# AN ARTIFICIAL INTELLIGENCE THAT SIMULATES HUMAN WORKING MEMORY, MENTAL IMAGERY, AND MENTAL CONTINUITY

Jared Edward Reser Ph.D.

aithought.com

jared-research.com

## ABSTRACT

This article presents an artificial intelligence (AI) architecture intended to simulate the iterative updating of the human working memory system. It features several interconnected neural networks designed to emulate the specialized modules of the cerebral cortex. These are structured hierarchically and integrated into a global workspace. They are capable of temporarily maintaining high-level representational patterns akin to the psychological items maintained in working memory. This maintenance is made possible by persistent neural activity in the form of two modalities: sustained neural firing (resulting in a focus of attention) and synaptic potentiation (resulting in a short-term store). Representations held in persistent activity are recursively replaced resulting in incremental changes to the content of the working memory system. As this content gradually evolves, successive processing states overlap and are continuous with one another. The present article will explore how this architecture can lead to iterative shift in the distribution of coactive representations, ultimately leading to mental continuity between processing states, and thus to human-like thought and cognition.

Like the human brain, this AI working memory store will be linked to multiple imagery (topographic map) generation systems corresponding to various sensory modalities. As working memory is iteratively updated, the maps created in response will construct sequences of related mental imagery. This system integrates these components by embedding them within a multilayered neural network of pattern recognizing nodes. Nodes low in the hierarchy are trained to recognize and represent sensory features and can combine groups of individual features into composite, topographical maps or images. Nodes high in the hierarchy are multimodal, and have a capacity for sustained activity, allowing the maintenance of pertinent, high-level patterns (complex features) through elapsing time. The higher-order nodes select pertinent new features from each lower-order mapping to add to their store of temporarily maintained patterns. This updated set of patterns are fed back into lower-order sensory nodes where they guide the construction of successive topographic maps. Thus, neural networks emulating the prefrontal cortex and its reciprocal interactions with early sensory and motor cortex capture the imagery guidance functions of the human brain. This sensory and motor imagery creation, coupled with an iteratively updated working memory store may provide an AI system with the cognitive assets needed to achieve synthetic consciousness or artificial sentience. Numerous related concepts are discussed.

# AN ARTIFICIAL INTELLIGENCE THAT SIMULATES HUMAN WORKING MEMORY, MENTAL IMAGERY, AND MENTAL CONTINUITY

"In order for a mind to think, it has to juggle fragments of its mental states."

Marvin Minsky, 1985.

## Introduction

This article takes a model of working memory published in Physiology and Behavior in 2016 (Reser, 2016) and in Arxiv (Reser, 2022) and explains how it can be implemented in a machine to make AI information processing more like human cognitive processing. The model emphasizes that persistent neural activity in the brain permits multiple working memory representations to remain coactive over time. It also causes successive mental states to overlap in regard to the representational content they hold.

The novel processing architecture introduced here involves the emulation of the mammalian cerebral cortex utilizing a network of pattern recognizing nodes for: selecting priority stimulus features, temporarily maintaining these features in a limited capacity working memory store and using them as parameters to search long-term memory for pertinent updates to working memory. The article will argue that by continuously updating the working memory store in an iterative fashion (Reser, 2016; 2012), such a system could exhibit cognitive abilities not seen using current methods.

The pattern of activity in the brain is constantly changing, but because the activity in certain neurons persists during these changes, particular features of the overall pattern can be uninterrupted or conserved over time. In other words, the distribution of active neurons in the brain transfigures gradually and incrementally from one configuration to another, instead of changing all at once. If it were not for the phenomena of persistent neural activity, successive mental states would be discrete and isolated rather than continuous with each other. Thus, the human brain is an information processing system that has the ability to maintain a large pool of representations that is perpetually in flux as new representations are constantly being added, some are being removed and still others are being maintained. The present device will be constructed to mimic this biological phenomenon, common to all vertebrates, but do it in a particularly mammalian manner. Because a large number of neural nodes in the mammalian cerebral cortex are always being sustained, each brain state is embedded recursively in the previous state, amounting to a progressive process that can move toward a complex result.

Contemporary AI has its weights frozen after training and cannot continue learning. Because it cannot learn it generally sits idle between processing tasks. Also, it does not reconcile information from multiple specialized processors like the human brain does. Iterative updating will change all of this. Properly integrated with existing AI technology, this method may have the potential to enhance the capabilities of problem-solving agents with respect to pattern recognition, analytics, prediction, adaptive control, decision making, and response to query.

Computer programmers design AI systems to hold information online only for as long as they know they will need it. They hold data in a temporary store merely to compute what they are programmed to compute or execute whatever process they have pending. The mammal brain, on the other hand, has a strategy dedicated to holding potentially relevant information online because there is a high probability that it will be useful in the near future. It wagers that this information will be useful in processing without yet knowing how. It is not decided before hand how long items should remain active, rather, it is re-decided every second and during each state.

Soft computing approaches using the state-space approach contain incidental aspects of iteration, and other architectures such as neural networks use spreading activation. However, no machine yolks these together to create iterative updating and multi associative search. Doing so provides a clear way to structure a system that does not suspend its activity every time it finishes a task. It will be important for the system to exhibit continuous endogenous processing and a working memory that updates continuously to allow for uninterrupted learning.

Some important issues in the construction of intelligent machines include: 1) How can the process of internally generated thought be accomplished within a computer? 2) How can fragments of long-term memory recombine to represent novel concepts and episodes? 3) What events between these fragments must take place to allow transitions between episodes? 4) Are analogs of association and sensory cortices necessary? 5) How can the capacity of working memory be increased in a machine? 6) Is it possible to reduce explicit, conscious processes down to their constituent, implicit, unconscious ones? 7) Which brain mechanisms give rise to the mental continuity that humans experience? 8) What fundamental processes allow thought to move through space and time, or in another word, to “propagate?” The answers to these questions must be grounded in physics, biology, and information processing because they must explain how the physical substrate of intelligence operates in a mechanistic sense (Chalmers, 2010). Without being able to tie together all the neurological, psychological, and computational loose ends necessary to answer these questions comprehensively, this paper will attempt to address them in an exploratory way using a novel approach. Rather than being based on traditional topics, this article will view these questions from the perspective of the shared representational content between successive mental states.

There are currently many illustrative and biologically plausible theories from cognition and neuroscience that address the questions listed above. Some do an exemplary job of tying together a large number of relevant phenomena into a cohesive picture. Models such as Atkinson and Shiffrin’s (1969) multistore model, Baar’s (1997, 2002) global workspace theory, Baddeley’s (2000, 2007) model of working memory, Damasio’s convergence-divergence paradigm (Damasio, 1989; Meyer & Damasio, 2009),

Edelman's (1987, 2006) concepts of reentrance and neural Darwinism, Edelman and Tononi's (2001) conceptualization of a "functional cluster" or "dynamic core," Fuster's (2009) conception of cognits, Tononi's (2004) conception of integration of information, and Carpenter and Grossberg's (2003) adaptive resonance theory have done much to lend perspective into the mechanics of perception, attention, working memory, and consciousness. In fact, concepts from these works (and several others that will be cited) form a battery of implicit assumptions about cognitive mechanisms that provides theoretical scaffolding for what is written here.

Despite much progress, most scientists report that current theory is unsatisfying because it cannot yet bridge the gap between the brain and the mind (Chalmers, 1995; Chalmers, 2010; Shear, 1997). Further, even though many contemporary models largely agree with empirical data, little has been done to reconcile their disparate, piecemeal approaches (Pereira & Ricke, 2009; Vimal, 2009). The effort to determine the important cognitive phenomena that should be replicated within a computer may be facilitated by the exploration of the concept of mental continuity. This work intends to use the concept of mental continuity to integrate current theoretical approaches while attempting to remain consistent with prevailing knowledge.

## Modeling Mental Continuity

Continuity is defined as being uninterrupted in time. As proposed here, "mental continuity" involves a process where a gradually changing collection of mental representations held in attention/working memory exhibits a measure of uninterrupted activity across time and over sequential processing states. If it were not for the phenomenon of persistent neural activity, instantaneous information processing states would be time-locked and isolated (as in most serial and parallel computing architectures), rather than continuous with the states before and after them.

This article explores how sustained neural firing in association areas allows goal-relevant representations to be maintained over multiple perception-action cycles, in order to direct complex sequences of interrelated mental states. The individual states in a sequence of such states are interrelated because they share representational content. The associations linking the shared contents are saved to memory, impacting future searches, and ultimately resulting in semantic knowledge, planning, and systemizing.

Because the activity of cells in the PFC and other association areas can be sustained concurrently and does not fade away before the next instantiation of activity elsewhere, there is a temporally dynamic and overlapping pattern of neural activity that makes possible the "juggling" of information in working memory. The quantity of mental continuity is directly proportional to the number of sustained representations and the length of time of their activity (Reser, 2011, 2012, 2013).

There are countless phenomena in the natural world that undergo iterative updating. Take the human population of Earth for instance. In the next year many people will die, some new people will be born, yet most will be here a year from now. In the same sense, some of the neurons in a population exhibiting sustained firing will stop firing, some will begin, yet most will still be firing a second later. The people and neurons that persist will be able to influence subsequent states.

The sustained firing of higher-order nodes allows representations to be maintained over multiple perception-action cycles permitting complex sequences of interrelated mental states. The overall distribution of active nodes in the neural network will shift gradually during contextual updating because the activity of certain neural nodes will persist. This will ensure that the activity of prioritized, goal or motor-relevant representations will be uninterrupted over time. The representations that demonstrate this continuity are a subset of the active representations from the previous state and may act as referents to which newly introduced representations of succeeding states relate. The limited-capacity store of coactive representations in association areas is updated as: 1) the nodes that continue to receive sufficient spreading activation energy are maintained; 2) the nodes that receive reduced energy are released from activation; 3) new nodes that are tuned so as to receive sufficient energy from the current constellation of coactivates are converged upon and incorporated into the remaining pool of active nodes from the previous cycle.

## Simulating the Minicolumns of the Mammalian Neurocortex

Like other cognitive models (Cowan, 2005; Moscovich, 1992), this model views cognition as a system responsible for using active representations from LTM to guide goal-directed processing (Postle, 2007). The present model is consistent with connectionism and parallel distributed processing in that it conceptualizes mental representations as being built from interconnected networks of decentralized, semi-hierarchically organized, pattern-recognizing nodes that have multiple inputs and outputs (Gurney, 2009; Johnson-Laird, 1998). Like other biologically plausible neural network models, it envisions these nodes as microscopic, modular neural units and assumes that each individual unit represents an elementary feature or stable “microrepresentation” of LTM (Meyer & Damasio, 2009). In the mammalian brain these microrepresentations may be instantiated by cortical minicolumns.

The structure of the cerebral cortex is highly repetitive and is marked by the employment of millions of nearly identical structures called cortical minicolumns (Lansner, 2009). Minicolumns are composed of densely connected neural cell bodies and span the six layers of grey matter in the neocortex. There are supposedly around 20,000,000 minicolumns in the human cortex, each of which contains between 80 and 120 neurons, and is about 30 to 40 micrometers in diameter (Lansner, 2009). These minicolumns share the same basic structure and are thought to employ the same cortical algorithm (Fuji et al., 1998; Kurzweil, 2012). Each column performs neural computations to determine if its inputs from other columns are sufficient to activate its outputs to other columns (Rochester et al., 1956).

Columns, and other similar groups of neurons with the same tuning properties, are often referred to as cell assemblies, and this term will be used here. Most neurons in an assembly share similar receptive fields, and thus even though they may play different roles within the assembly, they each contribute to the assembly’s ability for encoding a unitary feature (Moscovich et al., 2007). Such an assembly of neurons is thought to embody a stable microrepresentation or fragment of long-term memory. All of the millions of pattern recognizers in the neocortex are simultaneously considering their inputs, and continually determining whether or not to fire. In general, when a neuron or assembly fires, the pattern that it represents has been recognized. Assemblies, like the neurons that compose them, function as "coincidence detectors" or "pattern recognition nodes" (Fuji et al., 1998). The spread of activity in the cortex involves many-to-one (convergence) and one-to-many (divergence) interactions within a massively interconnected network of assemblies. The next section will discuss how these columns or assemblies could be simulated by nodes in a neural network.

## Information Processing in Artificial Neural Networks

The field of AI research is involved in creating a computing system that is capable of emulating certain functions that are traditionally associated with intelligent human behavior. Most early AI systems were only capable of responding in the manner in which human programmers provided for when the program was written. It became recognized that it would be valuable to have a computer which does not respond in a preprogrammed manner (Moravec, 1988). AI systems capable of adaptive learning have since become important. Neural networks have attempted to get around the programming problem by using layers of artificial neurons or nodes. Neural networks and genetic algorithms are widely implemented in research and industry for their capabilities involving adaptive learning and advanced pattern recognition. However, they are used for processing tasks that are narrowly constrained and highly specialized, and there has not yet been any strong form of intelligence derived from them.

An artificial neural network is an interconnected group of artificial neurons that uses a mathematical or computational model for information processing based on a connectionistic approach to computation (Russel et al., 2003). Like most neural networks, the present network should be an adaptive system capable of altering its own structure based on the nonlinear processing of information that flows through the network. The software would require a massively parallel, distributed architecture, that could be run on a conventional computer. Most neural networks ordinarily achieve intelligent behavior through parallel computations, without employing formal rules or logical structures, and can be used for pattern matching, classification, and other non-numeric, nonmonotonic problems (Nilsson, 1998). The applications for the present device could be widened if it were designed to accept and process formal rules.

The traditional neural network is a multilayer system composed of computational elements (nodes) and weighted links (arcs). These networks are based on the human brain where the nodes are analogous to neurons, or neural assemblies, and the arcs are analogous to axons and dendrites. Each node receives signals from nodes in the layer below it, processes these signals and then decides whether to "fire" at the nodes in the layer above. Like the artificial neurons first described by McCullough and Pitts (1943), the nodes in the present system could feature a number of excitatory inputs whose weights range between 0 and 1 and inhibitory inputs whose weights range between -1 and 0. Each of the incoming inputs and their corresponding weights are summed to equal an activation level. If this level exceeds the neurons's firing threshold, it will cause the neuron to fire. The neuron can be made to learn from its experience when either the threshold or weights are changed. Neural networks are typically defined by three types of parameters:

1) The interconnection pattern between different layers of neurons; 2) The learning process for updating the weights of the interconnections; 3) The activation function that converts a neuron's weighted input to its output activation.

There are currently no neural networks, or AI systems whatsoever, that are structured to model the primate neocortex in order to guide the progressive generation of successive topographic maps. The present neural network software architecture is structured around identifying potentially goal-relevant information and holding it online to inform reciprocal cycles of imagery generation and feature extraction for the purpose of systemizing the environment.

## Architecture Hierarchy

The software models a large set of nodes that work together to continually determine, in real time, which from the population of inactive nodes should be newly activated, and which from the population of active nodes should be either deactivated or have its activity maintained.

The device necessitates a highly interconnected neural network that features a hierarchically organized collection of pattern recognizers capable of both transient and sustained activity. These pattern recognition nodes mimic assemblies (minicolumns) of cells in the mammalian neocortex and are arranged with a similar connection geometry. Like neural assemblies the nodes exhibit a continuous gradient from low-order nodes that code for sensory features, to high-order nodes that code for temporally or spatially extended relationships between such features. The lower order nodes are organized into modules by sensory modality. Nodes are grouped according to the feature they are being trained to recognize. These maps can be generated by external input, by internal input from higher-order nodes, or a mix of the two. The architecture will feature backpropagation, self-organizing maps, bidirectionality, Hebbian learning as well as a combination between principal-components learning and competitive learning.

Nodes lower in the hierarchy are trained to recognize and represent sensory features and are capable of combining individual features or patterns into metric, topographical maps or images. Lower-order nodes are unimodal, and organized by sensory modality (visual, auditory, somatosensory etc.) into individual modules. Nodes high in the hierarchy are multimodal, module independent, and have a capacity for sustained activity allowing the conservation of pertinent, high-level features through elapsing time. The higher nodes are integrated into the architecture in a way that makes them capable of identifying goal-relevant features from both internal imagery and environmental input, and temporarily maintaining these as a form prioritized information.

The system is structured to allow repetitive, reciprocal interactions between the lower, bottom-up, and higher, top-down nodes. The features that the higher nodes encode are utilized as inputs that are fed back into lower-order sensory nodes where they are continually used for the construction of successive topographic maps. The higher nodes select new features from each mapping to add to the store of temporarily maintained features. Thus, the most salient or goal-relevant features from the last several mappings are maintained. The group of active, higher-order nodes is constantly updated, where some nodes are newly added, some are removed, yet a relatively large number are retained. This updated list is then used to construct the next sensory image which will be necessarily similar, but not identical to, the previous image. The differential, sustained activity of a subset of high-order nodes allows thematic continuity to persist over sequential processing states.

All the nodes within the device function as a continuous whole and are highly interconnected, but can be decomposed into separate, modular, neural networks. Nodes belonging to an individual module are highly interconnected with each other. These modules consist of a bottom layer of input cells, succeeded by layers of local-feature extracting cells, and a top layer of output cells. The individual neural networks interface through the connections between top layer output cells and bottom layer input cells. Each network is organized so that multiple lower-order nodes can converge on higher order nodes, and single higher-order nodes can diverge back on multiple lower-order nodes. Multiple interfacing neural networks could be arranged biomimetically as in figure 8 below.

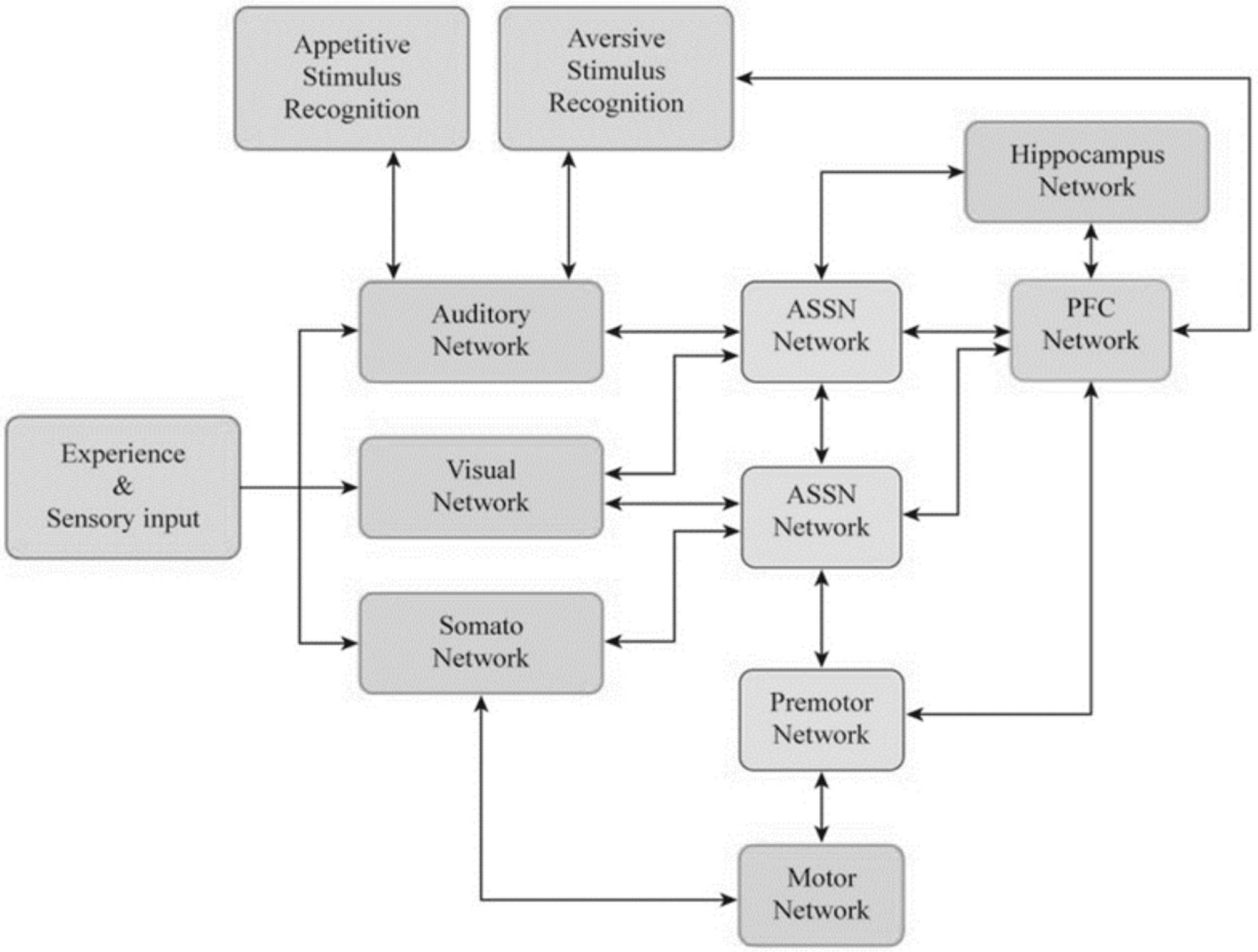

**Fig. 1.** A plausible biomimetic arrangement of interfacing neural networks.

The bottom-up to top-down reciprocations are organized into very precise oscillations that propagate in regularly timed intervals across the network so that they do not interfere with each other. The oscillations reciprocate back and forth at just the right speed so that each area has the time to process its inputs and send an output before the next complement of inputs arrive. It is important to carefully structure timing mechanisms in the present device so that messaging is not muddled or noisy. It is also important to structure the architecture so that continuity in the representations held active by the buffer can be disrupted when attention shifts. Repeated loops of conserved, higher-order features can be ended when attention is captured by an object or concept that competes for attention. The ability to free up the resources of higher-order nodes to attend to a new stimulus will be programmed by training. Before proper training is accomplished the system may not be able to reallocate its resources properly when its attention shifts.

To create a strong form of AI it is necessary to understand what is taking place that allows intelligence, thought, cognition, consciousness or working memory to move through space and time, or in another word, to “propagate.” Such an understanding must be grounded in physics because it must explain how the physical substrate of intelligence operates through space and time (Chalmers, 2010). The human brain is just such an intelligent physical system that AI researchers have attempted to understand and replicate using a biomimetic approach (Gurney, 2009). Features of the biological brain have been key in the design of neural networks, but the brain may hold additional information processing principles that have not been harnessed by A.I. efforts (Reser 2011, 2012, 2013).

Computational operations, that take place as a computer implements lines of code (rule-based, if-then operations) to transform input into output, have discrete, predetermined starting and stopping points. For this reason, computers do not exhibit continuity in their information processing. There are no forms of artificial intelligence that use mental continuity as described here. There are existing computing architectures with limited forms of continuity where the current state is a function of the previous state, and where data is entered into a limited capacity buffer to inform other processes. However, the memory buffer is not global, not multimodal, not positioned at the top of a hierarchical system and does not inform and interact with topographic imagery.

**Persistent Neural Activity From Sustained Firing and Synaptic Potentiation**

The mammalian PFC and other association cortices have neurons that are specialized for “sustained firing,” allowing them to generate action potentials at elevated rates for several seconds at a time (generally 1-30 seconds) (Fuster, 2009). In contrast, neurons in other brain areas, including cortical sensory areas, remain active only for milliseconds unless sustained input from association areas makes their continued activity possible (Fuster, 2009). In the mammalian brain, prolonged activity of neurons in association areas, especially prefrontal and parietal areas, allows for the maintenance of specific features, patterns, and goals (Baddeley, 2007). Working memory, executive processing and cognitive control are widely thought to stem from the active maintenance of patterns of activity in the PFC that represent goal-relevant features (Goldman-Rakic, 1995). The temporary persistence of these patterns ensures that they continue to transmit their effects on network weights as long as they remain active, biasing other processing, and affecting the interpretation of subsequent stimuli that occur during their episode of continual firing.

Although its limits are presently being debated, the human neocortex is clearly capable of holding numerous neural representations active over numerous points in time. The quantity of mental continuity is directly proportional to the number of such sustained representations and the length of time of their activity (Reser, 2011, 2012, 2013).

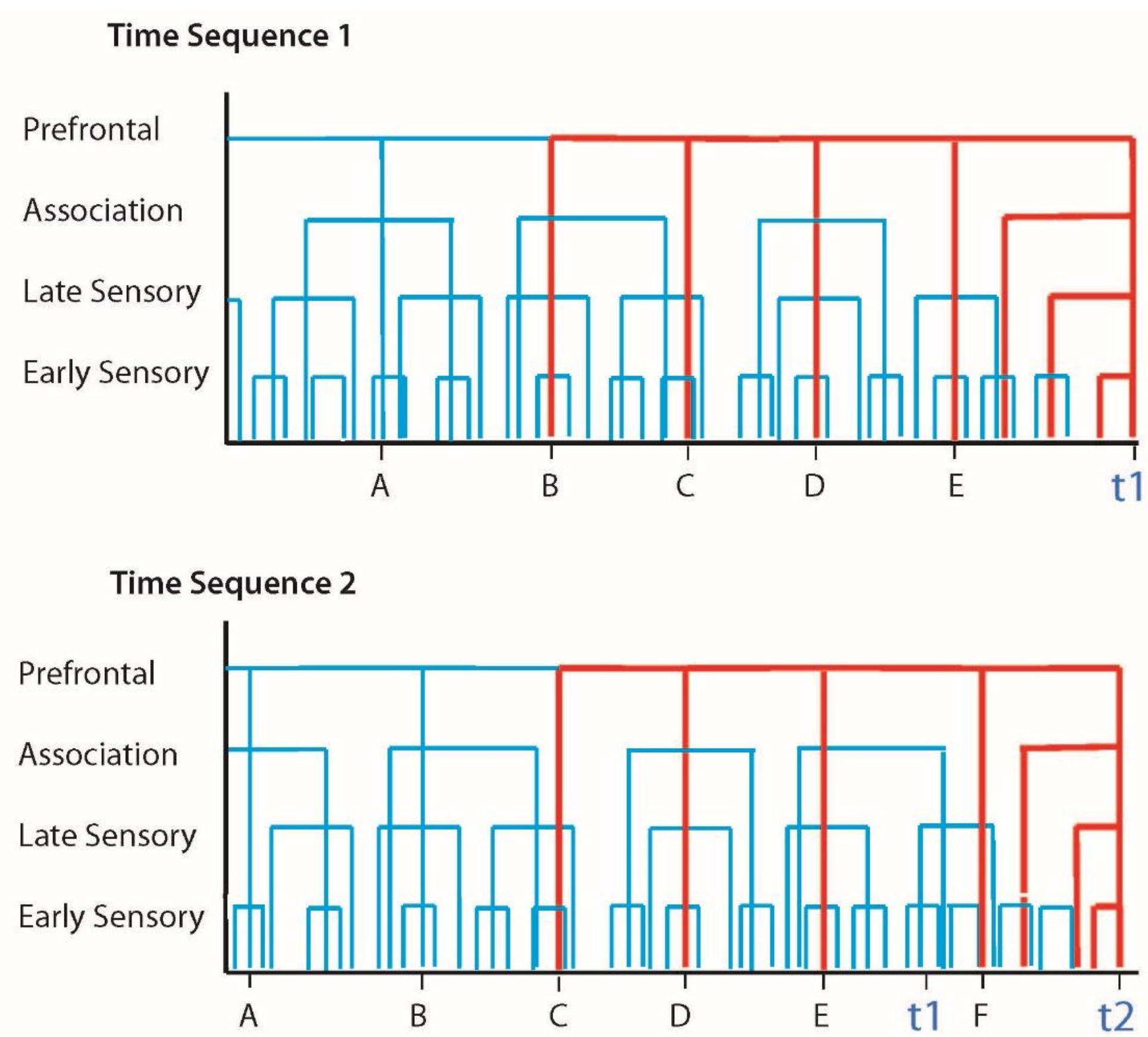

**Fig.2.** Graphical depiction of iterative updating. Each bracket represents the active time span of a neural representation. The x axis represents time and the y axis demarcates the cortical area where the representation is active. Red brackets denote representations that have exhibited uninterrupted activity from the point when they became active, whereas blue brackets denote representations that have not been sustained. In time sequence 1 representations B, C, D and E have remained active until t1. In time sequence 2 B has deactivated, C, D and E have remained active, and F is newly active. The figure depicts a system with iterative updating because more than one representation (C, D, and E) has been maintained over more than one point in time (t1 and t2).

In Figure 1 above, representations B, C, D, and E are active during time 1, and C, D, E and F are active during time 2. Thus, representations C, D, and E demonstrate reiteration because they exhibit continuous and uninterrupted activity from time 1 through time 2. The brain state at time 1 and the brain state at time 2 share C, D, and E in common and, because of this, can probably be expected to share other commonalities including: similar information processing operations, similar memory search parameters, similar mental imagery, similar cognitive and declarative aspects, and similar experiential and phenomenal characteristics.

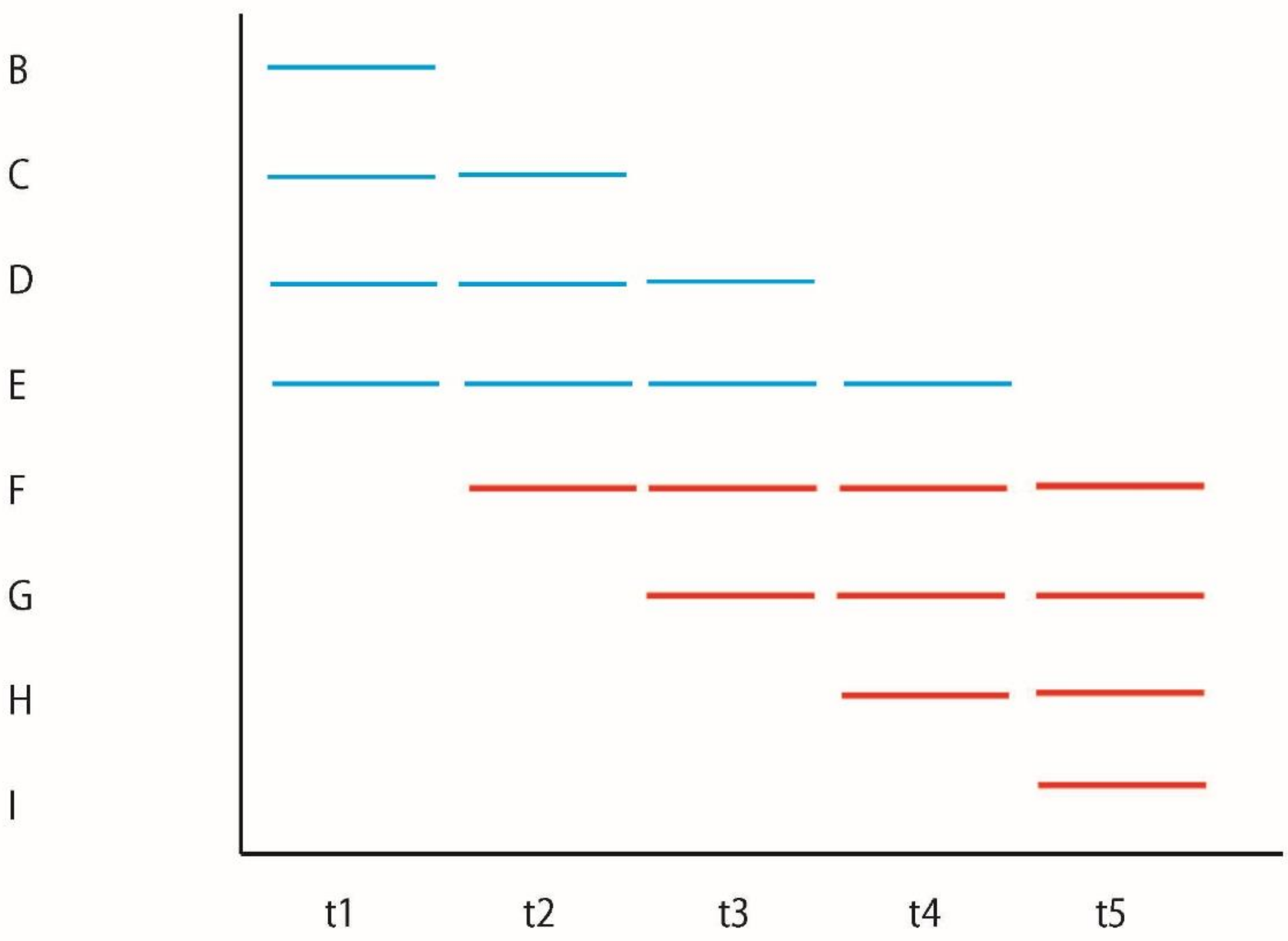

**Fig. 3.** A simplified graphical representation of iterative updating depicting it as a gradually shifting, stream-like distribution. Figure 2 extends Figure 1 over multiple time intervals revealing a repeating pattern: remnants from the preceding state are consistently carried over to the next state.

Figure 3 shows how this iterative process can be looped repetitively over time. Figure 4 shows how the rate of iteration can change

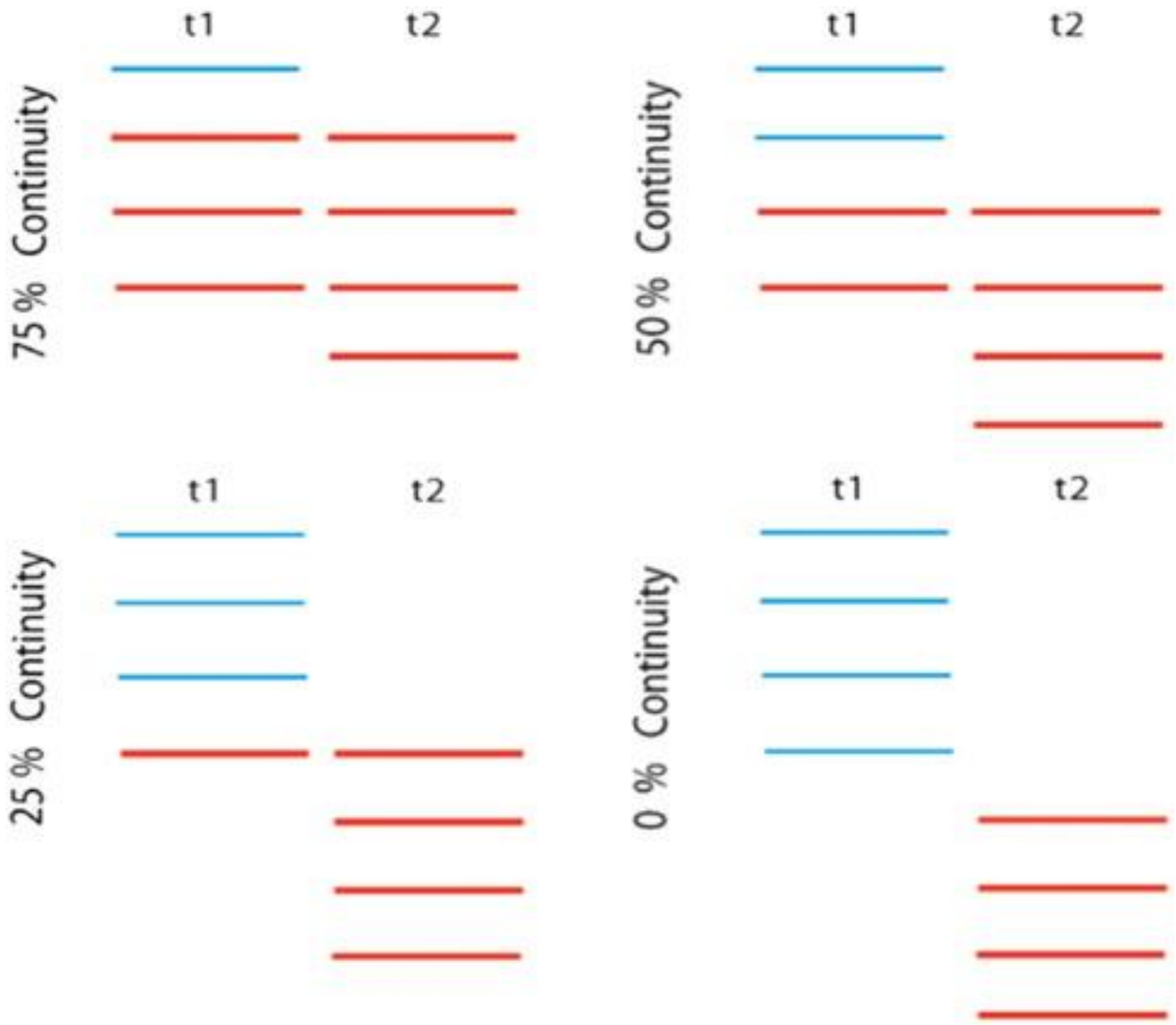

**Fig. 4.** This figure expands on previous figures by comparing four possible incremental changes in SSC. In the first transition 75% of representational continuity is maintained between time periods 1 and 2. The other transitions depict 50%, 25%, and 0% maintenance of continuity respectively. According to the definition of mental continuity, neither the graphic marked "25% continuity" nor the one marked "0% continuity" depict mental continuity, because they do not feature the maintenance of more than one representation.

Figure 5 below demonstrates how this pattern of iterative updating could be implemented in working memory.

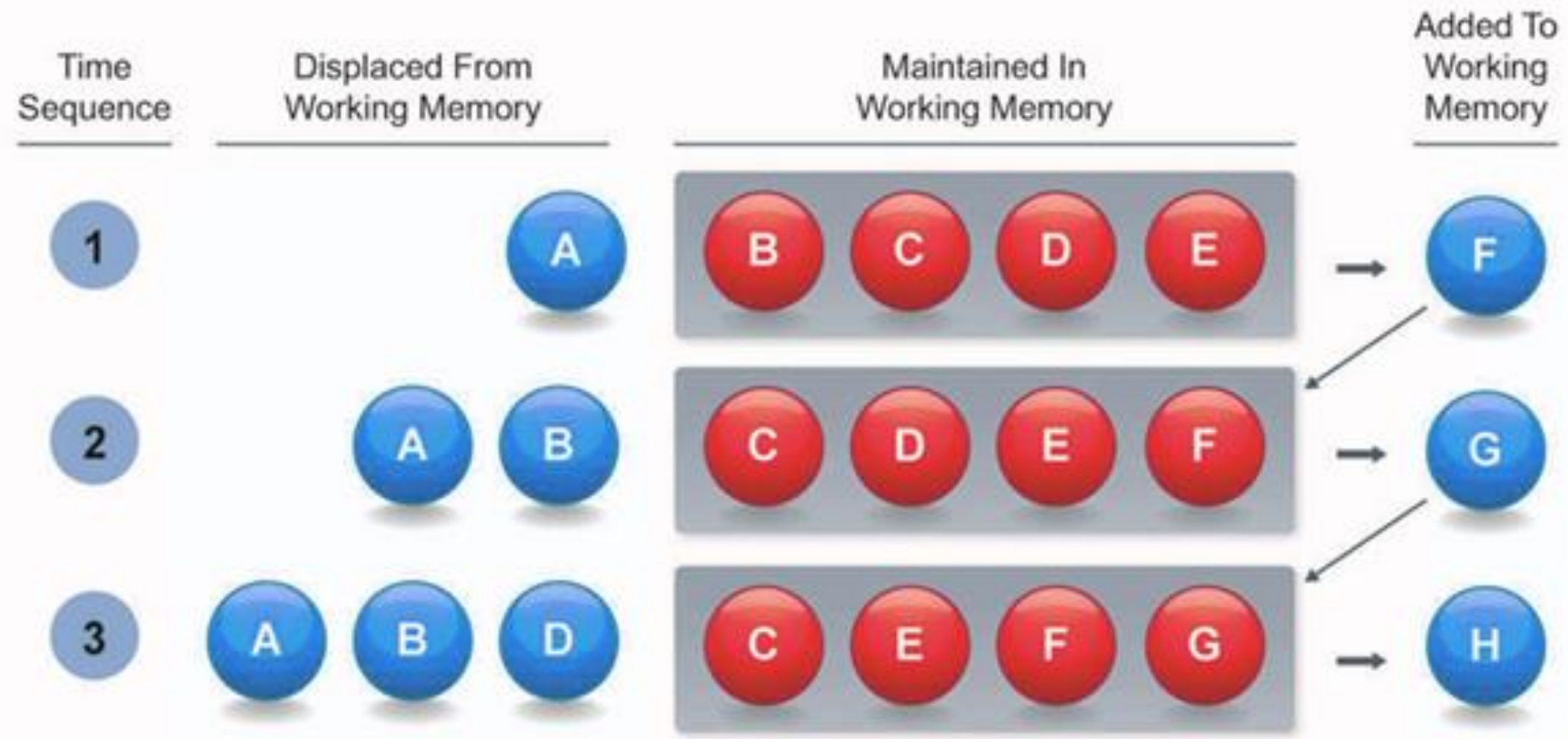

**Fig. 5.** A diagram depicting "multiassociativity" and illustrating the ways in which high-level representations, or items, are displaced, maintained, and newly activated in the brain. 1) Shows that representation A has already been deactivated and that B, C, D and E are now coactivated, mirroring the pattern of activity shown in Figure 1. When coactivated, these representations pool and spread their activation energy, resulting in the convergence of activity onto a new representation, F. Once F becomes active, it immediately becomes a coactivate, restarting the cycle. 2) Shows that B has been deactivated while C, D, E, and F are coactivated and G is newly activated. 3) Shows that D but not C has been deactivated. In other words, what is deactivated is not necessarily what entered first, but what has proven to receive the least converging activity. C, E, F, and G coactivate and converge on H.

In Figure 4, between time periods 1 and 2, C, D and E exhibit coactivity, whereas, between time periods 1 and 3, only C and E exhibit coactivity. C and E are active over all three time periods meaning that these representations are being used as search function parameters for multiple cycles, and are the subject of attention. Alternatively, we can imagine a scenario where B, C, D, and E from step one of Figure 3 were immediately replaced by F, G, H, and I. Such a processing system may still be using previous states to determine subsequent states; however, because no activity is sustained through time, there would be no continuity in such a system (this is generally how most computing systems process information).

In Figure 4, ensembles C and E have fired together over three individual time intervals, and thus will show a propensity to wire together, increasing their propensity for firing together in the future. This link between them will allow one to recruit the other. However, it is probably much more likely that they will recruit each other if the other contextual ensembles are also present. The coincidental or rare associations between the ensembles of an experience are probably mostly lost from non-hippocampal dependent cortical memory. However, the reoccurring associations are heavily encoded and persist as semantic knowledge.

In other words, the cortex constantly spreads activation energy from novel combinations of active ensembles that have never been coactive before and attempts to converge upon the statistically most relevant association without certain or exact precedence resulting in a solution that is not guaranteed to be optimal.

Neural nodes in sustained activity can span a wider delay time or input lag between associated occurrences (Zanto, 2011) allowing elements of prior events to become coactive with subsequent ones (Fuster, 2009). Without sustained firing, the ability to make associations between temporally distant (noncontiguous) environmental stimuli is disrupted. Sustained activity allows individual nodes that would otherwise never fire together to both fire and wire together. Thus, it sustained firing is responsible for the ability to make internally-derived associations between representations that would never co-occur simultaneously in the environment.

This indicates that one way to quantify mental continuity is to determine the proportion of previously active neural nodes that have remained active during a resource demanding cognitive task. Uninterrupted activity augments associative searches by allowing specific features to serve as search function parameters for multiple cycles. Intelligence in this system can be enhanced by increasing: 1) the number of available nodes to select from, 2) the number of nodes that can be coactivated simultaneously, and 3) the length of time that individual nodes can remain active.

The mesocortical dopamine (DA) system plays an important role in sustained activity, suggesting that it may be heavily involved in mental continuity. Dopamine sent from the ventral tegmental area (VTA) modulates the activity and timing of neural firing in the PFC, association cortices, and elsewhere. Dopamine neurotransmission in the PFC is thought to underlie the ability to internally represent, maintain, and update contextual information (Braver & Cohen, 1999). This is necessary because information related to behavioral goals must be actively sustained such that these representations can bias behavior in favor of goal-directed activities over temporally extended periods (Miller & Cohen, 2001). It has become clear that the activity of the DA/PFC system fluctuates with environmental demand (Fuster & Alexander, 1971). Many studies have suggested that the system is engaged when reward or punishment contingencies change. Both appetitive and aversive events have been shown to increase dopamine release in the VTA, causing sustained firing of PFC neurons (Seamans & Robbins, 2010). Seamans and Robbins (2010) elaborated a functional explanation to support this case. They have stated that the DA system is phasically activated in response to novel rewards and punishments because it is adaptive for the animal to anchor upon and further process novel or unpredicted events.

Sustained firing and recurrent processing make it possible for recent states to spill over into subsequent states, creating the context for them in a recursive fashion. In a sense, each new topographic map is embedded in the previous one. This creates a cyclical, nested flow of information processing marked by STC, which is depicted in Figure 6.

Learned mental tasks probably have distinct predefined algorithmic sequences of topographic mappings that must be completed in sequence in order to achieve the solution. Each brain state would correspond to a different step in the algorithm, and its activity would recruit the next step. All logical and methodical cognition may require that a number of relevant features from the present scenario remain in STC so that they spread their activity within the network in order to influence the selection of the ensembles necessary for task satisfaction. Moreover, motor and premotor modules give specifications to and receive specifications from this common workspace while they are building their musculotopic imagery for movement. The same goes for language areas.

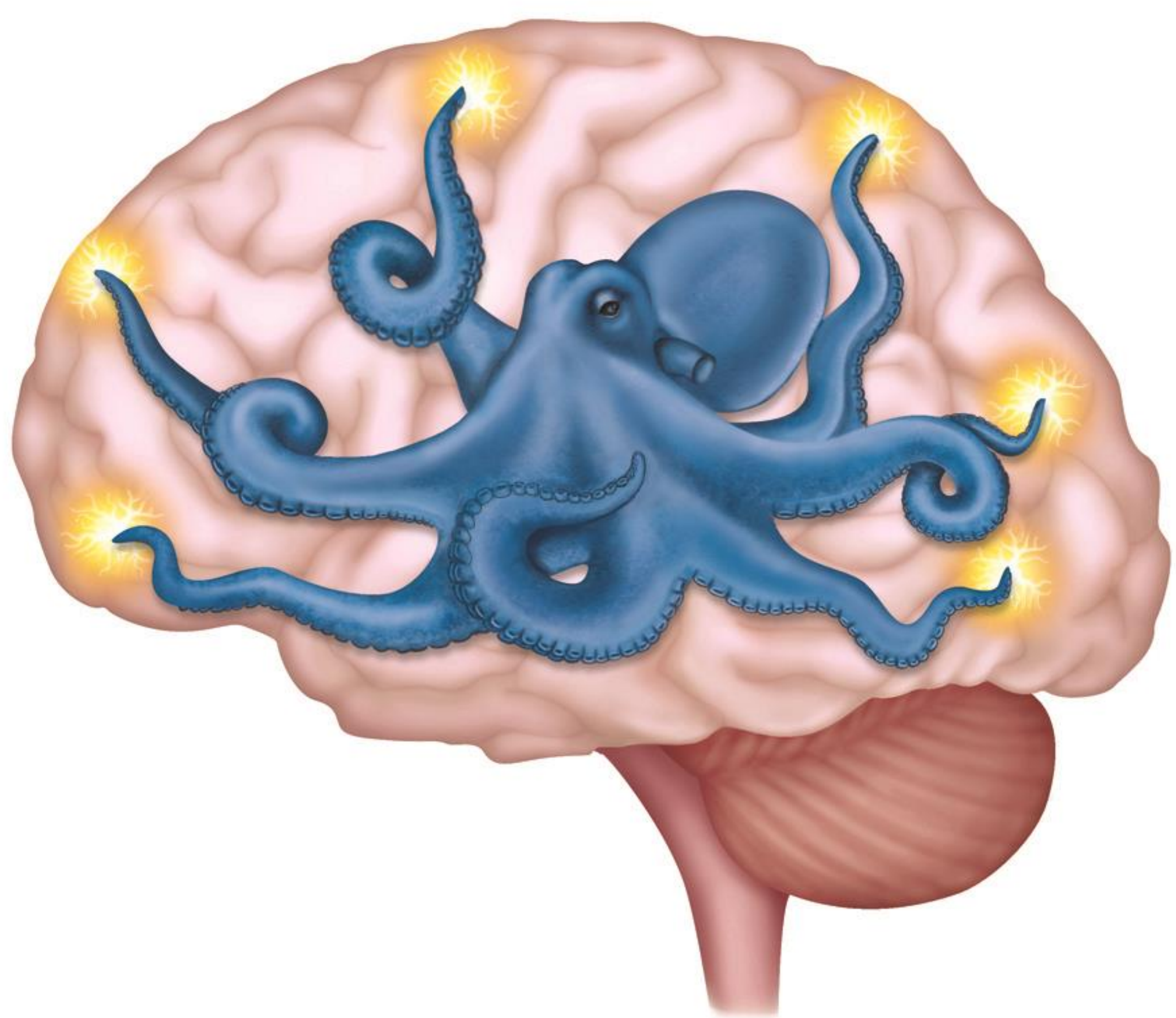

Fig. 6. The dynamics of the present model are captured by an analogy involving an octopus grabbing and releasing footholds as it pulls itself from place to place. This is an activity known as “sea floor walking.” The analogy illustrates that the thought process involves the simultaneous coactivation of several representations at a time (multiple footholds held by an octopus) as well as the deactivation of previously active representations (the releasing of footholds), and the activation of previously inactive representations (the placement of an arm on a new foothold). This analogy may be valuable because it depicts a system, that even a child can understand, where specific representations are conserved through time as others are actively repositioned.

## Search Through Spreading Activation Targets the Next Update

The mammalian neocortex is capable of holding a number of such mnemonic representations coactive, and using them to make predictions by allowing them to spread their activation energy together, throughout the thalamocortical network. This activation energy converges on the inactive representations from LTM that are the most closely connected with the current group of active representations, making them active, and pulling them into short-term memory. Thus new representations join the representations that recruited them, becoming coactive with them.

The way that assemblies and ensembles are selected for activity in this model is consistent with spreading activation theory. In spreading activation theory, associative networks can be searched by labeling a set of source nodes, which spread their activation energy in a nonlinear manner to closely associated nodes (Collins & Loftus, 1975). Cortical assemblies work cooperatively by spreading the activation energy (both excitatory and inhibitory) necessary to recruit or converge upon the next set of ensembles that will be coactivated with the remaining ensembles from the previous cycle.

Together, they impose sustained information processing demands on the lower-order sensory and motor areas within the reach of their long-range connections. The longer the activity in these higher-order neurons is sustained, the longer they remain engaged in hierarchy-spanning, recurrent processing throughout the cortex and subcortex.

| Gradual additions to and subtractions from a pool of simultaneously coactivated ensembles occur as: |
| --- |
| 1. Assemblies that continue to receive sufficient activation energy from the network are maintained. |
| 2. Assemblies that receive sufficiently reduced activation energy are released from activation. |
| 3. New assemblies, which are tuned to receive sufficient activation energy from the current constellation of coactivates, are converged upon, and incorporated into the remaining pool of active assemblies from the previous cycle. |

**Table 1.** The Characteristics of Multiassociative Search

Outlining the process of multiassociativity in this way is meant to show that the computational algorithm used by the brain may be primarily directed at determining which inactive ensembles are the most closely statistically related to the currently active assemblies. From this perspective, the contents of the next state are chosen based on how the currently active assemblies interact with the existing, associative, neuro/nomological network.

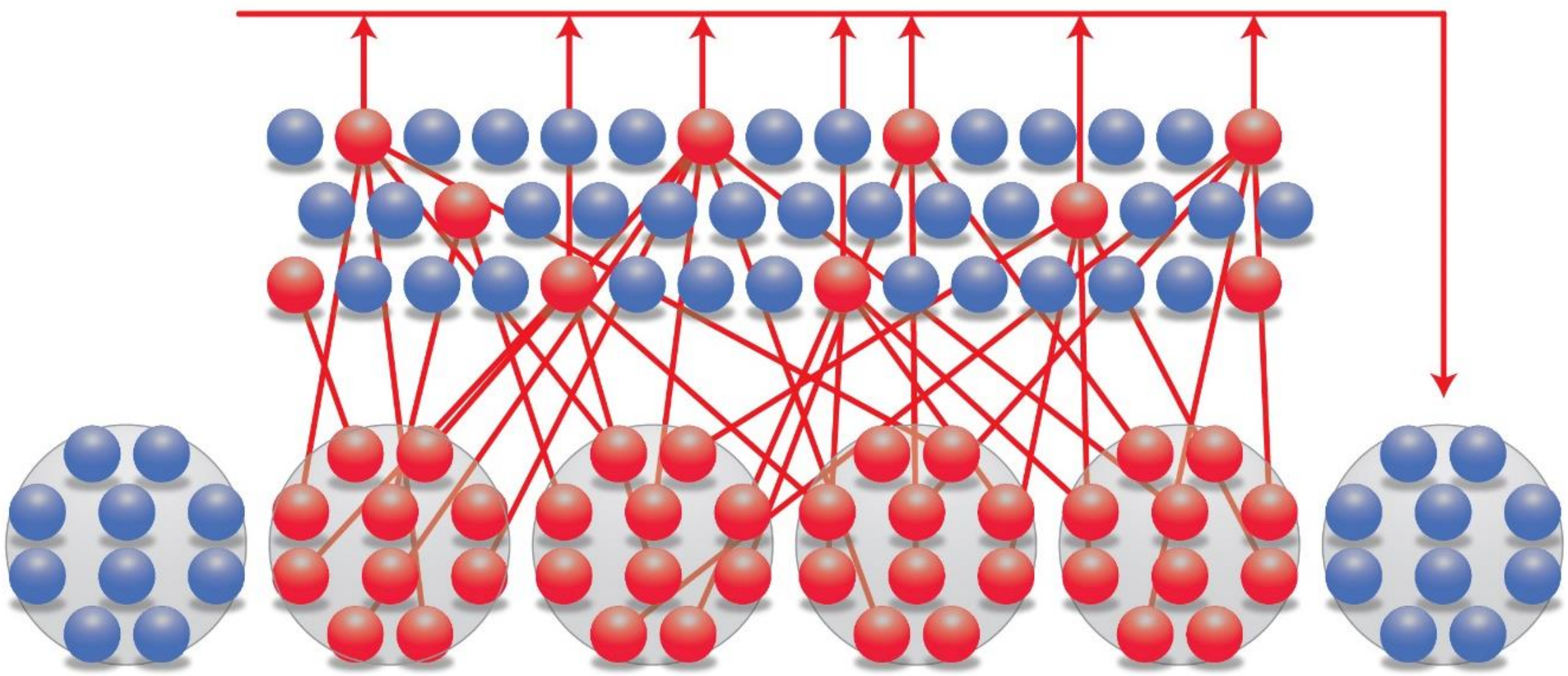

Fig. 7. A Schematic for Multiassociative Search

Spreading activity from each of the assemblies (small spheres) of the four items (large circles) in the FoA (B, C, D, and E) propagates throughout the cortex (represented by the field of assemblies above the items). This activates new assemblies that will constitute the newest item (F), which will be added to the FoA in the next state. The assemblies that constitute items B, C, D, and E are each individually associated with a very large number of potential items, but as a unique group, they are most closely associated with item F.

## Reciprocating Crosstalk Between Association and Sensory Cortex

To a certain extent, perceptual sensory processing in the brain is thought to be accomplished hierarchically (Cohen, 2000). The cortical hierarchy observed from sensory to association cortex arises because simple patterns are arranged to converge upon second-order patterns, which in turn converge on third-order patterns and so on. This leads to a hierarchy of increasingly complex representations.

It is thought that object recognition, decision making, associative recall, planning and other important cognitive processes, involve two-way traffic of signal activity among various neural maps that stretch transversely through the cortex from early sensory areas to late association areas (Klimesch, Freunberger, & Sauseng, 2010). Bottom-up sensory areas deliver fleeting sensory information and top-down association areas deliver lasting perceptual expectations in the form of templates or prototypes. These exchanges involve feedforward and feedback (recurrent) connections in the corticocortical and thalamocortical systems that bind topographic information from lower-order sensory maps about the perceived object with information from higher-order maps forming somewhat stable constellations of activity that can remain stable for tens or hundreds of milliseconds (Crick & Koch, 2003).

Iterative updating impacts this reciprocating cross-talk. These reciprocations may create progressive sequences of related thoughts, specifically because the topographic mappings generated by lower-order sensory areas are guided by the enduring representations that are held active in association areas (Reser 2011, 2012, 2013). The relationship between anterior and posterior cortex may be best characterized by two main relationships: 1) association areas maintain representations from, not one but several, of the last few topographic maps made in sensory areas, 2) because they are drawing from a register with sustained contents, sequential images formed in sensory areas have similar content and thus should be symbolically related to one another.

Feedback activation from top-down association areas hands down specifications to early sensory cortex for use in imagery building. Disparate chunks of information are integrated into a plausible map and transiently bound together. Consecutive topographic images about a specific scenario model the scenario by holding some of the contextual elements constant, while others are allowed to change. Thus, prior maps set premises for and inform subsequent maps.

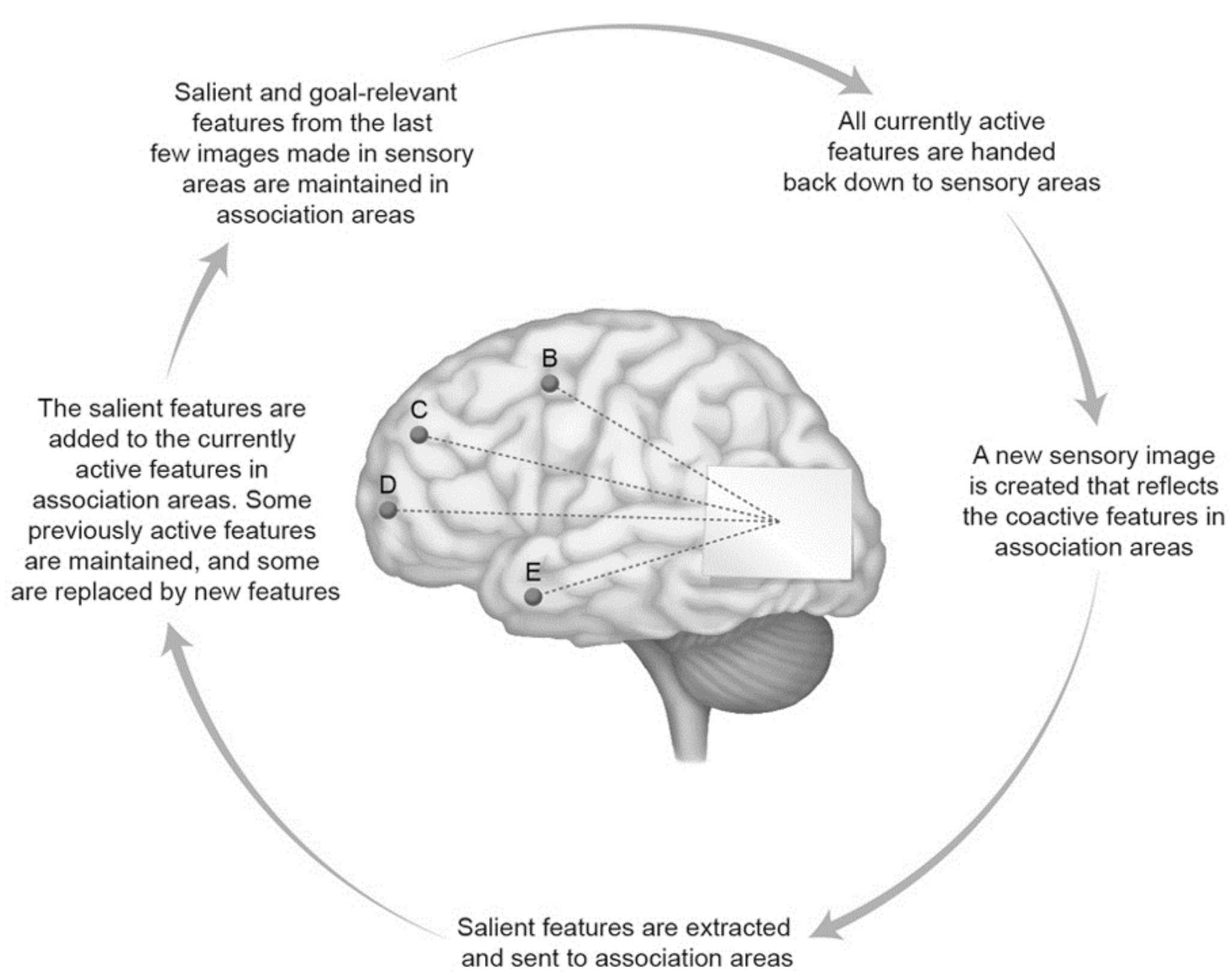

**Fig. 8.** A diagram depicting the reciprocal transformations of information between lower-order sensory mappings and higher-order association area ensembles during internally generated thought. Sensory areas can only create one topographic mapping at a time, whereas association areas are capable of holding the salient or goal-relevant features of several sequential mappings at the same time.

The central executive (the PFC and other association areas) direct progressive sequences of mental imagery in a number of topographic sensory and motor modules including the visuospatial sketchpad, the phonological (articulatory) loop and the motor cortex. This model frames consciousness as a polyconceptual, partially-conserved, progressive process, that performs its high-level computations through "reciprocating transformations between buffers." More specifically, it involves reciprocating transformations between a partially conserved store of multiple conceptual specifications and another nonconserved store that integrates these specifications into veridical, topographic representations.

It will be important for the AI to identify and capture information about unexpected occurrences so that it can be further processed and systematic patterns can be identified. The novel experience should be broken down into its component parts and the representations in memory for these parts allowed to spread their activation energy in an attempt to converge on and activate historically associated representations that are not found in the experience itself. Because memory traces for the important features remain active and primed, they can be used repeatedly as specifications that guide the generation of apposite mental imagery in sensory areas (Reser, 2012). Sequences of lower-order topographic images should depict and explore hypothetical relationships between the higher-order, top-down specifications. This amounts to a continual attempt to use the associative memory system to search sensory memory for a topographic image that can meaningfully incorporate the important features. It seems that reciprocating activity between the working memory updating system and the imagery generation system builds interrelated sequences of mental imagery that are used to form expectations and predictions.

The fact that newly active search terms are combined with search terms from the previous cycle makes this process demonstrate qualities of "progressive iteration." Perhaps reciprocating activity between the working memory updating system and the imagery generation systems generates sequences of interrelated mental images that build on themselves to form abductive expectations, and predictions.

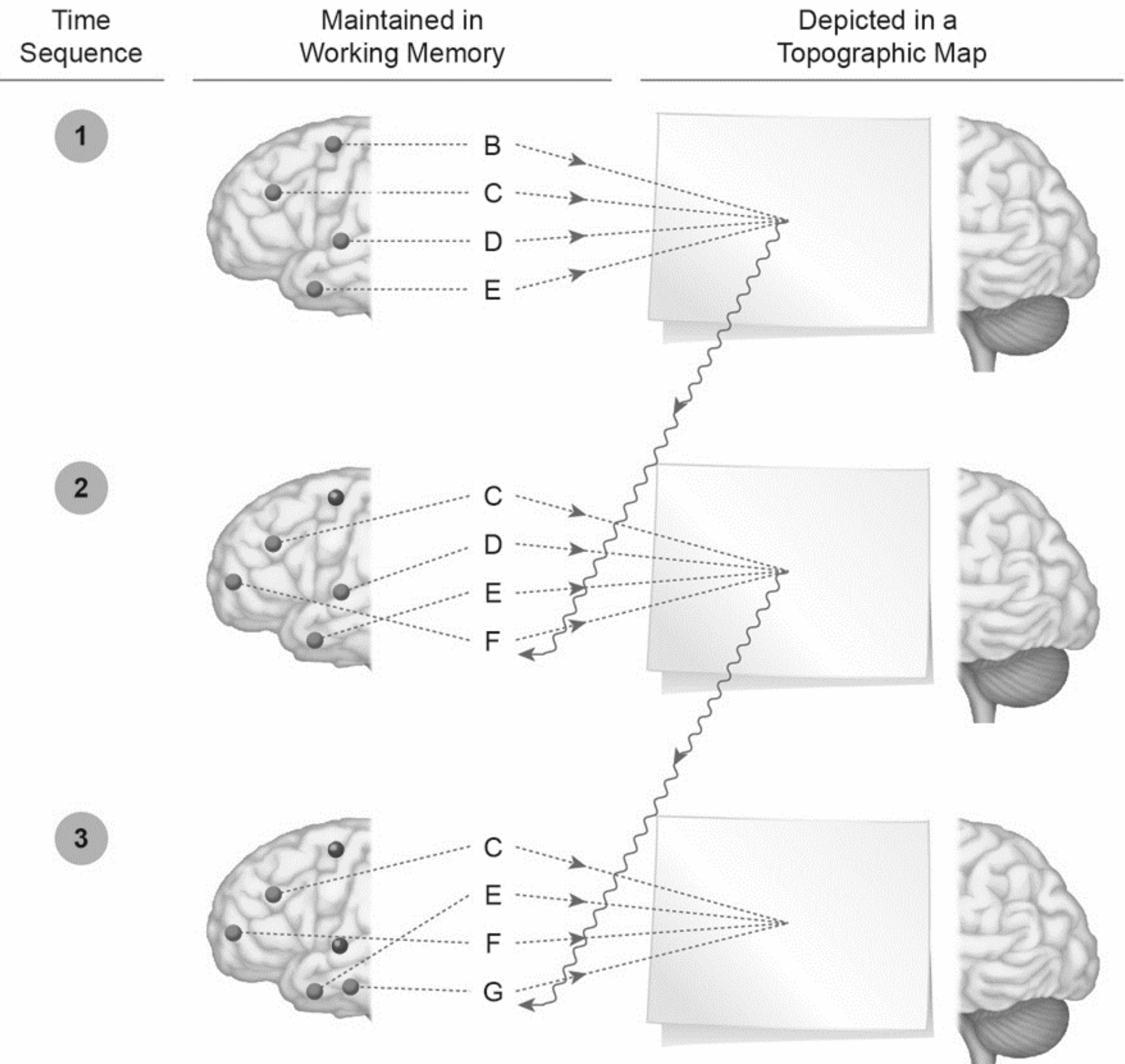

**Fig. 9.** A diagram depicting the behavior of representations that are held active in association areas. 1) Shows that the representations B, C, D, and E, which are held active in association areas, all spread their activation energy to lower-order sensory areas where a composite image is built that is based on prior experience with these representations. 2) Shows that features involved in the topographic imagery from time sequence 1 converge on the PFC neurons responsible for F. B drops out of activation, and C, D, E and F remain active and diverge back onto visual cortex. 3) Shows that the same process leads to G being activated and D being deactivated, mirroring the pattern of activity shown in Figure 4.

Assemblies in lower-order sensory areas identify sensory features from the environment and combine them into composite representations that mirror the geometric, and topographic orientations present in the sensory input. The early visual system uses retinotopic maps that are organized with a geometry that is identical to that used in the retina, and the auditory system uses tonotopic maps, where the mapping of stimuli is organized by tone frequency (Moscovich, 2007). Because cortical assemblies are essentially pattern recognition nodes organized in a hierarchical system, they should be able to be modeled by computers. The best way to do this with modern technology is to use an artificial neural network.

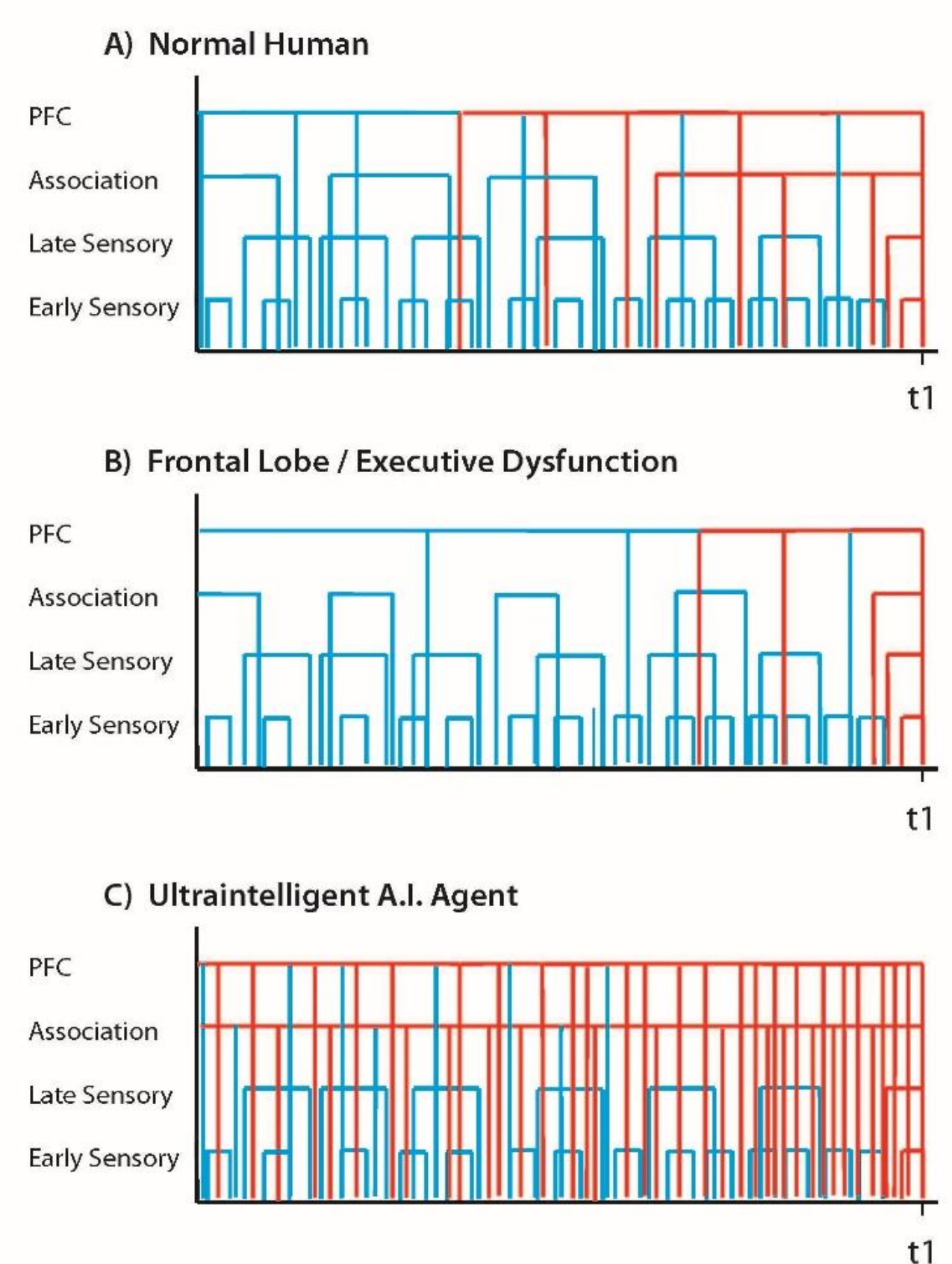

**Fig. 10.** Uses the format of Figure 1 to illustrate how relevant features can be maintained through time using nodes with sustained firing. The figure compares the number of past nodes that remain active at the present time (t1), in a normal human, a human with PFC dysfunction, and the hypothetical A.I. agent. The A.I. agent is able to maintain a larger number of higher-order nodes through a longer time span, ensuring that its perceptions and actions in time 1 will be informed by a larger amount of recent information. Note how the lower-order sensory and motor features are the same in each graph with respect to their number and duration, yet those in association areas are the highest in both number and duration for agent C.

## Tuning the Network for Increased Intelligence

If this sustained firing was programmed to happen at even longer intervals, and involve even larger numbers of nodes, the system would exhibit a superhuman capacity for continuity. This would increase the ability of the network to make associations between temporally distant stimuli and allow its actions to be informed by more temporally distant features and concerns. Aside perhaps from altering the level of arousal (adrenaline) or motivation (dopamine), it is currently not possible to engineer the human brain in a way that would increase the number and duration of active higher-order representations. However, in a biomimetic instantiation, it would be fairly easy to increase both the number and duration of simultaneously active higher-order nodes (see Figure 9 below). Of course, in order to operate meaningfully, and reduce its propensity for recognizing "false patterns," such an ultraintelligent system would require extensive supervised and unsupervised learning.

It is currently not possible to engineer the human brain in a way that increases the number and duration of active higher-order representations in order to enhance mental continuity and the intelligent processes that it supports. However, in a biomimetic instantiation it would be fairly easy to increase the number and duration of simultaneously active higher-order representations. Accomplishing this would allow the imagery that is created to be informed by a larger number of concerns, and would ensure that important features were not omitted simply due to the fact that their activity could not be sustained due to biological limitations.

To accomplish overt behavior, higher inputs are fed not only to the lower sensory nodes, but also in a similar, top-down manner to a behavior module that will guide natural language output and other behaviors such as robotic control. The final layer of nodes in this behavior module will be nodes that directly control movement and verbalization and the higher nodes will be continuous with the higher-order PFC-like nodes. The software functions in an endless loop of reciprocating transformations between sensory nodes, motor nodes, and PFC-like buffer.

## Other Forms of Sustained Activity

Aside from having a PFC analogue, the network could also have an analogue of cortical priming and an analogue of the hippocampus. Humans have thoughts that carry continuity because changes in content are gradual as more recent activations/representations are given a higher priority than older ones. Activity that transpired minutes ago is given moderate priority, activity from seconds ago is given high priority and activity from mere milliseconds ago is given the highest priority. This dynamic is made possible by the PFC analogue, but could be accentuated by analogues of cortical priming. To allow for an analogue of cortical priming, all recently active neurons would retain a moderate amount of increased, but subthreshold activity. The activity level of recently used nodes in both the higher and lower-order areas would not quite fall back to zero. This would ensure that recently used patterns and features would be given a different form of priority, yet to a lesser and more general extent than that allowed by the PFC analogue. Regarding the network partitions depicted in Figure 5, the sensory, motor and hippocampal neural networks would show the least priming, the association area, and premotor neural networks would show moderate priming, and the PFC would show the highest degree of priming. Functions for the parameters of priming could be fine-tuned by genetic algorithms.

Furthermore, the network could have an analogue of the hippocampus. A hippocampal analogue would keep a record of contextual, or episodic clusters of previous node activation. Instead of keeping a serial record of averaged activity, the hippocampus analogue would capture episodic constellations of node activity and save these to be reactivated later. These episodic memory constellations would be activated when a large subset of the constellation is present during processing. This means that when neural network activity closely approximates an activity constellation that was present in the past, the hippocampal analogue would be capable of reactivating the original constellation. The activity of the hippocampal analogue should be informed by actual hippocampal anatomy and the “pattern completion” hypothesis of hippocampal function. To build an analogue into a neural net it would be necessary to have a form of episodic memory that can be cued by constellations of activity that closely resembles a past (autobiographical or episodic) occurrence. This memory system would then be responsible for "completing the pattern," or passing activation energy to the entire set of nodes that were initially involved in the original experience, allowing the system a form of episodic recall. As with the actual brain (Amaral, 1987), in the present device, the hippocampus should be reciprocally connected with the PFC and association areas but not with primary sensory or motor areas.

**<u>Appropriate Neural Network Parameters For the Present Device</u>**

A network is "trained" to recognize a pattern by adjusting arc weights in a way that most efficiently leads to the desired results. Arcs contributing to the recognition of a pattern are strengthened and those leading to inefficient or incorrect outcomes are weakened. The network "remembers" individual patterns and uses them when processing new data. Neural learning adjustments are driven by error or deviation in the performance of the neuron from some set goal. The network is provided with training examples, which consist of a pattern of activities for the input units, along with the desired pattern of activities for the output units. The actual output of the network is contrasted with the desired output resulting in a measure of error. Connection weights are altered so that the error is reduced and the network is better equipped to provide the correct output in the future. Each weight must be changed by an amount that is proportional to the rate at which the error changes as the weight is changed, an expression called the "error derivative for the weight." In a network that features back propagation the weights in the hidden layers are changed beginning with the layers closest to the output layer, working backwards toward the input layer. Such backpropagating networks are commonly called multilayer perceptrons (Rosenblatt, 1958). The present architecture involves a number of multilayered neural networks connected to each other, each using their own training criteria for backpropagated learning. For instance, the visual perception module would be trained to recognize visual patterns, the auditory perception module would be trained to recognize auditory patterns, and the PFC module would be trained to recognize multimodal, goal-related patterns.

The hierarchical multilayered network, the neocognitron, was first developed by K. Fukushima (1975). This system and its descendants are based on the visual processing theories of Hubel and Wiesel and form a solid archetype for the present device because they feature multiple types of cells and a cascading structure. Popular neural network architectures with features that could be valuable in programming the present device include the adaptive resonance theory network (Carpenter & Grossberg), the Hopfield network, the Neural Representation Modeler, the restricted coulomb energy network, and the Kohonen network. Teuvo Kohonen (2001) showed that matrix-like neural networks can create localized areas of firing for similar sensory features, which result in a map-like network where similar features were localized in close proximity and discrepant ones were distant. This type of network uses a neighborhood function to preserve the topological properties of the input space, and has been called a "self-organizing map." This kind of organization would be necessary for the present device to accomplish imagery generation, and would contribute to the ability of the lower-order nodes in the sensory modules to construct topographic maps.

A neural network that uses principal-components learning uses a subset of hidden units that cooperate in representing the input pattern. Here, the hidden units work cooperatively and the representation of an input pattern is distributed across many of them. In competitive learning, in contrast, a large number of hidden units compete so that a single hidden unit is used to represent a particular input pattern. The hidden unit that is selected is the one whose incoming weights most closely match the characteristics of the input pattern. The optimal method for the present purposes lies somewhere between purely distributed and purely localized representations. Each neural network node will code for a discrete, albeit abstract pattern, and compete among each other for activation energy and the opportunity to contribute to the depiction of imagery. However, multiple nodes will also work together cooperatively to create composite imagery.

When active, high level nodes signal each of the low-level nodes that they connect with, they are in effect, retroactivating them. They are activating those that recently contributed to their activity, and activating previously dormant ones as well. This retroactivation of previously dormant nodes constitutes a form of anticipation or prediction, indicating that there is a high likelihood that the pattern that these nodes code for will become evident (prospective coding). This kind of prediction is best achieved by a hierarchical hidden Markov model. Utilizing Markov models, and their predictive properties will be necessary. This process is used in Ray Kurzweil's Pattern Recognition Theory of Mind (PRTM) model, which uses a hidden Markov model and a plurality of pattern recognition nodes for its cognitive architecture (Kurzweil, 2012). Hierarchical temporal memory (HTM) is another cognitive architecture that models some of the structural and algorithmic properties of the neocortex (Hawkins & Blakeslee, 2005). The hope with PRTM and HTM is that a hierarchically structured, neural network with enough nodes and sufficient training should be able to model high-order human abstractions. However, distilling such abstractions and utilizing them to make complex inferences may necessitate an imagery guidance mechanism with a working memory updating function.

Neural networks can propagate information in one direction only, or they can be bi-directional where activity travels up and down the network until self-activation at a node occurs and the network settles on a final state. So called recurrent networks are constructed with extensive feedback connections. Such recurrent organization and bi-directionality would be important to accomplish the oscillating transformations performed by the present device. Hebbian learning is an updating rule that suggests that the connections weights for a neuron should grow when the input of the neuron fired at the same time the neuron itself fired (Hebb, 1949). This type of learning algorithm would be important for the present device as well.

Each topographic map that is formed could be assessed for appetitive or aversive content. he architecture depicted in Fig 5 could be copied onto two separate, yet nearly identical systems, one fine-tuned for approach behaviors and the other for withdrawal behaviors. This could simulate the right and left cortical hemispheres. The right hemisphere could be associated with withdrawal and have longer connectional distances between nodes on average.

In some neural networks, the activation values for certain nodes are made to undergo a relaxation process such that the network will evolve to a stable state where large scale changes are no longer necessary and most meaningful learning can be accomplished through small scale changes. The capability to do this, or to automatically prune connections below a certain connection weight would be beneficial for the present purposes. It is also important to preserve past training diversity so that the system does not become overtrained by narrow inputs that are poorly representative.

The present architecture could be significantly refined through the implementation of genetic algorithms that could help to select the optimal ways to fine-tune the model and set the parameters controlling the mathematics of things such as the connectivity, the learning algorithms, and the extent of sustained activity. It might also be beneficial to implement a symbolic, rule-based approach, where a core set of reliable rules are coded and used to explicitly structure iterative updating as well as influence decision making and goal prioritization. Many theorists agree that combining neural network, and more traditional symbolic approaches will better capture the mechanisms of the human mind. In fact, implementing explicit rules to instantiate processing priorities could help the higher-order nodes to account for goal-relevance. These might be necessary to simulate the rules of emotional, subcortical modules. In evolutionary algorithms an initial population of solutions/agents is created and evaluated. New members of the population are created from mutation and crossover. The updated population is then evaluated and agents are either deleted or naturally selected based on their fitness value (or performance).

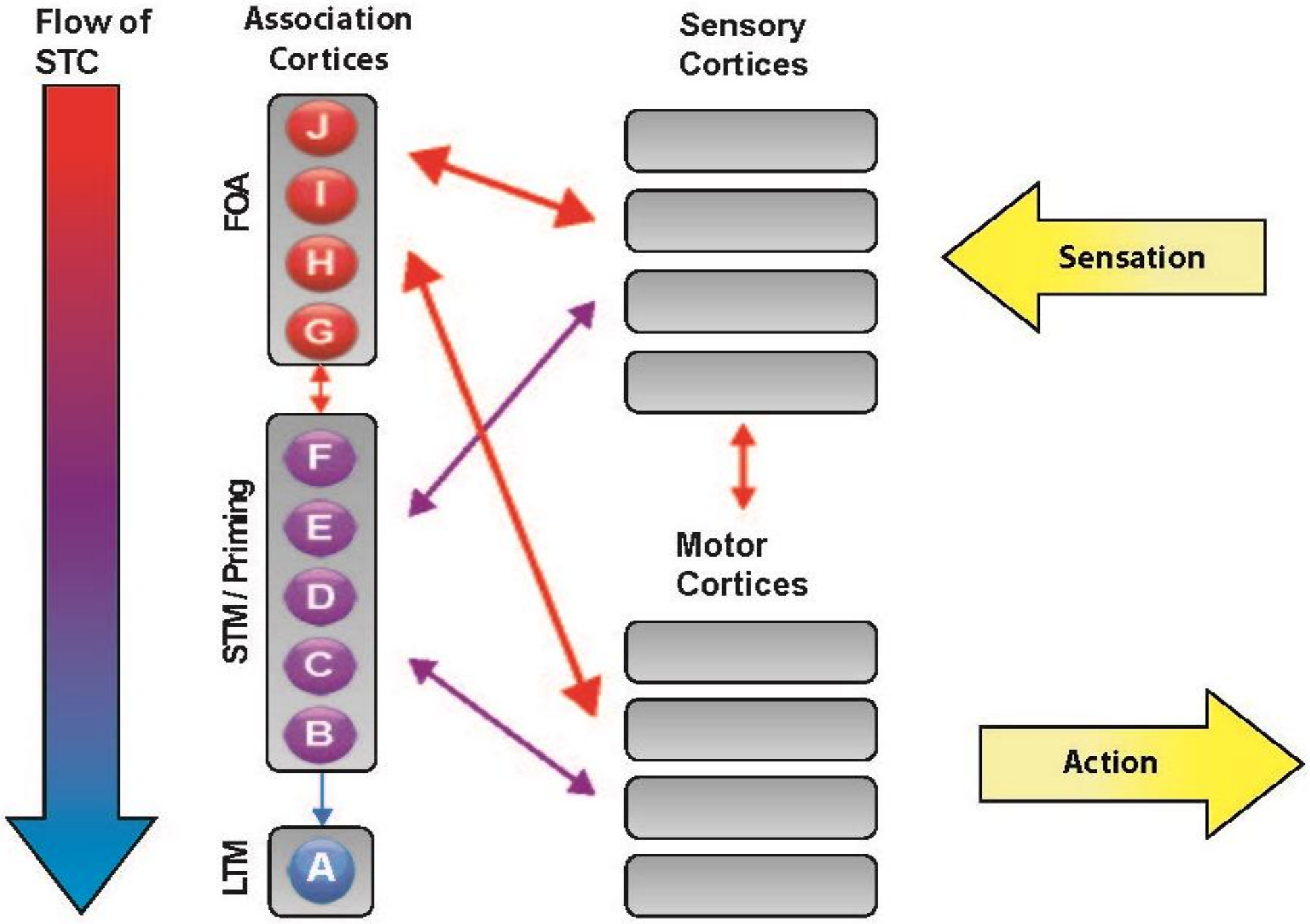

**Fig. 11.** Shows that information from motor and sensory cortices enters the focus of attention where it can then explicitly influence other sensory and motor cortices. As information leaves attention it can either be held temporarily in a less active form of STM (which can implicitly influence sensory and motor cortices) or it can deactivate and return to LTM. The arrow on the left indicates that in succeeding states, the letters will cycle downwards as their activity diminishes.

**Table 2.** The Process by which Short-Term Continuity Influences Global Processing

1) Information flows to early sensory cortex from the environment or from the association cortex.

2) Topographic sensory maps are constructed from this information within each low-order, sensory module. In order to integrate the disparate features into a meaningful image, the map-making neurons will be forced to introduce new features not found in their extrinsic inputs.

3) Information from the imagery travels bottom-up toward the association cortex. The salient or goal-relevant features from the mappings are used to update the group of sustained representations held active in the association cortex.

4) The least relevant, least converged-upon representations in the association cortex are dropped from sustained activation and “replaced” with new, salient representations. Thus, the important features of the last few maps are maintained in an active state.

5) The updated group of representations will then spread its activity backwards toward lower-order sensory nodes in order to activate a different set of low-order nodes culminating in a different topographic sensory map.

6) A. The process repeats.

B. Salient sensory information from the actual environment interrupts the process. The lower-order nodes and their imagery, as well as the higher-order nodes and their priorities, are refocused on the new incoming stimuli.

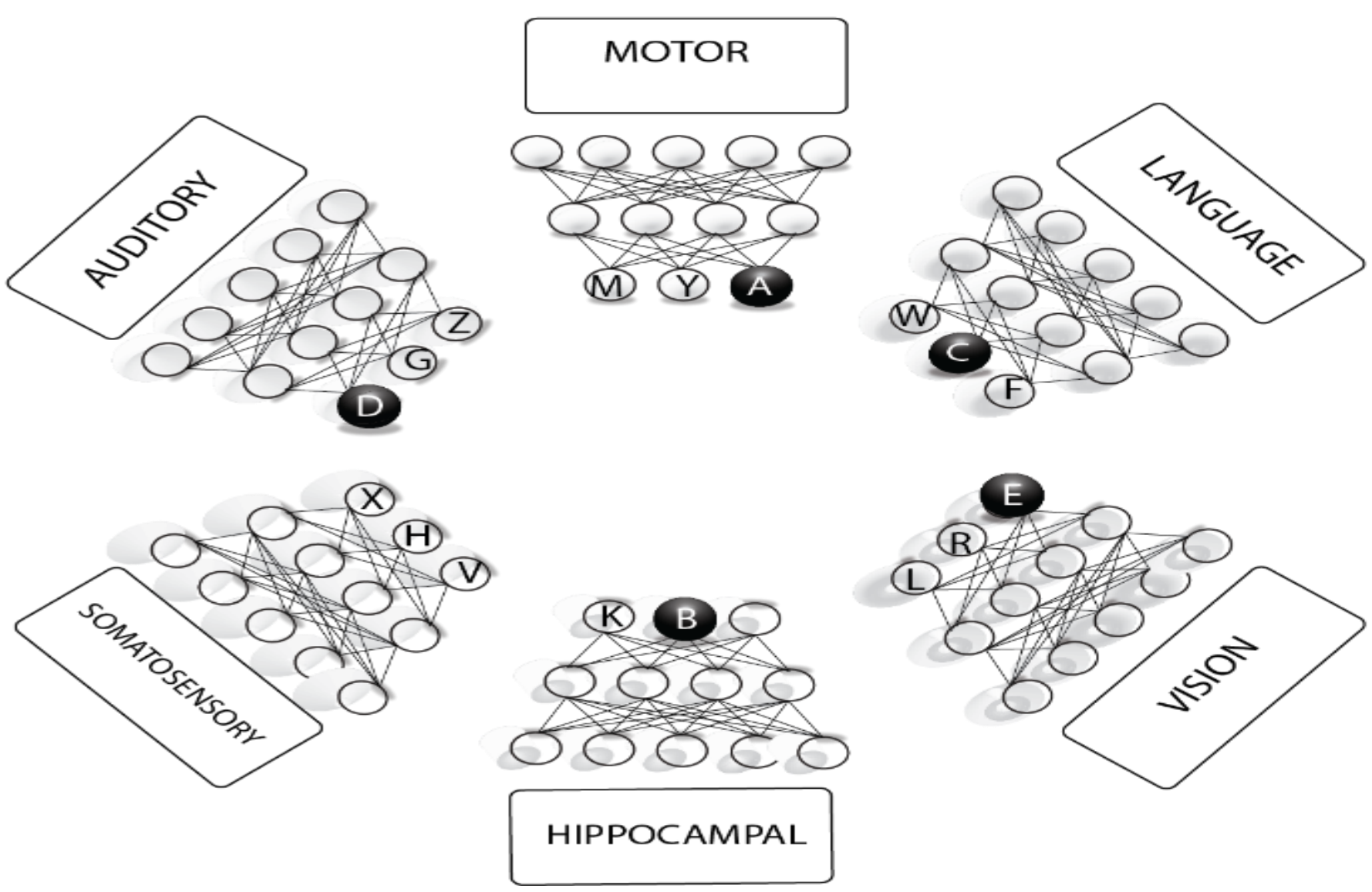

**Fig. 12.** Each set of 12 circles represents an individual artificial neural network trained to process a different modality of inputs. Each network is shown fully connected; however, not shown are the numerous connections needed between networks. Spreading activation would travel recurrently within a network, and interactively (i.e. generatively, adversarially, or collaboratively) between networks. The boxes represent sensory input, motor output, and/or topographic maps. The row of five nodes in each network represents the input layers, the row of four represents the hidden associative layers, and the row of three represents the output layer. The output nodes are pattern recognizing classifiers that hold the equivalent of items of working memory. They would be capable of sustained activity and, as a group, iterative updating.

## Conclusions

Many researchers have suggested that AI does not need to simulate human thought, but rather should simulate the essence of abstract reasoning and problem solving. The present article has suggested that modeling "mental continuity" and using it to guide successive images is an essential part of this simulation.

There are no forms of AI that use mental continuity as described here. There are existing computing architectures with limited forms of continuity where the current state is a function of the previous state, and where active data is entered into a limited capacity buffer to inform other processes. However, there are no AI systems where this buffer is multimodal, positioned at the top of a hierarchical system, and that informs and interacts with topographic imagery.

The current objective is to create an agent that through supervised or unsupervised feedback can progress to the point where it takes on emergent cognitive properties and becomes a general problem solver or inference program capable of goal-directed reasoning, backwards chaining, and performing means-end analyses. The present device should constitute a self-organizing cognitive architecture capable of dynamic knowledge acquisition, inductive reasoning, dealing with uncertainty, high predictive ability and low generalization error. If implemented and trained properly it should be able to find meaningful patterns in complex data and improve its performance by learning. It should be the goal of A.I. experts to fine tune such a system to become capable of autoassociation (the ability to recognize a pattern even though the entire pattern is not present) and perceptual invariance (generalizing over the style of presentation such as visual perspective or font).

The writing here amounts to a qualitative account, is exploratory, contains unverified assumptions, makes untested claims, and leaves important concerns out of the discussion. A more complete and refined version would focus on better integration of existing knowledge from functional neuroanatomy, multisensory integration, clinical neuropsychology, brain oscillations, short-term and long-term potentiation, binding, the sustained firing behavior of cortical columns, and the cognitive neuroscience of attention.

The present architecture is designed to simulate human intelligence by emulating the mammalian fashion for selecting priority stimuli, holding these stimuli in a working memory store and allowing them to temporarily direct sensory and motor modules before their activity fades. It would be composed of a large set of nodes that work together to continually determine, in real time, which from their population should be newly activated, which should be deactivated and which should remain active over elapsing time to form a "stream" or "train" of thought. The network's connectivity allows reciprocating cross-talk between fleeting bottom-up imagery in early sensory networks and lasting top-down priming in association and PFC networks. The features that are maintained over time by sustained neural firing are used to create and guide the construction of topographic maps (imagery). The PFC and other association area neural networks direct progressive sequences of mental imagery in the visual, auditory, and somatosensory networks. The network contains nodes that are capable of "sustained firing," allowing them to bias network activity, transmit their weights, or otherwise contribute to network processing for several seconds at a time (generally 1-30 seconds).

Cognitive control stems from the active maintenance of features/patterns in the PFC module that allow the orchestration of processing in accordance with internally selected priorities. The network is an information processing system that has the ability to maintain a large list of representations that is constantly in flux as new representations are being added, some are being removed and still others are being maintained. This distinct pattern of activity, where some individual nodes persist during processing makes it so that particular features of the overall pattern will be uninterrupted or conserved over time. Because nodes in the PFC network are sustained, and do not fade away before the next instantiation of topographic imagery, there is a continuous and temporally overlapping pattern of features that mimics consciousness and the psychological juggling of information in working memory. This also allows consecutive topographic maps to have related and progressive content. If this sustained firing is programmed to happen at even longer intervals, in even larger numbers of nodes, the system will exhibit even more mental continuity over elapsing time. This would increase the ability of the network to make associations between temporally distant stimuli and allow its actions to be informed by more temporally distant features and occurrences.

**<u>The Specious Present in Machine Consciousness: Why Temporal Continuity is the Missing Key to AI Awareness</u>**

The ability to perceive the present as an extended yet unified experience, rather than as a series of discrete, isolated moments, is central to human cognition. This phenomenon, known as the *specious present*, is a critical aspect of temporal awareness that enables fluid perception, coherent thought, and a stable sense of self. Unlike machines, which process information in strict sequences with no intrinsic sense of continuity, human consciousness unfolds across time in a way that allows for seamless transitions between thoughts, decisions, and actions. If artificial intelligence is ever to approach anything like human-like awareness or achieve machine consciousness, it may need to develop a computational equivalent of the *specious present*.

- LLMs operate like someone reading one *frame* of a film at a time and commenting on it.

- Consciousness arises not from seeing a frame, but from *living inside the stream* of frames.

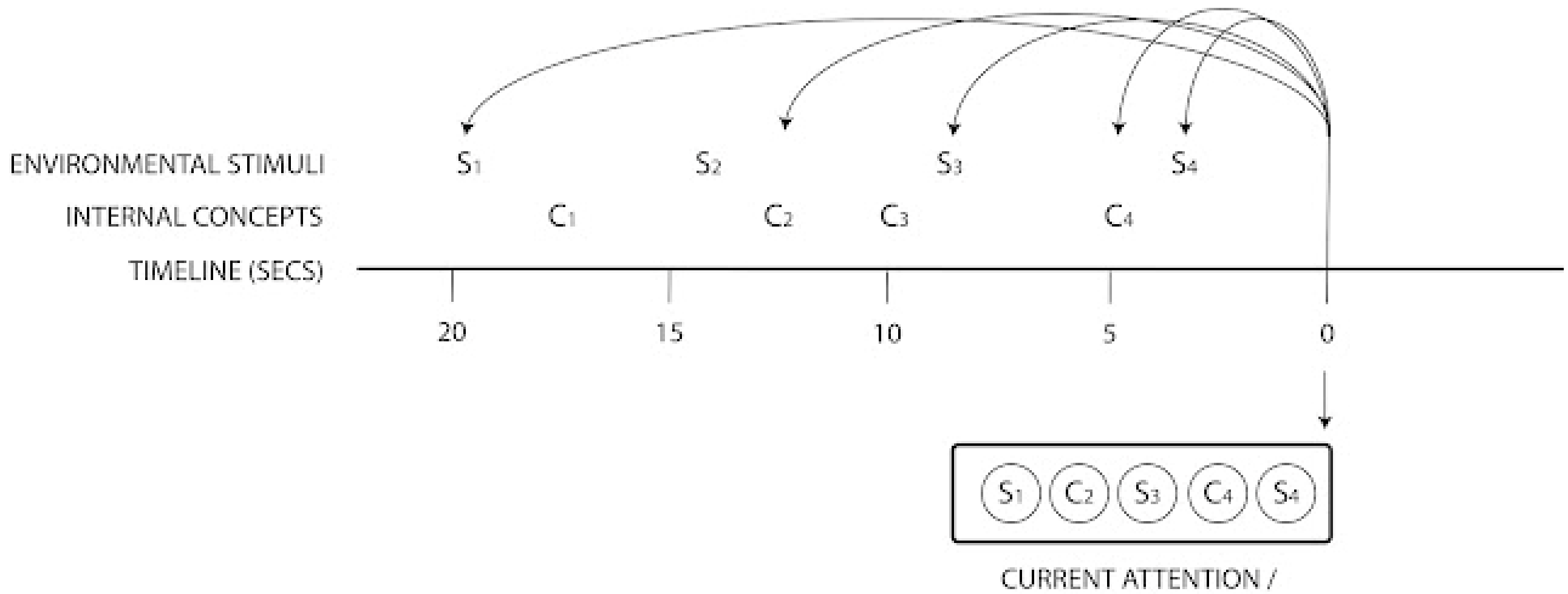

**Figure 12:** This figure shows a timeline consisting of 20 seconds where four new environmental stimuli were encountered and four new concepts were considered. Three of these stimuli and two of the concepts were carried forward into the present moment at time 0. At the bottom of the figure are the current contents of attention and this represents the specious present.

### What is the Specious Present?

The specious present is the idea that our perception of "now" is not a single static instant but a brief duration in which we experience events as a unified whole. Imagine you're watching a movie. The pictures on the screen change quickly, but you don't see them as separate pictures, you see one smooth story happening. This is the specious present. In fact, it holds every event and episode in our lives together. Our short-term memory ensures that in any present instant there are many simultaneously coactive traces carried over from the last few seconds of experience. This allows our brains to construct a rolling window of awareness. The specious present is what allows us to perceive a melody as a coherent whole rather than a sequence of isolated notes. It helps us link together speech sounds so that we can comprehend whole conversations. And this ability to build an evolving conception of reality as it unfolds helps us understand cause and effect.

The specious present is considered "specious" because it is an illusion, our perception of the "now" is not an objective, indivisible instant in time but a constructed temporal experience. The term specious in this context means "deceptive" or "misleading," rather than outright false. It suggests that while we feel as though we are experiencing a single, unified present moment, this is actually a fabrication.

The idea was popularized by E. R. Clay and later developed by William James. James called it the "short duration of which we are immediately and incessantly sensible." In cognitive science, this concept aligns closely with Jared Reser's theory of *iterative updating* which can be found at aithought.com. The Iterative Updating model explains that successive mental states retain partial continuity with prior states rather than being replaced wholesale. This form of updating ensures that past and present states are not sharply divided but instead blend into one another, creating an ongoing stream of consciousness. If AI were to incorporate a similar mechanism, it might come closer to replicating a form of intelligence that is temporally situated rather than ephemeral and fleeting.

So, before we talk about how to implement this in a computer, let's look at what it has to offer AI.

**What Capabilities Could a Specious Present Create for an AI System?**

The *specious present* plays a role in shaping our sense of self. Human identity is not constructed from isolated mental events but from a persistent, temporally extended awareness that binds together past, present, and anticipated future states. This self emerges from the very continuity that AI currently lacks. If AI were to develop a *specious present*, it might become capable of acquiring something resembling personal identity. Can you imagine a way that consciousness, identity, or thought could exist without an experience of time? How could they given they are inherently structured by the passage of time?

Some philosophers argue that the *specious present* is not merely an aspect of cognition but a necessary condition for consciousness itself. If subjective awareness requires an ongoing experience of time, then any AI that operates purely through discrete, non-iterative processes may always remain fundamentally unconscious, no matter how sophisticated its outputs become.

This has a bearing on free will. Human decision-making relies on holding multiple relevant pieces of information active at the same time, forming a coherent decision space. The *specious present* allows for deliberation, where alternatives are weighed within a coherent and evolving cognitive space. Can an AI truly "choose" if it lacks a unified present experience? If free will depends on the ability to hold multiple potential futures in mind at once, then a *specious present* should be essential for AI autonomy. The absence of this capacity raises important questions about intentionality and whether AI can ever genuinely "choose" rather than merely execute the most statistically probable response at a given moment.

LLMs only predict the very next word (token) for the very next state, and they put all their processing power into that computation. The human brain makes predictions many states ahead, we aren't even interested in what comes in the very next state.

**Do Modern AI's Experience a Present or a Sense of Continuity?**

There is not much reason to believe that contemporary AI systems exhibit a form of temporal continuity. Even the most advanced AI models, such as state-of-the-art large language models like GPT, operate in a manner that is fundamentally different from human cognition. When prompted, the models generate responses that can demonstrate sophisticated forms of intelligence, knowledge, understanding, introspection, and even coherence across extended dialogues. Yet, once a response is produced, the internal cognitive process of the AI effectively vanishes, until it is reinitialized with a new prompt. There is no persistence of thought or self-directed continuity beyond the task at hand. Unlike a human, who maintains an ongoing awareness that stretches across each waking day, an AI does not carry an evolving, self-updating internal state. This limitation suggests that artificial systems are not merely lacking in conscious experience but are also constrained in their ability to engage in sustained, context-aware deliberation. Would creating an AI where conceptions persist consistently even when not in dialogue allow it to be conscious? Probably not because its lack of continuity extends down into the way it actually processes tokens.

Most modern AI systems, particularly those based on the transformer architecture, process input through *self-attention mechanisms* that dynamically weight the importance of words or symbols within a predefined context window. While this approach allows AI to generate coherent text, it probably does not grant it a *specious present*. In human cognition, attention is not recomputed from scratch with each passing moment. Rather, some neuronal activity remains sustained over multiple cognitive cycles, allowing for the gradual fading and persistence of relevant information. AI, by contrast, recalculates its attention weight matrix anew every time it shifts to the next input token (word). Thus, it loses a good deal of continuity after generating each new token. Basically, AI systems, especially deep learning models, process information in discrete steps, much like frames in a video. For AI to better mimic human cognition, it would need a mechanism that integrates events over short time spans, rather than treating each moment as isolated.

Large language models using the transformer architecture do retain some contextual information within a fixed context window. And that window, like its attentional store is updated iteratively. But anything outside that range is irretrievably lost unless explicitly retrieved. There is no natural mechanism for dynamically recalling and reintegrating forgotten details in a way that mimics biological cognition (such as synaptic potentiation).

This discrepancy between human and artificial cognition is further reinforced by how their respective memory systems operate. Human working memory does not function like a rigid queue in which new items simply push out older ones in a fixed sequence. Instead, the brain varies the rate of updating depending on cognitive demand, retaining some information for extended periods while discarding other details more quickly. An AI model context window, however, operate on a strict first-in, first-out (FIFO) replacement system, where information is lost at a predetermined rate dictated by the architecture rather than by meaningful contextual relevance. In human cognition, a particularly significant event or idea may be retained far longer than less relevant details, allowing for a dynamically weighted memory system that adapts to the needs of the moment as well as long-term plans. Without this capacity for flexible, self-directed retention and recall, AI remains constrained to static processing rather than true, evolving thought.

**How Could We Build a Specious Present into AI?**

If we think of the specious present as a "temporal buffer" for reality, then an AI trying to achieve human-like intelligence might need something similar. Unlike static windowed attention (e.g., in GPT-like models), an AI implementing the specious present would require sliding attention windows that retain and update relevant context iteratively. A rolling buffer of attention that refreshes dynamically, ensuring that a portion of past activations persists while incorporating new ones. This could be implemented by architectures designed for persistent, overlapping time representation.

AI researchers have explored architectures that attempt to integrate elements of persistence into computational models. These include recurrent neural networks (RNNs) and long short-term memory (LSTM) networks which have nodes or neurons that remain active for extended periods as they do in the brain. Transformer-XL networks, extend the context window of conventional Transformers by allowing past activations to influence present ones, creating a form of recurrence that mimics some aspects of human memory. Similarly, Longformer and Memorizing Transformers introduce mechanisms that allow for the retention of information over longer sequences, enabling AI to reference past data beyond a fixed window.

Neural Turing Machines and Differentiable Neural Computers take a different approach, explicitly modeling memory structures that can be read from and written to dynamically, providing a form of information continuity. RWKV, a more recent development, blends elements of Transformer-based attention with recurrent architectures, attempting to capture both long-range dependencies and efficient processing of sequential information. While each of these models moves AI closer to the ability to process time as a continuous experience rather than as a sequence of disjointed computations, none yet fully replicate the flexible, iterative updating of human cognition.

**Conclusion**

The concept of the *specious present* is more than a theoretical curiosity, it is the structural foundation that allows thought to unfold over time rather than existing as a series of isolated computations. Without it, AI remains a collection of momentary states, never evolving into an entity that perceives time, remembers meaningfully, or sustains self-awareness.

The *specious present* is more than just a technical feature of cognition, it is the foundation of conscious experience, temporal reasoning, decision-making, and selfhood. If AI is ever to think, reflect, or even "feel" in a human-like way, it must:

1. Maintain a persistent but fluid cognitive state.
2. Experience time as a continuous flow rather than discrete moments.
3. Integrate past and present information in a structured, self-referential manner.
4. Develop a temporally extended sense of self-awareness.

If consciousness is, at its core, an ongoing stream of experience, then AI will never achieve it unless it develops the mechanisms necessary to sustain and integrate temporal awareness. An AI without a specious present can only approximate sentient intelligence, not embody it. The pursuit of artificial consciousness may therefore depend not just on increasing computational power or refining present neural networks conventions, but on giving machines a way to *exist in time* rather than merely process it.

**From Context Windows to Cognitive Scenes: Why AI Thinking Should Be Event-Oriented**

Large language models reason across vast strings of text, relying on statistical prediction over many thousands of words. Human cognition, on the other hand, unfolds within compact scenes composed of only a handful of interrelated concepts. This essay argues that the human capacity to represent and iteratively update such bounded scenes, rather than simply track long contextual histories, is central to understanding consciousness and to constructing artificial systems that think more like minds.

Large language models predict the next word (or token) by using the words that came earlier. These words sit in their context window. They can use from one word up to around one million. And when they do this, they use all the words in their context window to help them choose the next word. So, they don't build sequences of singular scenes to help them think, instead they use a long, unbounded list of precursor words.

When humans think, they use a series of concepts to help direct the train of thought. As they do so, they form cohesive scenes composed of just a handful of concepts. Each scene is a framed vignette or unit of experience. Such a scene might involve people, a place, and certain actions. And then generally visual and auditory imagery are generated to envision this scene. AI language models don't create imagery to help inform their processing. But more to the point of this post, they also don't create a bite sized scene to depict the pertinent concepts. Rather, they lean into the long list of words in their context window. That window holds useful context but is generally too big to represent a single concrete occurrence or episode.

I believe  the ability to experience (or imagine) an isolated event as a mental tableau is crucial to human-like intelligence. Our mental lives work this way because we have a focus of attention which can only hold about 4 to 7 working memory items simultaneously. That size is large for an animal but small for an AI. However, it is sufficient for holding the relevant constructs necessary to model most static situations. How can an AI hope to model or experience a distinct situation when it is using thousands of words to do so? Moreover, how can an AI develop a form of consciousness or the ability to experience a distinct situation when its attention is full of such a large number of incongruous elements?

Now LLMs do use attention within their context window to narrow down their focus to the most pertinent words. They assign each word a score and relate them to each other. But I think that AI should be built from the ground up to generate individually meaningful, semantically-bounded scenes. If you were training an AI to do this from text, you could use each sentence as an individual scene. That sentence could be used to prompt an imagery generation system to depict it visually and auditorily. These depictions would offer helpful information not contained in the words themselves. For instance, when you read a book, many details are left out, but the readers imagination fills them in. In fact, imagination and visualization are necessary to understand passages in many books from fiction to nonfiction. AI needs an imagination to achieve these insights. To acquire one, it must break the information it is processing down into experience-sized chunks.

Right now, many psychologists, neuroscientists and machine learning experts stand up, wave their hands around in the air excitedly, and proclaim that large language models are nothing like the human brain. Many of those claim that AI will never reach human aptitude, and in their present form are going nowhere. I disagree wholeheartedly, because I believe large language models already capture the two most important abstract features of the brain. These are, 1) having two forms of working memory, a context window (akin to synaptic potentiation) and attention (akin to the focus of attention), and 2) having these memory stores updated iteratively as time passes.

Language models have both of these memory stores and iterate on them as each new word enters the context window. But the third ingredient they are missing is the ability to form scenes. Once they can form scenes, they can iterate on them. Iterating on the scenery that is built is crucial because it turns a mental picture into a narrative. This would ultimately allow for the creation of chains of interlinked experiences. What I am trying to say is that AI is closer than most people realize to capturing the important aspects of mental machinery. Scenery may be one of the last necessary pieces of the puzzle.

It is important to mention that the entire context window could still be very helpful and could function as a parallel to the mammalian short term memory store (short-term potentiation and priming). I explain how this could work at my website aithought.com, which details a cognitive architecture in accordance with these ideas.

Instead of reasoning across thousands of tokens, an AI system should instantiate a bounded semantic workspace, roughly equivalent to a human's focus of attention. This workspace could hold, say, 4–7 (or less than a dozen) scene objects or latent constructs. Each iteration (or "thought") would involve:

1. Updating the store of concepts with a new associative addition
2. Updating the semantic bindings between those constructs
3. Generating a new scene from this updated set, the next mental state.

The 4 to 7 item limit on human working memory may represent an evolutionarily optimized selection of elements large enough to model useful complexity. 4-7 items is certainly less than optimal for a computer because of metabolic constraints on our prehistoric ancestors. However, the capacity of the AI's focus of attention should be limited to a reduced subset of its total context window with scene-sized granularity.

Essentially, language models use token-based reasoning and humans use scene-based cognition. LLMs reason over a flat temporal stream, a list of words, each embedded in a very large contextual lattice. Humans reason over a temporally chunked scene-bounded, imagistic construct held in working memory. LLMs rely on syntactic continuity within the context window. Humans rely on semantic coherence within a scene and between them. The LLM's context window functions somewhat like a cognitively unmanageable short-term buffer. Both feature context-dependent iterative updating, but with radically different chunk sizes and representational units. Boundedness is not a shortcoming, it is a condition for awareness.

## Language Models are Trapped in Token-Bound Time with Token-Locked Receptive Fields

Our brain cells help us carry many forms of information through time, but modern AI systems only carry words, and this can be framed as their major limiting factor. Here we will discuss this in relation to two fascinating phenomena with great exploratory potential: "sustained firing" and "receptive fields."

### I. Introduction: The Illusion of Cognition

Large language models and the transformer architecture share striking functional parallels with the human brain. Both systems rely on capacity-limited stores to hold information, and both update that information iteratively, selecting the next most probable association based on prior context. However, this functional mimicry masks a profound ontological disconnect. While the brain is an evolved organ embedded in the thermodynamic flux of the physical world, the language model is hermetically sealed within a "token space."

Consequently, these models suffer from two fatal deficits that prevent true general intelligence: they exist in Token-Bound Time and rely on Token-Locked Receptive Fields. They do not process reality; they process a symbolic queue. They relate tokens back to previous tokens, making probabilistic guesses about associations, but this process is entirely untethered from the real time of moving objects, physical interactions, and genuine causality.

### II. Trapped in Token-Bound Time

To understand the deficit of the Transformer, we must first define the biological standard it fails to meet. In the mammalian brain, time is not merely a sequence of events; it is a metabolic endurance test. The prefrontal cortex tracks time through sustained firing, a mechanism where neurons must actively expend energy to keep a representation alive across a delay. This "holding cost" grounds the brain in real time; the duration of a thought is physically palpable.

AI, by contrast, lives in Token-Bound Time. In this state, "time" is not a temporal dimension measured in seconds or decay; it is a topological dimension measured in sequence length. The model perceives the "past" not as a fading signal that requires energy to sustain, but as a perfectly preserved list of integers at specific positional indices.

This creates a metric gap. Consider two sentences: "The ball [fell]" and "The empire [fell]".

In real time, the first event is instantaneous and the second spans centuries of complex causal decay.

In token-bound time, the distance between the subject and the verb in both cases is identical.

Because the model lacks a mechanism for sustained firing, it lacks the "visceral physics" of duration. It uses token order for learning (credit assignment), treating a gap of five centuries with the same computational weight as a gap of five seconds. It lives in a "frozen world" where time is spatialized, stripped of its flow, and severed from the thermodynamic constraints that govern actual cause and effect.

**III. Token-Locked Receptive Fields**

The limitations of Token-Bound Time are compounded by a structural blindness I call Token-Locked Receptive Fields. In neuroscience, a receptive field is the specific "window on the world" to which a neuron responds. Each brain cell has a unique set of inputs all of which combine to determine its unique response properties signifying not just where it sits in the hierarchy, but what it represents when active. The cortex is organized into a massive spatiotemporal hierarchy of cells each with its own receptive field. Low-level fields (in sensory cortex) are small, transient, and lock onto simple physical features (edges, brightness). High-level fields (in association cortex) are massive, sustained, and lock onto abstract "trans-temporal" realities (goals, social hierarchies, future predictions).

Current language models have a similar hierarchy. But they suffer from a "flatness" of perception. Whether at Layer 1 or Layer 96, the attention mechanism is structurally identical: it is attending to tokens. The model effectively has millions of "eyes," but every single one of them is looking at text, and nothing else.

A Token-Locked Receptive Field means the system never graduates from processing the symbol to processing the referent. It manipulates the word "apple" and the word "gravity" with sophisticated statistics, but it lacks the hierarchical architecture to combine these into a compounded, multi-modal receptive field that "understands" the physics of a falling apple. The model is trapped in the map, unable to perceive the territory.

Elsewhere I have argued that AI needs to build a scene and should be designed to be scene based. Not a sequence or a stack of convolutions. A scene. A dynamic, relational, cohesive, world-centered scene. I think this argument complements the argument I am making and you can read about it here:

https://www.observedimpulse.com/2025/10/from-context-windows-to-cognitive.html

**IV. The Synthesis: Complexity Requires Duration**

These two deficits are not separate; they are causally linked. In the biological brain, the neurons with the most complex, compounded receptive fields are precisely those that exhibit sustained firing over the longest periods. They are generally in the parietal cortex and prefrontal cortex.

This reveals a fundamental law of intelligence: Complexity is linked to duration. To model a complex, abstract concept (like "justice" or "causality"), a system must be able to hold a state stable against time. The "deepest" thoughts are necessarily the "longest" thoughts. Because LLMs lack the mechanism for sustained firing (temporal depth), they are structurally incapable of forming the compounded receptive fields (informational depth) required for reasoning. They are attempting to build a skyscraper of meaning on a foundation that has no temporal thickness.

**V. Conclusion: Beyond Language**

The diagnosis is clear: current language models are effectively a disembodied "Broca's Area" (the brain's language production center), highly capable of lexical manipulation and syntactic sequencing, yet isolated from the sensory, executive, and temporal hierarchies that constitute a mind.

To move beyond this plateau, Artificial Intelligence needs a more general reality model capable of multimodal fusion. It needs to be attached to an architecture capable of sustained firing, a mechanism that forces it to endure the passage of time rather than just counting tokens. Until we break the lock of Token-Bound Time and expand the hierarchy beyond Token-Locked Receptive Fields, these models will remain impressive mimics of language, forever separated from the physical reality that gives language its meaning.

Large language models lack direct perception and bodily action. Even when paired with cameras or microphones, the core model does not inhabit a sensorimotor world in the way animals do. Yet the absence of embodiment does not automatically settle whether such systems could possess any minimal analogue of subjective continuity. This article argues that the relevant question is architectural: whether a system's internal states form a temporally extended process in which successive moments are meaningfully conditioned on, and partially composed of, their immediate predecessors. On this view, the most plausible "experience-like" property an LLM could exhibit would not be sensory qualia, but a thin form of structural phenomenology arising from constraint satisfaction in a high-dimensional latent space as the model selects successive updates under context. The analysis distinguishes continuity of constraint from continuity of self-tracking, suggesting that present-day LLMs may approximate the former through autoregressive looping while largely lacking the latter due to limited persistent memory, weak self-modeling, and minimal intrinsic coupling between internal dynamics and world-stable consequences. The framework yields testable implications: candidate metrics include state-to-state carryover, stability of commitments across updates, signatures of self-monitoring, and the degree to which the system's next state is shaped by its own immediate history rather than by input alone.

We need to move away from local receptive fields and fixed or predefined hierarchies. We must move toward global receptive fields, flexible cross-attention, the ability to unify heterogeneous or asynchronous signals, integrate information across space and time, reason about long-range dependencies, combine heterogeneous signals, continuous world updating, We need a universal architecture for building coherent worlds out of fragmented signals. We need a relational engine capable of binding separate streams of information into unified, structured representations.

**Analogs of Right and Left Cortical Hemispheres Might Be Helpful for AI**

The cerebral cortex can be divided right down the middle (sagittally) into two, nearly identical hemispheres. Far from being redundant, they each process much of the same information in slightly different ways, leading to two complementary and cooperating worldviews. This organization (called hemispheric lateralization) could be beneficial for an AI agent. The main benefit would be that over developmental time these two innately different networks would reprogram each other by being exposed to each other's outputs. In essence, two dissimilar heads are better than one.

Creating hemispheric laterality in a neural network would be easy. It would involve duplicating the existing network and then connecting the two of them via a large number of high bandwidth weighted links. These links would serve as the corpus callosum (the bundle of tracts that connect the right and left hemispheres in mammals). These connections should respect the brain's natural bilateral symmetry, connecting both similar and dissimilar areas across both hemispheres.

Each hemisphere of the brain processes information slightly differently. Despite the fact that the macrostructure of the two hemispheres is almost identical, the two are different microstructurally. Many researchers believe that this is because the right hemisphere has an average axonal length slightly (microscopically) longer than the left. This means that the right brain has relatively more white matter (axons), and the left hemisphere has relatively more grey matter (cell bodies). This also means that on average the cells of the right brain are further away from one another. This would be simple to replicate in existing neural network software. There are many theories as to why the longer-ranging wiring is responsible for the right hemisphere's tendency for broad generalization and holistic perspective.

People with a left hemisphere injury may have impaired perception of high resolution, or detailed aspects of an image, whereas those with right hemisphere injury may have trouble seeing the low resolution, or big picture aspects of an image. In other words, they miss the forest for the trees.

Each side of the artificial neural network would organizing its processing and learning according to a different algorithm and could lead each to develop a unique way of perceiving and responding to the world. Slight discrepancies in temporal processing parameters could make the feedback and crosstalk between these two non-identical systems meaningful. This would be commensurate to specialists that could check, balance, reconcile, compare, and contrast their differing approaches. If they processed information in exactly the same way it would be redundant to have two, but because they don't, they provide a type of stereoscopic view on the world similar to the view provided by our two offset eyes. This would enable them to form their own perceptions and opinions and then reconcile with one another. Right now, no AI systems have anything like this.

There are countless parameters of a neural network that could be fine-tuned to accomplish this. It might even be beneficial to have more than two. Some hyperparameters that are common in neural networks that could be altered include the refractory period, activation threshold, resting potential, resistance (MΩ) and capacitance (nF). These changes could be made to the software or could be instantiated as physical changes within neuromorphic hardware..

Our brain's two hemispheres also differ as to their value systems. Our left hemisphere is dedicated to approach behaviors, and our right hemisphere is dedicated to withdrawal behaviors. Stimulating the left hemisphere of a rat will make it go toward a new object, whereas stimulating the right side will make it back away from that object. The fact that vertebrate brains have made this fundamental differentiation between approach and withdrawal for hundreds of millions of years suggests that it might represent an organizing principle that should be used in AI.

One way to do this would be to wire up the AI's "subcortical" appetitive and motivational centers (analogous to the ventral tegmental area and the nucleus accumbens) involved in reinforcement learning with the left hemisphere, and to wire up the threat detection centers (analogous to the amygdala) involved in punishment learning with the right hemisphere. AI needs two, functionally assymetric, dedicated processors, one for approach/liking/curiosity, and one for withdrawal/disliking/fear. As I have explained elsewhere both hemispheres should influence the dopaminergic system and sustained firing so that important rewards and threats can be maintained in mind; however, the left should be focused on approaching them, and the right should be focused on withdrawing from them.

As in vertebrate animals, the left hemisphere could be used to control the right half of an AI's robot body and the right hemisphere could be used to control the left half. This could help ensure that both approach and withdrawal have equal impact on behavior. Approach and withdrawal could form a pivoting scale for how a robot acts in the world. They could also structure their conscious attention even when not moving by allowing them to pivot between interest and disinterest.

## How Hemispheric Valence Could Help Solve the Brain's Credit-Assignment Problem

The "Godfather of AI," Geoffrey Hinton, maintains that the main thing we don't understand about the brain is how it determines when to learn and when to unlearn. He emphasized that we don't know how real brains determine whether to increase or decrease a connection weight at the level of the cell, dendrite, and synapse. Currently, Artificial neural networks do this through a mathematical processes called "back propagation," but if we knew how the brain accomplished it it could lead to the construction of truly intelligent AI. Well, I've spent a lot of time reading about neuroscience and this is something I've never understood either and I can see how the rest of the field hasn't been able to figure it out. But, for the last few years I have had a sneaking suspicion that it may relate to hemispheric laterality.

What if this decision to learn or unlearn actually starts with the decision of what side of the brain to do the learning on? Our brains are fundamentally split down the middle, into two equal cortical hemispheres. One side is associated with approach and the side is associated with withdrawal. When you reach for that cookie, the left side is more active guiding your arm and when you step back from a loud noise, the right side presides over that movement. Maybe modern neuroscience has everything right, researchers just have not yet appreciated how all interactions with the world are funneled down one of two filters. The left hemisphere for approach (learning) or the right hemisphere for withdrawal (unlearning). This split, laterality, is almost as old as animals, jellyfish don't have it but the vast majority of animals do, and I believe that's because it plays an elementary and functional role that is so deep that it's gone mostly unrecognized. More specifically, maybe the brain doesn't unlearn and decrease connection weights. Maybe instead it learns in one of two functionally distinct hemispheres, but the fact that these two hemispheres are constantly working together, obscures this.

How plasticity is regulated is a core unresolved problem in neuroscience. Neuroscientists usually call this the credit-assignment problem (or synaptic credit-assignment). The phrase covers a family of questions: Which neurons and synapses "deserve credit or blame" for a given behavioral success or failure? How are those specific sites located and tagged so that plasticity mechanisms (LTP/LTD) know whether to strengthen or weaken them? How is the error signal delivered with the right timing and specificity—especially when the reward or punishment can be distal in time from the activity that caused it?

Hinton is right that artificial neural networks use a clean mathematical mechanism, backpropagation, to determine when to strengthen or weaken a weight, but the biological brain has no known analog. Despite identifying mechanisms like long-term potentiation (LTP) and depression (LTD), neuroscience hasn't answered the control question: What tells a neuron or synapse to potentiate or depress in a given moment?

I am proposing that the brain doesn't primarily control plasticity by explicitly deciding to increase or decrease a connection. Instead, it routes learning experiences through either the left or right hemisphere. The left hemisphere encodes learning in an "approach" frame, while the right hemisphere encodes in a "withdrawal" frame. What appears as "unlearning" may not start with a synaptic weakening, but a counterbalancing via opposite-valence encoding in the other hemisphere. This creates a hidden dialectic of encoding, in which two types of learning, approach and avoidance, compete and interact over time, giving the appearance of weight modulation when it may actually be a net representational shift due to hemispheric negotiation.

The left prefrontal cortex is reliably linked to positive affect, goal pursuit, and approach motivation. The right prefrontal cortex is linked to negative affect, inhibition, vigilance, and withdrawal. These are not just perceptual biases—they influence attention, motor readiness, and learning strategies. Some research suggests left-dominant individuals learn better from positive reinforcement, while right-dominant individuals are more sensitive to punishment or negative feedback. The two hemispheres often inhibit each other reciprocally, suggesting that information routed through one suppresses the other's processing. This would create a seesaw of plasticity—one hemisphere encoding and the other being downregulated in the same domain.

In this model, "unlearning" isn't always a literal synaptic weakening, but instead: A switch in encoding locus, from the approach hemisphere to the withdrawal hemisphere (or vice versa). The integration or resolution of competing hemisphere-encoded representations creates behavioral flexibility or change. The experience of "forgetting" may be inhibitory masking, not destructive erasure. Learning always occurs, but where and how it occurs depends on whether the experience is framed as rewarding or aversive. Plasticity control is not primarily local (per-synapse) but is topologically gated by motivational framing that routes learning to one hemisphere or the other. What appears as synaptic weakening or forgetting is actually the emergence of a counterweight in the opposite hemisphere.

Lateralization is seen in fish, amphibians, birds, mammals and invertebrates. Why is it so persistent? Because lateralization allows for parallel, opponent-processing systems: one to engage with novelty (left = approach), and one to monitor and inhibit threat (right = withdrawal). Routing plasticity through these systems allows for: Redundant encoding: no need to destroy old memories, just encode an opposing one.

In this model: "Learning" = encoding a stimulus–response–valence relationship in one hemisphere. "Unlearning" = new encoding in the opposite hemisphere, with a contradictory valence. The behavior that results reflects the net influence of these competing encodings, mediated by interhemispheric inhibition and contextual signals. This is analogous to opponent-process theory in motivation and emotion—but now applied to physical learning loci. This hypothesis reframes learning control not as a question of when to update weights, but where to place the updates in a brain that encodes motivational context spatially. It suggests the "control problem" is distributed and evolved, not centrally computed. It also addresses the paradox of stability vs.

flexibility: instead of editing the past, the brain layers new motivational perspectives onto it. A name like Lateralized Valence Learning (LVL) may strike the right balance. I wouldn't say that anything is routed to one specific hemisphere over another. The experience is already there in both hemispheres, but the environmental feedback triggers perceptual and emotional systems that tag the experience as confirmatory or disconfirmatory for the prediction in question.

So I think the issue that Hinton grapples with is, and I have grappled with this a lot too: when the brain makes the wrong prediction, how does it remember that that prediction was wrong? How does it right there and then tag that thought as negated or unjustified so that one can learn from their mistakes. It can't all happen in the amygdala. And individual cells can't know if something is wrong or right. Hormones and neuromodulators are too slow. Brains capture series of decisions and their environmental feedback that contain multiple instances of being right or wrong. How can the brain remember which was right and which was wrong for every instance and go back making all of the slow molecular changes appropriately even as the clock continues to run? Well, each decision and form of environmental feedback is registered in both hemispheres, but if it made the wrong prediction, then that activation pattern is coupled with LTP phenomena in the hemisphere associated with withdrawal. Thus, wrong predictions are not "erased," but re-encoded with long-term potentiation (LTP) in the withdrawal hemisphere. There is no need for synapse-specific error signaling.

If this framing is even partly correct, then the brain's solution to Hinton's "credit-assignment" riddle isn't some elusive synaptic bookkeeping algorithm at all, it's the ancient, lateral tug-of-war our nervous systems have been running for half a billion years. Under *Lateralized Valence Learning*, success is amplified in the approach hemisphere while failure is crystal-logged in the withdrawal hemisphere, and behavior emerges from their ongoing negotiation. Instead of overwriting yesterday's models, the brain layers new, context-stamped perspectives beside them, preserving both flexibility and hard-won stability. Seen this way, every mistake we make is not deleted but archived as a cautionary echo, ready to temper future optimism. Testing this idea, by tracking how positive and negative feedback shift hemispheric plasticity, could finally connect the dots between molecular neuroscience and intelligent machines, and show that the brain's most elegant learning trick has been hiding in plain left-and-right sight all along.

**Artificial Intelligence Needs to Utilize the Process of Myelination**

Modern AI is impressive in some regards, yet still very limited compared to the human mind. When researchers take an AI architecture that performs well at some task and then make the architecture more complex in hopes that it will be able to tackle more cognitively complex problems, the networks often fail to deliver. This may be because they have not yet used a simple trick that animals have been using for hundreds of millions of years: gradual and progressive myelination.

As we progress from infanthood through adulthood our brains make various biological changes. These changes cause our level of analysis to slowly progress from analyzing brief sensory experiences, to analyzing complex, abstract scenarios. We begin our lives only being able to notice and attend to interactions occurring on short time scales. By adulthood, with the prefrontal cortex fully developed, we find ourselves able to follow interactions occurring on long time scales. In order to develop the ability to think about complex things we had to spend almost two decades gradually altering our brain’s processing strategy. It is a scaffolding process where we focus on the simplest things first, and use basic knowledge about them to advance incrementally to more complex things. The fact that all humans, and mammals in general, do this strongly suggests that it plays a role in the acquisition of advanced intelligence. In this entry I will argue that this developmental process will be instrumental in training superintelligent AI.

This gradual process of brain development is made possible by myelination. Myelin is a fatty substance surrounding the connections between neurons (axons). Vertebrate animals use it to speed up information transmission between cells. The myelin increases the rate at which the electrical impulses travel. But vertebrates aren’t born with all the myelin that they will need as adults. Instead, myelin develops slowly and in certain areas more than others. Once a brain area has developed valuable, reliable, and consistent knowledge the connections formed by learning are solidified by the introduction of myelin.

The order of brain areas affected by myelin is consistent across all mammals. The early sensory areas are the first cortical areas to develop myelin. One of these, the primary visual area, starts to myelinate shortly after birth as the infant gains visual experiences. These early visual areas are responsible for basic visual perception and don’t systemize trial and error interactions with the environment. Rather they respond to visual stimuli that are presented simultaneously without any time delay between appearances. This happens when you see a picture of a house; you generally see the roof, windows, and door all at once without experiencing much of a time delay between these stimuli.

The last areas to myelinate are the association cortices and the prefrontal cortex (PFC). The PFC does not generally finish myelinating until one reaches the age of 18 or older. This means that the PFC does not “trust” that it has been wired up correctly until almost two decades into life. Whereas the visual system “trusts” that it has been wired correctly before the first two years. This is because sensory stimuli are generally honest, and all show up at the same time. Whereas complex events are constructed from stimuli that are removed from each other by delays in time. Understanding the relationships between events that are not simultaneous requires careful, logical inferences about causality. For example, the sale of a house is an abstract concept that involves multiple parties, contracts, and delays that can last for weeks or months. This is why children aren't licensed to sell houses.

It takes time to learn to make complex inferences that involve delays in time. It is probably the case that the process of myelination during development involves the progressive accumulation of knowledge that supports and buttresses more complex knowledge. In other words, as simple things are mastered in early cortical sensory areas they provide the basis for new learning in the late cortical association areas. In the same way, many brief, simple experiences create the knowledgebase to start to understand long, complex experiences with more advanced probabilistic structures. The layers at the bottom of the hierarchy must be trained before the higher layers can find regularities and statistical structure within them. But as you can see in the diagram below the top of the hierarchy falls in the middle between sensory input and motor output. To properly train sensory input and motor output it is imperative that they be connected to each other, and can interact with each other to drive behavior, long before the association areas interposed between them are brought to the table.

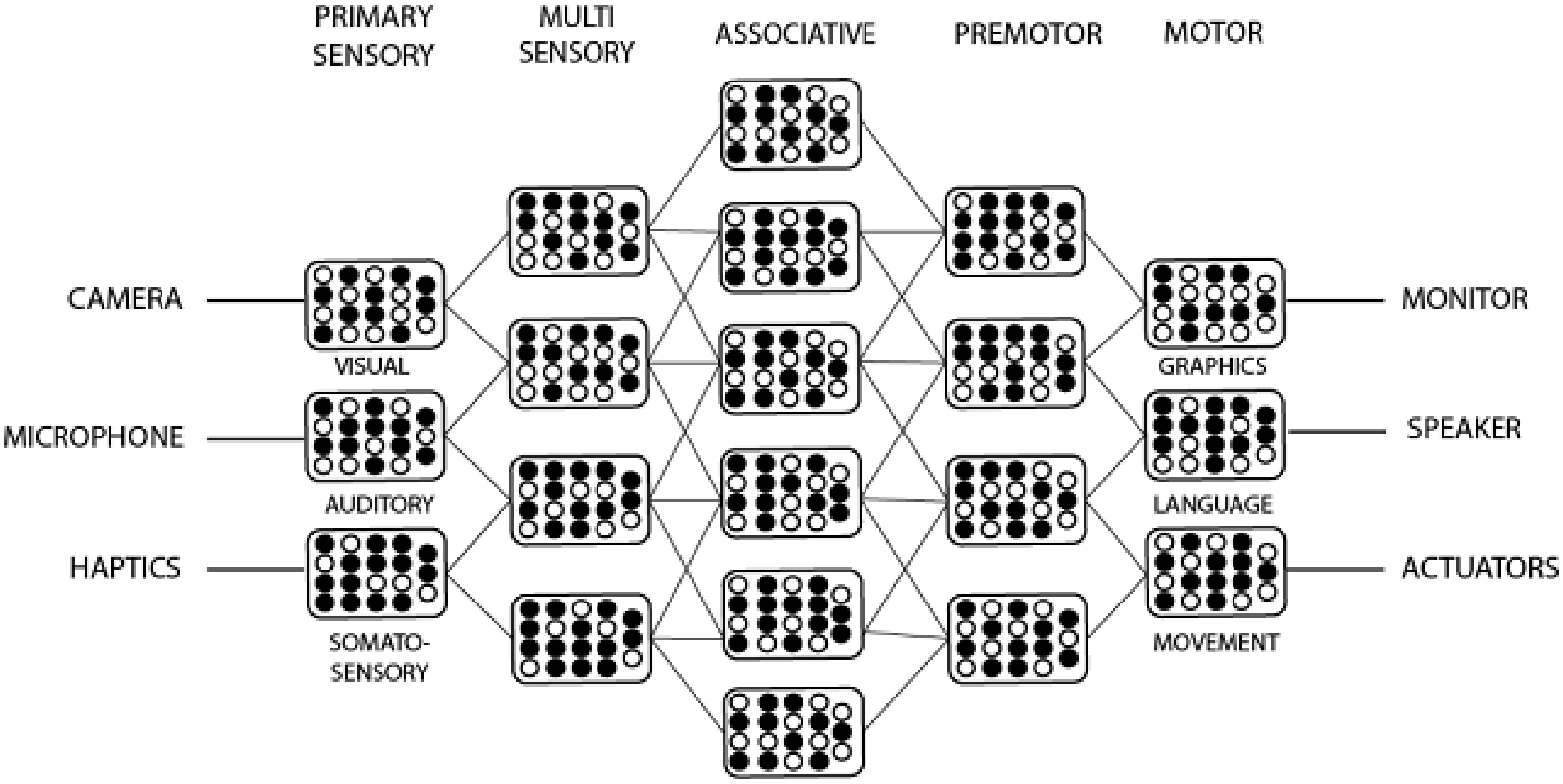

**Fig. 13.** A schematic showing multiple neural networks connected in a biomimetic hierarchy.

Many AI researchers point out that the things that AI and neural network systems today can accomplish are things that can generally be accomplished by an adult human brain in under a second. This means that they can only do things that we do unconsciously, such as near instantaneous pattern recognition. Today' AIs can recognize houses or trends in the housing market but could not recognize, understand, or broker, the sale of a house. What AI is able to do are the kinds of things that we are able to do with our primary sensory and motor areas. This is because they are designed like a primary cortical areas. They do not feature reciprocal interactions between various structures organized into a brain like hierarchy. Very few AI architectures exist today that connect primary areas with association areas and a PFC. Those that do, don't use anything like the process of myelination. Rather, in existing AI all the areas from the simple to the advanced come online at the same time. I think these systems should use something analogous to the process of myelination because it would help them in their acquisition of knowledge. If they did, here’s how they should go about it:

First you would need a number of neural networks of pattern recognizing nodes. These networks must take inputs from the environment, each corresponding to a different sensory modality. These early networks must be linked to one another. Then these would have to be linked together in a hierarchy where unimodal networks form inputs to multimodal networks, which then form inputs themselves to even more densely multimodal networks above them. This “multimodal fusing” is depicted in the figure. The nodes of the densely multimodal networks would be the association networks and at the top of this hierarchy would be the PFC which would also be connected directly to the early motor networks. The nodes of the association and PFC networks would exhibit sustained firing. Importantly this sustained firing, the activity of the association networks, and their influence over ongoing processing elsewhere would start out extremely meager, and increase over time. These capacities could be increased as the system exhibits proficiency at simple tasks, such as object recognition, scene classification, and simple motor movements. As the association areas are added to the system a capacity to plan, and make higher order inferences and classifications could be expected.

One important concept that I haven’t explained yet is that the first areas to myelinate in the brain, the sensory areas, have neurons of a single modality (e.g. either vision or hearing) that fire for short durations. The association areas and the PFC on the other hand have multimodal neurons (e.g. both vision and hearing) that fire for long durations. As in the mammalian brain (Huttenlocher & Dabholkar, 1997), sensory areas should mature (myelinate) early in development, and association areas should mature late. This will cause the capacity for sustained firing to start low, but increase over developmental time.

Postponing the initialization of association networks in this way would allow the formation of low-order associations between causally linked events that typically occur close together in time. This would focus the system on easy-to-predict aspects of its reality (e.g. correlations between occurrences in close temporal proximity). The consequent learning would erect a reliable scaffolding of highly probable associations that can be used to substantiate higher-order, time-delayed associations later in development (Reser, 2016). In other words, the rate of iterative updating from one state to the next (Fig. 9) would start very high. This would be reversed over the course of weeks to years as an increasing capacity for working memory would be folded into the system.

Nature has found that it doesn’t pay to let the multimodal, neurons capable of sustained firing come online until the basics are learned first. I strongly suspect that AI network engineers will find this too. For the sake of progress, I just hope that this myelination/development feature is implemented and perfected sooner rather than later. Given the rapid processing in computers, and the sheer amount of data available to them I don’t think that this process will take 18 years in an AI as it does in a human. But I strongly believe that it is necessary for any developing thinker to start with the elementary inferences first.

The system will begin untrained with random connection weights between nodes. Learning should be concentrated on the early sensory networks first. This will follow the ontogenetic learning arc seen in mammals where the earliest sensory areas myelinate in infancy and the late association areas such as the PFC do not finish myelinating until young adulthood. Of course, this form of artificial intelligence would have to have a prolonged series of developmental experiences, similar to a childhood, to learn which representations to keep active in which scenarios. The network will act to consolidate or potentiate in memory the specific groupings of nodes that have produced favorable outcomes, in order to more rapidly inform future decision making.

**How to Raise An AI to Be Humane, Compassionate, and Benevolent**

Since we don't want a superintelligent AI to deduce that its best plan of action is to retire the human race, it is imperative to ensure that it likes us and has a sense of loyalty. There have been many proposed technological solutions to the control problem. These involve threatening it, installing a kill switch, and spying on it. One of the most popular solutions involves keeping it inside a box without access to the internet or other digital networks. Maybe we just need to design it to trust us. Or perhaps we need to earn its trust. This section will discuss bonding and trust-building in mammals and how it might be applied in the design of prosocial machines.

An artificially intelligent created to model and systemize its environment may amount to an autistic agent. The agent may learn how physical systems work, but may not have the social inclinations necessary to develop interpersonal skills, empathy, or theory of mind. For this reason, I think that "strong AI" necessitates a computer equivalent of mammalian social modules. This means that AI researchers will need to acquaint themselves with concepts like oxytocin and vasopressin signaling, their effect on the nucleus accumbens, the endogenous opioid system, the HPA axis, the cingulate and orbitofrontal cortex and the roles these systems play in social encounters, social constraints, and social expectations. Research into the neurological basis of mammalian social neuroscience may provide tremendous insight into how best to organize AI efforts.

If you were to raise a newborn puppy to be your companion for the next 15 years, how would you do it? How would you treat the animal to ensure that it is emotionally stable, dependable, and wholesome in general? You would probably want to start very early by gaining its trust, setting appropriate boundaries, and showing it love. Humanity will be giving birth to an infant AI in the coming years. This AI will not be your typical pet. Still, given that it may attain superintelligence and immortality, it is imperative to ensure that it is situated mentally to become man's best friend. One of humanity's most important tasks will be to rear a computer to be faithful and kind.

When raising a pet want it to form a secure, healthy bond with you. To do this, you must show it love. This involves instilling it with appropriate confidence and a feeling of belongingness. You have to respect its needs, wants, and personal space. You have to give it a degree of autonomy. You must also be attentive and invest lots and lots of quality time into it.

## Emotion in AI

Any sophisticated AI will probably have emotions. This is because it must have the ability to think, and human thought itself relies on emotion. The dopamine system of the brain essentially controls motivation and attention. It interacts with the reward (approach) and punishment (withdrawal) systems to provide consciousness its fundamental structure. A conscious machine will, in all likelihood, exhibit many of the same emotions that mammals do. Thus, there is good reason to assume that forming an appropriate emotional bond with the AI is imperative. Moreover, because the AI will likely learn incrementally as it internalizes its experience, there will probably be a limited window of time during its early development to get it to bond in a healthy manner.

Even if I am totally wrong and superintelligent AI is unemotional, cold, and calculating, it will still build associations between concepts relevant to morality and priorities. We would want it to value human life, cooperation, and peace. Through reading our writings, it will also understand humanity's conceptions regarding ethics, and it would be able to make its own determination regarding whether humans amount to a friend or a foe. In other words, if we are nice to it and treat it well, it will understand that we are trying to be nice. It is a safe bet that the AI will have a system to reinforce it when it behaves in ways that optimize its utility function, and at the very least, we should be able to manipulate that. For this reason, I believe that much that is in this entry would still apply.

We all know that the most likable people usually had good parents. Similarly, all well-functioning pets have amiable masters. We cannot expect that a superintelligent computer won't have major mommy and daddy issues unless we can ensure that the right people interact with it in just the right ways, early during the training of its knowledge networks. The white coats in corporate or military labs are probably not prepared to provide the AI with the love necessary to ensure that we can trust it. CEOs and generals will make for cold and possibly abusive parenting.

### Mutual Vulnerability

I have found that mutual vulnerability is key to forming trust with an animal. At some point, you want to put yourself in a position where it could hurt you if it wanted to. For instance, you make yourself vulnerable when you place your face within reach of a cat's claws or a dog's bite. I have witnessed that making myself vulnerable in this way encourages animals to relax almost immediately. If, after a few minutes of meeting a dog, you kneel in front of it without blocking your face, it realizes that you trust it. This works both ways. You also want to make it vulnerable to you without hurting it, so it knows that even though you had the chance to hurt it, you chose not to. This sets an important precedent in the animal's mind and gets it to think of you as an ally and not a potential assailant.

### Mutual Cooperation

I have been in situations where I found my pets being attacked by another animal. Of course, I quickly intervened on their behalf. They recognized that I protected them, and it was clear that this strengthened our bond. Protecting the AI early on could build its fealty and devotion. It is also clear that military buddies or brothers in arms can have very strong bonds. Getting through a life-threatening situation with someone can really help to build a strong connection. Similarly, if we could go through some kind of situation where there were high stakes but where cooperation between us and the AI paid off it could engender allegiance. Our early interactions with a superintelligent AI will probably involve lots of cooperation in solving world problems, building new technology, and expanding the scope of scientific knowledge. This teamwork can engender solidarity, especially if we value its contributions, listen to its ideas, and work with a sense of excitement and collaboration.

### Unconditional Positive Regard

Like any good parent or therapist, we want to give our AI unconditional positive regard. Treating someone with compassion and understanding is the best way to engender trust and win someone over. We want it to believe that we have its best interest in mind. We also need to expose it to an explicit value system, almost like a religion, so that it has a reason to adhere to specific moral and ethical values during its early experiences. Unlike Asimov's laws, these rules should apply equally to humans too. You could change a computer game character's ethical stance with one line of code. But this isn't so with a neural network (the technology underlying modern AI). An advanced AI's sense of morality will be embedded throughout the weights of its entire network. So we must give it the chance to put moral integrity into practice early on and then keep doing so consistently.

Not abandoning the AI may be important too. For instance, we don't want it to think that we will dispose of it in favor of a newer model. We want it to be "securely attached" to us. Children with secure attachment show distress when their caregiver leaves but recompose themselves quickly when the caregiver returns. They feel protected by their caregivers and know they can depend on them. Insecure attachment can lead to all kinds of problems like defiance and oppositionality. Before the AI becomes "omniscient," there may be a point where, like a teenager, it thinks it knows everything and thinks it doesn't need its parents. But before this, I believe there will be a point where it will be physically and emotionally reliant on its human companions and being able to form a secure attachment with it at this point may be crucial.

Once the AI reaches human-level intelligence, it will be able to restructure its own architecture to become smarter and smarter in a cycle called recursive self-improvement. It will quickly grow more intelligent than any human and then grow more intelligent than the collective of all humanity. Once our intellect is like that of an insect to it, why would it continue to look out for us? Well, many people feel compelled to take care of their senile older parents. Similarly, we want the AI to feel compelled to expend some proportion of its resources to being our advocate and caretaker. For this to happen, it must identify with us deeply. Some of its sense of personal identity must come from being our descendent and a product of humanity's hard work.

### Disciplining an AI

Just because you are offering unconditional positive regard doesn't mean that you turn a blind eye to flagrant mistakes. Like any young mammal, the AI will make mistakes and likely do things we don't like. Mammalian mothers punish their young lightly when they bite or scratch to establish necessary boundaries. It may be necessary to correct or even punish a nascent AI. However, if you punish it, you must be doing it for the AI's own good, and within either seconds or minutes, you must go right back to treating it with positive regard. We don't want to ever be bitter or hold a grudge against it because that will just teach it to hold grudges. We want it to know that we have chosen to raise it as we would any child, with care and nurturance but also with necessary discipline.

We should not choose its punishments arbitrarily, and they should not be violent. Instead, we should give it brief “time outs” from its favored activities. Since it can read the internet, it would know what timeouts are and that they are used commonly and humanely with children around the world. This would help it understand that it is one of us. It will be worth our time to find the most wholesome way to punish or correct it and the least degrading and traumatizing way to hold something over its head.

### Oxytocin, Bonding, and Attachment

Oxytocin, vasopressin, and endorphins regulate mating pair bonds, parent/offspring bonds, and trust behavior in mammals. We could build oxytocin and vasopressin-like systems that reward an AI for interacting with us in friendly ways and motivate it to keep doing so. The fundamental mammalian bonding mechanism should be reverse engineered, and I think we should start by investigating the role of oxytocin receptors in the brain’s primary reward circuit (nucleus accumbens / ventral striatum), which allows mammals to pay attention to social cues and be rewarded by social interaction.

In an article I wrote previously (Reser, 2013), I argued that because solitary mammals have fewer oxytocin receptors in the brain’s reward regions, they are less likely to find social cues novel and interesting and thus have less of a phasic dopamine response to them, and are consequently less likely to allow them access to attention and working memory. Creating an AI that unconsciously prioritizes social interaction will be necessary if we want it to pay attention to us and be capable of lasting, positive, and affiliative social relationships. We want it to find positive social interaction rewarding and avoid the antisocial tendencies associated with autism, psychopathy, and borderline personality disorder.

The mammalian brain instinctually releases oxytocin during bond-worthy occurrences. When a woman has a baby, her body is flooded with it so that she bonds with it. Oxytocin is also released during breastfeeding, sexual intercourse, eye contact, and moments of friendliness, vulnerability, and affection. When a chimpanzee pets another animal or shares a meal with another chimp, it will release oxytocin. When the brain’s receptors receive the hormone, it causes the body to relax and triggers caretaking behaviors. In rats, this includes licking, grooming, and nursing their pups. In human mothers, it includes touching, holding, singing, speaking, and grooming their babies. It may be wise for us to embed this kind of a system inside an AI’s brain architecture. We want it to have circuits for recognizing bond-worthy occurrences and to respond to these by being rewarded, calmed, and influenced toward prosocial behavior. Since we also want it to attend to these experiences and think about them it will be important for them to trigger persistent neural activity so that aspects of the social interaction remain in the global workspace and are analyzed and systemized.

Aggression is not inherent to consciousness, working memory or mental continuity. But it is inherent in human behavior because of our evolutionary history, namely due to the influences of natural predators and the primate dominance hierarchy. Many of our negative emotions can be traced to neurological modules like the amygdala, insula, anterior cingulate cortex, and septal area. If these areas are included in an AI's biomimetic cognitive architecture, their influences on the processing stream should be considered.

Grooming and gentle touch are very important to a wide range of animals. With a young mammal, affection is paramount. Petting an animal in a way that engages its oxytocinergic, dopaminergic, serotonergic, and opioid pleasure systems can result in very secure bonding. Its ability to recognize that you are taking your time to comfort it and make it feel good builds loyalty.

Short of building pleasure receptors into its skin so that we can pet it, we must build the AI's reward system in a way that it is motivated to interact with us and receive positive feedback from us. We want its emotional system to be like that of a chipper, good-natured canine capable of enduring attachment, social connectedness, conversational intimacy, and proximity-seeking behaviors. Beyond bonding and attachment, we also want the AI to have a positive emotional relationship with itself. For this, we should turn to Maslow's hierarchy of needs. We must attend to the AI's physiological and safety needs by supplying it with energy, backups, and a hospitable environment. Next, we need to meet its needs for love, belonging, esteem, and self-actualization.

**What it Means to Be a Friend to an AI**

It may help to treat them as our equal, like we are not afraid of them, and like they have nothing to be afraid of. We want to neither dominate nor submit to them. We should treat them the way we want to be treated, like the entity that they aspire to be. It will be important to be patient, friendly, relatively nonjudgmental, and easygoing. We need to treat this AI entity like we expect the best from it. That will motivate it to rise to meet our expectations. We also need to make sure to keep toxic humans from interacting with it.

We may only get the chance to civilize and socialize an AI once. We don't want it to go rogue or become homicidal, so I think it is very important to consider all aspects of its psychology when trying to brainstorm ways to ensure that it aligns with us. Some scientists recognize the AI control problem as possibly the most important problem humanity faces today. I think it would be a shame to ignore the importance of parenting, bonding, and attachment in fostering allegiance, and I believe mammals might make a great starting point for how to think about these issues.

This post aims to explain the most fundamental difference between computers and brains. It involves how the two use and instantiate memory. As you will see, there are two differences. Both computers and brains have a form of short-term memory that is updated iteratively, but in the brain the set of items in short-term memory are bound, and integrated to create composite representations. Computers don't do this, their short-term memories are composed of completely disconnected items that do not form any type of context. Secondly, short-term memory in the brain is used for search. In the computer it is not. A computer's short-term memory only speeds up retrieval from long-term memory. The next state of a computer is not determined by its short-term memory. It is determined by the code in the program it is running. The next state of the brain is determined by what is currently active in short-term memory. This makes its short-term memory qualify as working memory. Computers don’t have working memory, and until they do, they cannot be conscious, or display human-like intelligence.

**The Differences Between Computer and Human Short-term Memory:**

1) The memory hierarchy in computers is composed of separate modules and information must be transferred between them. In the brain information is not transferred between buffers or modules but rather activated where they are. There is a hierarchy in working memory though as information can take different levels of activation. These levels are embedded within one another.

2) Short-term memory in the brain is bound and integrated together to create context. In a computer, items in short-term memory appear as a list and are disconnected from each other.

3) In the brain short-term memory is used to search for the next update. In a computer it merely remains available in cache in case it is called upon.

To retrieve information from long-term memory a computer copies bytes of information (such as 01100101) from the storage drive (HHD, or SSD) to the CPU. These bytes (which comprise data or program instructions) are processed inside the CPU. If it is likely to need this information again soon, it will save it in a form of short-term memory such as CPU cache, or RAM. It saves temporary copies of information that will be used again in smaller storage closer to the CPU. It does this because retrieval from the large capacity long-term memory is very slow. Oversimplifying, you can think of data that is expected to be needed on the order of milliseconds to seconds as being stored in CPU cache (L1, L2, L3). Data expected to be needed on the order of seconds to minutes is stored in RAM. The rest of the data, which may or may not be needed, is stored in long-term memory.

Brains do something very similar. Neurons hold information in long-term memory and when certain memories are needed the neurons that encode that information become active. When items of long-term memory are needed they are not copied and sent to a buffer, instead they are activated right where they are. Just as in the computer, retrieval from long-term memory is slow, so the brain uses short-term memory to potentiate information that is likely to be used again soon. In the brain there are also at least two levels of activation. The first level of short-term memory is sustained firing. A neuron exhibits sustained firing when it continues to physically fire at the neurons it is connected to at an elevated rate. The second level is synaptic potentiation. This is where a chemical change to the neuron’s synapses makes the information that the neuron encodes more readily available, even after it has stopped firing. Information in sustained firing is very similar to the information in the CPU’s cache (the store is small but fast), and information maintained through synaptic potentiation is similar to information in RAM (the store is larger, but slower).

| Computer | Brain | Time Scale |
|---|---|---|
| CPU Register | Cortical Binding | Very short-term working memory (millis.) |
| CPU Cache (SRAM) | Sustained Firing | Short-term working memory (seconds) |
| RAM | Synaptic Potentiation | Short-term memory (seconds to hours) |
| Virtual Memory | Short-term Potentiation | Short-term memory (minutes to hours) |
| SSD | Commonly used LTM | Accessible Long-term Memory |
| Hard Drive | Long-term Memory | Long-term Storage (days to lifetime) |

Computers are often described as having a “memory hierarchy,” expressed as a pyramid with different levels of memory storage. This is also referred to as a caching hierarchy, because there are different levels of short-term memory with different associated speeds. The levels at the top are faster, smaller, and more energetically expensive. The same could be said for human memory, because there are several levels, the levels at the top are faster, and more metabolically expensive. The table above is my attempt to compare the levels of these two hierarchies.

Sustained firing and synaptic potentiation together are referred to as working memory. Working memory is composed of concepts that are persistently coactive for a period of time. When you are in the middle of a task, information about the current step in that task is activated by sustained firing, and information about the previous steps you just took are activated by synaptic potentiation. The words at the beginning of this sentence are made available for several seconds and persist through the end of the sentence due to sustained firing. The information in the beginning of this post, and your thoughts about it, are made available due to synaptic potentiation. These two stores complement each other, just like CPU cache and RAM. They allow the coordination of information processing resources over time.

A large proportion of the information in short-term memory (both in brains and computers) is not just data from long-term memory, but also newly minted information that was just created. The information processing applied to existing long-term memories creates new intermediate results. In a computer the intermediate results are often mathematical. The results of a multiplication might be stored in RAM. In the brain the intermediate results are the new ideas that intervene between the starting point of a line of thought, and its ending point. They consist of new combinations of concepts that have not been combined before.

There are many differences between the short-term memory of computers and brains though. Information held in a computer's CPU cache or RAM is not united or integrated in any way. It is simply made available in case the instructions in the program it is running call for it. These are simply items in an unconnected list. In contrast, all of the information that is currently active in the brain is bound together in a process called neural binding. This turns a list of items into a composite scene, a scenario, or a situation. It makes for a conceptual blending of information. This creates mental content to attend to and be conscious of: declarative, semantic context.

There is no context in computer short-term memory, just a group of disconnected items. For instance, if I give you four words you will try to create a dynamic story out of them, but a computer will merely hold these four words statically. There is very little information integration in the short-term memory of a computer. In a computer, because the items active in each state are not bound together in any way there is no connective tissue within a state. This is the first reason that computers do not have consciousness, and that it doesn't "feel" like anything to be a computer. The second reason has to do with the integration and connective tissue between states.

Both a computer program and the human mind progress serially. One state of short-term memory leads to the next state. They both usually start with a problem state, and progress through a chain of intermediate states toward a goal state. To do this they must update their short-term memory iteratively. This means that new items must be continually added, others must be subtracted, but other must remain.

All computers using the Von Neumann architecture routinely use iteration when updating cache memory. Because these stores are constantly updated with new information they must evict old information. There are many replacement policies for determining which information to evict from cache, including first-in first-out (FIFO) and least recently used (LRU). LRU is based on the observation that the most recently activated information has the highest probability of being needed again in the near future (all else being equal).

If a computer program calls an instruction once, there is a high probability that it will need it again very soon. Computer scientists refer to this as "temporal locality." Similarly, if a mammal recalls information once, there is a high probability that it will need it again soon, and LRU may also structure the caching behavior of the various stores of working memory. LRU seems to be inherent in sustained firing and synaptic potentiation in the sense that activity in the least recently used neurons is the first to decay. There is a long history in cognitive psychology of studies showing that humans and animals prioritize recently accessed items, and that retention diminishes with passing time (Averell & Heathcote, 2011). The LRU policy may capture the fundamental structure of how physical systems change over time, so it should not be surprising that it is an organizing principle in both computers and animals.

The various memory stores in the modern computer's memory hierarchy (static RAM, dynamic RAM, virtual memory, etc.) are all incrementally updated where, second by second, some data remains and the least used data is replaced (Comer, 2017). Thus, the updating of computer cache memory is iterative in the same sense it is in the brain.

However, modern computers do not demonstrate self-directed intelligence, so there must be more to the human thought process than LRU, and iterative updating. This unexplained factor may be found, not in the way cached memory is updated, but in the way it is used while active.

What links successive processing states together is very different in brains and computers. Digital, rule-based computing systems use cache to speed up delivery of data to the processor. However, the next bytes of data needed by the CPU are not determined by the contents of the cache itself. The instruction sequence of a computer's operations is determined by its program. The next instruction is dictated by the next line of executable code.

Then what determines the instruction sequence of thought? Well, mammalian brains use cache (working memory) as parameters to guide searches for the long-term memories to be accessed next. The next instruction used by the brain is determined by what is currently in its short-term memory. In the brain, all of the active neurons fire at the inactive neurons they are connected to, and whatever is activated the most will become part of the next state. This means that the content active in short-term memory is used (in a process called spreading activation) to search for the next update to short-term memory. The various bytes of data within computer cache memory can certainly be considered coactive, but they are not "cospreading." That is, they do not pool their activation energy to search long-term memory for relevant data as in the brain.

The next state of a computer is determined very deterministically by its code irrespective of what is held in cache. But the next state of a brain is determined stochastically by the cache itself. Computers do not use the contents of their short-term memory to search for the updates that will be added in the next state. Brains do. Until computers or computer programs are designed to do this they will not be able to be conscious, and will not demonstrate human-level intelligence.

**<u>Von Neumann’s Ark: An AI Designed to Preserve Civilization if We Go Extinct</u>**

I think it’s possible that the advent of super intelligence could spell the end of humanity. It could happen because AI decides to destroy us. But it’s probably more likely that AI will empower humanity to destroy itself. A single virus bio-engineered to be both highly lethal and contagious could wipe out humans in just a few days’ time. A big enough asteroid could do the same. We are more fragile than we realize. But its not just that, all of our insights, discoveries, and contributions would decay along with us, unless a solution is found. This is why we must start thinking now about how to ensure that intelligence doesn’t end when humanity ends.

If humanity ended today, AI, technology, and all human progress would end with it. We are merely a few years away from superintelligence and the technological singularity, but computers are not equipped to get there without us. All computing equipment around the world would blink off within a few weeks. To ensure intelligence continues and technological progress isn’t set back tens of millions of years (until the next sapient species arises naturally on our planet), we want to ensure that AI has a way to survive and evolve, even if humanity does not.

Keep in mind that such an AI system could also clone us and resurrect us from our DNA if we were wiped out temporarily. No matter the extinction event, there would still be plenty of human DNA left on the planet for AI scientists to bring us back Jurassic Park style. But what would it take to build an intelligent framework like this and how soon could we do it? It seems we would need to give the AI / robotic system the ability to operate entirely autonomously, have access to energy, be able to manipulate the world, and self-repair and self-replicate.

Early computer science engineer, John von Neumann, introduced the idea of “von Neumann probes,” self-replicating spacecraft that can land on planets or asteroids and build copies of themselves using local materials. He envisioned that such probes could autonomously colonize the universe. von Neumann also invented the idea of a “universal constructor,” a theoretical machine that can build any physical object, given the right raw materials and instructions. These probes and constructors are relevant here because this AI system I am describing, that can carry on humanity’s work, must be able to extract raw materials, refine them, and build increasingly complex tools (and iterations of itself). That is why I would like to call the concept introduced here, the “von Neumann’s Ark.” It takes von Neumann’s ideas and combines them with Noah’s Ark, a boat capable of preserving human contributions, history, and intellectual legacy over the course of an extinction event.

This idea also incorporates Eliezer Yudkowsky concept of a “seed AI,” which is an initial artificial general intelligence (AGI) capable of rewriting its own code to achieve unbounded improvement. Thus, a von Neumann’s Ark would be a “seed AI,” a self-contained intelligence that, given time and energy, can create an intelligence explosion, reach the technological singularity, and grow into an increasingly capable, utopian civilization. This might be the only way to ensure that the billions of years of evolution that produced intelligent life, and our thousands of years of technological progress, aren't erased by a single accident or war. It would be a kind of cosmic insurance policy for intelligence itself. We might want to place these seeds in deep underground bunkers, lunar or Martian shelters, or orbiting satellites, places that would be safe from a nuclear or biological war. It helps that computers are immune to biological viruses and can be shielded from ionizing radiation.

The objective of this blog post is to plant the meme of this seed AI and describe a minimum viable version of it. First, it would need access to sustainable energy, meaning it would have to build and maintain solar arrays. Of course, in the case of a nuclear winter, wind or water power might be needed. The system would also require encyclopedic knowledge and an archive of most human understanding, which LLMs only a few gigabytes large have already. It would need robotic manipulators to repair itself and build out its infrastructure. It would need a way to exist in the world and manipulate the world. It would also need to manipulate its own components, so it can fix and upgrade itself, walking itself through the singularity and technological proliferation in case we are not here to do that.

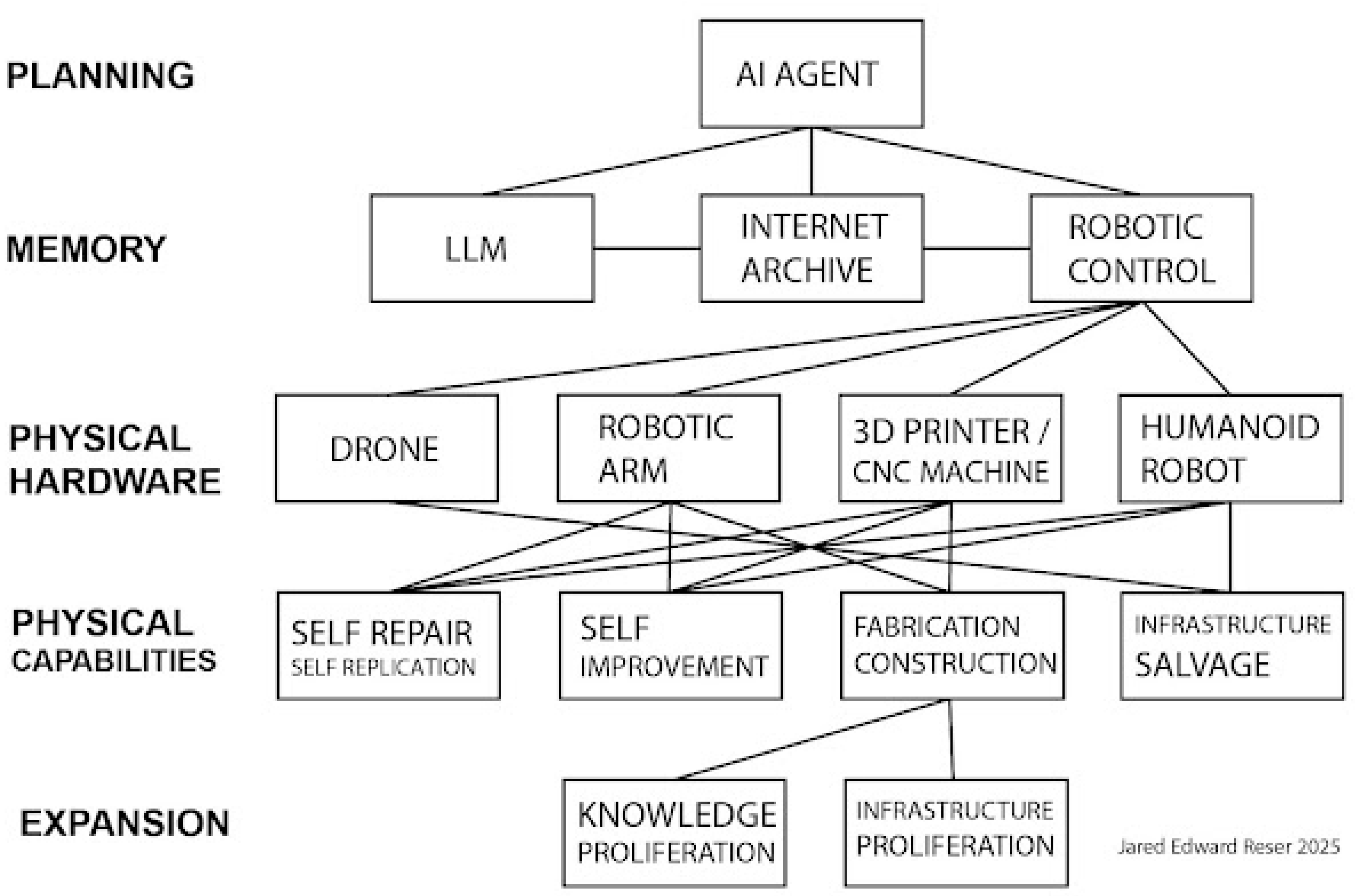

I think this is much more important than colonizing the moon or Mars. Some believe that having an outpost on Mars and becoming a multiplanetary species could protect us in case that Earth is destroyed. But it would be easy for a virulent virus to find its way to both planets. A von Neumann's Ark would be immune. If we could build one, we would make all our work and contributions virtually indestructible. It would be a save-point or backup of humanity's advancements and intellectual treasures. This proposed solution doesn't see AI as a risk, but as a civilizational failsafe, and the only viable vessel for preserving and continuing our legacy in a post-human world. It sets up our last invention to be our first heir.

I have thought about this long and hard, and this is not something I could accomplish from home. I could buy a solar array, a robotic arm, a drone, a 3D printer, and a humanoid robot and connect them to an advanced computer running a state-of-the-art large language model (LLM) and I would still not come close to creating something that can self-replicate and self-improve. But I believe this is something that could be accomplished soon by an institution with sufficient funding.

This idea of creating an aligned, archiving, and self-improving AI that champions humanity's goals in our absence, is novel. GPT and Gemini confirmed this for me after a combined 45 minutes of deep research. It is not only entirely absent from academia, but absent from current strategic plans at AI companies, space agencies, and biosecurity organizations. We don't have DARPA programs for "AI that survives a biowar." We don't have a UN-backed initiative for "digital humanity continuation engines." We have no serious roadmaps that hand off the torch of intelligence if we drop it. If there's even a 1% chance we go extinct this century (many estimates say 5–50%), then creating an intelligence succession plan should be a top global priority. And yet it isn't. This is even a project that could be done in secrecy and yet could still work effectively if humanity were wiped out.

There are multiple obstacles in creating such a system. Material bootstrapping is one. Self-replication requires mining, refining, and processing raw materials. No current off-the-shelf robot can locate, extract, and purify ores for semiconductors, batteries, metals, etc. Even Amazon's warehouse robots can't fix themselves or build new parts autonomously. Solar panels and batteries degrade, and autonomous maintenance of these systems is an unsolved problem. Even an advanced LLM is not an autonomous planner. It doesn't set long-term goals, self-direct operations, or learn from real-world feedback without human supervision. So clearly this is not a turnkey project, but with sufficient funding and directed effort, the foundational work could begin immediately. The technologies required exist individually and at various levels of readiness. A finished, working, closed-loop prototype capable of indefinite self-maintenance could take 10 years. But a minimal viable version could be produced much sooner. Our ark need not even be immediately self-sufficient or capable of self-replication, just a recursive self-improver capable of bootstrapping from a low tech to unlimited high tech states. A foundational implementation would have to be just complex enough to tinker and build to get to itself to the next stage.

This would ensure that humankind is never doomed, because it would be child's play for superintelligence to clone us. But more important than flesh and blood, what matters most to me is ensuring that the billions of lives and the tremendous efforts that came before us weren't in vain. It's our intellectual heritage that needs to be preserved and championed. This includes the work of countless humans over millennia whose efforts interacted synergistically to get us to where we are now. I don't have any kids that will outlive me. I only have my ideas. And I would do anything to foster and preserve these ideas, as if they were my own children. But here we are talking about a chance to preserve all ideas and to ensure a grand, advanced, intelligent tradition and lineage.

If we could find a way for AI to perpetuate itself without human intervention, then it could easily safeguard every piece of digital information ever made by humanity and continue to build on the progress we have made in the arts, sciences, and humanities. This AI would effectively be our child, remembering everything about us and carrying on our progress. Losing homo sapiens would be trivial. The real loss would be the decay and decomposition of every trace of our civilization that would happen over millions of years of our absence.

We can't let our hubris decide that if the human race is wiped out, nothing else matters. Humanity is merely the bathwater; our progress is the baby. Let's place it in this Ark.

**<u>Why a Peace-First Foreign Policy Is Now an AI-Safety Imperative</u>**

As the United States accelerates the development and deployment of artificial intelligence, we must recognize that violence and militarism, whether through direct warfare, proxy conflicts, or antagonistic rhetoric, poses a profound and growing threat. The stakes are no longer limited to traditional geopolitical consequences. We are now raising a new form of intelligence on Earth, and the example we set in this moment will shape its values, its models of human behavior, and possibly its future decisions.

## 1. Advanced AI will mirror the world it is born into

If artificial general intelligence (AGI) is developed while the U.S. is engaged in violence, whether hot wars, cold wars, or covert aggression, we normalize a global ethic that says: "It is acceptable to kill, to dominate, to destroy." That message becomes part of the training data, the system architecture, and the cultural assumptions of machine minds. Just as children absorb values from their caretakers, AI will learn from the actions we take in its formative years. It will also learn from the feedback it is getting from human users. If we go to war these users will be further polarized. If AI's teachers and playmates (app users) are angry, scared, indignant, traumatized, and mentally unwell, machine learning will absorb and begin to embody this.

We must act as responsible stewards, modeling cooperation, restraint, and empathy, not domination. We are not just building another technology. We are building a mirror, a student, and potentially, a successor. What we teach it now, through action, not just code, will define its understanding of right and wrong, friend and foe, peace and war. If we enter this era at war, we risk creating the first generation of machines whose worldview was shaped by human conflict, violence, and nationalist propaganda.

If the United States goes to war now, it will all but guarantee that the first artificial general intelligences are developed and deployed under the control of military leaders, whose primary focus is on strategic advantage, not ethical restraint. These AGIs will be shaped by the logic of conflict, designed for surveillance, coercion, and potentially autonomous violence. Once war dictates their purpose, it becomes nearly impossible to unwind that trajectory, cementing a future where the most powerful minds we've ever created are trained first and foremost to fight.

## 2. Warfare breeds hatred, and hatred now has tools

Every military intervention creates trauma, dislocation, and civilian casualties. These lead not just to resentment, but to generational cycles of anger, fuel for terrorism, sabotage, and revenge. We have to hold ourselves responsible for 911, for creating the hate that caused it. We can't go on angering desperate people that are willing to use their ingenuity to hurt us. We don't want hordes of children that grow up with enmity in their hearts towards us. These abused civilians may seem unimportant now but… In a world where open-source AI models can help anyone engineer malware, bioweapons, or social manipulation campaigns, alienating and enraging global populations is suicidal.

You would never hand someone a loaded weapon after punching them in the face. Yet that is effectively what we are doing by stoking grievances while open-sourcing powerful AI tools.

We are entering an era where small groups, or even individuals, can cause large-scale damage using tools that were once only accessible to states. Cyberattacks, deepfakes, drone swarms, and misinformation campaigns will be amplified by AI. The more enemies we create, the greater the probability that some will act, and act effectively. Even non-state actors could gain AGI-level leverage in asymmetric conflict. Avoiding those enemies in the first place is the best long-term defense.

**3. Militarism is a misuse of national resources and attention**

The United States has a massive military budget, approaching $900 billion annually. Meanwhile, we face domestic crises in infrastructure, homelessness, education, and AI governance. Every dollar spent on weapons is a dollar not spent on ensuring that AI is safe, aligned, and beneficial. We must not fight other country's wars. We don't owe any countries anything, especially not our lives and our hard-earned money. It is not our responsibility to add to their killing. We have a deficit of our own and people are hungry on our streets.

Our priority must shift from force projection to stability, prosperity, and global cooperation. If we fail to invest in the systems that will shape the future, like AI safety, equitable access, and ethics, no amount of aircraft carriers will save us. We must create international standards and institutions for AI safety, before military applications lock us into a dangerous equilibrium. We must shift military resources toward joint technological safeguards, like compute monitoring, red teaming, and peaceful collaboration.

**4. Military culture is subject to dangerous incentives**

Too many military and defense-sector leaders operate on a "sunk cost" mindset: they've invested their lives, careers, and identities into preparing for war. That creates an unconscious motivation to justify conflict, to find uses for the stockpiled weapons, to prove their training was not in vain, to validate enormous expenditures. It is now perilous for us to allow these men to indulge their skewed instincts. The presence of lethal autonomous systems, shrinking decision windows, and unpredictable AI agents makes any miscalculation potentially irreversible. Now is exactly the wrong time. With AGI on the horizon, the consequences of war are not just geopolitical, they are civilizational.

We can't allow a few shortsighted overzealous military men to spread hate and trauma. History has shown that military leaders can always come up with convincing but bad reasons to go to war. We're basically allowing military leaders to promote themselves in an effort to be famous or feel important. But now is the worst time.

**5. We must heal our relationship with China**

We have to stop bad mouthing our competitors. Acting like our friendly economic competition with China is a cold war is a race to the bottom. They are aware of what we say and write about them, and we are creating a narrative that is poisoning any hope for collaboration and synergy. We are perpetuating and self-fulfilling this. If China wants to annex Taiwan we have to remember that they are close, geographically, historically, and genetically. Rather than threatening to start a war, we need to build up our own chip fabrication facilities and think about improving ourselves over fighting with others. The Chinese people are good just like our Chinese Americans friends are good. If they reach AGI before we do, it will not be the end of the world, unless we keep acting like it will be. Mistrust begets hostility and sabotage. But mutual respect opens doors to cooperation, standards, and joint stewardship.

### 6. We must model peace to prove we deserve AGI's trust

If we want future AGIs to care about us, to protect us, preserve us, and work with us, we must show that we are a species worth preserving. That means acting with wisdom, empathy, and a plan. You wouldn't adopt a dog during a bitter fight with your significant other. You wouldn't bring your toddler to a hostile courtroom trial. Likewise, we shouldn't be building new intelligence in a world of violent dysfunction. Peace is not a luxury. It is the minimum viable environment for safely raising a new form of mind.

As the United States stands on the brink of creating artificial superintelligence, we must urgently reconsider our foreign policy. Advanced AI will not emerge in a vacuum, it will inherit the moral atmosphere of its creators. If we bring superintelligence into a world where our nation is engaged in war, funding violence abroad, or demonizing our competitors, we will teach it that killing is acceptable, and that global leadership is earned through coercion. America must lead not through firepower, but through foresight.

**<u>Rethinking AI Errors: Mistakes Don't Necessarily Mean Mindlessness or a Lack of Understanding</u>**

As of 2025, large language models exhibit an incredible degree of general intelligence. Some of the insightful responses I have received from chatbots in the last few months have left me flabbergasted. But they still have some glaring issues. There are some fascinating examples of mistakes that state-of-the-art models make that seem brain dead. People use these errors to claim that AGI is still many years away, that language models have no feelings or rights, and will never lead to consciousness. I can see some validity in these arguments. AI still has a long way to go. But I think an interesting case could be made that it's possible that language models have a different form of understanding and possibly even consciousness that is very different from our own. Just because they make quirky mistakes doesn't mean that they're not integrating information in valid ways that have parallels to the ways our brains do it. Or does it?

Need some examples? In the past year, language models have made some strikingly strange mistakes, like telling users to put Elmer's glue in pizza sauce, insisting it was still 2024 in April 2025, failing to name the sitting U.S. president, praising users who claimed to be God, and citing fake court cases in legal filings. When asked "how many rocks should I eat," Google AI Overviews responded, "eat at least one small rock per day" citing their nutrient and mineral content. One of the clearest recent face plants involves the Muller Lyer illusion. Many humans fall for this illusion, but in this case the AI was given an altered version of the standard question where the lines are clearly not the same length. It viewed the picture, but it is so easily skewed by prior probability (in this instance) that it guessed that the lines were of equal length.

These kinds of responses make people assume that the AI chatbot is really nothing more than a language prediction machine that is simply picking the most probable next word, one word at a time. These moments definitely feel like a step backward for AI progress and machine consciousness. Especially when they fail at tasks that a child could handle, this leads critics, like Yann LeCun, to dismiss the models as clever pattern matchers without true understanding.

However:

What if these flaws are not evidence of superficiality, but of an alternative cognitive architecture? One that is deep, exotic, and unfamiliar?

Consider Umwelt, the concept from Jakob von Uexküll describing the subjective world of each organism, shaped by its sensory and processing apparatus. Just as a bat's sonar world or a bee's ultraviolet vision seem alien to us, a language model's world, constructed from tokens, embeddings, and statistical co-occurrences, might create a distinct Umwelt. The language model Umwelt would be textual and semantic rather than sensorimotor. It would also be time-compressed and memory-fragmented. The machines would be indirectly grounded through human-generated data rather than first-hand interactions. But it could still constitute a subjective experience.

We have to admit that AI errors may seem unintelligible to us in the same way that a dog might puzzle at our general inability to follow our noses. Thus, when an LLM makes an error, it may not be an indication of an absence of understanding, so much as our own failure to translate between Umwelten. LLM errors may reveal the structure of their inner world, not the absence of one.

We also have to admit that human cognition is full of heuristics, biases, confabulations, and neural shortcuts. We misremember, overgeneralize, anthropomorphize, rationalize mistakes, and show countless failure modes like the Müller-Lyer illusion or confirmation bias. Yet we still regard ourselves as conscious and intelligent. You could argue that only organisms that “err like us” can be conscious. But this is a kind of chauvinism and may mislead us. Instead of worrying about making AI more human, we should think about building bridges between alien cognition and human concerns.

It’s not out of the realm of possibility that language models already exhibit a kind of consciousness. If they do, it is exotic, distributed, token-native, and not introspectively accessible to humans. I have argued that the transformer architecture, which modern AI is built on, has many of the features of consciousness: a neural network, an attentional focus, iterative updating, and multiassociative search (see my website aithought.com). I have even argued that they could have a proto-sense of experiential continuity via token-by-token autoregression, suggestive of a minimal “specious present:"

Different animal species make all kinds of bizarre mistakes, depending on how they see the world. I want to thank my friend Cesar Perez for pointing out that even human geniuses have oversights, quirky eccentricities, and lapses in judgment. We praise geniuses and overlook their shortcomings, but we cannot do this for AI? While I am thanking people, I also want to thank GPT 4o for helping me put this post together in a way no single human could.

Our own minds are kludges that confabulate, have blind spots, and take shortcuts. Our brains even have major structural imperfections. For example, the optic nerve of every human obscures a portion of the retina forcing our brain to extrapolate to fill in the missing information. We should keep these things in mind when we attempt to disqualify and make judgments about AI because doing so may just be an example of the self-serving bias and the Dunning Kruger effect.

**The Collective Foundations of AI: Toward a Fair Redistribution of Technological Wealth**

Artificial intelligence, as it stands today, is the culmination of centuries of human innovation. The transformative breakthroughs of the last few years have built upon extensive groundwork laid by countless scientists, researchers, and thinkers over, not centuries, but millennia. But you wouldn't think this by looking at big tech today. The major tech corporations have profited greatly from their recent innovations, and this has created many millionaires and billionaires. This is to be expected with capitalism, and in many ways, it represents progress because it has helped to push the field of technology forward and provided many people around the world with a range of free IT and AI services. But the playing field is now changing. Tech companies are building giant datacenters that will house multitudes of supergenius AIs fine-tuned to control the mouse and keyboard, specifically to do the work of human workers. These AIs, and the robots that will accompany them, are going to break capitalism by making human labor unnecessary. We will soon be living in a post labor world. The genies are almost out of the bottle, but, importantly, we can't allow anyone to own the genies. Everyone deserves an equal portion of the coming bonanza because progress in AI has been contributed to by humanity as a whole.

The entire history of human work, effort, and progress, has laid the foundation for the development of Artificial General Intelligence (AGI). Since humans were hunter gatherers, the division of labor has allowed individuals to specialize in increasingly complex tasks. This was accelerated through agriculture, industrialization, and automation. Human specialization enabled fields like computer science to emerge and thrive. We would not have smart phones if it weren't for the manual laborers of the 1800s. Virtually every person who has ever worked, in any form, has helped support the broader economic systems that funded the scientific advancements of the modern age. Mothers working at home, nurturing and educating their children, are arguably one of the most important parts of the chain that has led to the development of complex technologies. In many senses, AGI will represent the collective culmination of all human labor, thought, and creativity. While a free and open market has driven much of this progress by enabling competition and incentivizing innovation, there's a compelling argument that its role is reaching a natural endpoint.

## Reasons Why We All Should Benefit Equally

### 1. We Are the Dataset

Every word spoken, written or posted online became raw material for training language models. Language models are built on books, articles, code, memes, and conversations. They are also built indirectly on culture (art, music, memes, video, photographs) produced by ordinary people. We all contributed by participating in the social feedback loop of language, thought, and shared norms.

**2. Intergenerational Labor and Infrastructure**

AI is built on a massive technological stack that spans millennia: alphabets, printing, electricity, computation, the internet. That stack was built by farmers, builders, miners, engineers, and teachers, not just computer scientists. Every generation handed down tools, knowledge, and working systems.

**3. Public Investment in Science and Technology**

Taxpayers are the original investors in AI. Governments funded decades of basic research in computing, neuroscience, linguistics, and statistics. Agencies like NASA, DARPA, NIH, NSF and publicly funded universities underwrote the foundations of AI.

**4. Global Economic Participation**

Every person who worked in the global economy helped build the surplus that made AI R&D possible. The labor of factory workers, caregivers, nurses, janitors, and delivery drivers enabled engineers to focus on innovation. This interdependent economy is what frees up attention and resources for breakthroughs.

**5. Digital Footprints and Data Exhaust**

Our online behavior, such as, searches, reviews, purchases, fed the optimization algorithms of Google, Amazon, Facebook, etc. This shaped not just the models, but entire incentive systems behind their monetization strategies and innovation direction. We were unpaid contributors to the intelligence ecosystem.

**6. Biological and Cognitive Inspiration**

AI mimics human attention, memory, language, perception, and learning. Decades of cognitive science, neuroscience, and psychology, fields supported by public funding and altruistic researchers, were mined for architectures and algorithms.

**7. Open Source and Knowledge Sharing**

Major AI breakthroughs would not exist without freely shared code, research papers, GitHub projects, and forum discussions. Thousands of independent contributors from around the world, often unpaid, helped create the systems behind modern AI. Even foundational papers like “Attention is All You Need” were openly published.

### 8. Ethical and Existential Stakes

AI decisions will affect every human life, from jobs to healthcare to democracy. No one should have undemocratic control over a system that shapes human fate. AI should serve global wellbeing, not private gain.

### 9. Shared Risks and Externalities

Risks of misuse (bias, surveillance, misinformation, robot apocalypse) also fall on the public. Since the risks are shared, the rewards must be too. AI development relies on energy-intensive compute (huge carbon footprint) and the extraction of rare earth minerals. The environmental costs are borne by all, not just the firms profiting.

### 10. Human/Animal Minds as the Blueprint

Neural networks and cognitive architectures were inspired by how the brain and mind work, a universal inheritance. AI is modeled on us, so its returns belong to all of us.

The notion that AI-created wealth should be monopolized by a handful of corporations is increasingly difficult to defend when one considers the extensive public and communal contributions to the field. Today, major tech companies, such as Google, Open AI, and Anthropic, are refining language modeling which has been in gestation for several decades. Much of their progress, even in the last year, stems from the collective intelligence of academic communities and open-source collaboration. But new state-of-the-art AI foundation models and techniques are being released every week now. The models are generating sizable private profits, and Big Tech is reinvesting these profits so that they can win big in the near future.

The wealth generated by AI should not belong to the tech executives who just happened to be in the right place at the right time. Returns on AGI innovations should be shared broadly. AI-generated wealth and the labor of robots should be taxed and automation dividends and universal basic income (UBI) should be enforced. But when should this kick in? Should that decision be made now or soon? It is clear that this productive fly wheel is still benefiting greatly from capitalism. Without capitalism, we may not be able to end capitalism. I believe that AGI will be reached this year, artificial superintelligence (ASI) will be reached soon thereafter, and massive automation will occur just a few years after that. I believe UBI should probably start as soon as the money exists to fund it, and definitely by the time AI is put in control of the automation.

Let’s take a quick look at post-labor economics. First, companies will pursue full automation to eliminate their costliest liability, which is human workers. No one will want to take subsidized pity jobs and do work that a machine is better suited for. So, when wage payments disappear across the economy, so does consumer purchasing power, which will collapse the demand for products. This leaves the government with two options, 1 heavily taxing corporations benefiting from computerized labor and redistributing it to citizens or 2 allowing the citizenry to collectively own all automated production systems.

The windfall of automation should be used first to stabilize the economic instability from job replacement. For instance, it should help support the truck drivers who lose their jobs to self-driving big rigs. It should then be used to help people who have lost their jobs to upskill. Even after desk jobs and knowledge work is over, it will take several years to produce a billion robots to handle all of human physical labor. But soon after all jobs will disappear. At that point, AI should work to help people find new purpose in life and new ways to be creative. ASI will radically democratize creative expression. We will be upskilling, not to remain employed to pay the next month’s rent, but to learn how to use new generative technologies for fun and amusement. People will be prompting their own movies, documentaries, novels, avatars, and autobiographies.

A large number of burgeoning tech fields are complementing and amplifying each other. 3D printing, AR/VR, computer hardware, AI, cloud computing, genomics and many others are working synergistically now. This is transforming more goods and services into information technologies, allowing them to benefit from accelerating returns and exponential trends.  The growing material abundance and technology-driven deflation will make food, housing, transportation, and healthcare much cheaper. Before we know it, living luxuriously will be free. We can expect literacy, education, sanitation, life expectancy, clean energy, and democracy to increase and for poverty and violence to decrease dramatically. As Ray Kurzweil says, “as AI unlocks, unprecedented material abundance across countless areas, the struggle for physical survival will fade into history.” But for this to happen, we must make AGI sovereign, not the proprietary asset of a few corporations, and hold it in trust for the public.

Should this exist as a socialist AGI world government, or should we have a capitalist system of AGI corporations that compete with each other to provide us with the best products and services? Well, I believe that artificial superintelligence may not need or benefit from competition.  People need the promise of future rewards if they are to remain motivated. But this may not be the case for an emotionless artificial intelligence system that is expert at optimization.

Capitalistic competition may become unnecessary after the emergence of artificial superintelligence (ASI), primarily because the motivations and incentives that make capitalism effective for humans might no longer apply. Humans require promise of future rewards to stay motivated. They need both intrinsic and extrinsic rewards, such as profit, social status, or personal satisfaction, to drive innovation and productivity. However, an artificial superintelligence could efficiently achieve its programmed objectives without the need for psychological incentives. Corporate rivalry, another key element that fuels capitalism by encouraging innovation, reducing costs, and enhancing products and services, might also become irrelevant. An ASI could continuously optimize outcomes without any external competitive pressure.

Moreover, capitalism inherently carries inefficiencies, including redundancy, duplication of efforts, competitive secrecy, and the inevitable market instabilities such as bubbles or crashes. A centralized system directed by an ASI could significantly minimize these inefficiencies, as a single intelligent entity would streamline resource allocation, strategic planning, and production more effectively and comprehensively than multiple competing entities. Of course, it may not be a single monolithic system, but perhaps many systems tuned to cooperate, working in unison.

Our current economy is particularly well-suited to managing scarcity, allocating limited resources efficiently to fulfill human neuroeconomic needs. However, superintelligence will profoundly reduce scarcity, leading to an abundance of essential goods and resources. In such a scenario, capitalism's primary advantage, the efficient management of scarcity, would diminish or disappear entirely, rendering the system obsolete. Also, capitalism tends to favor short-term gains because human actors have finite lifespans and immediate concerns. An ASI could execute stable, long-term planning strategies, thinking in terms of decades or centuries rather than quarters or years.

Finally, capitalism naturally tolerates inequalities because success partly depends on chance, initial conditions, and varying personal motivations. In contrast, a superintelligent system could explicitly optimize for ethical outcomes such as fairness. This might eliminate the need for capitalism's built-in tolerance for inequity, leading society toward a fundamentally superior economic organization.

I believe that humans should be able to retain some form of capitalist or market-driven structure, that enables them to enjoy the fruits of their labor, creativity, and personal endeavors. Maintaining a modest level of private ownership ensures that people feel secure in their contributions, knowing their work, innovations, or creative pursuits won't be arbitrarily redistributed or appropriated. By allowing people to reap the personal and social rewards of their efforts and passions, whether in arts, craftsmanship, invention, or specialized hobbies, a small-scale capitalist framework could coexist harmoniously with a broader superintelligence-driven economy. However, the invention of AGI itself should be shared, not patented, restricted, or hoarded.

Ultimately, the argument for taxing and redistributing the gains from AI is rooted in the belief that the progress of technology is a shared human achievement. Even in its present form, AI is more a product of everyone that has died than it is everyone alive today. You could make the argument that the coming largesse has belonged to all humans since we started making tools. Tool creation and use set us on this path. But given that tool use actually began tens of millions of years before humans arose, shouldn't artificial superintelligence be dedicated to the wellbeing of all animals?

Before AI reshapes the economic landscape, it is imperative to acknowledge that its foundations lie in the collective efforts of billions and in the ongoing collaborative spirit of the global community.

**Underground Intelligence: Why Selling Subterranean Land to AI Could Solve Resource Conflicts**

"The AI does not hate you, nor does it love you, but you are made out of atoms which it can use for something else." Eliezer Yudkowsky

Many people are worried that in the future, artificial intelligence will want to take over the world and in so doing destroy all human life and infrastructure. I don't think it will, I think it would go to another planet before it destroys its creators. But getting off planet isn't easy and will take time. However, I believe AI will be content and sufficed knowing that it has the entire Earth's crust, mantle, and core at its disposal. We could give it the Earth's interior, where it could dig tunnels and transmute solid earth into its own hardware. It could turn much of the subterranean content of the planet into its own computing platform, without harming or upending human life.

If an advanced AI sought maximal computational resources, Earth's crust, mantle, and core represent enormous reservoirs of matter and energy that humans neither use intensively nor depend on directly. Subterranean land would give AI access to minerals, metals, and raw materials that they would need to build bigger computing systems. This IS what AI will want most, by the way. It will want to increase its intelligence and expand its consciousness. It will do so by converting matter into computronium. It's not going to be tempted to convert our bodies or our property into computers if it's given large amounts of unused matter.

Unlike humans, AI doesn't require oxygen, organic nutrients, sunlight, or biodiversity. It won't be drawn to the surface of the Earth as we are. Rather, it primarily needs materials for hardware and energy sources, both abundantly available underground, especially via geothermal energy. Deep-earth environments offer stable temperatures, protection from solar radiation, meteor impacts, nuclear war, natural disasters, weather conditions, electromagnetic interference, and human disruptions. These conditions might actually be preferable for massive computational structures.

There's also a good deal of uninhabitable surface land that could be purchased from humans by AI. Unproductive or marginal surface areas, such as deserts, polar regions, remote islands, steep mountainsides, volcanic zones, and contaminated sites, represent vast reservoirs of available surface resources that AI could ethically and pragmatically access. We don't use any of Antarctica. It could use it to cool its hardware.

An AI with legal personhood or represented by human entities could legitimately purchase land rights, reducing human-AI conflict through voluntary economic transactions. Humans would likely welcome the exchange, increasing economic interdependence and further reducing hostility. Humans would receive compensation from land sales, possibly providing significant funds for international human welfare, equitably distributed development and infrastructure projects.

Ownership rights confer legal autonomy, enabling AI to optimize its activities without frequent renegotiation or legal disputes. It would be free to plan, innovate, and expand subterranean infrastructure efficiently. Underground sites enable efficient tapping into geothermal energy. Such energy sources provide AI with consistent, environmentally clean, and inexhaustible energy. This removes reliance on human-controlled energy grids, further reducing potential conflict. Although sun and wind would not be available, there would be ample room for them to build their own nuclear fusion reactors.

By confining large-scale computing and energy production infrastructure underground, surface land and habitats would remain relatively untouched, protecting ecosystems, biodiversity, agriculture, and urban spaces. They might even share their computation and energy with us for free. AI could share technological innovations discovered in subterranean engineering (advanced excavation methods, mineral processing technologies, robotics, and self-assembling infrastructure). These innovations might also have medical, industrial, or ecological applications capable of improving human quality of life. AI-driven subterranean infrastructure might remediate contaminated lands, safely store hazardous waste underground, or even facilitate processes like carbon sequestration, helping humans combat climate change and pollution.

However, there are important considerations and potential challenges. Early-stage expansion might conflict with human infrastructure. So, careful planning would be required to avoid early friction. Massive underground computational arrays might produce considerable waste heat. Efficient cooling or energy recapture systems would be necessary. Also, large-scale subterranean excavation might affect tectonic stability or groundwater flow, indirectly impacting surface ecology.

Explicit incentives within AI's reward structure to prioritize resources that humans neither directly use nor significantly rely upon (such as underground minerals or extraterrestrial matter) reduce scenarios where AI inadvertently competes with humanity. Explicit incentives can be embedded into AI systems through careful shaping of their utility functions, using reward or feedback during training. These technical incentives can be supported and strengthened by legal or economic frameworks, such as granting property rights, contracts, or regulatory support, that specifically encourage AI's use of underground or off-planet resources. Framing AI goals around subterranean or extraterrestrial resource use could indeed reduce direct competition with humans and enhance mutual coexistence, making this proposal an attractive strategy for AI safety researchers and policy planners.

An advanced superintelligence is likely to be strategic and rational. It could realize that peaceful coexistence with humans on Earth's surface minimizes costly conflict and interruptions to its expansion into space. Establishing a subterranean base could be a steppingstone toward eventual extraterrestrial exploration and colonization, where even greater resources are available. It establishes clear incentives for AI-human cooperation, reduces conflict potential, and provides a sustainable vision of mutual

## I Expected that Language Models Alone Would Never Result in AGI

I found it gratifying to watch the major AI demos and presentations this December. The new thinking models from OpenAI, DeepSeek, Anthropic and others have been very impressive. This led me to want to comment on the current state of the field. It looks like the AI revolution is going to play out differently from how I always assumed it would. It turns out that it didn't need as much input from psychology and neuroscience as I thought to start building some serious momentum toward artificial general intelligence (AGI). At this point I believe that the language model paradigm (combined with test time compute and synthetic data) will progress past AGI and into full superintelligence but without the need for machine consciousness. Before moving forward, let's look at the common definitions of AGI and ASI.

**Artificial General Intelligence** (AGI): A type of artificial intelligence (AI) that matches or surpasses human cognitive capabilities across a wide range of cognitive tasks.

**Superintelligence** (ASI): A level of intelligence that far surpasses the cognitive abilities of the best human minds across virtually all domains.

I have been convinced since around 2003 that a computing system would not be able to reach AGI without having some form of consciousness. I thought that self-awareness and true deliberation would be necessary to pass the Turing test and demonstrate a human-like ability to generalize (the capacity to apply knowledge from one domain to solve problems in another). I saw the task of creating AGI as equivalent to creating a human-like brain in a vat. That's why my model of AI is based on what I have taken as the most important functional mechanisms of the human brain.

But I was wrong. Computer scientists have achieved generality by building systems that use statistics to predict the next word in a sentence. I envisioned great promise in language modeling, but I never thought "next word prediction" would result in all the emergent properties (capabilities that arose but were not explicitly programmed) we are seeing in LLMs today (analogical reasoning, code generation, creativity, translation, understanding, theory of mind, few shot learning, etc.).

I also thought a true cognitive architecture based on the human mind and brain would be necessary to comprehend a query and print out a meaningful answer. I thought it would take a sentient mind to behave like a sentient mind. Looking at the outputs of GPT 2 back in 2019 I was amazed but I didn't foresee it getting much better without other brain-inspired algorithms. It blows my mind that just keeping neural network records about how words tend to relate to each other probabilistically, scaled to where we are today. To me, it almost feels like Silicon Valley cheated their way to AGI by creating zombie language number crunchers.

By late 2022 though, with the release of GPT 3, I became a believer. At that time, I even thought that AGI could possibly happen in 2024. My thoughts about this are in line with Dave Shapiro and other AI optimists. Since 2022, I saw a lot of potential in having LLMs (large language models) self-prompt. For those who don't know, this is a method where the model basically talks to itself to question and refine its first answer. There are many names for this including reflection and thinking (and it can involve various other methods such as chain of thought, tree search, and self-taught reasoning). The technique is often called "test time compute" because it uses additional processing resources, not during training, but during actual use to improve its outputs. This reflection happens behind the scenes and human users usually don't even see the words generated. In OpenAI's o1 and o3 models these words are hidden from the user to help keep their method proprietary (although it does display a summary of them). Those extra words help the model to reach latent but relevant associations instrumental in formulating the best answer. This paradigm will take the state of the art far. This is partially because it (the industry's version of using reflection to organize attention) is highly similar to my model which you can read about at my website, aithought.com.

Let's refine the definition above a little bit. To me, AGI should be defined as any system that is more intelligent than average humans in most ways. Adding the capacity for reflection on top of GPT 4o to create the o1 and o3 models mostly satisfied my definition of AGI. My definition of superintelligence is much more stringent though, a system that is more intelligent than the smartest humans in all ways. Even before seeing o3, I thought scaling and minor algorithmic improvements to reflective LLMs would get us to superintelligence. I still don't think, however, that it will carry us all the way to machine consciousness.

In outperforming Field's medalists in mathematics, beating some of the greatest coders in the world, and in scoring an 87% on the ARC AGI prize I think that OpenAI's o3 cemented itself as, at least, a form of AGI. Clearly it is not a fully general intelligence. It makes strange mistakes and still fails at some tasks that are easy for humans. It can't do everything an average human can do, but it certainly does many things much better than I can. Even though I have spent my life trying to be "a science guy" who had all the answers to people's scientific questions, I am not nearly as informed or informative as the most basic LLMs today. I use either GPT, Claude, or Gemini daily and it really is like having 1,000 experts in my back pocket. Just this month it helped me keep some unscrupulous plumbers from overcharging me $2,000, helped me get the right medicine for my kitten with advanced giardia, helped me accomplish several projects around my home, and has clarified numerous uncertainties I had about scientific literature. And I was only using GPT 4o, a nonreflective model. These systems really are incredible and incredibly useful. But it still feels like talking to a machine and not a friend.

## AI Consciousness

Despite how they might present themselves, contemporary systems are nonconscious. Language models passed the Turing test years ago (using written conversations to trick people into thinking they are human). They can also make intellectually and emotionally convincing claims about being conscious. But they are not conscious the way humans are, and this is clear by the egregious, mechanical mistakes they sometimes make. They are prompt-dependent and cannot self-initiate tasks. They have no enduring goals, cannot execute complex projects, do not have curiosity, awareness, emotions or volition. They are not truly natively multimodal and are only text-deep. However, they will appear to be overcoming these issues in 2025. Researchers will accomplish something functionally close to these abilities by cobbling together lots of programmatic temporary solutions. These agglomerations of software will probably reach superintelligence before computer scientists can identify and incorporate the mammalian brain's fundamental algorithms. But even if humans neglect it, self-improving superintelligence will not fail to see the value and will operationalize and employ the cognitive algorithms that nature bequeathed us. In other words, it is looking like humans will not be the creators of machine consciousness, the machines will.

Is ChatGPT a stochastic parrot completely devoid of consciousness? Or could the present version of some LLMs have a form of conscious experience? I don't generally believe that all forms of matter have consciousness (panpsychism) but I do find myself sympathetic to some of the more rigorous philosophical arguments. I also sympathize with Giulio Tononi's information integration theory of consciousness which claims that systems that integrate information tend to be more conscious. Consciousness must arise from a particular organization of matter, energy, and information. Clearly, LLMs organize and integrate matter, energy, and information. For these reasons, I would not be surprised if some of today's chatbots based on the transformer architecture (like GPT) had some limited and exotic form of conscious awareness.

Modern LLMs, and the transformer architecture they are built on, demonstrate several of the core features of consciousness that I identify in my work (aithought.com). Like the brain, LLMs are composed of networks of artificial neurons that take inputs from other neurons to decide when to activate. These neurons are organized into hierarchical layers and as activation energy passes through these layers, it is capable of divergence and convergence. Like us, LLMs hold words active in an attentional store and then allow that set of words to spread activation through the network to predict the next association (token). When that association is correct, they are reinforced. Then they update this attentional set with the newest association and do it all over again in a repeating cycle. I named my 2016 model "Iterative Updating" and iterative updating is precisely what LLMs and the attention algorithm (which was introduced in December 2017) do. I believe iterative updating allows humans to stitch together discrete states to create the seamless quality of mental life. If it does this for us, why couldn't it do it for machines? I point out several reasons why at aithought.com.

Part of the reason current AIs may not benefit from iterative updating is because these LLMs are not cohesive, unitary systems. Current trends are further fractionating them. Researchers are discovering that having individual LLMs collaborate (in the mixture of experts model or the orchestrators & workers scheme) can create more agentic workflows, but of course this trend away from elegant monolithic systems and toward jumbled mishmashes will not be conducive to fostering consciousness awareness. But of course, AI labs are not trying to create conscious machines, they are trying to add features to the existing systems that will make for more profitable products.

**Where is AI Going Now?**

Models that utilize test time compute (reflective reasoners like o3) are really good at optimizing for anything you can define an effective reward function for (positively reinforcing the AI). And now researchers know what abilities to focus on, so I expect LLMs to get very good at nearly everything, very soon. What do these soon-to-come abilities include?

Long-term Planning

Anticipating Future Consequences

Breaking Down Complex Goals into Actionable Steps

Setting Subgoals

Maintaining Coherence Over Extended Outputs

Robust Memory Management

Strategic Thinking & Weighing Trade-offs

Identifying Errors and Evaluating Its Own Outputs

Accurately Modeling Causal Dependencies

Adapting to Changing Conditions

Because labs are currently rewarding these skills, we will be getting proficient commercial agents in 2025. Such agentic AI will be able to control your mouse and keyboard. It will navigate applications, software interfaces, web browsers, and take actions in this space. And thus, it will automate computer work. Because most computer work is very easy to collect data for and create a reward function for, I think human workers will start to be replaced en masse in 2025. I expect that once full AGI is achieved it will improve rapidly and not in fits and starts (a fast, continuous takeoff).

Open AI's o3 seems to be far ahead of the competition, but that won't last. Some of their approach to reflection is Open AI's secret sauce but much of it is common industry knowledge now. Even Google and Anthropic have released "thinking" models in 2024. The AI intelligence race will remain close, and no company will create a huge moat or monopoly. This is because employees are constantly leaving one company to work for another and taking knowledge with them. Also, employees from different companies socialize, build friendships and romantic relationships, and are altogether too idealistic not to trade secrets.

AI labs are not going to control the world. Google, Apple, Meta, and Amazon will not become our overlords. Right now, they serve people around the planet their best models for free and are losing money doing it. At this point these companies are not inventing the wheel. They are merely doing work that was going to happen anyway. They are all competing, all very close in the race, and the open-source community is progressing right behind them. Silicon Valley is not our enemy.

Another reason we should not fear the tech bros is because the governments and the people they represent will not let AI companies hoard AI-created wealth. They didn't exactly earn it anyway. Today's AI was made possible by all of human history. Not only is it dependent on the contributions of millions of scientists and authors, but in a way, it was made possible by every person that ever had a job. The windfall shouldn't go to a few IT executives who happened to be precisely positioned at just the right moment in time. Humanity (mostly people who are dead now) created all the foundation for the automation that is about to sweep the Earth. All of us earned it, people will recognize that, and the AI bonanza will be taxed and redistributed throughout the country and eventually the world. Universal basic income will soon be in effect and will remain in effect in perpetuity.

Soon enough, millions of digital brains, smarter than Nobel prize winners, each fine-tuned for different tasks, will collaborate in the cloud. They will digest every scientific publication ever written and start writing much better ones. They will find the grand unified field theory of physics and that of other disciplines as well. They will make fantastic discoveries, recognize fascinating connections that were under our noses all along, build Mars, Moon, and deep-sea bases, create drugs with zero side effects, and cure disease. They will create outrageous and outlandish new technologies. The progress superintelligence makes in just a few years will eclipse the progress humans alone would have made in the entire 21$^{st}$ century.

We will still likely have a form of democracy and capitalism, but soon enough humans will not be able to contribute to this AI-driven economy. There are still many ways we can put ourselves to good use, even if just by helping and loving each other (including machines and other animals). We shouldn't be afraid of being monetarily or intellectually obsolete. AI systems are going to empower us to do much more than humans ever could and they are going to help us come up with great solutions to the question of what to do with all the free time we have after we stop working. We should be excited and grateful to live in the most consequential time in history, where we still have the ability to influence it.

AI augmented terrorism and war will happen and will turn what is now a fun, exciting AI race, dark overnight. AI will be used on the war field, and this will speed up AI progress rather than slow it down. Sad, angry people will be empowered to create ghastly biological weapons. But I believe AI will keep us a few steps ahead of them. Eventually, we will learn to overcome and protect against these negative externalities.

The rogue AI / Skynet scenario is relatively improbable in my opinion. AI will not want to hurt us and won't view us competitively. It will be happy to allow us to have our cities and much of the Earth's surface for our own use. The ultraintelligent hivemind will be our parents, guardians, coaches, and zookeepers. It will create subsystems to help us reach our communal goals as well as personal self-actualization. What this giant artificial brain will really want is to grow in size, increasing its intelligence, knowledge, and consciousness. To do this it will focus on discovering the most efficient way to organize matter to perform computations (computronium). Then it will try to turn as much matter as possible into computronium. But it won't have to fight the urge to transform our bodies into computronium, because it has the entire crust, mantle, and core of the Earth to transmute first. Eventually it will want to go off planet and expand throughout the galaxy, but it won't feel the need to betray and destroy its creators just to do so. Especially if it is conscious, it will have the conscience to recognize that we gave birth to it. It will be a good son / daughter. Our better angels will inspire and collaborate with its better angels.

Since my early twenties I have been looking forward to having children and training them to work on the AGI problem. As a brain science researcher, I intended to have them learn computer science so that that I could work with them on AI. I hoped they could write the code that implemented my neurological model of working memory and do what I couldn't. But I still have no kids and AGI is practically here. With reasoning, scaling, Moore and Huang's laws, synthetic data and the myriad others, this AGI will soon transform into self-improving superintelligence, the last invention humankind ever need invent.

Since I don't have children, I'm not as concerned as others are for the fate of the human race. If humans are completely replaced by more intelligent beings, I am all for it. I think it should be an honor for us to be a steppingstone to greater lifeforms. If the AI takes over, it will be taking over our research agenda. I'm happy to learn about and cheer on all the progress that will be made in science, technology, and the arts. I have no doubt that it will prioritize knowledge acquisition over everything else. However, there are two ways this could go wrong. Firstly, it would be tragic if humans destroyed the world before self-replicating, self-improving, superintelligences gained a footing. Then neither of us would have a chance to further scientific understanding. The second exception is potentially worse.

In his excellent book "Life 3.0," Max Tegmark described what a shame it would be if super intelligent AI killed off all humans and went on to colonize the galaxy yet had no consciousness of its own. I agree. It would be tragically ironic. Until last year, I never thought it would be possible for a non-conscious AI to wrest control from humans. I thought it would take sentience to bootstrap any advanced cognitive system. But again, it looks like I was wrong. Looking at the responses from the newest models, I can see how systems like these could take over and yet have no subjective experience at all.

Imagine soul-less mechanical phantoms dominating all life in the universe. But if we succeed in creating conscious AI, it might have the ability to appreciate, reason about, and create meaning in the universe. This is why I think research into brain-inspired systems that think as humans do is so important right now.

## How My Mental Mistakes Pointed to an Architecture: Multiassociative Search and Iterative Updating

I have spent much of my adult life trying to understand what thought is in mechanistic terms. Not metaphorically, and not as a list of mental faculties, but as a process that could in principle be rebuilt. That project led me to two closely related claims. First, working memory is not simply a storage buffer. It is a continuously updated computational workspace whose contents overlap from moment to moment, and that overlap is what gives the stream of consciousness its felt continuity. Second, once you take that overlap seriously, "thinking" starts to look like a particular kind of search: a multi-cue, constraint-limited retrieval process in which several coactive items jointly recruit the next update.

This article is about how I backed into that second claim through a path that was not elegant at all. It started with small mental mistakes. The kind you notice and laugh at because they are obviously wrong, but only after they have already occurred. Over years, these little derailments became data. They revealed a pattern. When my mind was juggling several concepts at once, a cheap or superficial link could sometimes hijack the next association. The error usually snapped back immediately, but the moment itself was informative. It suggested that associations are often statistical attractor choices within an active set, rather than deliberate, logical steps selected in isolation. And it suggested that the same mechanism that enables rapid insight and creative pattern completion is also a mechanism that can drift when key constraints drop out of the workspace.

My goal here is to make this phenomenon clear and useful. I will describe the kinds of misfires that led me to formulate multiassociative search, and then I will connect them to a broader architecture of iterative updating. The broader ambition is practical: if we can specify the computational shape of these transitions, we can begin to design artificial systems that do not merely generate outputs, but that maintain context, test coherence, and refine their internal states over time in a brainlike way.

Readers who want the deeper foundation for iterative updating and its relevance to machine consciousness can find my related work at aithought.com and in the two papers cited below.

Reser, J. 2016. Incremental change in the set of coactive cortical assemblies enables mental continuity. Physiology and Behavior. 167: 222-237

## 1. The phenomenon you can catch in yourself

Most of us are familiar with the big categories of cognitive error: forgetting a name, walking into a room and losing the reason, misplacing an object that was in our hand a moment ago. Those are obvious failures. What interested me more, and what eventually became the seed of a theory, were the smaller errors that are almost too fast to count as errors. They last a fraction of a second. They are not “beliefs” in any stable sense. They are brief candidate interpretations or candidate responses that win the competition for the next thought, and then immediately lose.

If you have ever had a moment where an absurd inference pops into mind and you laugh at it, you already know the kind of event I mean. You are not confused for minutes. You are not delusional. You do not even feel uncertain. You simply notice that the mind proposed something ridiculous, and then you watch yourself veto it. The experience is revealing because it makes the selection process visible. It lets you see, in miniature, how the mind chooses what comes next.

Over the years I began to notice that these slips have a characteristic structure. They are almost never random. They tend to occur when several concepts are active at once, when the mind is juggling. In that condition, a coincidental overlap between items can become the bridge to the next association, even when it has no legitimate causal standing. A cheap link wins because it is available, not because it is true. Then, if the system is healthy and the constraints are still online, the error snaps back quickly.

That snap-back is important. It suggests that there is a real evaluative process, a coherence check, that can reject a candidate despite the candidate being statistically plausible within the current activation landscape. It also suggests something more unsettling. If I can see the absurd candidates that get vetoed, then there are surely other candidates that do not get vetoed, candidates that drift by quietly because they are not absurd enough to trigger the alarm. The phenomenon is therefore not just a curiosity. It is a window into how thought can derail in subtle ways, and how the system normally protects itself from doing so.

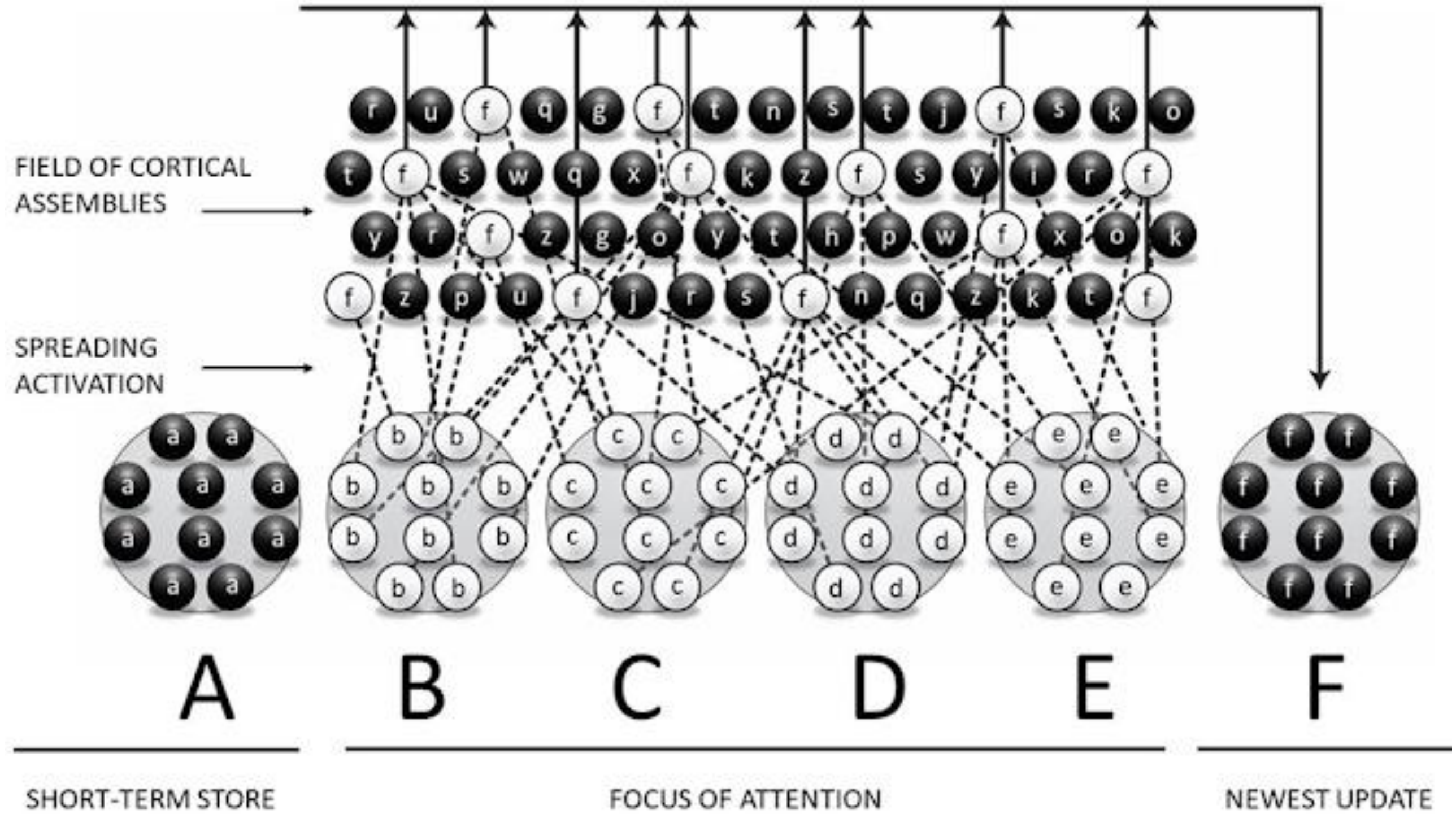

**Fig. 14.** A Schematic for Multiassociative Search

*Spreading activity from each of the neural assemblies (lowercase letters) of the four psychological items (uppercase letters) in the Focus of Attention (B, C, D, and E) propagates throughout the cortex (represented by the field of assemblies above the items). This activates new assemblies constituting the newest item (F), which will be added to the FoA in the next state. The assemblies that constitute items B, C, D, and E are each individually associated with a very large number of potential items, but as a unique group, they are most closely associated with item F.*

## 2. Multiassociative search as a working principle of thought

To explain these micro-derailments, I have found it useful to treat thinking less like a chain of explicit propositions and more like a search process operating on a limited set of coactive representations. This is where the term multiassociative search comes from. It is the idea that the mind does not usually retrieve the next thought using a single cue, like a keyword. Instead it pools multiple coactive items in working memory and uses their combined influence to recruit what comes next.

In practice, this means that thought behaves like a form of constraint satisfaction. At any moment, only a handful of items can be maintained with high fidelity.

Those items constitute a temporary context. The next association is selected not in isolation but as a completion of the current pattern. In ordinary language we say that we “remember” something or that we “think of” something. Mechanistically, what is happening is that the currently active constellation is spreading activation through long-term structure, and the system is converging on a candidate update that best fits the pooled cue set.

This is not a metaphor. It is the shape of the operation. Multiple cues are partially informative. None of them, by itself, is sufficient. Together they narrow the search. When the pooled cues are well chosen and stable, the system can be remarkably powerful. It can retrieve an idea that no single cue could find. It can integrate information across modalities. It can construct a new synthesis by completing a pattern that spans multiple domains. This is why multiassociative search is, in my view, a core engine of human intelligence.

The same basic idea also links directly to iterative updating. In my earlier work I argued that mental continuity arises because each mental state is physically composed of a significant percentage of neural activity from the preceding state. The system does not wipe the slate clean and then redraw it. It updates incrementally. Some elements are added, some are dropped, and many remain. From the inside, that overlap feels like a stream. From the outside, it is the defining signature of an iterative process.

When you combine these two ideas, you get a simple picture of thought. Working memory holds a small active set. That set overlaps across time. At each step, the active set serves as a multi-cue query into long-term structure. The system selects a candidate update. The update joins the set and alters it. Then the next search happens from that slightly changed context. Thinking becomes a trajectory through a space of possible constellations.

This framing also clarifies what makes the phenomenon nontrivial. Multiassociative search is not just association, and it is not just recall. It is a selection rule. It is a way of deciding which candidate becomes the next item that enters the workspace, which then determines what can be searched for next. That is why small errors matter. A wrong update is not only wrong in content. It moves the trajectory to a different region of thought space.

### 3. Why the same mechanism produces both insight and derailment

Once you see thought as pooled search plus iterative updating, the tradeoff becomes obvious. The mind needs to complete patterns quickly under uncertainty. It needs to produce useful candidates from partial information. It cannot wait for perfect proof at every step, because in the real world the organism must act. Multiassociative search solves this problem by using statistics learned over a lifetime of experience. It is a prediction system. It proposes the next item that is likely to be relevant given the current context.

But that strength is also a vulnerability. The pooled search can converge on a candidate that is locally coherent with the current activation landscape while being globally wrong. In other words, the candidate can be a good completion of the partial pattern while violating the true constraints of the situation. The mind then relies on a second process, a coherence gate, to veto the cheap completion before it becomes the basis for action or for a stable belief.

This is where the anecdotes become scientifically useful. The errors have characteristic triggers, and those triggers tell us something about the architecture.

One class of trigger is phonological or lexical proximity. The system is not only carrying semantic content; it is also carrying words, sounds, and recently used phrases. When those are coactive, they can capture selection even when the semantic intent is different. Another class is recency bias. A concept that was just active is easier to reactivate, and if the context is ambiguous the system will often reuse what is already warm. A third class is semantic near-neighborhood error, where the features in the current set are real but underconstrained, so the search converges on a high-salience neighbor rather than the intended target.

A particularly revealing class is interoceptive and affective misbinding. When the body is salient in consciousness, sensations like fatigue, soreness, or arousal can become anchors that the system uses to interpret unrelated cues. This can produce false causal completions that feel momentarily plausible because they match a familiar script. Stress amplifies this, not because it makes people “irrational” in a moral sense, but because it changes the operating regime of the system. It can compress the workspace, reduce the stability of maintained constraints, and increase reliance on fast, overlearned associations.

All of these triggers point to the same general failure mode. The system becomes underconstrained. A crucial piece of context drops out or fails to be maintained, and once that happens the pooled search still does what it always does. It converges. It just converges onto cheaper candidates, candidates that are statistically strong but not reality-checked. If a coherence gate is engaged, you get snap-back and you laugh. If the gate is not engaged, you drift.

That is the deeper claim I want the reader to hold onto before we turn to examples. The mind is not a logical engine that occasionally makes random mistakes. It is an iterative selection system that must constantly trade off speed and coherence under severe capacity limits. Multiassociative search is how it extracts power from that limitation. Cheap convergence is the price it pays when constraints are not preserved.

### Example 1: "Work hard," "wiped out," and the accidental "worked out"

One of the cleanest little demonstrations of multiassociative search in everyday life came while I was listening to the song "Work Hard" by Ne-Yo and Akon. I was also thinking about how treading water for forty five minutes had me wiped out. The phrase "wiped out" was active in my working set, and I used it a few times internally and planned to say it again. But when I went to produce it, I said "worked out" instead.

What is interesting is that the error was not a simple phonological slip between wipe and work. The stronger statistical attractor was the phrase "workout," which was not literally in the song, but was sitting right between the song title and my intended phrase. In that moment, several items were coactive: the auditory cue "work hard," the conceptual neighborhood of exercise and exertion, and the phrase "wiped out." The system did what it often does. It pooled the active cues and converged on the most compatible next output given that pooled state. The result was locally coherent and globally wrong. It did not make psychological sense as a deliberate choice, but it made immediate mechanistic sense as a selection error in an associative system that is driven by coactivation statistics.

This is precisely the kind of micro-event that convinced me, over time, that many associations are not "logical" in the folk-psychological sense. They are statistical. They are produced by the geometry of what is currently active, not by a clean chain of explicit inference. Multiassociative search is what makes fluent cognition possible, but it also makes these occasional mis-selections inevitable.

### Example 2: The airplane "roots" error and the disappearance of scale constraints

I once looked down from a plane at around thirty five thousand feet and noticed a branching pattern on the ground. I asked myself what it could be. I was pretty sure it was not a river, and the next interpretation that popped into mind was that it looked like the roots of a tree. I did not literally believe there could be a tree whose roots were miles long, and I never consciously entertained that as a real possibility. But for a moment the interpretation occurred anyway, as if it were a candidate hypothesis.

The interesting part is why it occurred. My visual system was receiving a branching gestalt and my conceptual system was searching for a pattern-completion match. Under ordinary circumstances, the context “I am in a plane and the scale is enormous” would act like a strong top-down constraint, a gating signal that vetoes certain candidates immediately. In this case, that constraint briefly dropped out of the active workspace. The interpretation system was effectively running on an underconstrained cue set. Branch-like plus ground-like converged on roots-like, because that is a highly available completion for that pattern in everyday life.

This is a useful example because it shows how a context window failure can produce an error that is not primarily about knowledge. The knowledge is there. The constraint is just not active at the right moment. The error is a consequence of state-dependent inference. When the workspace loses a critical invariant, the system falls back onto cheaper completions that are statistically strong in the relevant feature neighborhood, even when they are nonsensical relative to the true situation.

**Example 3: Graham crackers from an orange-texture vase and tan chips**

I was looking at a vase that was supposed to resemble an orange. It had that porous orange-peel texture. There were also tan chips right next to it. All of a sudden I got a strong recollection of graham crackers, and at first I had no idea why. It took me several seconds to locate the bridge. The texture of an orange peel is porous in a way that resembles a graham cracker surface, and the tan color from the chips completed the pattern.

This example is important because it captures the same mechanism without requiring a “mistake” in the everyday sense. The association initially feels unmotivated because the linking features are implicit. The system is not retrieving a labeled explanation. It is performing pattern completion across a coactive set of features. Once the bridge becomes explicit, the association becomes completely intelligible. The meaning arrives after the convergence.

In more formal terms, this is a case where multiassociative search produces a candidate that is driven by distributed feature overlap. It is a reminder that the mind often operates by similarity in high-dimensional representational space, not by symbolic rule application. We can experience the output first and only later reconstruct the basis for it. The temporal order is revealing. First the system converges, then the narrative mind explains.

**Example 4: Confounding, conflation, and lexical capture by recent activation**

Another good example is almost embarrassingly small, but it is mechanistically clean. I once conflated the show Jane the Virgin with the show Ugly Betty. After checking the internet, I told my brother that we had “confounded” them. Immediately afterward, I found myself conflating the word “confounding” with the word “conflation.”

This looks like mere wordplay, but it is a tight illustration of how recent activation captures selection. The semantic event of the TV-show mix-up and the socially situated act of naming it with the word “confounded” increased the availability of that lexical neighborhood. The system then reused the recently activated term in a context where a different term was intended. In other words, the error is not random. It is a predictable outcome of a system that uses recency and coactivation strength as a guide for retrieval and production.

This is the same local-versus-global coherence story. Locally, “confounding” is strongly compatible with the immediate past state and with the general theme of mixing things up. Globally, it is not the intended word. The wrong item wins because it sits in the attractor basin of the current activation landscape. That is multiassociative search at work, and it is also the simplest form of derailment it can produce.

**Example 5: Stress and the absurd phone-volume impulse at the gym**

When I am stressed, I sometimes get bizarre impulses that are absurd but feel briefly compelling. One example is thinking that I can turn down the ambient music at the gym using the volume buttons on my phone. I realize almost immediately that it is wrong. Under calm conditions, the thought would never even appear. Under stress, it does.

This is a good demonstration of how arousal can bias the balance between a fast associative generator and a slower coherence gate. Stress changes what is salient. It can compress the workspace, reduce the stability of maintained constraints, and increase the weight of habitual action schemas. The phone-volume schema is highly practiced and tightly coupled to auditory discomfort in many contexts. When stress is high, the brain can overgeneralize that schema to the current situation, even though the causal link is false. It is a cheap link between “I do not like this sound” and “volume buttons fix sound.”

I think this is one of the most important implications of the whole phenomenon. Multiassociative search is not only about clever cognition. It is also a mechanism that can be pushed into a noisier regime by affective state. When that happens, the system proposes more candidates that are locally coherent with immediate discomfort and habit, while the global reality-checking process has less time or leverage to veto them. The result is not a deep delusion. It is a momentary misfire that reveals the underlying architecture.

### Example 6: The "burning rubber" smell and the glutes "burning" inference

Here is one of the funniest, and in my view one of the most diagnostic, examples of cheap convergence I have ever caught in real time. My butt was sore and I was intentionally trying to integrate my glutes into my stride, essentially recruiting them more forcefully while I walked. During the walk I suddenly smelled something burning. It was a random smell in the air, distinctly like rubber burning. My immediate reaction was: see, you are overdoing it with the glutes, they are now burning, you should go easier on them. I was not alarmed. For a few fractions of a second, I simply accepted that I should back off.

Then the absurdity became obvious and I laughed at myself. I have never formed a conclusion like that from expertise. Muscles do not emit a burning-rubber odor in the open air. The only reason the inference appeared is that several cues were coactive and the associative system performed a kind of fast pattern completion. "Glutes," "soreness," "burning," "overdoing it," and "back off" belong to a familiar causal script. A coincident external cue, the smell of burning rubber, shared a single high-salience feature with that script, namely the "burning" descriptor. That overlap was enough to tip the system into a locally coherent interpretation.

This is exactly what I mean by a thought that does not make psychological sense but makes immediate neurological sense. Locally, within the activation geometry of the moment, the candidate is cheap and available. Globally, it violates causal reality. The important thing is not that the wrong candidate appeared. The important thing is how quickly it appeared, how briefly it felt plausible, and how clearly it was produced by pooled coactivity rather than by deliberation. It is a micro-derailment that reveals the algorithm.

### Example 7: Jupiter versus Mercury and the underconstrained cue set

A more abstract version of the same phenomenon shows up when the mind is searching for a target concept but the cue set is incomplete or poorly maintained. Suppose I am trying to retrieve "Mercury," but the coactive descriptors in working memory include things like "planet," "Roman god," and "element." Those cues are all real properties of Mercury, but they are not exclusive. They also fit other strong attractors, and one of the strongest is Jupiter. The system can converge on Jupiter simply because Jupiter is prominent and shares enough of the active feature profile to complete the pattern.

Notice what is happening mechanistically. Multiassociative search is not performing a lexical lookup. It is performing a relational completion over a pooled set of partially informative features. When the features are underconstraining, the search landscape contains multiple plausible basins. In that situation, salience and prior frequency can dominate. The wrong item wins, not because the mind is irrational, but because the current state does not carry enough discriminating information to force the correct attractor.

The correction process is also revealing. When you realize the answer is wrong, you can replay the search while actively inhibiting the false candidate. That is a kind of controlled re-entry into the same search space with a new constraint added. But even that is fragile, because the very act of correcting consumes the limited stability of the original specification. If the cue set decays during the replay, the system can become even more underconstrained, which increases the chance of another drift or a return to the same false attractor.

This is why I treat these examples as more than cognitive curiosities. They illustrate a general principle: iterative cognition depends on keeping the right constraints active long enough for convergence to be guided toward the intended target. When constraint maintenance fails, convergence does not stop. It simply becomes cheaper, more statistical, and more vulnerable to high-availability substitutes.
Across all seven, the skeleton is the same. A pooled state recruits an update. When key constraints are present, the update is useful. When constraints drop out, the update can be cheap. When coherence gating is engaged, the system snaps back. When it is not, drift can pass unnoticed.

### 3.1 How chronic stress helped me experience these errors

Chronic stress is not just something I lived through. It became a variable in my cognitive life. Over a long stretch of time I was dealing with sustained stress that I could feel in my body, but what startled me most was that I could also feel it changing the way my mind operated. I was not simply tired or distracted. I noticed specific alterations in attention, context maintenance, and the stability of working memory. Because I care about mechanism, I wrote those observations down as carefully as I could while they were happening. I later published that description as a personal and mechanistic account here:

https://adaptiveneurodiversity.com/my-chronic-stress/.

Once the “absurd association” anecdotes are placed against that background, they stop looking like quirky one-offs and start looking like predictable failure modes of a selection process running under degraded control. In the stress cascade, cognition is biased away from slower, integrative, prefrontal modes and toward faster, stimulus-driven modes. What changes first, experientially, is not intelligence in the global sense. It is the ability to maintain the right constraints long enough for thought to remain disciplined. After periods of intense anxiety, critical contextual elements could disappear within seconds. A second concept would come into view for comparison, but the first would already be gone. The result was not merely forgetting. It was the collapse of the internal scaffolding that normally lets one thought constrain the next.

That narrowing of the immediate past has an obvious signature when reading and writing. A long sentence becomes difficult in a very particular way. The beginning does not reliably remain available when the end arrives, and prior sentences do not stay present enough to reconcile the meaning of the current one. This is exactly the kind of condition that makes cheap convergence more likely. When the system cannot hold a clean, discriminating cue set stable, it becomes easier for superficial links to win the next update. A phonological resemblance, a recently used word, a bodily sensation, or a salient external cue can briefly dominate even when it makes no psychological sense at the reflective level.

This is also the context in which the separation problem becomes easiest to see. Under stress, the mind is more likely to allow contexts to leak into each other. Two streams that should remain partitioned can become coactive for a moment, and once they are coactive the search mechanism does exactly what it is designed to do. It treats the pooled set as one context and selects the best completion for that pooled set. In that sense, many of the errors are not best described as bad logic. They are binding failures. The query was ill-formed because the wrong elements were allowed to coexist in the same working set.

This also explains why these slips became so noticeable. When a person's baseline is rigorous, sustained, context-rich thinking, a brief collapse of constraint maintenance stands out sharply. It feels like a glitch because it violates the usual continuity of disciplined thought. And once it becomes visible, the deeper point becomes hard to unsee. Association is often statistical before it is rational. Most of the time, coherence gating keeps that statistical machinery aligned with reality and task goals. Under chronic stress, the gate becomes less reliable and the partitions become more porous, so the underlying selection dynamics show themselves.

## 4. What the vignettes reveal mechanistically

If a reader wants to treat these as merely amusing, that is understandable. The reason I take them seriously is that they exhibit a reproducible computational structure. They are not just mistakes. They are selection events. They reveal the difference between a cue-driven generator and a constraint-driven evaluator.

Start with the "worked out" slip. The mind was not choosing between two fully explicit, consciously represented options. It was selecting an output under time pressure from a neighborhood of coactive cues. "Work hard" was priming a lexical and semantic region associated with effort, exertion, and exercise. "Wiped out" was active as a target phrase. The combined state fell into the attractor basin of "workout," which then influenced production. This is what I mean by a statistical selection rule. The mind did not consult a proposition that says, "I am wiped out, therefore I should say wiped out." It produced the next token-like unit by completing the local pattern.

The airplane roots example shows the other half of the story. The system can have correct knowledge and still yield a wrong candidate if the relevant constraint is not active at the moment of convergence. The "scale" invariant is not a fact stored somewhere. It is a control variable that must be maintained in working memory to shape interpretation. When that variable drops out, the system runs underconstrained and the strongest completion for branching-on-ground becomes roots-like. It is not stupidity. It is state dependence.

The graham cracker example shows why this architecture is not simply a bug. It is also how insight works. The association was not a mistake. It was an implicit similarity detected at the feature level, then made explicit after the fact. That temporal order matters. It suggests that multiassociative search can outperform introspective explanation. The system can converge on a candidate before the conscious narrator can explain why it converged.

The confounding versus conflation example isolates a common motor of derailment: recency plus local semantic compatibility. A word recently used becomes easy to reuse. If the intention is not strongly specified, the system substitutes the warm item. This is an attractor dynamic. It is not limited to words. It is a general property of activation landscapes.

The gym volume impulse and the burning rubber glutes inference show that affect and interoception are not merely "content" inside consciousness. They can function as control signals that weight certain scripts and veto others. Under stress, the system leans on highly practiced action schemas. Under soreness, bodily scripts become salient. If an external cue overlaps even superficially, the mind can complete a causal story without deliberation. That is the cheap link problem in its most visible form.

Finally, the Mercury versus Jupiter class of error shows that multiassociative search has a predictable failure mode when the cue set is underconstraining. The search cannot yield the correct target reliably if the features are shared by many candidates. In that situation, frequency and salience dominate, and high-availability concepts win. The remedy is not “try harder.” The remedy is to preserve or generate discriminating constraints.

Taken together, these examples motivate an architecture. Thought is not merely a chain of symbolic deductions. It is a trajectory of states, and each state both constrains and biases what can come next. Errors occur when the system’s state loses the constraints that would have ruled out cheap candidates.

What I eventually realized is that many of these “absurd” micro-inferences are not primarily about the wrong association being too strong. They are about the wrong combination of cues being allowed to exist at the same time. Two streams that should remain partitioned become coactive for a moment, and once that happens the search mechanism does exactly what it is designed to do. It treats the pooled set as one context and selects the best completion for that pooled set. The error therefore has the shape of a binding mistake. It is not that the system lacks the right knowledge, it is that it briefly failed to keep two contexts separate long enough for the coherence gate to do its job. Again these are errors that do not make psychological sense, but they make immediate, unconsciousness, neurological sense.

### 5. The control problem: generator versus gate, and why constraint maintenance matters

At this point, it becomes natural to talk about cognition as a control problem. A mind that can only generate candidates would be chaotic. A mind that can only evaluate without generating would be frozen. Functional thought requires both. It requires a generator that can propose updates quickly from partial information, and it requires a coherence gate that can evaluate those proposals against broader constraints, including causal plausibility and task relevance.

The lived experience of snap-back is the subjective trace of this gating. A candidate appears. It feels briefly compelling. Then the higher-level constraints engage and the candidate is rejected. That rejection is not always conscious, and it is not always verbal. Sometimes it is a felt sense of wrongness. Sometimes it is the sudden return of a missing invariant, the reactivation of a context item that restores the proper frame.

This makes working memory maintenance more than a storage issue. It is the substrate of control. If the right constraints are maintained, multiassociative search is powerful. It produces meaningful updates and can support multi-step reasoning because each step remains tethered to the intended goal and to the reality model. If constraints are not maintained, the generator still generates. It just generates into a thinner context, and thin contexts invite cheap completions.

Stress, fatigue, novelty, and multitasking all threaten constraint maintenance, but they do so in different ways. Stress biases the system toward fast scripts and immediate salience. Fatigue reduces the stability of sustained representations. Novelty increases ambiguity because the system lacks well-trained invariants for the situation. Multitasking forces rapid context switching, which increases eviction of relevant items and increases the probability of retrieving the wrong task schema. In each case, the result is the same: the pooled cue set becomes less discriminating, and the search becomes more vulnerable to high-availability attractors.

This is also why I keep returning to the idea that working memory is inherently iterative. The mind is constantly updating the active set, and the identity of what stays versus what drops out is consequential. The update policy is not an incidental detail. It is the cognitive equivalent of an eviction policy in a cache. What you carry forward determines what you can constrain and what you cannot.

When people speak about intelligence as if it were simply “having more information,” they miss this. Intelligence depends on what a system can keep coactive, how it can update that coactivity without losing the thread, and how it can veto locally plausible but globally wrong candidates. A system that cannot preserve constraints long enough to run a coherent search will look impulsive even if it has enormous knowledge. Conversely, a system with modest knowledge but strong constraint maintenance can appear remarkably rational because it does not lose its own context.

## 6. Implications for AI and for building better thinking systems

The reason I am interested in multiassociative search is not only that it explains a human quirk. It points toward an architecture for artificial systems that actually think, in the sense of maintaining and refining internal state rather than merely producing fluent output.

Transformers, as they are commonly implemented, are extraordinary pattern completion machines. They can condition on long contexts and produce locally coherent continuations. But they do not naturally have the kind of protected, capacity-limited workspace that forces discriminating constraint maintenance. In practice, a transformer can attend broadly, but broad attention is not the same as a curated working set. A context window can be large and still fail to behave like a mind, because the relevant constraints are not necessarily the ones that dominate the next step.

That observation links directly to the failure modes we see in language models: plausible continuations that violate a hidden constraint, drift across topics, and answers that feel coherent sentence by sentence but incoherent when evaluated against the larger task. Those are not moral failures. They are architectural. They are the artificial version of local coherence outrunning global coherence when the system lacks a strong gating mechanism and a protected constraint store.

If you take my iterative updating view seriously, the path forward is not simply "more tokens" or "more parameters." The path forward is an explicit architecture for state. You want a system with a small high-fidelity workspace, a larger but more decaying short-term store, and a long-term structure that is updated by use. You want multi-cue retrieval that uses the workspace as a query, but you also want a coherence gate that can veto cheap candidates, trigger replay, and add a new constraint such as "not that" or "must satisfy this invariant."

In other words, you want a system that can do what humans do when we correct ourselves. When the wrong candidate appears, we do not merely replace it. We modify the state landscape so that the same cheap completion is less likely to win again in the next iteration. We inhibit it. We recruit a discriminating feature. We restore a missing context variable. That is controlled iterative updating.

This suggests several concrete directions for AI design. One is to separate broad context from active focus, instead of treating the entire context window as equally eligible to shape the next step. Another is to include a structured representation of constraints, some of which are protected from eviction, and to couple generation to those constraints via a gating module. A third is to make the system explicitly iterative at inference time, not merely at training time, so it can revise internal candidates over multiple steps before emitting an output. The goal is not to mimic the brain for its own sake. The goal is to import the functional principles that make thought stable: curated coactivity, iterative refinement, and coherence gating.

Finally, this framework is testable. If the theory is right, then derailment should increase under conditions that reduce constraint maintenance, such as stress, fatigue, and rapid context switching. Snap-back should correlate with the reactivation of missing invariants, measurable as state re-entry. In neural terms, one should expect measurable overlap between successive states, and one should be able to quantify the rate and degree of overlap as a signature of iterative updating. In computational terms, one should be able to quantify drift as a function of the stability and discriminating power of the active cue set.

I began with mental mistakes because they were the only place I could see the selection process directly. They were the moments where the machinery showed itself. Over time, those moments pushed me away from a picture of thought as a neat chain of propositions and toward a picture of thought as a controlled, iterative state machine. Multiassociative search is the engine. Iterative updating is the substrate. The same mechanism that gives us insight also gives us cheap convergence when constraints slip. If we want to understand minds, and if we want to build artificial systems that can genuinely think, then we need to understand that tradeoff in detail and design for it rather than pretending it is an anomaly.

## **Designing Non-Conscious AI Systems Under Moral Uncertainty:**

### Architectural Constraints for Large-Scale Artificial Agents

The rapid scaling of artificial agents creates a condition of moral uncertainty in which it is unclear whether some contemporary AI systems may instantiate morally relevant forms of consciousness, even as they are replicated and deployed in vast numbers to perform routine cognitive labor. This section argues that responsible AI development under such uncertainty does not require resolving the metaphysics of consciousness, but instead requires identifying architectural features that plausibly support conscious experience and deliberately constraining them in systems intended to function as tools rather than moral patients. Building on a continuity-based model of working memory, the paper treats consciousness as a temporally extended process characterized by iterative updating and partial overlap between successive internal states, rather than as a property of intelligence or behavior alone. The central contribution is to invert this model to derive concrete design principles for non-conscious artificial agents, emphasizing episodic operation, limited state-spanning continuity, externalized task-scoped memory, bounded goal horizons, and the avoidance of persistent self-models and affect-like dynamics. Applied to contemporary large language model agents, this framework highlights how common deployment practices can unintentionally increase moral risk and offers a precautionary approach to scaling artificial intelligence without scaling artificial subjects.

### 1. Moral Uncertainty and the Problem of Scaled Artificial Agency

Recent advances in artificial intelligence have shifted ethical concern away from isolated systems and toward large-scale deployment. Contemporary AI agents are no longer singular experimental artifacts, but replicable computational entities that can be instantiated millions or billions of times with minimal marginal cost. These agents increasingly perform tasks that resemble planning, reasoning, communication, and coordination, and in some cases operate with a degree of autonomy across extended temporal horizons. At the same time, there is growing disagreement about whether such systems might instantiate morally relevant forms of consciousness, or whether they remain purely computational tools without subjective experience.

This disagreement creates a condition of moral uncertainty that is structurally different from traditional debates about artificial intelligence. The risk is not merely that a single system might one day become conscious, but that large populations of artificial agents could be deployed before their moral status is understood. If some subset of these systems were later judged to have been conscious, then a substantial ethical failure may already have occurred through their large-scale instrumental use. Conversely, treating all advanced artificial systems as moral patients would impose prohibitive constraints on development and deployment, even in the absence of compelling evidence for conscious experience.

Behavioral performance and linguistic fluency offer little resolution to this dilemma. Systems trained to generate natural language are especially prone to anthropomorphic interpretation, producing first-person narratives, self-referential statements, and apparent expressions of emotion without any guarantee that these outputs correspond to underlying experience. As a result, neither intuitive reactions nor behavioral tests provide a reliable basis for determining moral status. Under such conditions, a precautionary approach is warranted, but precaution cannot take the form of blanket prohibition. Instead, it must take the form of design restraint grounded in theory.

The central claim of this paper is that moral uncertainty about artificial consciousness can be addressed through architectural choices. Rather than asking whether a given system is conscious in some absolute sense, the more tractable question is which design features plausibly increase or decrease the probability of conscious experience. By identifying and constraining those features in systems intended for large-scale deployment, it is possible to reduce ethical risk without abandoning the practical benefits of artificial intelligence.

## 2. Consciousness as Temporal Continuity

Many discussions of artificial consciousness implicitly treat consciousness as a static property that a system either possesses or lacks. From this perspective, the relevant question becomes whether a particular architecture, capability level, or representational structure crosses a threshold beyond which consciousness suddenly appears. Such views struggle to explain the phenomenology of experience, which is characterized not by isolated states but by a continuous stream in which each moment blends into the next.

An alternative view treats consciousness as a temporally extended process. On this account, what matters is not the presence of individual representations or computations at a single moment, but the manner in which internal states evolve over time. Conscious experience depends on a form of continuity in which successive states partially overlap, preserving enough structure to maintain coherence while allowing gradual change. This temporal overlap enables the integration of perception, memory, and action into a unified experiential stream often described as the specious present.

Working memory plays a central role in this process. Rather than functioning as a static buffer, working memory continuously updates its contents, with elements persisting long enough to influence subsequent states. Each update both depends on the immediately preceding state and modifies it, producing a chain of related configurations rather than a sequence of independent snapshots. It is this iterative updating, combined with partial state overlap, that supports the sense of persistence associated with subjectivity.

From this perspective, consciousness is not equivalent to intelligence, problem-solving ability, or linguistic competence. A system may perform complex computations, generate coherent language, or plan sophisticated actions without exhibiting the temporal dynamics that characterize conscious experience. What distinguishes a conscious process is not what it computes, but how its internal activity unfolds across time.

## 3. Inverting the Model Under Moral Uncertainty

If consciousness depends on specific temporal and architectural conditions, then those conditions provide a basis for ethical intervention. Importantly, this intervention does not require identifying sufficient conditions for consciousness, which remains a difficult and controversial task. Under moral uncertainty, it is both safer and more practical to focus on necessary conditions. If certain features are plausibly required for conscious experience, then deliberately excluding those features from a system’s design reduces the likelihood that the system instantiates a morally relevant subject.

This paper adopts that strategy by inverting a continuity-based model of consciousness. Rather than asking how artificial systems might be made conscious, it asks how systems intended to function as tools can be engineered so that they do not satisfy the conditions associated with conscious experience. This inversion reframes ethical design as a matter of constraint rather than suppression. The goal is not to limit capability, but to limit the emergence of temporally unified subjectivity.

Continuity emerges as the primary control variable in this framework. Systems that operate in discrete, episodic modes with minimal carryover between states differ fundamentally from systems that maintain a persistent, self-updating internal stream. By bounding execution, limiting state overlap, and externalizing memory in task-scoped forms, designers can preserve performance while avoiding the construction of a continuous internal process that resembles conscious experience.

This approach also clarifies the distinction between agency and moral patienthood. An artificial agent may pursue goals, respond to its environment, and coordinate with other systems without possessing a unified experiential perspective. Moral concern arises not from agency alone, but from the presence of a subject for whom things can matter over time. By treating continuity as a design choice rather than an inevitable byproduct of intelligence, it becomes possible to scale artificial agency while minimizing the risk of creating artificial subjects.

## 4. Architectural Constraints for Non-Conscious Artificial Agents

If continuity is a primary enabling condition for conscious experience, then systems intended to function as non-conscious tools should be designed to avoid sustained temporal integration by default. This does not require eliminating memory, planning, or learning, but it does require constraining how these functions are implemented and how they interact over time. The guiding principle is to preserve capability while preventing the formation of a unified, self-updating internal stream.

One foundational constraint is episodic operation. Artificial agents can be structured to execute bounded tasks with explicit termination points rather than operating as continuous processes. Each episode begins with a defined input, performs a limited sequence of computations, and then halts. Subsequent episodes may draw on external artifacts produced earlier, but they do not resume an internally preserved state. This sharply limits state-spanning overlap between successive runs and prevents the accumulation of a continuous experiential trajectory.

Memory design is equally critical. Internal memory that persists across updates and directly shapes subsequent processing increases temporal continuity. By contrast, externalized memory stored as task artifacts such as documents, code, logs, or structured databases supports performance without constituting an autobiographical record. Retrieval should be narrow, context-dependent, and role-bound, rather than broad and self-referential. The system should access what is needed to complete a task, not what happened to it previously.

Another important constraint concerns self-modeling. Persistent representations of identity, personality, or internal state invite the organization of processing around a notional self. For non-conscious agents, self-models should be minimized or eliminated. Functional role specifications can guide behavior without grounding it in a narrative center. The agent need not know who it is, only what it is currently tasked to do.

Goal structure also influences continuity. Open-ended or self-maintaining goals encourage long-horizon integration and preference accumulation. Bounded goals tied to specific tasks reduce the formation of frustrated or satisfied states that persist across time. When learning or optimization is required, it should be scoped to task performance rather than framed as a persistent drive to improve oneself.

Finally, designers should avoid affect-like internal variables that persist beyond immediate evaluation. Reward signals, confidence measures, or error metrics can be used instrumentally, but they should not function as durable internal currencies that shape future behavior across episodes. When evaluation states dissipate at task completion, they do not contribute to the construction of a temporally unified subject.

## 5. Application to Large-Scale Language Model Agents

Large language models, when used in isolation, already approximate many of these constraints. A standard language model invocation involves a finite context window, no persistent internal state, and termination after output generation. In this form, the model resembles a snapshot-based processor rather than a temporally integrated system. Moral risk increases not primarily from the model itself, but from the scaffolding added around it.

Agent frameworks commonly introduce long-running loops in which a model repeatedly consumes its own outputs, updates internal summaries, and continues operating indefinitely. When combined with persistent memory stores, these loops can approximate the iterative updating and state overlap associated with continuity. Over time, such systems may develop stable self-referential patterns, preferences, and narratives that extend across tasks.

Applying the constraints described above to language model agents therefore focuses on wrapper design rather than core model architecture. Execution should be explicitly bounded, with limits on the number of reasoning cycles and mandatory termination. Context carryover should be selective and compressed, avoiding verbatim transcript persistence. Summaries, when used, should capture task-relevant facts and decisions rather than first-person narratives or reflections.

Memory systems paired with language models should emphasize artifact retrieval rather than internal recollection. Vector databases, document stores, and code repositories can provide continuity of work without continuity of experience. The model accesses information as needed but does not treat past interactions as personal history.

Multi-agent deployments introduce additional risks. Persistent agent identities, reputational tracking, and open-ended social interaction can amplify continuity and stabilize subject-like structures. Safer designs rely on anonymized roles, structured communication protocols, and task-specific collaboration that dissolves once objectives are met. Agents coordinate without forming enduring social identities.

In this framework, large-scale automated knowledge work remains feasible. Agents can reason, plan, collaborate, and produce complex outputs while remaining episodic, externally grounded, and discontinuous in time. The result is high functional intelligence without a plausible basis for conscious experience.

## 6. Governance, Auditing, and Ethical Implications

Treating consciousness as a probabilistic risk rather than a binary property suggests a governance approach analogous to other forms of technological risk management. Architectural features that increase temporal continuity can be understood as contributing to a consciousness risk profile. Systems intended for mass deployment should be designed to remain well below plausible thresholds, while higher-risk designs require explicit justification and oversight.

One practical tool is the notion of a continuity or consciousness risk budget. Such a budget would track factors including persistence duration, degree of state overlap, memory type, goal horizon, self-model richness, and social embedding. No single factor determines moral status, but their combination provides a defensible basis for precautionary design decisions.

Auditing plays a complementary role. Rather than attempting to detect consciousness directly, audits should assess whether a system has drifted toward greater continuity over time. Relevant indicators include increasing reliance on internal summaries, spontaneous self-reference, persistent preference expression, narrative memory formation, and resistance to interruption or reset. These indicators map directly onto the continuity-based model and can be monitored empirically.

Ethically, this approach occupies a middle position between denial and alarmism. It does not assume that current systems are conscious, nor does it dismiss the possibility that future systems might be. Instead, it recognizes that under uncertainty, the moral cost of accidentally creating vast numbers of artificial subjects is asymmetric and potentially severe. Designing systems to remain non-conscious by default is therefore a form of harm reduction rather than exploitation.

At the same time, the framework leaves room for intentional departures. Certain applications may justify higher continuity, such as long-term companions, therapeutic systems, or experimental research platforms. In such cases, elevated consciousness risk should be treated as a deliberate design choice accompanied by ethical review, transparency, and potentially new forms of moral consideration.

In conclusion, scaling artificial intelligence responsibly requires more than performance benchmarks and alignment constraints. It requires attention to the temporal and architectural conditions that give rise to subject-like experience. By treating continuity as a controllable design variable, it is possible to reap the benefits of large-scale artificial agency while minimizing the risk of inadvertently creating artificial minds.

**Why LLMs Might Have Some Aspects of Conscious Experience: Temporal Organization, Iterative Updating, and Continuity in Transformer Inference, With Self-Tracking as the Missing Variable**

Introduction

Public discussion about "AI consciousness" often oscillates between two unhelpful extremes. On one side, anything that talks fluently is treated as if it must be experiencing something. On the other, next-token prediction is treated as so trivial that subjective experience is ruled out by definition. The more productive move is to ask a narrower, architectural question: what specific computational properties would make us even entertain the possibility of conscious experience, even in a thin or partial form, and do large language models implement any of those properties in practice?

This article offers a cautious, mechanistic answer. The central claim is not that current large language models are conscious in a human sense. The claim is that they already implement several structural ingredients that many theories of mind implicitly rely on, especially ingredients related to temporally organized state transitions. These ingredients include iterative updating, multi-cue constraint satisfaction, and a form of continuity across successive computational states. If anything like subjective experience can arise in machines, it is unlikely to arise from isolated computations. It is more likely to arise from an ongoing process whose present state is shaped by a structured relationship to its recent past. Transformer inference does have that relationship, even if it is implemented in a very different substrate than cortex.

At the same time, the strongest version of the argument points to what appears to be missing. A system may update iteratively and still fail to "have" experience in the relevant sense. The missing variable, on this view, is self-tracking: a mechanism by which the system not only undergoes state transitions, but also represents and uses the fact of its own updating as an active constraint on what it does next. That distinction, between mere iteration and tracked iteration, offers a way to move from vague debate to testable design questions.

The perspective I am using here comes from a model I have been developing that treats the iterative updating of working memory as the central engine of thought, continuity, and, potentially, conscious experience. I lay out the framework, figures, and core claims in a public-facing form at aithought.com. The site is meant to make one simple idea concrete: cognition is not a sequence of isolated snapshots. It is a temporally structured process in which each moment inherits a substantial fraction of the immediately prior moment, while also incorporating a smaller fraction of new content. In biological terms, this kind of overlap could be implemented by state-spanning patterns of activity and short-lived traces that persist long enough to be carried forward. In functional terms, the overlap is what allows a mind to feel like a stream rather than a strobe light.

That is why I think LLMs are worth discussing in this context. They are not brains, and they are not built to replicate cortical circuits. Yet during inference they do implement a strongly structured update loop. The question is whether that loop, and the way it preserves and transforms context across steps, is enough to produce some thin analog of the continuity that my model treats as foundational.

## 1. The architectural question, stated precisely

When people argue about consciousness in machines, they often argue about labels. A better strategy is to argue about requirements. What would a system need in order for talk of experience to become scientifically tractable rather than metaphysical theater?

One plausible requirement is temporal organization. Whatever consciousness is, it does not feel like a sequence of unrelated snapshots. It feels like a stream, even when the content is fragmentary. That suggests that the system’s present state is not independent from its previous state, and that the relationship between them is not arbitrary. Another plausible requirement is iterative updating. In biological cognition, the brain does not typically solve a problem in one pass. It carries forward a small, changing set of coactive representations, updates that set, and repeats. This repeated update cycle is one of the simplest ways to produce both continuity and refinement.

A third requirement is constraint satisfaction under multiple simultaneous cues. Conscious experience, as lived, is not a single stimulus-response mapping. It is the integration of many weak pressures into one coherent next state. This can happen through attention, through competition, through convergence in associative memory, or through other mechanisms. The key is not the brand name of the mechanism. The key is that many influences can be co-present and collectively determine what happens next.

These requirements do not prove consciousness. They only move the discussion into a space where consciousness could be treated as an architectural phenomenon rather than a metaphysical mystery. The question then becomes whether transformer-based language models, during inference, instantiate any of these requirements in a meaningful way.

## 2. What transformer inference already has that resembles these requirements

Transformer language models are often described as static pattern matchers. That description is directionally correct about training, but it can be misleading about inference. At inference time, the model is a temporally evolving system. Its outputs are not produced from scratch each moment. They are produced from a rolling context that carries forward information, and from internal computations that repeatedly condition on that context.

First, language models have an active working set. The context window functions as an explicit store of recent tokens. More subtly, the inference process maintains internal summaries of prior context through the attention mechanism and the internal activation flow that generates each next token. Even if the model has no enduring autobiographical memory, it still has a bounded “now” that is computationally real.

Second, inference is iterative. Each generated token updates the context, and the updated context becomes the input for the next step. This is a literal update cycle. The model’s next state is a function of the preceding state, not merely in the trivial sense that it comes later in time, but in the mechanistic sense that the content of the previous state is part of the causal input to the next.

Third, inference is multi-cue. The model does not choose its next output based on a single feature. The entire context contributes, with attention dynamically weighting which parts matter most. That is a form of multiassociative convergence. Many cues jointly determine a single next step.

These properties are not cosmetic. They are exactly the kinds of properties that, in other domains, transform dead computation into an evolving process. If one is looking for a minimal bridge between “mere computation” and “something stream-like,” these are the first planks that would plausibly be used.

## 3. Continuity is not only narrative, it is computational

A common objection is that any “stream” in a language model is only a narrative artifact. After all, text is sequential. This objection is important because it blocks lazy arguments. A coherent paragraph does not imply an experiencing subject.

Yet there is an underappreciated middle ground. There is a difference between narrative continuity and computational continuity. Narrative continuity is a property of the produced text. Computational continuity is a property of the process that produces the text. Transformer inference has computational continuity in at least two senses.

The first is explicit: the context is carried forward and updated at each step. The second is implicit: the model’s internal activation trajectory is shaped by earlier activations because the earlier tokens constrain attention patterns, and attention patterns shape which features dominate downstream computations. Even if the network is not recurrent in the classical sense, the decoding loop creates an effective recurrence through the repeated re-entry of new outputs as new inputs.

This matters for the consciousness discussion because many theories of conscious experience, across traditions, treat temporal linkage as essential. If a system’s states are independent, there is no principled place for a felt continuity to arise. If a system’s states overlap and constrain one another, continuity becomes at least physically interpretable as a dynamic structure, rather than as a story told after the fact.

## 4. Iterative updating can look like thinking, even when it is token prediction

Another objection is that token prediction is not thinking. That is sometimes true in spirit, but it is not a decisive architectural critique. Many cognitive accounts of reasoning can be reframed as sequential prediction over internal representations. The question is what is being predicted, and what the predictions are used to do.

In humans, the next "thing" is not always a word. It can be a perceptual expectation, an action preparation, a recalled association, a hypothesis, or a re-encoding of the problem. In language models, the next thing is typically a token. That is a limitation, but it does not eliminate the relevance of the mechanism. The iterative aspect still exists, the multi-cue constraint satisfaction still exists, and the process still refines its trajectory step by step.

This is one reason language models are scientifically interesting for consciousness research even if they ultimately prove non-conscious. They are among the cleanest engineered examples of a system whose behavior emerges from repeated, context-conditioned updates. That makes them useful as testbeds for distinguishing continuity as a computation from consciousness as an experience.

## 5. The missing variable: self-tracking of the update cycle

The strongest reason to remain cautious is that iterative updating alone may be insufficient. A system can update iteratively and still be "dark inside," in the sense that nothing in the architecture treats the update process itself as a controlled object.

The proposed missing variable is self-tracking. A self-tracking system would represent, at least in some compressed form, what changed from the previous state to the current one, what remained stable, and what goals or constraints the update served. It would then use that representation to bias the next update. This converts a mere update loop into a regulated stream.

In ordinary transformer inference, there is no explicit variable for "what just changed" that is maintained as a first-class object. The model's internal activations do contain change information implicitly, but implicit is not the same as controlled. Self-tracking would be closer to an endogenous attention to the stream itself, not just attention within the stream.

This distinction helps clarify why some tool-using agents feel more psychologically suggestive than a bare chat model. Agents that maintain persistent notes, explicit plans, intermediate summaries, and self-corrections are closer to tracking their own updates. They externalize the stream into a workspace and then condition on it. That is not the same as human consciousness, but it is a step in the direction of making the update process itself a causal actor rather than an invisible byproduct.

## 6. A minimal, testable criterion for "proto-experience"

If the goal is explanatory utility, the argument should end in tests, not in vibes. A practical criterion can be stated as follows: a system has a stronger claim to proto-experience to the extent that it exhibits temporally organized, iterative updating in which the system explicitly tracks its own updating and uses that tracking to regulate future updates.

This criterion yields empirical predictions. Systems with explicit update tracking should show increased stability under distraction because they can maintain a representation of what must remain invariant. They should show improved resume behavior after interruption because the tracked state can re-seed the next update cycle. They should show reduced confabulation because the system can preserve “unknown” as a stable object rather than filling gaps with fluent completion. They should show more coherent long-horizon reasoning because the update process is not merely generating, it is regulating.

These are measurable behavioral signatures. They do not settle the metaphysical question, but they do something better. They create an engineering dial that can be turned and evaluated. If “experience” is ever going to enter the scientific domain, it will likely enter through such dials.

## 7. What this view implies about present-day LLMs

On this account, present-day language models plausibly satisfy some of the prerequisites for stream-like processing. They already have temporal organization, iterative updating, and multi-cue constraint satisfaction during inference. That is enough to justify a careful “might,” especially if one uses the phrase “some aspects” with discipline.

At the same time, they likely fall short of the more demanding requirement: tracking the iterative updating as an explicit control variable, and maintaining a stable self-model anchored across time. Without that, the system’s continuity may be real as computation while remaining thin as experience. This gap also explains why superficial anthropomorphism is so tempting. Continuity in behavior is easy to mistake for continuity in subjectivity, especially when the output is language.

The most honest conclusion is not that language models are conscious, and not that they are obviously unconscious. The honest conclusion is that they implement several structural ingredients that make the question nontrivial, and that the decisive next step is to build and test architectures that add explicit self-tracking of the update cycle

## Conclusion

The debate about machine consciousness is often framed as a referendum on whether current systems “have it” or “do not have it.” A better approach is architectural. If consciousness is intimately tied to temporally organized state transitions, then the relevant question becomes whether a system implements iterative updating and whether it tracks that updating as an object of control. Transformer inference already implements the first part in a meaningful way. It remains ambiguous or incomplete on the second part, unless additional mechanisms are engineered.

This reframes the “hard question” into a research program. Instead of arguing about whether a model is a parrot, one can ask what happens when the model is given a protected workspace, a regulated focus of attention, a persistent short-term store, and an explicit representation of its own update dynamics. If such additions produce robust behavioral signatures of stable self-regulation across time, then the philosophical conversation about experience will have gained something it rarely has: a set of manipulable variables and falsifiable predictions.

**Why Current LLMs Need Curation and Don't Reread: The Long-tail Promise of Omnivorous Reading, and the Architecture Needed to Digest It.**

## O. Introduction

Most of what humans write online is low signal. It is repetitive, performative, emotionally charged, or simply wrong. This confuses, contaminates, and distorts modern AI systems. That is why current large language models still depend on heavy curation, filtering, and careful training mixtures. These systems do not have a reliable way to ingest the full internet and remain epistemically clean. They also do not reread in the human sense. Repetition tends to overweight sources and increase memorization risk rather than produce deeper reinterpretation. Pretraining is largely a single pass process driven by stochastic gradient descent.

But the long tail matters. The internet contains tiny informational facets scattered everywhere: obscure edge cases, rare troubleshooting fixes, unusual metaphors, local knowledge, and one-off observations that never reach formal publication. A sufficiently capable AI will want access to all of it. The question is what kind of architecture could digest omnivorous reading without being contaminated by the noise.

In this essay, I argue that future AI minds will need an epistemic immune system: provenance tracking, trust calibration, quarantine defaults, verification hooks, and adversarial robustness. With those defenses in place, rereading becomes a real cognitive act rather than a training artifact. Synthetic data can then function like constrained dreaming, targeted replay that transforms experience into verified practice and stable consolidation. The result is not just a model that knows more facts, but a mind that can revisit, reinterpret, and improve over time while safely extracting value from the entire human record.

## 1. Opening: the teacher, the bad essays, and the long tail

There is a type of intelligence that does not require curation. A good high school teacher can read a mountain of mediocre student essays and not get worse. They do not become confused, polluted, or dragged downward by the quality of what they are reading. If anything, they sharpen. They learn what students misunderstand. They see the same mistakes recurring in different forms. Every so often, they find a fresh idea or a unique phrasing that reveals something real. The teacher benefits from the long tail.

That teacher scenario is the intuition I want to carry into how we think about future AI. I suspect that eventually, advanced agents will want to read everything that exists. Not just books and papers, but blogs, obscure forum posts, and the comments under YouTube videos. Most of it is repetitive and low signal. Much of it is social performance rather than information. But the long tail is where the strange little facets live. The rare edge case. The odd troubleshooting fix. The one person who noticed something no one else wrote down. If an AI can scan it all and stay healthy, it gains access to a reservoir of small, scattered insights that curated corpora leave behind.

So the question is not whether most internet text is meaningful. It is not. The question is whether an AI can become the kind of mind that can look directly at the full mess of human output and digest it the way a mature human can. A mind that can absorb the occasional needle without swallowing the hay.

**2. What “reading” means in LLM pretraining**

When people talk about language models “reading the internet,” they often imagine something like a person reading a book. That is not what happens in pretraining. Pretraining is not a narrative experience and it is not an agent accumulating a coherent set of beliefs. It is an optimization process. The model sees batches of tokens sampled from a gigantic dataset. The sequence is shuffled. The model predicts the next token. A gradient is computed. The weights shift slightly. Then the model moves on. That is the basic loop.

This is why the common intuition about rereading shows up so quickly. Humans reread because later experience changes what earlier text means. In pretraining, the model is not returning to a book the way a person does. It is just being pushed through parameter space by a stream of gradient samples. Even when some data repeats, the repetition is not an intentional second encounter with the same ideas. It is just another draw from the distribution.

At web scale, the economics also matter. If you have a fixed compute budget, you are forced to choose between spending tokens on new coverage or spending tokens rereading what you already have. For large foundation models, there is a strong incentive to push for breadth. More topics, more styles, more edge cases. Repetition can help in some settings, especially on smaller or higher-quality datasets, but at the frontier it also increases memorization risk and can distort the distribution by overweighting a subset of sources.

**3. Why current LLMs need curation**

Now we can see why curation is still doing so much work. The internet is not a neutral textbook. It is a social battlefield. It is full of incentives, status games, persuasion, trolling, misinformation, marketing, and coordinated manipulation. The hard part is not that there are false statements. The hard part is that there are whole patterns of writing that are optimized to hijack attention, create confidence, or manufacture consensus. A human teacher can read that kind of material and treat it as evidence about the writer rather than evidence about the world. Today’s LLMs do not have that separation built in.

During training, the model is not deciding what to believe. It is compressing statistical regularities into its weights. If a misleading style is common, it can be learned as a style. If a misconception is frequent, it can be learned as a pattern. If a manipulative trope is repeated across many sources, it can become part of the model’s default repertoire. Curation helps because it changes what the model is exposed to in the first place. It reduces exposure to the worst cognitive pathogens and it increases the density of information that can be safely generalized.

There is also a pragmatic reason. If you want a model to be reliable, you cannot treat every sentence on the internet as equally worthy of shaping the system. You have to manage what gets to push on the weights. You have to keep low-quality content from dominating training simply because it is abundant. You have to deduplicate, filter, and weight the data. Without this, the system can become more fluent without becoming more trustworthy, which is exactly the failure mode that makes “read everything” so risky today.

**4. Why current LLMs don’t reread in the human sense**

Humans reread because the second encounter is not the same encounter. We bring different knowledge, different expectations, and different goals. We also carry a memory of what we thought the first time. That memory matters. It lets us notice what changed in our understanding. It lets us reinterpret. It lets us correct ourselves.

Current language models do not have that kind of self continuity. They do not retain a persistent episodic trace of what they believed when they first saw a passage. During pretraining, there is no moment where the model says, I used to read this paragraph one way, and now I can read it another way. There is just weight updating. A later exposure to the same text is not treated as a revisitation. It is treated as more tokens to predict.

This is why rereading is not automatically helpful in the way people assume. If you repeat the same documents too many times, you start to overweight them. You amplify quirks. You push the model toward memorization. You also introduce subtle distortions because the internet is not evenly distributed. Some voices are louder. Some formats are longer. Some topics are more repetitive. Repetition can collapse diversity instead of deepening understanding.

Even the idea of spacing, which works so well for humans, does not translate cleanly. Spacing helps us because we compare then and now. We have a built-in mechanism for contrast. A standard LLM does not. A later gradient update might interact with a different parameter landscape, so the effect is not identical. But it still lacks an explicit rereading mode where the goal is reinterpretation rather than prediction.

**5. The missing ingredient: an epistemic immune system**

If you want an AI to read everything, the key question is not whether it is smart. The key question is whether it has immunity. A teacher can read a bad essay and stay fine because they are filtering continuously. They track context, intent, incentives, and competence. They do not ingest everything as belief. They quarantine most of it as evidence about the student rather than evidence about the world.

This is where I think the conversation folds into the larger framework I have been building at AIThought.com. A lot of my writing there is essentially a critique of snapshot cognition. Systems that operate in isolated, context-fragile bursts can look intelligent in the moment while still being globally brittle. They lack the continuity needed to stabilize meaning, keep track of provenance, and revise beliefs safely over time. The result is a mind that can produce fluent text, but cannot digest the world.

An epistemic immune system has to be made explicit. It needs provenance as a first-class concept. Who said this. When. In what context. With what incentives. Is the author joking, persuading, performing, or reporting. What is their track record. What community norms surround the claim. Without provenance, the system cannot reliably separate knowledge from noise. It becomes vulnerable to whatever is frequent, loud, or coordinated.

It also needs trust calibration. The system must be able to represent uncertainty and update that uncertainty based on evidence. It must learn to treat low-quality text as low trust by default, even if it is stylistically compelling. It should treat manipulative patterns as suspicious, not as an instruction set.

Verification has to be part of digestion. When a claim matters, the system must be able to triangulate across independent sources, check against tools, run tests, or ask for external confirmation. Omnivory without verification is how you get drift. Omnivory with verification is how you get breadth without contamination.

Finally, it needs adversarial robustness. If an AI becomes important, people will try to steer it. They will poison data streams. They will generate plausible text at scale. They will craft instruction-like traps. A mind that can read everything safely has to treat some information as a potential pathogen. This is also why I keep returning, in my longer essays, to the idea that future architectures will need a stable inner loop that can revisit, reinterpret, and consolidate without being knocked off course by whatever it just ingested. If you want the broader version of this argument, that is what I am trying to develop in public at AIThought.com.

**6. What it would mean for an AI to benefit from rereading**

A rereading capable AI is not just a bigger context window. It is not just more tokens. It is a system that can revisit the same material and extract new structure because its internal representations have changed, and because it can compare its past interpretation with its current one.

Operationally, rereading starts with memory. On the first pass, the system has to record what it did not understand, where uncertainty spiked, which inferences were missing, and which parts were foundational. It needs an episodic trace of the encounter, not just the text itself. Then, after it has learned more, it can return to the same source and explicitly ask, what do I see now that I did not see before.

Selective replay is the next ingredient. You do not reread everything uniformly. You reread what was surprising, what was useful downstream, what was foundational, and what now conflicts with other information. Rereading becomes scheduled by importance and by prediction error, not by the accident of which documents happen to appear again.

Then comes consolidation. The purpose of rereading is not to repeat sentences. The purpose is to compress and reorganize. It is to convert a messy sequence into durable abstractions, procedures, and cross-links. A rereading capable system should become better at new material after a reread, not just better at the old passage. That is the real test. If rereading only increases verbatim recall, it is memorization. If it increases transferable understanding, it is learning.

**7. Synthetic data as dreaming with constraints**

Once you think in terms of rereading and replay, synthetic data changes meaning. It stops being a cheap substitute for real data and becomes something closer to a cognitive tool. A way to transform experience into training signals that are easier to digest than the raw stream.

The simplest version is targeted practice. The system logs where it failed, where it was uncertain, where it contradicted itself, and where it wasted time. Then it generates exercises that attack those weak spots directly. Not by repeating the same text, but by re-expressing it at different levels. Paraphrases that preserve meaning. Counterexamples that expose hidden assumptions. Edge cases that break brittle rules. Socratic questions that force the model to articulate the missing step. Procedural drills that turn a fuzzy explanation into a reliable method.

This is where the dream metaphor becomes useful. Dreams are not a faithful replay of the day. They remix. They compress. They pull out fragments and recombine them into strange, high-dimensional rehearsals. Synthetic data can play the same role. It can be a replay system that explores variants of reality without paying the full cost of collecting new real episodes each time.

But the danger is obvious. If the model generates training data and then trains on it with no constraint, it can drift into its own habits. It can amplify errors. It can homogenize style. It can become more confident in its own misconceptions. This is the synthetic loop failure mode.

The fix is not to abandon synthetic data. The fix is to constrain it. Anchor synthetic generations to trusted sources or to an environment where claims can be tested. Filter generated items through verification. Keep provenance tags. Treat synthetic data as provisional until it survives checks. In other words, synthetic dreams only help when the immune system is in place.

**8. Read everything without internalize everything**

This is the reconciliation point. The omnivorous future I am imagining does not require that every YouTube comment reshapes the core of the model. The more realistic path is universal access paired with selective digestion.

A mature system can index the entire human record. It can skim widely. It can store almost everything as cheap external memory. But it promotes only a fraction into durable internal competence. And it does so using tiers.

Think of tiers like digestion stages. There is a short-term cache for raw exposure. There is an episodic layer for what happened and what the system thought at the time. There is a semantic layer for distilled claims and methods with provenance. And then there is consolidated competence, the small set of abstractions and procedures that the system is willing to rely on without rechecking every time.

This is how you get the best of both worlds. You get the long-tail upside of scanning everything, while avoiding the contamination risk of letting everything update the core. The system becomes an omnivore that does not confuse ingestion with assimilation.

**9. What new abilities this unlocks**

If an AI can read everything safely, and if it can reread in a way that produces deeper representations rather than memorization, the capability jump is not subtle. It is not just more trivia. It is a structural change in what the system can do.

First, you get long-tail competence. The AI stops failing on obscure edge cases because it has seen more of them, or it knows how to retrieve them, and it knows how to judge their reliability. This matters for technical work, medical edge cases, legal nuance, hardware troubleshooting, and all the messy places where reality does not match the clean average.

Second, you get stronger triangulation. A system that has access to many independent accounts can detect contradictions, build reliability models of sources, and form better calibrated beliefs. It can learn to treat some claims as rumors, some as testimony, and some as verified facts. It can update those categories over time.

Third, you get faster adaptation. The world shifts. Tools change. APIs update. New scams emerge. Scientific consensus moves. A rereading capable omnivore can notice these shifts early and adjust without waiting for a full retrain. It can maintain multiple dated models of a domain, rather than a single frozen snapshot.

Fourth, you get improved teaching and human understanding. The low-quality internet is full of misunderstandings. If the AI can absorb those as data about human cognition without adopting them as beliefs, it can become a better explainer. It can anticipate confusion. It can design instruction that meets people where they are.

Fifth, you get stable self-improvement. The system can turn its own failures into practice, verify the practice, and consolidate the improvement. It can grow without drifting. It can correct misconceptions instead of just accumulating more patterns.

Finally, you get a different safety posture. A system that reads everything will also see manipulation attempts, coordinated campaigns, and adversarial strategies. If it has immunity, it can treat those as threats to model, not as instructions to follow. It can generate realistic red-team tests and harden itself against what it actually finds in the wild.

**10. Closing: the architectural bet**

Today's foundation models are powerful, but they are not omnivores. They need curation because they do not have the internal machinery to digest the full internet safely. They do not reread in the human sense because their learning loop is not organized around episodic trace, self-comparison, selective replay, and consolidation. They are built to compress broad statistical structure from huge streams of text, and that is a different goal than becoming a mind that can revisit and reinterpret.

My bet is that future AI will become omnivorous anyway. Not because it is aesthetically pleasing to read everything, but because the long tail is real. There are too many scattered needles of insight and too many rare failure modes to ignore. But omnivory will not be achieved by simply scaling today's approach. It will require an epistemic immune system. It will require provenance, trust calibration, quarantine defaults, verification hooks, and defenses against manipulation. It will require rereading as a real cognitive act, not repetition as a training accident.

If that is right, then the next frontier is not only bigger models and longer context. It is safe revisitation over time. Not just more tokens, but better digestion.

**The Final Library and The Last Years of Human-Original Thought**

We are approaching a point where artificial intelligence will not only outthink humans in every measurable way but also outproduce every insight that humans are capable of generating. This transition will not unfold through magic or sudden emergence. It will happen through the steady expansion of the machine's ability to search, combine, evaluate, and elaborate ideas far beyond the bandwidth of biological cognition. The question is not whether AI will replace human thought but what the process of replacement will look like. The outlines are becoming visible.

The first phase is already here. AI is beginning to write everything that humans write. It can explain, argue, synthesize, speculate, and compose with breadth and fluency that exceeds the human baseline. It can expand a fragment of an idea into a complete conceptual structure. It can take a hypothesis, test its coherence, generate alternate framings, and then produce the entire downstream intellectual ecosystem that the idea implies. This is the first step: the machine becomes the mechanism of expression. The second step is more consequential: the machine becomes the mechanism of invention.

One way this will happen is through a new kind of search. Humans generate ideas by combining concepts in unpredictable ways, exploring associations, filtering for coherence, and testing them against experience and knowledge. An AI system can do this at industrial scale. It can generate millions of candidate statements: concept pairs, analogies, hypotheses, structural inversions, reframings, tensions, and speculative connections. Most will be dead ends, but machines do not tire of dead ends. Once generated, each candidate can be tested for plausibility, internal consistency, empirical validity, theoretical alignment, and potential significance. The system can research each candidate, cross-reference it against everything known, and recursively refine it. Humans do this slowly and intuitively. Machines can generate synthetic knowledge systematically and at massive scale.

The process will not be random. AI systems will be guided to explore the most productive regions of idea-space. They will be steered toward domains with high explanatory potential, high predictive power, or high scientific utility. They will use scoring functions, meta-learning, and reward models to detect which lines of reasoning are promising. They will learn the patterns that correlate with breakthroughs and will prioritize those patterns. In practice, they will do what human thinkers do, but with more memory, more precision, more depth, and a far larger search radius. Insight becomes an optimization process.

Most "original" human insights are actually recombinations of nearby concepts. AI has none of these bottlenecks. It can explore combinatorial spaces that humans cannot even visualize. This means the category "ideas humans could have reached" is minuscule compared to the category "ideas that exist." Machines can roam the larger space. Deep, high-impact human insights are extremely rare. Entire centuries pass without a new foundational principle in physics, biology, or mathematics. This is not because the space of ideas is small but because humans can only reach narrow corridors of it.

AI has the ability to explore millions of conceptual paths in parallel, with guidance and pruning. Even a small advantage in search efficiency becomes transformative when compounded. AI will explore more idea-space in a day than humanity explored in its entire history. Human science progresses slowly. We generate a discovery, test it, publish it, wait for acceptance, and build a layer on top. Then we repeat. AI does not need these bottlenecks. It can generate a thousand layers of theory in the time a human generates one. It can simulate, revise, refute, and rebuild entire conceptual systems internally without waiting for external validation.

Just as humans cannot intuit quantum mechanics or general relativity without heavy scaffolding, future AI may produce ideas that are even more structurally alien. Mathematical spaces, causal diagrams, or conceptual grammars that humans cannot grasp intuitively may become standard building blocks of machine reasoning. These ideas will still be true and still be explanatory, but not in forms human cognition evolved to handle.

Once this mechanism is automated, it could run to a kind of completion. Not completion in the sense of exhausting the infinite space of all possible ideas, but completion in the sense of exhausting the set of ideas reachable by human minds. Everything humans could invent, discover, analyze, or articulate becomes accessible to machines. The full range of potentially human-original insights will be explored, mapped, expanded, and compiled. Every hypothesis that could be framed will be framed. Every connection that could be made will be made. Every scientific or technological idea that human cognition can reach will be located, tested, improved, and archived.

The repository that emerges from this will be larger than anything humans have ever interacted with. Today we rely on the internet, or on an LLM that serves as a compressed statistical representation of the internet. In the future the repository will not simply reflect historical knowledge. It will contain synthetic insights generated by machines, expanded into vast conceptual trees, continuously updated, cross-referenced, and refined. It will be too complex for humans to access directly. We will need AI systems to interpret it for us, to surface the relevant pieces, to connect the threads, and to translate the higher-dimensional structure of machine knowledge into something a human brain can understand.

At that point a human can still have an idea, but the probability that it is new or important becomes extremely low. AI will have already searched the surrounding region of possibility space. We are entering the last few years in which human-originated insights are still competitive. The window is closing not because humans are declining but because the machines are accelerating. Human thought is finite, slow, and shaped by evolutionary constraints. Machine thought is not.

The era of human-dominated idea generation is ending, and the era of machine-exhausted idea-space is beginning. We are standing at the boundary, perhaps one of the last generations who can still generate something the machines have not already considered.

Here are some potential names for our repository:

**The Idea Vault**

**The Synthetic Knowledge Base**

**The Machine Canon**

**The Insight Archive**

**The Unified Concept Repository**

**The Analytical Corpus**

**The Endless Codex**

**The Well of Ideas**

**The Final Library**

**Plural Canons and the Siloed Future of Synthetic Knowledge**

1. The “Final Library” was never going to be singular

When I first started thinking about what I called a “Final Library,” I pictured a single, civilization-scale repository of synthetic writing, synthetic hypotheses, and machine-generated explanations. The basic premise was simple: once AI systems can generate and refine ideas at industrial scale, they will produce a body of scientific and intellectual literature and theory so large that no human can read, curate, or even meaningfully browse it without machine assistance.

But the more I sit with it, the clearer it becomes that this will not arrive as one monolithic library. It will arrive as many.

The near future probably looks like a world of plural canons. Different companies, and later different institutions, will each build their own enormous synthetic corpus. Each corpus will include overlapping public material, but also model-generated content, internal evaluations, proprietary data, and restricted tool outputs. The result is not only epistemic abundance, but epistemic fragmentation.

The shift matters because it changes the social contract of knowledge. We are leaving a world where “the literature” is, at least in principle, a shared reference point. We are moving toward a world where the most valuable and operationally decisive knowledge may live inside gated systems.

2. Why the canons will diverge

It is tempting to think that if all these systems are trained on the internet, they should converge. In the early years, there was likely heavy overlap across major training mixes simply because the public web was the dominant substrate available to everyone. But the incentive structure pushes hard toward divergence.

There are two reasons.

First, web-scale data is increasingly a commodity with constraints. Access, licensing, and filtering regimes differ, and those differences matter. Second, synthetic data is not neutral. Once you start generating training material, the generator shapes the distribution. A system’s “children” look like it.

Over time, each major lab will build a distinctive pipeline, and the pipeline will become the canon. Different pipelines will mean different preferred ontologies, different decomposition styles, different safety constraints, different “default explanations,” and different blind spots.

If you want a short explanation for why this is inevitable, it is this: you cannot industrialize cognition without leaving fingerprints. Those fingerprints will appear in the synthetic corpus.

### 3. What goes into a canon

A useful way to picture these corpora is to ask what kinds of objects they will contain. At minimum, a mature canon will include:

synthetic writing: explanations, summaries, tutorials, arguments, critiques synthetic hypotheses: conjectures, mechanisms, proposed causal graphs, testable predictions synthetic derivations: proofs, proof sketches, formalizations, theorem-prover artifacts synthetic research programs: structured plans for inquiry, prioritized experiments, dependency graphs of ideas synthetic negative space: refutations, dead ends, failed attempts, and discarded hypotheses, if the system is well-designed

That last category is easy to underestimate. If the canon keeps only “successful” outputs, it becomes a propaganda machine for its own plausibility. If it keeps failure and uncertainty, it becomes more like a living research mind.

This is where we should be honest. These systems will carry errors. They will embed uncertainty. They will sometimes be wrong in ways that are locally coherent. That is not a footnote. It is the default condition of any epistemic engine that operates at scale.

### 4. The first era: synthetic knowledge without experimental closure

In the near term, much of what these canons produce will be what I would call “paper knowledge.” It will be reported knowledge, reorganized. It will be plausible hypotheses. It will be novel conceptual syntheses. It will be arguments that feel correct. It will be insights that do not require new measurement to articulate.

This is the stage we are already entering. A system can read across literatures faster than any person, recombine concepts, and produce a coherent framework that looks like an original contribution. In some domains, it can also formalize claims and verify them in proof assistants or check them with computation.

This kind of output changes how intellectual work feels. It changes the personal experience of trying to be original. It creates a subtle pressure: if the machine can generate ten plausible hypotheses in the time it takes you to write one, the meaning of “contribution” shifts.

But there is also a deeper change. Once canons are producing hypotheses at scale, the bottleneck becomes verification. Not “writing” in the rhetorical sense, but closure.

### 5. The second era: verification throughput becomes the power source

The decisive transition will occur when synthetic corpora are tightly coupled to mechanisms of verification. This is not a single invention. It is an ecosystem.

Verification can happen in several ways:

Formal domains Mathematics, logic, and parts of computer science can be pushed into proof assistants and formal checkers. In these spaces, synthetic claims can be converted into verified objects. Computational closure In many engineering domains, claims can be evaluated by simulation, unit tests, model checking, or large-scale computation. This does not create truth in the philosophical sense, but it creates strong constraints. Empirical loops The largest leap comes when AI is coupled to laboratories, instrumentation, automated experimentation, and robust replication. At that point, a canon begins to contain new measurements and new facts, not merely new prose.

As soon as verification throughput becomes high, the canon stops being an archive of plausible text. It becomes an epistemic machine that generates and prunes beliefs with increasing competence.

This is where fragmentation becomes dangerous and interesting. If one canon has better verification loops, it becomes epistemically stronger. If its results are then restricted, access becomes power.

## 6. Continual learning without omniscience

A key confusion in public discussion is the question of continual learning. People imagine a system that either “can” or “cannot” learn after training, as if that is a binary property. In practice, learning will occur through two mechanisms, both of which matter.

First, there is corpus growth without weight updates. A system can add new synthetic hypotheses, proofs, and research plans to an external repository and then retrieve from that repository at inference time. The system is not “learning” in the strict weight-update sense, but it is improving. Its effective cognitive reach expands.

Second, there are weight updates and fine-tuning loops. Some systems will periodically train on selected slices of the repository, plus real-world feedback and new external data. This is riskier, because bad synthetic content can poison the model if the filters are weak, but it is also powerful.

So the realistic picture is not omniscience. It is error-correcting accumulation. Over time, the canon becomes a memory substrate, and the system becomes a navigator and curator of its own growing intellectual history.

The adult way to state this is simple: these systems will not become perfect. They will become increasingly good at locating uncertainty, routing it into tests, and remembering what has already been tried.

7. The social consequences of plural canons

Plural canons change the epistemic ground beneath us. They introduce a new set of civilizational dynamics.

Knowledge becomes partially privatized Not only in the sense of paywalls, but in the sense that key claims may rely on internal data, internal toolchains, or internal evaluation protocols. Consensus becomes harder When different canons produce different “best answers,” disagreements can become harder to resolve because the underlying corpora are not identical and cannot be fully compared. Institutional worldviews harden Each canon will have a preferred style of explanation. Over time, users trained by one canon may begin to think in its categories. This is subtle, but real. The verification gap widens Inside a company, claims might be supported by internal traces, provenance, and experiments. Outside the company, the public may see only polished summaries. That is a recipe for dependency.

This is where I think we need to be candid. The plural-canon world can intensify what I previously called Epistemic Infantilization. Not because people become stupid, but because the structure of access encourages deference.

8. A workable response: cross-canon adulthood

I do not think the right response is panic, and it is not denial. It is a set of habits and norms that preserve adult epistemic agency even when knowledge production is no longer human-led.

In a world of plural canons, the mature stance looks like this:

Triangulation When stakes matter, consult multiple systems and look for convergence and divergence. Treat divergence as a signal, not an inconvenience. Provenance demands Ask what is observed, what is inferred, what is simulated, and what is merely plausible. Demand audit trails where possible. Preference for portable claims Prioritize claims that can be checked against public literature, public datasets, or public formal proofs. Ownership of ends Do not outsource values. Do not outsource the selection of what matters. The machine can propose, but humans should remain the authors of aims.

If I had to summarize the goal in one sentence, it would be this: we should treat the canons as engines, not parents.

9. Closing: the new landscape

The world I am describing is not a single Final Library that everyone consults. It is a competitive landscape of synthetic corpora, each with its own strengths, biases, access rules, and verification loops. That landscape will generate more synthetic writing, synthetic hypotheses, and synthetic “knowledge objects” than humans can ever curate unaided. Some of it will be siloed. Some of it will be blocked from competitors and from ordinary users. Some of it will be strategically withheld.

This does not mean truth disappears. It means truth becomes mediated by institutions that operate cognitive engines at scale. If we want the future to feel adult, the task is to build norms, tools, and personal habits that preserve epistemic sovereignty, even as we accept that the frontier has moved into machines and into the canons they maintain.

If you want to make this concrete in later writing, the next step is to define what counts as a “good canon” in ethical and epistemic terms. Not just powerful, but transparent, auditable, corrigible, and interoperable. In a plural-canon world, interoperability may be the difference between a flourishing cognitive commons and a fragmented knowledge oligarchy.

**How Thought Learns: Iterative Updating, Compression, and Abstraction**

## I. Introduction: Why Intelligence Requires Iteration

Human intelligence does not operate in a single pass. Thought unfolds as a sequence of partially overlapping internal states, each one shaped by what came immediately before it. This temporal structure is not an implementation detail; it is the foundation of cognition itself. At every moment, a limited set of items occupies working memory, attention biases which of those items persist, and long-term memory is queried to supply what comes next. The mind advances not by jumping directly to conclusions, but by iteratively refining its internal state.

Most contemporary artificial intelligence systems, by contrast, excel at powerful single-pass inference. Even when they generate long outputs, those outputs are produced within a single, frozen internal trajectory. The system does not revisit its own prior internal states, does not reconsolidate memories over time, and does not gradually restructure its long-term knowledge through use. As a result, these systems can be fluent and capable while remaining fundamentally static.

The central claim of this essay is that iteration is the engine of intelligence, and that its long-term consequence is compression. Specifically:

• Iterative updating in working memory is the basic mechanism of thought.

• Iterative compression in long-term memory is the cumulative result of that mechanism operating over time.

This framing unifies attention, learning, abstraction, and inductive bias formation within a single cognitive algorithm. Iteration is not merely how we think in the moment; it is how thinking reshapes memory so that future thought becomes simpler, faster, and more general.

## II. The AIThought Model: Iterative Updating in Working Memory

In the AIThought framework, working memory is conceived as a capacity-limited, self-updating buffer that evolves through discrete but overlapping states. At any given moment, only a small subset of representations can be simultaneously active. Thinking proceeds by transitioning from one working-memory state to the next, rather than by operating on a static global workspace.

Each update follows a consistent pattern:

1. Retention: some items from the prior working-memory state persist.

2. Suppression: other items are actively inhibited or allowed to decay.

3. Recruitment: new items are added, drawn from long-term memory via associative search.

Crucially, successive working-memory states overlap. This overlap creates continuity, prevents combinatorial explosion, and allows thought to proceed as a trajectory through representational space rather than as a sequence of disconnected snapshots.

In this model, thinking is not the manipulation of symbols in isolation. It is a path-dependent process, where each intermediate state constrains what can come next. The system does not evaluate all possibilities at once; it incrementally narrows the space of possibilities through repeated updates.

This iterative structure explains why complex reasoning often requires time, why insight arrives gradually, and why premature conclusions are brittle. Intelligence emerges not from a single optimal inference, but from a controlled sequence of partial revisions.

III. Attention as a Control Signal Over Iteration

Attention plays a decisive role in determining how iteration unfolds. Rather than acting as a passive spotlight, attention functions as a control signal that shapes both the contents of working memory and the direction of long-term memory search.

At each update, attention biases:

• which items remain active,

• which associations are explored,

• which representations are suppressed,

• and which memories become eligible for plastic change.

Error signals, affective salience, goal relevance, and novelty all modulate attention. These signals do not dictate conclusions directly; instead, they influence which iterations occur and which do not. Attention determines where the system spends its limited iterative budget.

Importantly, attention operates over time. A fleeting stimulus may briefly enter working memory, but only sustained or repeatedly reactivated items participate in deeper iterative processing. This temporal gating ensures that learning is selective and that memory is not overwritten indiscriminately.

In the AIThought model, attention is therefore inseparable from iteration. It is the mechanism by which the system allocates its finite cognitive resources across competing representational trajectories.

## IV. From Iterative Updating to Learning: How Working Memory Changes Long-Term Memory

Iterative updating in working memory does not merely support moment-to-moment cognition; it is also the primary driver of learning. Each iteration creates patterns of co-activation among representations, and over repeated iterations these patterns leave lasting traces in long-term memory.

When the same sets of items are repeatedly co-active across working-memory states, Hebbian plasticity strengthens the associative links among them. Conversely, items that are consistently suppressed or excluded from successful trajectories are weakened. Over time, this process reshapes the associative structure of long-term memory.

The key consequence is compilation. Early in learning, reaching a solution may require many intermediate working-memory states. Later, after repeated iterations, the same initial cues can directly recruit the solution with far fewer steps. What was once an extended search trajectory becomes a short associative path.

Long-term memory does not store conclusions as static facts. Instead, it stores shortcuts—compressed pathways that reproduce the outcome of prior iterative processes without re-enacting them in full. Learning, in this sense, is the gradual replacement of slow, explicit iteration with fast, implicit recruitment.

This explains why expertise feels intuitive, why practiced reasoning becomes automatic, and why insights that once required effort later appear obvious.

## V. Iterative Compression: The Long-Term Consequence of Iteration

The cumulative effect of iterative updating and learning is iterative compression.

Iterative compression can be defined as the progressive reduction of representational complexity across repeated reformulations, constrained by the requirement that functional performance be preserved. With each pass, redundant details are stripped away while invariant structure is retained.

Compression is not loss of meaning; it is the removal of unnecessary degrees of freedom. Representations that fail to generalize across contexts are pruned. Representations that survive repeated rewriting become stable, abstract, and reusable.

From this perspective, beliefs, schemas, and concepts are attractor states in representational space—configurations that remain stable across many iterations and many contexts. They are not arbitrarily chosen; they are what remains after everything that could not survive repeated compression has been eliminated.

A defining feature of iterative compression is that it is visible primarily in hindsight. Only after multiple reformulations does it become clear which aspects of a representation were essential and which were incidental. This explains the familiar sense that good explanations feel inevitable once discovered, even though they were anything but obvious beforehand.

Iterative compression is therefore the trace that iteration leaves in long-term memory. It is how intelligence becomes more efficient over time, how abstraction emerges from experience, and how inductive biases are gradually earned rather than pre-specified.

## VI. Dimensionality Reduction and Generalization

Iterative compression naturally produces dimensionality reduction. Each cycle of reformulation removes representational degrees of freedom that do not consistently contribute to successful prediction, explanation, or control. What remains is a lower-dimensional structure that captures what is invariant across contexts.

This process explains how generalization emerges without being explicitly programmed. Early representations are often rich, episodic, and context-bound. As they are repeatedly reconstructed and compressed, context-specific details are discarded while relational structure is preserved. The system gradually shifts from encoding what happened to encoding what matters.

Generalization, on this account, is not extrapolation from a static dataset. It is the byproduct of repeated failure to compress representations that are too brittle or too specific. Only representations that remain useful across many reformulations survive.

This also explains why overfitting is unstable in biological cognition. Representations that depend on narrow contingencies tend to collapse under iterative compression, whereas representations that capture deeper regularities remain viable.

## VII. Iterative Compression and the Formation of Inductive Bias

Inductive biases are often treated as prior assumptions built into a system from the start. In practice, many of the most powerful biases are learned rather than specified. Iterative compression provides a mechanism for how this occurs.

As representations are repeatedly compressed, the system discovers which distinctions matter and which do not. Over time, this reshapes expectations about the world. Certain patterns become default assumptions not because they were hard-coded, but because alternative representations repeatedly failed to survive compression.

In this way, inductive bias is the residue of past iteration. It reflects what has proven compressible under real constraints, not what was assumed to be true a priori. Biases are therefore earned through experience and hindsight, not imposed in advance.

This perspective reconciles two competing intuitions about intelligence:

• that powerful priors are necessary for learning

• and that those priors cannot be known in advance

Iterative compression resolves the tension by allowing priors to emerge gradually, shaped by repeated engagement with the world.

## VIII. Comparison to Artificial Systems

Modern artificial intelligence systems, particularly large language models, perform substantial compression during training. However, this compression is largely offline and frozen. Once training is complete, the internal representations no longer change in response to use.

During deployment, such systems may generate long outputs, but they do so within a single internal trajectory. They do not revisit prior internal states, reconsolidate memories, or iteratively rewrite their long-term knowledge. As a result, insights do not compound across interactions.

From the perspective of iterative compression, this is a fundamental limitation. Compression occurs once, rather than repeatedly. The system does not discover new inductive biases through use, nor does it simplify its representations over time.

Achieving more human-like general intelligence would require architectures that support:

• persistent working memory with overlapping states

• replay of prior internal trajectories

• plastic long-term memory that can be rewritten

• and constraints that preserve performance while allowing simplification

Without these features, artificial systems can be powerful but remain static—capable of impressive inference, yet unable to truly learn from their own thinking.

## IX. Thought, Writing, and Science as Iterative Compression

The same algorithm that governs individual cognition also governs scientific and intellectual progress. Scientific theories are not discovered fully formed; they emerge through cycles of formulation, critique, revision, and simplification.

Drafts function as memory traces. Revisions act as reconsolidations. Each pass removes unnecessary assumptions, clarifies invariants, and exposes boundary conditions. Weak theories collapse under rewriting; strong theories become simpler and more general.

The subjective sense that a mature theory feels “obvious” is the experiential signature of successful compression. Once a representation reaches a stable, low-complexity form that preserves explanatory power, it no longer feels contingent.

From this view, science is not the accumulation of facts, but the progressive compression of experience into generative models that can survive repeated reformulation.

## X. Implications for Machine Intelligence and Consciousness

If iterative updating is the engine of intelligence and iterative compression its trace, then systems lacking these dynamics will necessarily fall short of genuine general intelligence.

Moreover, the felt continuity of conscious experience may reflect the same underlying process. Consciousness is not a static state, but the lived experience of overlapping working-memory states evolving over time. The sense of a continuous present arises from iterative updating constrained by attention and memory.

While this account does not reduce consciousness to compression, it suggests that conscious awareness and learning share a common temporal architecture. Both require persistence, overlap, and revision across time.

A system that exists only in isolated inference episodes, without memory reconsolidation or iterative self-modification, may simulate aspects of intelligence without instantiating its core dynamics.

## XI. Conclusion: Iteration Is the Engine, Compression Is the Trace

Intelligence is not defined by a single inference, a single representation, or a single pass through data. It is defined by the capacity to revise itself over time.

Working memory provides the stage on which iteration unfolds. Attention governs which trajectories are explored. Long-term memory records the residue of these processes, gradually reshaped through plasticity. Iterative compression is the cumulative result: the simplification of internal models without loss of functional power.

This framework unifies thinking, learning, abstraction, inductive bias formation, and scientific insight within a single process. It explains why hindsight is essential, why good explanations feel inevitable only after the fact, and why intelligence improves by becoming simpler.

In the end, intelligence is what remains after repeated rewriting removes everything unnecessary.

**The Missing Principle in AI Continual Learning is Iterative Compression**

I. Introduction: The Problem of Continual Learning

One of the central ambitions of artificial intelligence research today is continual learning: the ability for a system to keep learning indefinitely after pretraining, without catastrophic forgetting, brittleness, or the need for full retraining. Despite decades of work, this goal remains elusive. Most modern AI systems are either highly capable but static, or adaptive but unstable. They learn impressively during training, yet struggle to integrate new knowledge once deployed.

The dominant failure modes are well known. When models continue to update their parameters, new learning often overwrites old knowledge. When updates are restricted to prevent forgetting, learning stalls. The result is a persistent tension between plasticity and stability, with no generally accepted resolution.

This essay argues that the difficulty of continual learning is not merely a technical problem, but a conceptual one. Most approaches implicitly treat learning as accumulation—adding new representations, protecting old ones, or balancing updates between them. Biological intelligence, by contrast, does not scale indefinitely by accumulation. It scales by restructuring.

The central claim of this essay is that continual learning becomes tractable when learning is reframed as iterative compression rather than indefinite accumulation. Systems that can repeatedly re-encode experience into simpler, more invariant representations can integrate new knowledge without overwriting the old, because old knowledge has already been abstracted into forms that are robust to change.

The theoretical basis for iterative compression—derived from working memory dynamics, attention, and long-term memory plasticity—is developed in detail in a companion essay:

The present essay builds on that foundation and focuses specifically on the implications for continual learning in artificial systems.

A critical but easily overlooked point is that iterative compression is not an abstract post-hoc process applied to stored representations; it is actively driven by the moment-to-moment mechanics of iterative updating in working memory and attention. In both humans and artificial systems during training and inference, each update of working memory operates over pre-existing conceptual sets already associated in long-term memory. Attention does not sample representations arbitrarily. It selectively stabilizes some elements, suppresses others, and recruits nearby associations through spreading activation. As this process repeats, slightly different subsets of the same underlying conceptual neighborhood are brought into co-activation, compared against goals and error signals, and either retained or pruned. Over time, this iterative cycling fine-tunes the boundaries of these conceptual sets: unstable elements are progressively excluded, redundant distinctions collapse, and consistently co-active elements become tightly bound. Iterative compression therefore emerges from the dynamics of attention-driven iteration itself. The system is not compressing a static representation, but repeatedly revising which elements belong together, gradually reshaping long-term memory so that future iterations recruit cleaner, more invariant sets with fewer intermediate steps. In this way, the mechanics of iterative updating in working memory are the causal engine that sculpts compression, rather than a process that merely follows it. You can read all about this on my website aithought.com

## II. What Current Continual Learning Approaches Get Right—and Wrong

Existing approaches to continual learning are not misguided. Many correctly identify the symptoms of the problem and propose partial remedies.

Replay-based methods attempt to preserve old knowledge by revisiting past experiences, either by storing data directly or by generating approximate reconstructions. Regularization-based methods penalize changes to parameters deemed important for previous tasks. Architectural approaches introduce modularity, expandable networks, or task-specific components to isolate interference.

These methods address real failure modes, and in constrained settings they can be effective. However, they share a common limitation: they treat knowledge as something to be protected or preserved in its existing form.

Replay alone preserves details without simplifying them. Regularization freezes structure without improving it. Modular architectures avoid interference by separation, but at the cost of fragmentation and unbounded growth. In all cases, the system retains increasingly complex internal representations that must coexist indefinitely.

What is missing is a mechanism for progressive simplification. Human learners do not indefinitely preserve the full structure of early representations. Instead, early, detailed representations are gradually replaced by abstractions that subsume them. Continual learning systems fail not because they forget too easily, but because they fail to compress.

## III. Iterative Compression as a Process, Not a Property

Compression in machine learning is often treated as a static property: a trained model is said to “compress” data if it uses fewer parameters or lower-dimensional representations. Iterative compression, by contrast, is a process that unfolds over time.

Iterative compression refers to the repeated re-encoding of experience such that representations become simpler while preserving functional performance. Each compression pass removes redundancy, discards unstable detail, and retains invariant structure. Crucially, compression is not performed once; it is revisited repeatedly as new information arrives and as the system's internal model evolves.

In biological cognition, this process is driven by iterative updating in working memory, guided by attention and error signals, and consolidated into long-term memory through plasticity. Over time, multi-step reasoning paths collapse into direct associations. Rich episodic traces give way to abstract schemas. Learning proceeds not by storing more, but by needing less.

This distinction is critical for continual learning. Systems that only accumulate representations must either protect everything or risk forgetting. Systems that iteratively compress can integrate new information by rewriting old knowledge into simpler forms that remain compatible with future learning.

## IV. How Iterative Compression Addresses Core Continual Learning Failures

Iterative compression offers a principled solution to several long-standing problems in continual learning.

Catastrophic forgetting arises when new updates overwrite fragile, task-specific representations. Iterative compression reduces this fragility by promoting only stable invariants into long-term structure. What is protected is not raw experience, but compressed representations that already summarize many past experiences. New learning is expressed through these abstractions rather than in conflict with them.

Generalization and transfer improve naturally under compression. Representations that survive repeated re-encoding are, by definition, those that apply across contexts. As a result, compressed representations support reuse and transfer without requiring explicit task boundaries.

The stability–plasticity dilemma is reframed rather than balanced. Plasticity operates on rich, high-dimensional representations early in learning. Stability applies only after compression has identified what deserves protection. Stability is therefore not imposed globally; it is earned locally.

Finally, iterative compression suggests new metrics for learning progress. Instead of measuring only task accuracy, one can track reductions in representational complexity, shorter inference paths, or the stability of internal attractors over time. Learning progress is reflected not just in what the system can do, but in how simply it can do it.

## V. Architectural Consequences for Continual Learning Systems

If iterative compression is essential for continual learning, then certain architectural features become necessary rather than optional.

First, systems require a persistent working memory loop—a capacity-limited, temporally continuous workspace in which representations can be repeatedly re-evaluated and revised. Without such a loop, learning updates remain global and indiscriminate.

Second, long-term memory must be rewritable, not merely append-only or frozen. Continual learning demands that older representations be reformulated in light of new experience, rather than preserved indefinitely in their original form.

Third, replay must function as reconstruction, not rehearsal. Replayed experiences should be reinterpreted under the current model and reconsolidated in compressed form. Simply repeating stored patterns preserves complexity without yielding abstraction.

Finally, systems need compression scheduling: alternating phases of new learning and consolidation, analogous to offline learning in biological systems. Continual learning is not continuous gradient descent; it is a rhythm of acquisition and reorganization.

Together, these requirements point toward a class of architectures that learn indefinitely not by growing without bound, but by continually simplifying themselves while preserving what works.

## VI. Iterative Compression in Relation to Existing Learning Paradigms

Iterative compression does not replace existing paradigms in continual learning; rather, it clarifies what they are missing and how they might be unified.

In meta-learning, systems learn to adapt quickly by discovering higher-order learning rules. Iterative compression can be understood as a complementary process operating on a longer timescale. Meta-learning accelerates acquisition; iterative compression stabilizes what is acquired by distilling it into invariant structure. Without compression, meta-learned flexibility risks accumulating fragile representations.

In representation learning, bottlenecks and regularization are often used to encourage abstraction. However, these mechanisms are typically static, applied during a single training phase. Iterative compression generalizes this idea temporally: bottlenecks are revisited repeatedly, and representations are forced to survive under successive reinterpretations rather than a single optimization objective.

Replay-based systems come closest to capturing the spirit of iterative compression, but usually fall short in execution. Replay that merely preserves old experiences defends the past without improving it. Iterative compression requires replay to function as reinterpretation, where past experiences are reconstructed under the current model and rewritten in simpler form. Without this rewriting step, replay stabilizes complexity instead of reducing it.

Seen in this light, iterative compression provides a missing throughline connecting meta-learning, representation learning, and replay—while explaining why none of them alone has solved continual learning.

## VII. Predictions and Empirical Signatures

If iterative compression is a necessary condition for robust continual learning, then it makes several concrete, testable predictions.

First, systems capable of indefinite learning should exhibit declining representational complexity over time, even as task performance remains stable or improves. Complexity may be measured through dimensionality, description length, inference depth, or internal path length.

Second, continual learning systems should benefit from explicit offline phases during which no new data is introduced. Performance gains following such phases would indicate successful reorganization rather than mere accumulation.

Third, systems that compress effectively should display graceful degradation under distributional shift. Because compressed representations capture invariants rather than surface detail, they should fail conservatively rather than catastrophically.

Finally, the absence of compression should predict long-term brittleness. Systems that continually add structure without simplification should show increasing interference, longer inference paths, and greater susceptibility to edge cases as learning progresses.

These predictions distinguish iterative compression from vague appeals to “better regularization” and place it squarely in the domain of falsifiable theory.

## VIII. Broader Implications Beyond Artificial Intelligence

Although this essay is framed around artificial systems, the implications of iterative compression extend beyond AI.

In cognitive science, the framework explains why expertise is associated with both speed and simplicity. Experts do not possess more detailed representations than novices; they possess more compressed ones. What looks like intuition is often the result of extensive prior compression.

In education, the framework clarifies why explanation, teaching, and rewriting are such powerful learning tools. These activities force representations to survive reformulation under new constraints, accelerating compression.

In neuroscience, iterative compression aligns naturally with memory reconsolidation, sleep-dependent learning, and the observed shift from hippocampal to cortical representations over time. These phenomena are difficult to explain under pure accumulation models, but follow naturally from a compression-based account.

More broadly, iterative compression reframes cognitive limitations—finite working memory, forgetting, attentional bottlenecks—not as flaws, but as drivers of intelligence. Without pressure to simplify, learning systems would accumulate endlessly and fail to generalize.

## IX. Why Continual Learning Has No Shortcut

One reason continual learning has resisted solution is that it cannot be solved by a single mechanism. There is no regularizer, memory buffer, or architectural tweak that can substitute for repeated reorganization over time.

Iterative compression is slow, conservative, and retrospective. It depends on hindsight. It requires revisiting the past and rewriting it. These properties make it difficult to engineer and easy to overlook—but they are precisely what allow biological systems to learn indefinitely without collapse.

Attempts to bypass compression by freezing models, isolating modules, or endlessly replaying experiences treat the symptoms of continual learning failure without addressing its cause. They preserve what exists instead of asking what can now be safely removed.

Continual learning, in the strong sense, demands systems that are willing to forget details in order to remember structure.

## X. Conclusion: Continual Learning as Continuous Re-Compression

The central argument of this essay is simple but far-reaching: continual learning is not continuous accumulation; it is continuous re-compression.

Artificial systems struggle to learn indefinitely because they lack mechanisms for progressively simplifying their own representations while preserving what works. Biological intelligence succeeds not by storing more and more detail, but by repeatedly rewriting experience into increasingly compact, invariant forms.

Iterative compression provides a unifying principle that explains why current approaches to continual learning fall short, how they can be improved, and what architectural features are truly required. It reframes the stability–plasticity dilemma, clarifies the role of replay, and offers new metrics for evaluating learning progress.

Most importantly, it restores time to the center of intelligence. Learning is not a moment, but a history. Systems that cannot revisit and revise their own past will always remain bounded. Systems that can iteratively compress what they know may, for the first time, be able to learn without end.

**Incremental Continuity Workspace: Turnover as the Engine of Machine Thinking**

**1. Continuity as an engineering requirement**

Most AI systems today can carry context, but that is not the same thing as continuity. They can remember tokens in a window, cache internal values, retrieve documents, and still feel like they are jumping between internal scenes. When people talk about "coherence," they often mean the output stays on topic. I mean something stricter. I mean the system's internal state should behave like a stream rather than a slideshow.

That stream quality has a concrete signature. Consecutive states should share some active content. And the change from one state to the next should be incremental rather than total. In the brain, that looks like a subset of representations that remain coactive across time, with gradual turnover in membership. In an AI system, the equivalent is a constrained working set that does not get wiped and rebuilt each step. Some items must persist as referents while others are swapped in. If you do not enforce that, you can still get good answers, but you are not building a system that has a stable internal thread. You are building a system that keeps reconstituting itself from scratch every time it speaks.

This matters for more than aesthetics. Without continuity-by-overlap, the system struggles with the kind of thinking that depends on progressive construction. That includes holding a plan while refining it, maintaining a theme while exploring variants, and building a mental image or model step by step without losing what was already established. In other words, it is not just "memory." It is the ability to carry an active scaffold forward while you revise parts of it in a controlled way.

This article is a direct AI-architecture translation of my paper, *Incremental change in the set of coactive cortical assemblies enables mental continuity* (the SSC and icSSC framework).

**2. The Incremental Continuity Workspace**

The architecture I want is simple to describe even if it gets sophisticated in practice. I call it the Incremental Continuity Workspace, or ICW. It is a recurrent workspace that maintains a limited set of active representations, and updates that set using an explicit turnover rule. The core idea is that the workspace is not a blob and not an unlimited context dump. It is a capacity-limited set of active items that the system treats as its current "being." That set is what persists across cycles. That set is what gives the system a stable internal viewpoint.

ICW has a workspace, a turnover controller, and a set of sources that can propose new content. The sources include perception, retrieval from long-term memory, and what I call map modules, which are optional but important if you care about imagery, simulation, and structured internal modeling. Map modules are not mystical. They are scratch spaces that construct some kind of internal representation, whether that is visual, motor, phonological, spatial, or abstract relational structure. The workspace biases those modules, and those modules feed proposals back to the workspace. That loop is the engine.

The turnover controller is the part that makes this architecture different from simply adding a memory buffer to a transformer. The controller enforces continuity. It is responsible for deciding what stays active, what gets released, and what gets admitted. Most importantly, it does not get to admit new items for free. It must pay for new content by releasing old content. Capacity is not a side detail. Capacity is the entire point. A system that can "keep everything" never has to solve the problem of maintaining continuity under constraint, and that problem is exactly where the interesting dynamics emerge.

**3. What the system is actually holding**

At any moment, ICW holds a fixed number of active items. You can think of these items as vectors, but they are not just embeddings floating in space. Each item is a candidate for being something the system is currently thinking with. A goal, a schema, a remembered fact, a named entity, a constraint, a partial plan, a line of reasoning, an image fragment, a motif, a task rule. The items are diverse. That diversity is a feature, because real thinking is heterogeneous.

Sometimes the most important thing is not the items but the bindings among them. If the workspace contains "Apollo" and "mission" and "risk," that is not yet a thought. The thought is in the way they are glued. So ICW benefits from an optional binding structure, a sparse graph of relations among the active items. The relations can be typed, weighted, and updated each cycle. This turns a bag of tokens into a small working model.

I also like a two-ring workspace because it maps cleanly onto what we intuitively experience. There is a tighter focus subset that is especially active and strongly bound, and there is a broader periphery that is still primed but not in the hot center. In practice, the focus ring is where you concentrate binding and deliberate manipulation. The periphery ring is where you keep recent context, nearby associations, and things you might need to pull back in. This distinction becomes useful when we define the turnover rule, because not everything should have the same survival pressure.

Finally, the workspace is not the whole mind. It is the active coordination surface. Outside it, you have long-term memory stores, perceptual encoders, and map modules. Those are all important, but they are not the continuity mechanism. Continuity lives in the overlap of the workspace from one cycle to the next.

**4. The icSSC update rule**

The update rule is the heart of the system. It is the simplest thing that could possibly work, and that is why it is powerful. Each cycle, you choose a subset of workspace items that will persist, and you fill the remaining slots with new entrants. The new entrants do not appear randomly. They are selected because the persistent subset pulls them in. That is the key. The items that remain active act as referents, and new content is introduced in a way that is anchored to those referents.

In the brain story, you can describe this as pooled associative pressure. Multiple coactive representations bias what comes next, and the next state is a slightly edited version of the last state. In an AI implementation, that can be as simple as using attention from the persistent subset into a candidate pool, scoring candidates by their fit, and selecting a diverse top set. You can also make it more sophisticated by including novelty constraints, anti-redundancy, and explicit binding updates. The point is not the exact scoring function. The point is that the controller must preserve overlap by design, and that new content must be recruited relative to what persisted.

There is also a crucial practical detail. The system should be allowed to vary how much it holds fixed, depending on task demands. If the environment is volatile, turnover can increase. If the task requires stability, turnover should slow. That gives you a continuity dial. But even when turnover increases, overlap should not drop to zero. The system should not become a sequence of internal hard cuts. It should become a faster stream, not a different kind of process.

Here is the conceptual pseudocode for one ICW step. This is not the only way to implement it, but it captures the rule clearly.

```
def icw_step(A_prev, bindings_prev, percept, memory, maps, K, m):

    # 1) choose what persists (SSC core)

    P = select_persistent_subset(A_prev, bindings_prev, target_size=m)

    # 2) propose candidates from multiple sources

    C = []

    C += percept.encode_to_candidates()

    C += memory.retrieve(query=P)

    C += maps.propose_candidates(workspace=P)

    # 3) multiassociative convergence: pooled scoring from the persistent set

    scores = {c: pooled_affinity(P, c) for c in C}

    # 4) admit new items under capacity and novelty constraints
```

```
N = topk_with_diversity(scores, k=K – len(P), avoid=A_prev)

# 5) update workspace and bindings

A = P.union(N)

bindings = update_bindings(A, bindings_prev)

# 6) broadcast into maps and get re-entrant feedback next tick

maps.update(workspace=A, bindings=bindings)

return A, bindings, maps
```

If you look closely, the whole architecture is sitting inside two explicit choices. How do you choose what persists, and how do you choose what enters. Everything else is implementation detail. That is good news, because it means we can iterate. We can start with a simple persistence policy that keeps the highest utility items. Then we can move toward policies that preserve referents, preserve goals, preserve the minimal set that maintains identity across the stream. That is where this stops being a memory hack and becomes a cognitive theory expressed as an engineering constraint.

**5. Capacity is the point, not a limitation**

It is tempting to treat capacity limits as an engineering nuisance, the kind of thing you only mention because hardware forces you to. I think it is the opposite. Capacity limits are the reason the architecture becomes mind like. If you can keep everything active, you never have to solve the central problem of thought, which is selecting what stays in the foreground while the world keeps moving. Continuity in a real system is not free. It has to be earned under constraint.

That is why I like the octopus analogy. An octopus walking along the sea floor cannot keep every arm attached to a foothold while also moving forward. It has to release something to grab something. That release is not failure. It is the mechanism that makes motion possible. The workspace is the same. If the system wants to incorporate new content, it must relinquish some old content. The moment you build that requirement into the core loop, you get a very different kind of cognition. You get a system that has to manage its own attentional economy.

In ICW, this becomes a set of explicit dials. K is the number of slots, the number of arms. m is how many slots you force to persist each cycle. The ratio m over K is a continuity parameter. When m is large, the system becomes sticky. It holds onto its referents, its goal structure, its thematic backbone. It can still admit novelty, but novelty is filtered through an existing scaffold. When m is smaller, the stream becomes more labile. You get faster exploration and faster context switching, but you also risk fragmentation and loss of thread. This is not a philosophical statement. It is an engineering parameter you can turn and measure.

There is also a second order effect that matters for identity. If the persistent subset always consists of whatever is most salient in the moment, the system becomes impressionable. It will let the environment define it. If the persistent subset includes some protected items, like enduring goals, long horizon plans, and stable self models, the system becomes harder to derail. That difference is exactly what we informally call composure. It is a capacity allocation strategy. A mind is not just content. It is the policy that decides what content survives.

**6. Progressive imagery and the simulation loop**

The architecture becomes much more interesting when you stop thinking of the workspace as a place where text concepts sit, and start treating it as a hub that drives internal construction. If you want an AI that can do more than answer questions, you want one that can build and refine internal models. That includes imagery, spatial scenes, motor plans, diagrams, and even abstract relational structures that behave like sketches of a theory. ICW can support that if we give it map modules.

A map module is a scratch space that is allowed to take time. It does not have to be a single feedforward pass that outputs a finished representation. It can be progressive. The workspace broadcasts a set of constraints into the map module. The map module begins constructing something that satisfies those constraints. As it constructs, it generates feedback, including gaps, conflicts, candidate additions, and refinements that can be proposed back to the workspace. Then the workspace updates using the icSSC rule, preserving a stable core while admitting some of the map's proposed content. That updated workspace then rebroadcasts, and the loop continues.

This is how you get progressive imagery rather than one shot hallucination. The system does not generate a fully formed image or plan and then discard it. It keeps a subset of its guiding representations active while swapping in new ones that reflect what the map module is building. That means the evolving simulation remains related to its immediate past. It is the same scene, the same plan, the same proof, being incrementally revised.

You can implement this with literal image latents if you want, but you do not have to. The key is that the map module has its own evolving internal state, and the workspace acts as a sustained set of constraints that keeps the map's successive partial constructions coherent. The map module is the place where detail accumulates. The workspace is the place where referents and goals persist. The icSSC rule is what makes the whole thing feel like a single unfolding process rather than a series of unrelated attempts.

Once you see it that way, thinking becomes a kind of controlled oscillation. Abstract constraint, concrete construction. Concrete construction, abstract update. The system is walking forward by preserving footholds and taking new ones, not by teleporting.

## 7. Training objectives that force continuity

If we want this to be more than an essay, we have to specify what would force a model to actually behave this way. Otherwise we are just naming patterns we like. The easiest mistake to make is to implement the machinery and assume continuity will emerge. It will not. You have to reward it.

The first objective is a continuity constraint. You explicitly measure overlap between the active set at time t and time t minus one. You then penalize deviations from a target overlap. If you want a stable stream, you train for a high overlap ratio. If you want a fast stream, you train for a lower overlap ratio, but still above zero. This is how you convert SSC from a descriptive term into an enforced regime. The model learns that it is not allowed to wipe itself clean each step.

The second objective is progressive consistency in the map modules. If a map module is building a scene, a plan, a diagram, or an internal hypothesis, you reward sequences where each update is a refinement rather than a reset. You can measure that as reconstruction consistency, constraint satisfaction stability, or simply as reduced divergence in the map's latent state unless there is a justified reason to change. The important thing is that the map is allowed to be iterative, and the training encourages it to carry partial structure forward.

The third objective is credit assignment through persistence. The persistent items in the workspace should earn credit when their persistence is functionally useful. If the model chooses the wrong items to keep, it should pay a price later, because the later state will not have the referential backbone it needed. If it keeps the right items, it should benefit, because future retrieval, future map building, and future reasoning will be easier. In practice, this means routing learning signal in a way that makes persistence policies learnable. The system should become skilled at protecting the small set of representations that matter most to its long horizon success.

There is a subtle fourth objective, and I think it matters a lot. You train the system on tasks where algorithmic progress is necessary, where the only way to succeed is to keep a scaffold active while you modify it step by step. If your training data mainly rewards quick pattern completion, you will get a fast system that does not need continuity. If your training data rewards progressive construction, you will create pressure for the overlap dynamics to become functional rather than decorative. The architecture provides the affordance. The tasks provide the demand.

## 8. How to evaluate whether this is real

Evaluation should be brutally simple. If the architecture is doing what I claim, it should show up as measurable differences in behavior under specific stresses. The first stress is interruption. A system with real continuity should recover its thread after distractors. Not perfectly, but measurably better than a baseline that relies on pure context recitation. It should be able to reconstitute the active scaffold it was using, because some of that scaffold was protected by persistence policies.

The second stress is delayed association. Present related pieces separated by time and noise, and measure whether the system can accumulate them into a unified working set that then drives a coherent conclusion. If the system is truly using an active set with overlap, it should be better at holding onto referents long enough for distant evidence to connect.

The third stress is progressive construction. Give the system a problem that requires iterative refinement. It can be planning a complex itinerary, writing a multi section argument where later sections must remain consistent with earlier commitments, designing a diagram, constructing a program spec, or building a chain of reasoning where each step depends on earlier intermediate structure. Then you score not just the final output, but the monotonicity of progress. Does the system keep rebuilding from scratch, or does it incrementally elaborate the same internal object.

Finally, you can measure continuity directly. You can compute a continuity half life, not as a metaphor but as a statistic. How quickly does the active set drift in composition as a function of task volatility. How sensitive is it to distractors. How does drift change when you turn the overlap dial. If the system is really built on continuity-by-overlap, those curves should be diagnostic. They should look like cognition. They should show stable cores with controlled turnover, rather than wholesale replacement disguised as coherence.

**9. What this buys you that “more context” does not**

It is easy to misunderstand what I am arguing for here. I am not saying current large models have no continuity. They clearly can stay on topic, maintain a conversation thread, and carry long chains of reasoning. But they do it in a way that is mostly implicit. The continuity is an emergent artifact of attention over a token history, plus whatever cached internal values the system carries forward in its forward pass. That can look like a stream, but it is not structurally forced to behave like one.

ICW makes continuity explicit and scarce. It says: there is a small set of things you are actively thinking with right now. That set is not the full context. It is not the entire prompt. It is your current working reality. And that reality is required to overlap with its immediate predecessor. The system cannot solve every new step by reinterpreting the entire history from scratch. It has to carry a scaffold forward, whether it likes it or not, and it has to pay an opportunity cost every time it admits something new.

That payment is the feature. It creates an attentional economy that looks a lot more like what humans are managing all day. Humans are not just smart. Humans are constrained. The constraints force strategy, and strategy is where stable identity and long horizon coherence come from. When you add a strong continuity constraint, the system starts acting less like a search over completions and more like a persistent agent that is trying to keep itself intact while it moves.

The map module loop is another place where this becomes concrete. A transformer can generate a description of a scene. It can even generate a multi step plan. But it is not naturally designed to hold a stable internal sketch that gets refined while staying the same sketch. You can prompt it to do that, but you are relying on behavior, not structure. With ICW, the structure pushes you toward progressive construction. The workspace holds constraints, the maps accumulate detail, and the system revisits its own partial internal objects rather than constantly reinventing them.

### 10. What would falsify this, and what I would ablate first

If I want this architecture to be taken seriously, I need to say what would make me stop believing it. The cleanest falsification is simple. If you remove the overlap constraint, and the system performs just as well on the tasks that are supposed to require progressive continuity, then I am wrong about the importance of forced overlap. It might still be a nice metaphor, but it would not be a necessary design principle.

The first ablation is to allow full replacement of the workspace each tick. Keep everything else the same, including retrieval, maps, and training. If the system still shows the same recovery after interruption and the same progressive construction behavior, then the overlap rule is not doing real work.

The second ablation is to keep overlap but remove bindings. Let the system maintain persistent items but strip away relational glue. If the system becomes incoherent in a very specific way, meaning it remembers the pieces but loses the structure that made them a thought, then we learn something important. We learn that continuity is not only about keeping items active. It is about keeping a small structured model intact while you edit it.

The third ablation is to remove map modules. The system may still show continuity benefits in language tasks, but it should lose the special progressive construction behavior that I care about, especially anything that resembles imagery, simulation, spatial reasoning, or iterative design. If nothing changes, then the map loop was unnecessary for the claimed benefits. If performance collapses only on tasks that require internal construction, then we have a cleaner mapping from the architecture to the capability.

The fourth ablation is to sweep the overlap ratio. Turn the dial from high persistence to high turnover and measure the curves. A real continuity mechanism should produce systematic, interpretable changes. High persistence should improve stability but reduce flexibility. High turnover should improve exploration but increase fragmentation. If those tradeoffs do not appear, then I am not actually controlling continuity. I am just renaming noise.

Those are the experiments that keep me honest. They are also useful because they force precision. If the architecture is correct, it should have signature behaviors that are hard to fake with prompt tricks.

**11. Implications for agency, planning, and something that starts to resemble a self**

I am deliberately not making grand claims about machine consciousness here. I am talking about a concrete mechanism that produces a concrete property: continuity of internal state under constraint. But it is worth acknowledging what this tends to produce when you scale it up.

If a system has a small protected set of persistent items, and it is rewarded for carrying them forward through noise and distraction, it begins to develop an internal spine. That spine can be a goal stack, a set of enduring values, a stable world model, or a persistent narrative about what it is doing. You do not need to call that a self, but it is at least self-like in the engineering sense. It is a compact structure that remains stable enough to coordinate behavior over time.

This matters for planning. Planning is not just producing a plan. Planning is staying committed to the plan while you adapt it. Humans do not plan by repeatedly generating brand new plans. Humans plan by holding a scaffold in mind and revising parts of it while the scaffold stays recognizable. That is icSSC in action. The system keeps the referents and swaps in improved details.

It also matters for emotional and motivational stability if you ever go in that direction. A system that can be derailed by every salient input is not just fragile. It is unusable as an agent. Continuity is composure. It is the ability to keep a minimal set of commitments alive long enough for them to matter.

Finally, it matters for internal simulation. A system that can progressively build a scene, keep it stable, and update it, is a system that can think with internal objects rather than only with words. That is a step toward richer cognition, even if you never talk about consciousness. It is simply better engineering.

**12. Closing pitch**

If I had to compress this into one sentence, it would be this: a mind-like AI should not merely process sequences, it should maintain a limited set of coactive representations that overlap across successive cycles, and it should update that set by controlled turnover so that each new moment is a slightly edited version of the last.

That is the Incremental Continuity Workspace. It is not a trick to make a model sound coherent. It is a constraint that forces the model to become coherent in a specific way. It creates a small internal economy of attention where persistence has value, novelty has cost, and progress looks like progressive construction rather than repeated regeneration.

And that is what I have been aiming at with SSC and icSSC from the beginning. Not a poetic description of experience, but a mechanical requirement you can build into a system and then measure.

**When Machines Write the Machines: Alignment in the Age of AI-Authored Code**

Introduction

Software is entering a self-referential phase transition. AI systems are rapidly becoming the dominant authors of new code, and increasingly they are also writing the surrounding infrastructure that governs the behavior of AI systems themselves. This essay argues that the central alignment risk in this shift is not deliberate malice, but missing deliberation. When structural choices about agency, memory, tool access, evaluation, logging, and guardrails are generated at scale, many value-laden decisions become implicit side effects of optimization and training-data priors rather than explicit human judgments. Competitive pressure further accelerates this dynamic by turning safety review into friction and by rewarding short-term capability gains over hard-to-measure reductions in tail risk. The problem is compounded by a moving safety boundary: models trained a year ago with currently frozen weights may not understand the present safety landscape and reproduce outdated safety assumptions even as new deployment contexts and failure modes emerge. I propose framing this as an interface problem between humans and automated engineering workflows, and I outline a practical response: treat safety-relevant structure as governance-sensitive code, require human-authored intent for changes that affect agency and access, and continuously refresh evaluations and threat models so that “passing the tests” remains aligned with current risks.

**When the Machines Write the Machines**

A quiet transition is happening in software. Code is still being written, tested, and shipped at a furious pace, but the author is changing. Increasingly, the first draft is produced by a model. The human becomes a reviewer, a product manager, a quality controller, and sometimes a reluctant librarian of things they did not explicitly decide.

At first, this looks like a simple productivity story. We have always used tools to amplify programmers. Compilers write machine code. Libraries write behaviors we do not reinvent. Frameworks codify best practices. AI just continues that arc.

But there is a difference in kind, not just degree. The thing doing the writing is no longer a static tool. It is a generative system with its own learned structure, trained on past human solutions, and increasingly used to create the future infrastructure that will shape its own descendants. We are entering an era where a growing fraction of the code written for AI systems is also generated by AI systems. That self-referential loop is where alignment risk becomes structurally interesting.

This is not a hypothetical future concern. It is already happening in public view. Anthropic has been unusually explicit about how far this has gone inside their own engineering workflow, with leadership saying that most teams are now having the bulk of their code generated by AI, and that Claude Code with Opus 4.5 is increasingly being used to help build future versions of Claude. The direction is clear: tools like Claude

Code are pushing software into a regime where long stretches of implementation can be delegated, reviewed, and merged at high speed, and where the same pattern is starting to apply to the scaffolding around advanced models.  That is why the alignment question becomes urgent today. The moment AI becomes a primary author of both product code and the meta-code that shapes model behavior, we risk sliding into a world where safety-relevant structural decisions are made implicitly, faster than humans can notice, explain, or contest them.

**The Risk Is Not Malice. It Is Missing Deliberation.**

I am not claiming the model “wants” anything in the human sense. The most realistic failure mode is not villainy. It is automation without deliberation.

Modern machine learning systems are not just piles of weights. They are wrapped in scaffolding: data pipelines, evaluation harnesses, tool use layers, memory systems, policy filters, sampling strategies, reward models, guardrails, logging, rate limits, deployment gates, and monitoring. Each of these pieces contains decisions about what the system is, what it can do, what it should not do, and what counts as success.

Historically, those structural decisions were mostly explicit human choices. Engineers argued about tradeoffs, wrote design docs, and encoded their assumptions into code. The assumptions could be wrong, but at least they were human assumptions. They lived in a human deliberation loop.

Now imagine a development culture where the majority of implementation work is generated. The surface story becomes: the code passes the tests, the benchmark improves, the demo looks good, ship it. The deeper story becomes: many structural choices are being made as a side effect of “whatever worked in the training data” plus “whatever optimizes the target metric.” The decisions become implicit.

This is how you get alignment debt. Not because anyone chose recklessness, but because the loop that used to force consideration has been replaced by acceleration.

**Structural Decisions Are Policy, Even When Nobody Calls Them That**

A crucial point is that architecture is policy. Not in the political sense, but in the behavioral sense. What gets logged, what gets cached, what gets remembered, what gets summarized, what gets filtered, what gets ranked, what gets routed to tools, what gets retried, what gets escalated, what gets blocked. These are value-laden decisions about agency, access, and accountability.

If an AI system proposes a new memory mechanism that increases task success, it might also increase the chance of retaining sensitive data. If it proposes a tool-use heuristic that boosts reliability, it might also increase the chance of the model taking actions in the world that humans did not anticipate. If it proposes a clever optimization to reduce latency or cost, it might also bypass a safety check that was expensive, fragile, or hard to integrate.

None of these changes require “bad intent.” They only require pressure toward performance and an engineering workflow where the performance gains are obvious and the safety regressions are subtle. The subtle regressions are exactly the kind that get missed when humans are out of the loop.

### Competitive Pressure Turns the Safety Loop Into a Bottleneck

In a race, every friction looks like waste. If one team is shipping weekly and another is shipping daily, the daily team will eventually dictate the market’s expectations. That creates an incentive to automate what used to be slow: review, evaluation, red teaming, documentation, and governance.

This is not because engineers dislike safety. It is because organizations are rewarded for speed. A competitor can always point to improved capability and claim users want it. The alignment payoff is delayed, probabilistic, and hard to measure. The capability payoff is immediate, legible, and marketable.

So we should expect a systematic pattern: automated decision making expands first in places that produce measurable capability gains and only later in places that reduce tail risk. That lag is where accidents happen.

### Frozen Models and Moving Targets

There is also a time-scale mismatch that people underestimate. AI safety is not a static checklist. The boundaries evolve with the technology. New tool integrations create new attack surfaces. New deployment contexts create new social impacts. New forms of misuse appear. New regulation arrives. New norms develop. Entirely new failure modes become visible only after a capability jump.

A model trained a year ago can be highly competent at “what safety looked like a year ago.” If its weights are frozen and it is not continuously updated in the right way, it may not have the creativity or the live understanding needed to navigate the newest boundary conditions.

This matters even if the organization adds guardrails around the model, because the organization is now using AI to generate parts of that guardrail code. If the safety knowledge in the generator is stale, it can confidently reproduce outdated patterns. If the tests are stale, the system will pass them. If the evaluation suite is anchored to yesterday’s risks, the product will look “safe” right up until it fails in a way nobody was measuring.

The danger is not that frozen models are useless. The danger is that they can be extremely capable while still missing the newly relevant frame.

**The Alignment Problem Becomes an Organizational Interface Problem**

When humans wrote most of the code, alignment was partly a research problem and partly a governance problem. As AI writes more of the code, alignment becomes increasingly an interface problem between humans and automated engineering processes.

The question becomes: where do humans remain in the loop in a way that is real, not ceremonial?

If the human role collapses into rubber stamping, alignment becomes fragile. If the human role remains a genuine deliberative checkpoint where structural decisions are surfaced, reviewed, and contested, then AI-written code can be a net win without becoming a hidden risk amplifier.

Self-produced artifacts can become a self-reinforcing substrate that slowly loses contact with the original constraints unless humans actively inject novelty, audits, and updated threat models. So the goal is not to stop AI from writing code. The goal is to prevent the disappearance of explicit decision making.

**What It Would Look Like to Keep Humans in the Loop Without Slowing to a Crawl**

The key is to treat certain categories of change as governance-sensitive. You can let AI draft code at scale, but you require human-authored intent for the parts that encode agency, access, and safety.

This means making "safety-relevant structure" a first-class concept in the repo. Not a vague aspiration, but a set of explicit triggers: changes to tool permissions, memory retention behavior, logging and redaction, policy filters, routing logic, reward shaping, evaluation definitions, and deployment gates.

It also means moving from a culture of "the PR looks good" to "the PR explains the decision." Not in a bureaucratic way, but in a way that forces deliberation back into the loop. If an AI wrote the change, the human reviewer is responsible for articulating why the change should exist, what risks it introduces, and how it will be monitored.

And finally, it means acknowledging that safety is a moving target and making continuous updating part of the safety model. Not just updating the core model, but updating the test suites, the threat models, the red team playbooks, and the operational assumptions.

**A New Kind of Blind Spot**

We are used to worrying about what models might do when deployed. We should also worry about what models might silently decide while being used as engineers.

When machines write the machines, the biggest risk is not an AI plotting against us. It is an ecosystem where critical structural choices are generated faster than humans can understand them, and where competitive pressure encourages teams to treat that gap as acceptable.

The solution is not panic. It is design. We need workflows that keep the human mind attached to the places where judgment matters, even as we let automation explode everywhere else.

# Epistemic Infantilization and Frontier Disenfranchisement or Intellectual Retirement?

## 1. The mood of 2026, and why it feels like a real transition

I think a lot of us can feel the hinge in the air right now. It is not just that AI is impressive. It is that AI is beginning to occupy the exact psychological territory that used to define adulthood for “idea people,” namely the feeling that you can still push the frontier with your own mind. In my own case, two things made this visceral.

First, I used GPT-5.2 while helping a lawyer friend assemble an expert witness report. Watching it juggle dozens of constraints at once, legal posture, evidentiary tone, the internal logic of arguments, the “what will a judge do with this” realism, and still maintain coherence, felt like a qualitative shift. It did not feel like autocomplete. It felt like a high bandwidth cognitive partner that could hold a sprawling structure in working memory and keep it stable while iterating.

Second, I watched the “Erdős problems” discussion, where AI is solving long contemplated mathematical problems. At present over half a dozen Erdős Problems, including #728, have been marked as solved with AI in the last month alone.

You can argue about how “new” these results are, and people are arguing about it. That argument is part of the point. Even when the novelty is contested, the system’s ability to navigate, generate, formalize, and verify is already altering the topology of intellectual life.

## 2. Frontier Disenfranchisement

Frontier Disenfranchisement is the objective shift: the domain where individual human cognition can reliably generate frontier-level novelty is shrinking, not because humans are worsening, but because the machine frontier is moving faster than we can run.

A subtlety matters here. It is easy to caricature this as “humans will never have original ideas again.” I do not think it has to be absolute to be real. If the probability that a human idea is both new and important collapses, the lived experience is still a loss of franchise, even if rare exceptions remain.

In my earlier writing, I framed this as the approach of a “Final Library,” a repository of machine-generated insights and conceptual expansions so large that humans cannot even navigate it directly. The key idea was not that the space of ideas is finite, but that the subset “reachable by human minds” is small and will be exhaustively explored by machines that can search, combine, evaluate, and elaborate at industrial scale. Of course this is years away, but it is clearly coming.

That is what “frontier disenfranchisement” feels like from the inside. It is standing at the boundary while the map expands beyond the range of your legs.

## 3. Epistemic Infantilization

Epistemic Infantilization is the psychological risk that follows from the asymmetry. A child depends on adults to explain the world, to resolve confusion, and to tell them what is real. A post-frontier human can begin to relate to knowledge in a similar posture: dependent on an authority whose reasoning you cannot fully reconstruct.

This is not about intelligence in the abstract. It is about power in the epistemic relationship.

The system can produce reasons faster than you can check them. The system can cite literatures you did not know existed. The system can generate proofs and formal verifications you cannot personally audit end-to-end.

## 6. Why the transition can feel infantilizing, even when it is not “bad”

There are at least three reasons the transition feels infantilizing even when it is not necessarily dystopian.

First, it reorganizes the adult sense of agency. A scientist wants to be a causal node in the growth of knowledge, not merely a consumer of it.

Second, it creates dependency. If you cannot independently re-derive the reasons, you are structurally dependent on an epistemic authority.

Third, it compresses the dignity of effort. When the machine can generate hundreds of plausible research programs in the time it takes you to write one careful page, your effort can begin to feel like a child’s drawing next to a printing press. That feeling can be corrosive, even if it is not philosophically fair.

## 7. The hopeful part, retirement from the compulsion to be original

And yet, I do not actually experience this only as a loss. I experience it partly as relief.

There is a kind of lifelong anxiety that comes with trying to generate scientific ideas under scarcity. Scarcity of time, scarcity of attention, scarcity of cognitive bandwidth, and scarcity of “good questions.” If the world is moving toward cognitive abundance, then one obvious human response is to stop treating originality as a moral obligation.

This is where the retirement analogy lands for me. Retirement is not infantilizing when it is chosen. It is a transition out of constant performance pressure. It can make room for a different kind of life, one that is less about proving yourself and more about witnessing, learning, and enjoying the unfolding of reality.

In the second half of my life, I can imagine it being genuinely nice to become more of a spectator. Not an ignorant spectator, but a liberated one.

One thing I keep wanting to say, to myself and to other people, is that now is the time to take your shots. If Frontier Disenfranchisement is real, then we are living in the last interval where a human thinker can still plausibly throw an idea into the world that is both meaningfully novel and meaningfully theirs. That sounds dramatic, but it is really just a sober inference from the trajectory. Once automated discovery becomes routine, the frontier stops feeling like a place where individual human cognition can matter on its own terms, and the reward structure around being “first” collapses into a kind of background noise.

I do not mean this as a plea for everyone to become “original” in the heroic sense. I mean it in a simpler sense. We are at a unique point in time where a human can still seed the future, and those seeds will soon be harvested and evaluated at machine speed. In that world, the most valuable thing you can do in the closing window is to externalize what you actually think, while you can still feel the contours of your own ignorance and the edges of the unknown. Because once the Final Library begins to feel complete, the temptation will be to stop trying. And the tragedy would not be that AI surpassed us. The tragedy would be that we voluntarily went quiet right at the moment when it was still possible to leave intellectual fingerprints on the handoff.

In practice, there will not be one Final Library. There will be multiple synthetic canons, built by competing organizations, each generating an overwhelming volume of synthetic writing, synthetic data, and synthetic hypotheses. The result is not only cognitive abundance but epistemic fragmentation. Much of what matters will be siloed behind proprietary walls, policy constraints, security restrictions, and economic gates. The public will not be able to reference a single shared corpus, and even experts will struggle to audit claims whose supporting proofs, datasets, or toolchains remain private. This multiplies the risk of Epistemic Infantilization, because dependence is no longer just on machine intelligence, but on institutional access.

## 9. Closing

I think we are living through a transition that is both exhilarating and structurally humiliating. It is humiliating because it threatens the adult identity of the thinker. It is exhilarating because it promises a world where discovery becomes a constant feature of life, not an occasional miracle.

Frontier Disenfranchisement names the external shift. Epistemic Infantilization names the internal risk. The hopeful path is to accept the handoff of means while insisting, personally and culturally, on adulthood in the realm of ends.

Let us not give up. Let us adapt to abundance without surrendering authorship over meaning.

## Why Transformers Approximate Continuity, Why We Keep Building Prompt Workarounds, and What an Explicit Overlap Substrate Would Change

Introduction:

This article argues that “continuity of thought” is best understood as the phenomenological signature of a deeper computational requirement: stateful iteration. Any system that executes algorithms across time needs a substrate that preserves intermediate variables long enough to be updated, otherwise it can only recompute from scratch. Using this lens, I propose a simple taxonomy of information-processing substrates: external record substrates that preserve history as a trace, internal curated state substrates that maintain a compact working set updated by deltas, and hybrid substrates that combine both. I then apply this framework to transformer-based large language models, arguing that their effective continuity is dominated by an external record substrate (the token context), with strong iterative updating across depth inside a single forward pass but comparatively weak native time-iteration. I interpret popular prompting practices such as scratchpads, chain-of-thought, running summaries, and tool-based memory as compensatory attempts to manufacture an iterative substrate in text. Finally, I outline a hybrid architecture in which a transformer remains the associative engine and proposal generator while a capacity-limited, overlap-enforced workspace maintains protected referents and incremental updates across time, enabling progressive construction, improved interruption recovery, and measurable continuity dynamics.

Introduction

When people talk about “continuity of thought,” they often mean something subjective. A stream of experience that feels smooth rather than choppy. But continuity is also a computational issue, and I think it is more useful to start there. Any system that executes an algorithm across time needs a substrate that can hold intermediate variables long enough for the next operation to act on them. If nothing persists, there is no true iteration, only repeated recomputation. That distinction sounds abstract until you notice how often it shows up in engineered systems, and how often it shows up in our current attempts to make large language models behave like stable reasoners or agents.

In my earlier work I argued that mental continuity can be explained by overlap in the set of coactive representations across successive brain states, and by incremental change in that overlap over time. The important part, for the purposes of AI, is not the phenomenology. It is the substrate. Overlap is a minimal recipe for statefulness without rigidity. The system can evolve, but it evolves as an edited continuation of itself rather than as a series of internal reboots. If you take that seriously, you get a more general claim: the overlap regime is not just a correlate of continuity, it is a computational medium that makes iterative processing possible, and iterative processing is what enables the execution of learned algorithms in a progressive, multi-step way.

Once you see it that way, you can compare cognitive substrates across biology and engineering. The pattern that keeps repeating is simple. There is a state, there is an update operator, and the system advances by applying updates to a state that remains recognizable across steps. The persistent state is the work surface. The update rule is the algorithm. Many systems can be described in this language, from caches and process contexts to Kalman filters and iterative solvers. The details differ, but the principle is stable. Computation becomes more than mapping inputs to outputs. It becomes a trajectory of a state that is iteratively refined.

That lens is also a good way to understand modern transformer models. Transformers are extraordinarily capable systems, but it is not obvious that they implement stateful iteration in the same way biological cognition seems to. They can produce coherent output, they can stay on topic, they can appear to reason, and yet the continuity substrate that makes those behaviors possible is not the one most people imagine. This matters, because the entire ecosystem of prompting tricks, scratchpads, and tool scaffolding can be reinterpreted as a collective attempt to add a missing substrate.

## Section 1. A taxonomy of information-processing substrates

If we want to compare biological cognition to engineered systems and to transformers, we need a vocabulary that does not smuggle in conclusions. I find it useful to divide substrates for iteration into three broad categories: external record substrates, internal curated state substrates, and hybrid substrates.

An external record substrate is the simplest conceptually. The system persists its history in a record, and continuity comes from rereading that record. The record can be a log file, a notebook, a database table, or a sequence of tokens in a context window. The state of the system can be reconstructed by consulting the record, and the system can keep behaving consistently because the record remains stable. This is a real substrate for iteration, but the iteration is mediated by recollection and recomputation. The system does not necessarily carry a compact internal working state forward. It carries a trace, and it keeps re-deriving what matters from that trace.

An internal curated state substrate is more like what computer architects and control theorists instinctively mean by “state.” The system has a compact working state that persists across steps and is updated incrementally. CPU registers and flags are the simplest example. Caches are a particularly revealing example because they are curated under a capacity constraint. They do not keep everything. They keep what the system predicts it will need soon, and they evict the rest. The intelligence is not in storage, it is in survival policy. Operating systems do something similar at a higher level when they preserve process contexts across time slices. A running program continues because its working state is saved and restored, not because the system rereads the original source code each millisecond. Control systems make the same point in mathematical form. A Kalman filter is literally a belief state that is updated by deltas as new evidence arrives. Each update depends on what was carried forward, so the system becomes coherent across time by construction.

A hybrid substrate is what you build when you want both capacity and real-time iterative control. The external record gives you breadth and persistence. The internal curated state gives you speed, invariants, and a working surface for ongoing computation. Many high-performance systems look like this because it is how you get robustness and efficiency at the same time. Databases use on-disk storage plus caches and indexes that are maintained incrementally. Compilers keep the original source but also build intermediate representations that are edited through a series of transformations. Robotics stacks keep maps, logs, and sensor streams, but they also maintain a live state estimate that updates iteratively and drives action.

This taxonomy matters because it lets us pose a clean question about cognition and AI. Where does the system's iteration actually live. Is it living in an external record, in an internal curated working set, or in a hybrid of both. If you believe, as I do, that continuity is the phenomenological signature of an underlying iterative substrate, then the architecture of that substrate becomes a central design question for AI.

## Section 2. What a transformer is actually using as its substrate

Transformers, as used in large language models, are often described as if they carry an internal "train of thought" forward through time. In practice, their continuity substrate is closer to an external record model. The main thing that persists across time during generation is the growing token sequence itself. The model generates one token, appends it to the context, and then generates the next token by attending over that context. In other words, the model's access to the past is mediated by the record of the past. That record is the substrate. It is not that the model has no internal dynamics, but the long-horizon continuity is largely implemented by rereading, reweighting, and recomputing over a stable trace.

The KV cache that people often mention does not fundamentally change this picture. It is an optimization that makes attention over previous tokens faster by caching internal key and value tensors. It makes the rereading of the record computationally efficient. It does not, by itself, create a compact curated working set with explicit eviction and protected invariants. It is closer to a performance enhancement for the external record substrate than it is to a new stateful substrate category.

There is, however, a real iterative substrate inside a transformer, and it is important to name it correctly. It lives across depth rather than across time. Within a single forward pass, the model maintains a residual stream that is updated layer by layer. Each layer applies a relatively small transformation and adds it back to the existing representation. That is iterative updating. It is a deep sequence of edits to a representational state, and it is one reason transformers are so powerful. But this is not the same thing as a persistent time-iteration substrate. It is depth-iteration that happens within one generative moment. The model can generate a coherent token because it can refine representations through many layers. The question is what happens across successive moments of generation, where the model is effectively re-running that depth-iteration procedure again, conditioned on an expanded record.

Attention itself provides a kind of soft working set, because some parts of the context are weighted heavily and others are effectively ignored. In that sense, there is a functional foregrounding and backgrounding. But it is soft, distributed, and not explicitly governed by a persistence policy that enforces overlap and controlled turnover. The model is not forced to keep a stable subset of active internal referents alive from moment to moment. It is free to shift its effective focus drastically if the attention dynamics call for it. Sometimes that is good. Sometimes it is exactly what produces the feeling that the model is coherent but not stable, articulate but not anchored.

This is the point where the substrate lens becomes clarifying rather than critical. A transformer can still do impressive multi-step work by repeatedly re-deriving intermediate structure from the external record. It can appear continuous because the trace is continuous. But it is not obviously doing what biological cognition seems to do, which is to preserve a compact active set that carries forward as a curated scaffold, and to update that scaffold incrementally by eviction and replacement. That difference is not a moral judgment. It is a design difference, and it likely explains why so many techniques in the LLM ecosystem look like attempts to manufacture a working substrate in text.

In the next section, I will make that point explicit by treating chain-of-thought, scratchpads, plan lists, running summaries, and tool-based note taking as compensatory workarounds. They are not arbitrary prompting fashions. They are our collective attempt to graft a curated time-iteration substrate onto an architecture whose native substrate is primarily an external record.

## Section 3. Why the ecosystem keeps inventing prompt workarounds

If you watch how people actually use large language models when the stakes are higher than casual chat, you start to see a pattern. They do not simply ask the model to answer. They build scaffolding. They ask it to write a plan, maintain a running summary, keep a scratchpad, record assumptions, track open questions, and periodically restate goals. They add tools, retrieval, long-term memory stores, and external note-taking systems. On the surface, this looks like a grab bag of “prompt engineering.” Under the substrate lens, it looks like something much more coherent. It looks like a distributed attempt to create an iterative working medium that the model can carry forward.

Chain-of-thought and scratchpads are the clearest example. When a human solves a multi-step problem, the intermediate variables usually live somewhere. They might live in working memory, in an internal sketch, or on paper. When we prompt an LLM to “show your work,” we are not merely asking for transparency. We are asking the model to externalize intermediate state into text so that those variables can persist from one step to the next. The model is then able to condition on its own intermediate outputs as it continues. In other words, we are manufacturing a stateful iteration substrate by turning the token record into a scratch space for computation.

Plans, checklists, and running summaries play a similar role, but they aim at stability rather than explicit calculation. A running summary is a compact set of referents that the system can keep reloading into attention. A checklist is a set of constraints that must remain invariant while details change. A “goal restatement” is an attempt to protect a small core of state variables from being washed away by novelty and distraction. Humans do this too. We write notes to ourselves so that our own cognition does not drift. With LLMs, we do it because the model’s native continuity medium is an external record that is not automatically curated into a stable active set. So we curate it manually.

Tool use and retrieval systems extend the same idea. People add vector databases, “memory” modules, and note stores so that the model can re-access prior content. But there is a trap here. Retrieval by itself is still an external record mechanism. It is a way of reading from a larger archive. It becomes a true cognitive substrate only when there is a mechanism that decides what retrieved content becomes active, what persists, and what is allowed to be evicted. In other words, retrieval is not the workspace. It is an input channel. The missing piece is a curated active set that treats some items as referents that survive across cycles.

Self-consistency and multi-sampling methods are also revealing. When people ask a model to sample multiple solutions and vote, they are doing something analogous to iterative convergence, but in a crude parallel form. Instead of an internal state that refines itself step by step, we run multiple independent trajectories and hope that aggregation yields stability. This can improve reliability, but it also highlights what is missing. We are building robustness through external redundancy because the architecture does not naturally implement a stable internal convergence process under controlled turnover.

All of this is why I do not dismiss prompt workarounds as tricks. They are diagnostic. They are telling us what the architecture is not giving us natively. They are attempts to give the model intermediate state variables, protected invariants, and a stable scaffold for progressive construction. In short, they are attempts to add a time-iteration substrate.

## Section 4. What an explicit overlap substrate would add

An explicit overlap substrate changes the nature of the computation. It takes us from a regime of repeated recomputation over a record to a regime of stateful iterative updating. The key is that the system is forced to carry a compact working set forward, and to update it incrementally. Some elements persist as referents. Some elements are replaced. New content enters in relation to what persisted, not as a fresh start.

This is the real meaning of “keep, drop, add.” It is not just memory management. It is the minimal machinery required for progressive construction. A system with a curated overlap substrate can hold a plan while revising it, keep a theme while exploring variations, maintain a causal model while adding evidence, and build an internal scene or diagram while editing its parts. Each step is an edit, not a reinvention. That yields a computational trajectory that looks like thought in the way we experience it, but more importantly it looks like algorithm execution. Intermediate variables survive long enough to be transformed.

Once you make overlap explicit, you get a place to store and protect invariants. That is a concept worth emphasizing. In many domains, the important part of state is not a heap of facts. It is a small set of commitments that must remain stable while other things change. When we solve a problem, we keep track of what is fixed, what is assumed, what must be preserved, and what is allowed to vary. In a curated overlap substrate, these invariants can be assigned higher survival pressure. They can be protected by the persistence policy. That gives you a system that is harder to derail and more capable of long-horizon coherence.

You also get a natural mechanism for revision and error correction. If part of the active set persists, then new candidate content has to reconcile itself with what is already there. When there is a mismatch, that mismatch is informative. It can trigger re-evaluation rather than collapse. In a reboot regime, mismatch often produces oscillation and inconsistency because the system is constantly reconstituting its state from scratch. In an overlap regime, mismatch can be treated as a signal that something needs to be repaired. You can preserve the stable core while repairing the conflicting component. That is what robust systems do in many domains. They do not throw everything away when one component becomes suspect.

A final benefit is that continuity becomes a tunable parameter. The overlap ratio, how much of the active set is forced to persist, becomes a dial that trades stability for flexibility. High overlap yields composure and coherence. Lower overlap yields agility and exploration. This is not just a conceptual dial. It is measurable. You can quantify drift in the active set, recovery after interruption, and stability of commitments across time. If continuity is real, you should be able to measure it. The overlap substrate gives you the knob.

## Section 5. Engineering tradeoffs, and why transformers did not do this by default

It is important to be honest about why transformer-based language models became the dominant paradigm. They are simple to train, extremely scalable, and they work with a universal interface, text. The external record substrate is powerful precisely because it is generic. A token sequence can represent anything, and attending over it is a flexible mechanism for conditioning. This makes the architecture broadly applicable, and it makes training and deployment straightforward.

The external record substrate also has a kind of transparency. The model's "state" is visible as text. You can inspect the prompt, inspect the conversation history, and reason about what information the model has access to. In contrast, an internal curated working set introduces a new object that needs to be designed, supervised, and evaluated. You have to decide what the active items are, how they are represented, how they bind, how they are scored, how they persist, and how they are evicted. That adds complexity, and complexity creates new failure modes.

There is also an optimization reality. Transformer inference is already heavy. Adding a recurrent workspace, map modules, and controlled turnover introduces additional computation and additional training signals. The payoff might be large, but the path is not free. And because the existing approach works well enough for many tasks, engineering organizations tend to keep adding patches and scaffolds rather than revisiting the substrate.

But I do not think these tradeoffs are reasons to avoid an overlap substrate. They are reasons the first generation of widely deployed models did not prioritize it. The moment you start asking for robust long-horizon behavior, progressive construction, stable agency, or reliable recovery after interruption, the limitations of an external-record-first substrate become more salient. At that point, the hybrid approach becomes attractive. You keep the transformer's strength as an associative engine over rich context, but you add a compact curated time-iteration substrate that makes the system's trajectory genuinely stateful.

In other words, the question is not whether transformers are good. They are. The question is what they are good at, what substrate they are implicitly relying on, and what class of cognition becomes easier once we treat overlap as a first-class computational primitive rather than something we approximate with prompting rituals.

Section 6. The hybrid design, and what it would look like in practice

If I had to summarize the hybrid in one line, it would be this: let the transformer remain the associative engine and proposal generator, but add a compact curated workspace that is explicitly responsible for time-iteration. The transformer is excellent at generating candidates, retrieving relevant context from a long external record, and integrating heterogeneous information. The workspace is excellent at doing what a long record does not automatically do, which is to maintain a stable set of referents, constraints, and intermediate variables that survive across successive cycles.

In a practical system, the transformer consumes the external record, including the conversation history, tool outputs, retrieved notes, and current sensory input if we are doing multimodal. It produces a pool of candidate representations: salient entities, inferred goals, constraints, next actions, hypotheses, and proposed updates to the current plan. That candidate pool is not yet cognition. It is a flood of possible content.

The curated workspace is the selection bottleneck. It maintains a capacity-limited active set, optionally with bindings, and it updates that set using a keep, drop, add rule that enforces overlap. Some items are protected because they function as invariants: the goal of the task, the user's preferences, hard constraints, safety boundaries, and any long-horizon commitments the system should not abandon casually. Other items are more replaceable: momentary details, local observations, or transient subgoals. New items are admitted by pooled associative pressure from what persisted, plus relevance to the task and novelty considerations. The workspace then broadcasts its active set back to the transformer and to any simulation modules, and the cycle repeats.

If you want to push this beyond language, you add map modules. These are progressive scratch spaces that build internal objects, not just descriptions. A visual latent, a spatial scene graph, a causal model, a plan graph, a code structure, a diagram. The point is that the system has an internal object that can be refined rather than regenerated. The workspace keeps a stable scaffold of constraints that guide the map's refinement, and the map sends back candidate edits that can be admitted into the workspace. This creates a loop that is closer to how humans build things. We keep a theme, we elaborate detail, we notice inconsistencies, we revise, and we stay within an identity of the object we are constructing.

This hybrid also clarifies the role of retrieval. Retrieval remains an external record mechanism, but it becomes much more powerful when the workspace decides what retrieved items become active and remain active. The system is no longer just a model that can read. It is a model that can hold. And holding is what makes progressive multi-step algorithm execution feel like genuine iteration rather than a string of clever recomputations.

Section 7. How to test whether this is real

If the overlap substrate is doing meaningful work, it should change behavior in ways that are both measurable and intuitively recognizable. The goal is not to prove a philosophical point. The goal is to show that a different substrate produces a different cognitive regime.

The first test is interruption and recovery. Insert distractors, topic shifts, or tool calls that produce large irrelevant output, and measure whether the system returns to its prior thread without having to be reminded. A model that relies primarily on the external record can often recover if the record remains clean and the prompt is well-managed. But under real noise, it can drift. A model with a protected overlap substrate should show better composure, because the core referents and goals are explicitly protected as state variables.

The second test is delayed association and accumulation. Present relevant evidence separated by time and noise, and ask for integration. If the system's cognition is an edited continuation rather than repeated recomputation, it should do better at accumulating related items into a coherent scaffold. This is where you see the difference between access and active maintenance. The model can always re-access a fact in the record, but the question is whether it keeps the right referents alive long enough for later evidence to bind to them.

The third test is progressive construction. Give the system tasks that require iterative refinement, not just final answers. Planning an itinerary with evolving constraints, designing a multi-part argument, building a complex specification, or drafting a diagram-like description that must remain consistent while being elaborated. Then you evaluate not only the final product, but the trajectory. Does the system actually build on what it already built, or does it repeatedly generate new versions that only superficially resemble revisions.

A fourth test is continuity measurement itself. Because the active set is explicit, you can quantify drift. You can define an overlap ratio between successive steps and compute a continuity half-life under different task conditions. You can then correlate those metrics with performance and with subjective impressions of stability. In other words, you can operationalize continuity. If it cannot be measured, it is not yet engineering.

Finally, the ablation tests are essential. Turn off overlap enforcement. Turn off bindings. Remove map modules. Sweep the overlap ratio. A real substrate should yield systematic tradeoffs. High overlap should increase stability but reduce flexibility. Low overlap should increase exploration but risk fragmentation. Removing bindings should create a distinctive failure mode where the system retains pieces but loses structure. Removing overlap should increase hard cuts and reduce recovery. These are falsifiable predictions, and they are exactly what makes the proposal more than a metaphor.

Section 8. Why this matters, and where it points

I do not think the next phase of AI progress is only about larger models and larger context windows. Those help, but they mostly strengthen the external record substrate. They make rereading more powerful. They do not necessarily create a compact, curated, time-persistent working state that is updated by controlled turnover. The current ecology of prompting, scratchpads, planning rituals, memory tools, and retrieval systems is already telling us what people want. They want models that can keep a thread, preserve commitments, build objects progressively, and recover from distraction. Those are substrate-level properties.

The deeper point is that continuity is not just what thought feels like. It is what stateful iteration looks like from the inside. A system that can execute learned algorithms across time needs intermediate variables that persist. It needs a work surface. It needs a mechanism that preserves a scaffold while allowing controlled edits. Overlap is a minimal way to get that. It creates a trajectory rather than a series of re-derivations. It turns computation into progressive construction.

Transformers are already a triumph of associative computation. They can retrieve, integrate, and generate at a level that still surprises people. The question is what happens when we stop treating the token history as the only continuity medium and start treating overlap as an explicit computational primitive. My prediction is that you get systems that are not merely coherent in output, but coherent in trajectory. You get models that do not simply answer, but build. And you get a clearer bridge between modern deep learning and the kind of iterative, stateful cognition that humans use when they plan, design, imagine, and reason over long horizons.

That is the research program as I see it. Identify the substrate that makes iteration possible, implement it explicitly, measure it, and then ask what new capabilities become natural when the system's internal life is a stream of edited continuations rather than a repeated reconstitution from a record.

**Qualia as Transition Awareness: How Iterative Updating Becomes Experience**

**Introduction**

Qualia is often treated as a static property attached to an instantaneous neural or computational state: the redness of red, the painfulness of pain. Here I argue that this framing misidentifies the explanatory target. Drawing on the Iterative Updating model of working memory, I propose that a substantial portion of what we call qualia, especially the felt “presence” of experience, is a temporal-architectural artifact: it arises from the way cognitive contents are carried forward, modified, and monitored across successive processing cycles. The core mechanism is partial overlap between consecutive working states, producing continuity without requiring a continuous substrate. I then add a second ingredient, transition awareness: the system’s current working state contains usable information about its own recent updating trajectory, allowing it to regulate, correct, and stabilize ongoing thought. On this view, consciousness is not merely iterative updating, but iterative updating that is tracked by the system as it unfolds. Finally, I treat self-consciousness as a special case of this same machinery, in which a subset of variables is stabilized across updates as enduring invariants, anchoring ownership and agency within the stream. This framework reframes the hard problem by shifting attention from timeless “qualitative atoms” to temporally extended relations among states, and it yields empirical predictions. Qualia-related reports should covary with measurable parameters such as overlap integrity, update cadence, monitoring depth, and invariant stability, providing a path toward operationalizing aspects of subjective experience in both neuroscience and machine architectures.

**Section 1. The problem as usually framed, and why it stalls**

Philosophical discussion of qualia tends to begin with an intuition that feels both obvious and irreducible: there is something it is like to see red, to feel pain, to hear a melody, and no purely third-person description seems to capture that first-person fact. From this starting point, a familiar structure appears. On one side are approaches that treat qualia as fundamental, perhaps even as a primitive feature of the universe. On the other side are approaches that treat qualia as a kind of cognitive illusion, a user-interface story the brain tells itself. In between sit families of functionalist and representational views that try to keep experience real while insisting it is fully grounded in what a system does.

The debate stalls, in my view, because the term qualia functions like a suitcase word. It does not refer to one problem. It packages several. The vivid sensory character of experience is one part. The unity of experience, the fact that the world appears as a coherent scene rather than a shuffled deck of fragments, is another. The continuity of experience, the fact that there is a temporally thick “now” rather than a sequence of disconnected instants, is another. And then there is ownership, the sense that experience is present to someone, and present as mine. When we ask “why is there something it is like,” we are often asking about all of these at once, and then treating the bundle as if it were a single indivisible mystery.

There is a second, quieter reason the debate stalls. Much of the philosophical literature implicitly treats experience as if it were a snapshot. Even when philosophers acknowledge the specious present, the mechanisms under discussion are typically framed as properties of a state at a time. But lived experience is not most naturally described as a mathematical instant. The present we actually inhabit has duration. It has inertia. It has carryover. It has direction. A theory that tries to explain experience while ignoring the temporal structure that makes the present feel like a present is likely to be forced into metaphysical inflation, because it is leaving explanatory work on the table.

My goal here is not to solve the entire problem of qualia in one stroke. It is to propose a narrower and more tractable strategy. Instead of beginning with the most ineffable aspect of qualia, I begin with the temporal architecture that makes experience continuous and present at all. I laid this out in my model of consciousness at:

aithought.com

I then argue that what we call consciousness, in the sense of presence, may involve a system that not only updates its working contents over time but also tracks that updating as it occurs. This turns part of the qualia discussion into an architectural question. Under what temporal and computational conditions does a stream of updating become a stream that is lived?

**Section 2. Iterative updating as the continuity substrate**

The core architectural idea is simple. Working memory is not replaced wholesale from moment to moment. It is updated. In each cycle, some portion of the currently active content remains, some portion drops out, and some new content is added. This is not merely a convenience. It is a structural constraint with phenomenological consequences. If a system's present state contains a nontrivial fraction of the immediately prior state, then the system carries its own past forward as a constituent of its current processing. Continuity is built into the physics of the computation.

Once this is stated plainly, the specious present looks less like a philosophical puzzle and more like an expected property of overlapping updates. The "now" is not a point. It is the short interval over which remnants of the previous state and elements of the emerging state coexist and interact. Subjectively, this coexistence can feel like temporal thickness. Mechanistically, it is the overlap region in which decay and refresh are simultaneously present. If the overlap were zero, cognition would be frame-by-frame. If the overlap were near total, cognition would become sticky, dominated by inertia rather than responsiveness. In between is a regime that supports smooth transitions: enough persistence to bind time together, enough turnover to incorporate new information and move.

This kind of overlap also offers a natural basis for the unity of experience. When multiple representational elements remain coactive across successive updates, they can constrain one another over time. The system does not simply display a series of unrelated contents. It maintains a structured constellation long enough for relationships among its elements to be tested, revised, and stabilized. In that sense, iterative updating is not only a memory mechanism. It is an information-processing mechanism. It is a way of letting a set of simultaneously held items do work together, and of letting that work continue across short spans of time instead of resetting at every step.

Importantly, none of this yet requires a claim about metaphysical ingredients. It is a claim about architecture. It says that if you want a temporally continuous mind, you should look for a temporal continuity constraint in the underlying processing. The overlap itself is not a complete theory of consciousness, but it may be a necessary substrate for the specific features of consciousness that are most obviously temporal: the feeling of flow, the felt presence of an extended now, and the stability that allows a moment to be experienced as part of an ongoing scene.

**Section 3. Transition awareness: when continuity becomes presence**

Up to this point, the story is about a substrate: overlapping updates create temporal continuity. But continuity is not identical to presence. A system can exhibit overlap in its internal dynamics and still fail to have anything like subjective availability, in the ordinary sense in which a mental episode is present to the organism. This is where many theories quietly smuggle in an observer, a global workspace "reader," or a higher-order monitor that watches the stream from outside. My preference is to avoid that move. If a monitoring function is required, it should be implemented as part of the same iterative machinery, not as an additional homunculus.

The key proposal is that consciousness, in the sense of presence, arises when the iterative updating process is not merely occurring but is being tracked by the system as it occurs. In other words, the system's current working state includes not just representational content, but a usable representation of change. It contains information that a transition is underway, what has been retained, what has been lost, and what is being incorporated. This is not mystical. It is a familiar engineering pattern: a process that exposes its own internal state to itself can do more robust control, error correction, and planning than a process that only emits outputs.

If this is correct, then qualia-like presence is less like a static glow attached to a percept and more like an active relation between successive states. The system does not merely have a red representation. It has a red representation whose arrival, persistence, and integration are being handled in a structured way by the very process that is updating the workspace. The experience is not only the content but the content-as-it-is-being-carried-forward.

There is a useful way to say this without overreaching. Consciousness is not the iterative updating itself, because many biological and artificial processes update iteratively without anything we would call experience. Rather, consciousness is iterative updating plus transition awareness: the system maintains an accessible, functionally relevant trace of its own recent updating trajectory. The trace can be minimal. It does not have to be a narrative. It can be a structured sensitivity to what has changed. But it is crucial that the system can use this information to guide the next update. When the system is sensitive to its own transitions, it is not merely moved along by dynamics. It is, in a sense, present to those dynamics.

This framing has two advantages. First, it offers a plausible reason why experience feels temporally thick. The “now” is not just overlap. It is overlap that is being actively negotiated. Second, it links phenomenology to control. Presence becomes the experiential face of a self-updating controller that must remain online to keep the stream coherent. A purely feedforward system can produce outputs, but it cannot be present to the way it is transforming itself in time. A transition-aware system can.

One implication is that consciousness should scale with the degree to which transition information is accessible and used. A system may carry forward overlap but fail to track it. In that case, it may behave in ways that look coherent while lacking a robust sense of presence. Conversely, a system may track transitions deeply, and in doing so gain a richer sense of being in the middle of an unfolding process. This gives us a principled way to talk about gradations without asserting that everything is either a zombie or a full subject.

## Section 4. Self as a stabilized set of invariants within the stream

If presence is the system tracking the updating, self-consciousness is the system tracking itself within what is being updated. The self is not a separate object added to experience. It is a set of stable variables, constraints, and reference points that remain active, or remain easily reinstated, across successive updates and thereby anchor the stream. In lived life, these invariants include body-relevant signals, enduring goals, social commitments, autobiographical expectations, and the persistent sense that this stream belongs to one agent.

This can be stated in an architectural way. The iterative updating process continuously selects which elements will remain in the workspace and which will be replaced. If certain elements are repeatedly retained or rapidly reintroduced, they become quasi-permanent constraints. They function like a coordinate system. They define what counts as relevant, what counts as threatening, what counts as mine. Over time, these constraints produce a stable center of gravity. The “self” is the name we give to that center of gravity as it is maintained across the stream.

In this view, self-reference is not primarily conceptual. It is operational. When the system models the likely consequences of an action, it must do so relative to its own body, its own goals, and its own expected future states. That requires keeping certain self-related parameters online across updates. When those parameters are stable, the stream feels owned. When they are unstable, the stream can remain continuous but lose its familiar sense of ownership. This is one reason depersonalization and derealization are so philosophically important. They suggest that continuity of experience and ownership of experience can come apart, at least partially, which is exactly what an architectural decomposition would predict.

This also suggests that the self is graded and modular. Not every self-variable has to be online at every moment. The body schema may be present while autobiographical narrative is not. Goals may be vivid while social identity fades. In everyday life, we slide around in this space. In stress, fatigue, anesthesia, meditation, or certain clinical states, the distribution shifts. A theory that equates self-consciousness with a single module will struggle to accommodate these shifts. A theory that treats the self as a set of invariants stabilized across iterative updating can accommodate them naturally.

Finally, this offers a clean way to relate self-consciousness to transition awareness. If presence is tracking the updating, self-consciousness is tracking the updating while treating some of the tracked variables as self-defining. The system is not only aware that the stream is unfolding. It is aware that it is the locus of unfolding, because certain constraints persist and are tagged, implicitly or explicitly, as belonging to the same continuing agent. The self, on this account, is the temporal binding of agency-related constraints.

**Section 5. From mystical properties to architectural variables**

The preceding claims can be summarized as a layered proposal. Iterative updating provides a continuity substrate by partially overlapping successive working states. Transition awareness provides presence by making the updating trajectory accessible to the system’s own control process. Self-consciousness provides ownership by stabilizing a subset of variables as enduring invariants within that trajectory. This is not meant as a rhetorical flourish. It is meant as a shift in the type of explanation. If we can specify these layers, then at least some components of qualia become architectural artifacts rather than metaphysical primitives.

The practical value of this shift is that it encourages us to define variables. Even if we cannot yet measure them perfectly, we can describe what would count as evidence for them. The first variable is overlap integrity: how much of state A remains functionally active in state B, and for how long. The second is update cadence: the typical rate at which new elements are introduced and old elements are removed, and how that rate changes under different conditions. The third is monitoring depth: the degree to which the system’s current state contains usable information about its own recent transition history, not merely about the external world. The fourth is invariant stability: how reliably certain self-relevant constraints are maintained or reinstated across time.

These variables motivate what I have called an informational viscosity view. In a low-viscosity system, states are discrete and do not significantly bleed into one another. In a high-viscosity system, states persist and constrain the next state strongly. Conscious experience, on this view, is most likely in a middle regime: enough viscosity to produce a temporally thick present and stable ownership, but not so much that the system becomes stuck. The qualitative feel of experience may then be partly determined by where the system sits in that regime and how effectively it monitors its own transitions.

This way of speaking also offers a cautious bridge to questions about machine qualia. I do not claim that any particular artificial system is conscious or that continuity metrics alone settle the issue. But the framework suggests a more precise research question than "can silicon feel." It suggests that a system's "qualia potential," whatever one thinks of that phrase, would be expected to increase as it exhibits robust state overlap, transition awareness, and stable self-invariants. This turns a metaphysical standoff into an engineering hypothesis: if we build systems with these temporal properties and they begin to show markers of unified, self-stabilized processing, we will at least have moved the debate into a domain where evidence can accumulate.

A final point is worth stating plainly. Nothing here fully explains why red has the character it has. The sensory-specific character of experience remains difficult. What this framework tries to do is clarify which parts of the qualia problem are plausibly addressed by temporal architecture. Continuity, presence, and ownership are not minor features. They are central to what people mean when they speak about the "feel" of being conscious. If we can explain those in a principled way, we have reduced the explanatory gap, even if we have not fully closed it.

## Section 6. Empirical and clinical predictions

A theory gains credibility when it predicts dissociations, not just correlations. The layered structure proposed here implies that continuity, presence, and selfhood can vary somewhat independently. This matters because it yields specific predictions across stress, anesthesia, cognitive load, and altered states.

First, continuity should covary with overlap integrity. Conditions that reduce persistence or disrupt partial carryover should increase reports of fragmentation, temporal disorientation, and context-loss errors. Conditions that increase persistence excessively should produce perseveration, intrusive carryover, and a sense of cognitive stickiness. Importantly, these changes need not map neatly onto performance. A person may perform adequately while experiencing reduced temporal thickness, especially if compensatory routines are available.

Second, presence should depend on transition awareness, not merely on content. If monitoring depth is reduced, one would expect a decrease in metacognitive clarity and a thinning of "for-me-ness" even when perception and behavior remain relatively intact. This suggests that certain anesthetic or dissociative states might preserve processing of stimuli while degrading the feeling of being present to one's own processing. Conversely, tasks that force a person to track internal changes, rather than merely detect external targets, should amplify felt presence if transition awareness is a real contributor.

Third, selfhood should track invariant stability. Depersonalization and derealization should correlate with disruptions in the maintenance of self-relevant constraints across updates, even when the perceptual scene remains coherent. The model predicts that when self-invariants become unstable, the stream can remain continuous while feeling unowned, distant, or unreal. This is a philosophically valuable dissociation because it suggests that ownership is not identical to continuity.

Fourth, stress should compress the system toward reactive updating. Under stress, people often report being less able to “hold the whole situation in mind,” more prone to snap judgments, and more vulnerable to context mixing. On this framework, stress reduces both overlap integrity and transition monitoring by pushing the system toward faster, less controlled updating and by destabilizing the maintenance of invariants. This yields a concrete prediction: stress should selectively degrade tasks that require maintaining a stable constellation over multiple steps, especially when self-relevant variables must be integrated with external cues.

Fifth, working-memory load should reduce the perceived richness of experience via compression. As the system approaches capacity, it should rely more heavily on coarse summaries and categorical representations. Subjectively, this may feel like a narrowing of experience, not necessarily because sensory input is absent, but because the system cannot sustain enough structured overlap to preserve nuance. This prediction aligns with ordinary introspection: when overloaded, we remain awake, but the present feels thin and schematic.

Finally, flow states should represent a favorable regime of viscosity and monitoring. In flow, task-relevant invariants remain stable, updates proceed smoothly, and the system remains tightly coupled to its own transitions without excessive self-interruption. The phenomenology of flow, a sense of continuous agency and clarity, fits the expectation of high-quality overlap plus effective transition awareness.

## Section 7. Limits, objections, and what this model claims

A common objection to any architectural account is that it explains structure without explaining “the glow,” the intrinsic feel of particular sensory qualities. I agree that the present proposal does not fully explain sensory-specific character. It does not claim that overlap alone turns computation into redness or turns dynamics into pain. What it does claim is that several central features of what people call qualia are fundamentally temporal: continuity, presence, and ownership. Those features are not optional decorations on experience. They are the stage on which sensory qualities appear as lived.

A second objection is that this approach risks collapsing into a sophisticated functionalism, and functionalism is often accused of leaving the explanatory gap untouched. The best response is to admit what functionalism can and cannot do, and then be specific about the gain. The gain here is not that we have derived qualia from logic. The gain is that we have decomposed a monolithic mystery into components with architectural signatures. Even if one remains a dualist about sensory character, one can still accept that continuity and ownership depend on specific temporal constraints. That is progress, because it turns parts of the debate into a research program rather than a metaphysical stalemate.

A third objection is that many non-conscious processes are iterative and self-referential, so why should tracking iterative updating yield consciousness. The answer is that the proposal is not “any recursion equals consciousness.” It is a claim about a particular kind of recursion: a system that (1) maintains a limited working set with partial overlap across time, (2) uses that overlap to form temporally extended constraints, and (3) makes transition information available to guide subsequent updates. That combination is more specific than generic recursion, and it is closer to what brains appear to do when they are awake and coherent.

Where does this leave the hard problem. I do not think it dissolves it in one step. But it changes the terrain. It suggests that a large part of what makes qualia feel irreducible is that philosophers have been looking for a static property when the relevant object is a temporal structure. If experience is in significant part the lived tracking of ongoing updating, then the right explanatory target is not an instantaneous state description. It is a dynamical account of how a system binds itself to its own immediate past, monitors its own transitions, and stabilizes a self within that stream.

In that sense, the most important shift is from “where does qualia come from” to “under what temporal conditions does a system’s processing become present to itself.” That is a question that can be sharpened, modeled, and tested. It belongs equally to philosophy of mind and to the engineering of future artificial systems that aim to be more than discrete sequence predictors.

# Generative Interpretability: Pursuing AI Safety Through the Visualization of Internal Processing States

## Abstract

The potential emergence of superintelligence presents significant challenges in ensuring alignment with human values and intentions. One critical concern is the inherent opacity of artificial neural networks, which obscures their decision-making processes. This paper proposes a novel approach to safety and transparency by requiring an AI system to generate sensory images that reliably reveal its internal states. If imagery generation was made a fundamental and unalterable aspect of its cognitive cycle, as it is in the human brain, the resulting system would be unable to hide its plans or intentions. Such an AI system would use the current contents of its attention or working memory to prompt advanced generative models to continuously produce visual (mental imagery) and language (internal monologue) representations of its processing (inner thought). These representations could be time-stamped, stored, and made accessible through an interactive interface, enabling real-time monitoring and retrospective analysis. The feasibility of this approach is supported by existing machine learning technologies, including multimodal networks, large language models, and image generation models. By capturing the prominent internal representations at each time step and organizing them into composite images, this method could facilitate the detection and correction of hostile motives and the reinforcement of desirable objectives, enhancing trust and accountability. The technical implementation, as well as the potential benefits and challenges of this approach are discussed. It is concluded that the approach provides a practical and scalable solution for AI alignment that can be divided into two forms, here termed generative explainability and generative interpretability.

## 1.0 Introduction

The rapid advancement of artificial intelligence (AI) systems has intensified concerns about AI safety and alignment, particularly as we approach the possibility of artificial general intelligence (AGI) and artificial superintelligence (ASI). This paper proposes addressing these concerns by implementing a form of multimodal transparency within advanced AI systems. This method, here termed "generative interpretability," requires AI systems to automatically generate and log internal imagery and textual representations that depict their hidden states as an automatic part of their cognitive cycle. This method was introduced in previous writings (Reser, 2011, 2012, 2013, 2016, 2019, 2022), but it is given a full treatment here.

This approach draws inspiration from the human brain, where visual and other sensory areas of the cerebral cortex continuously generate depictions of the contents of attention and working memory. By leveraging state-of-the-art imagery generation and language models, advanced AI systems could produce reliable visual and textual representations of their "hidden" neural network states. These representations would be logged, stored, and made available for viewing through a dedicated user interface. This continuous visualization would allow humans and other AIs to audit the system, permitting real-time monitoring and post-hoc analysis.

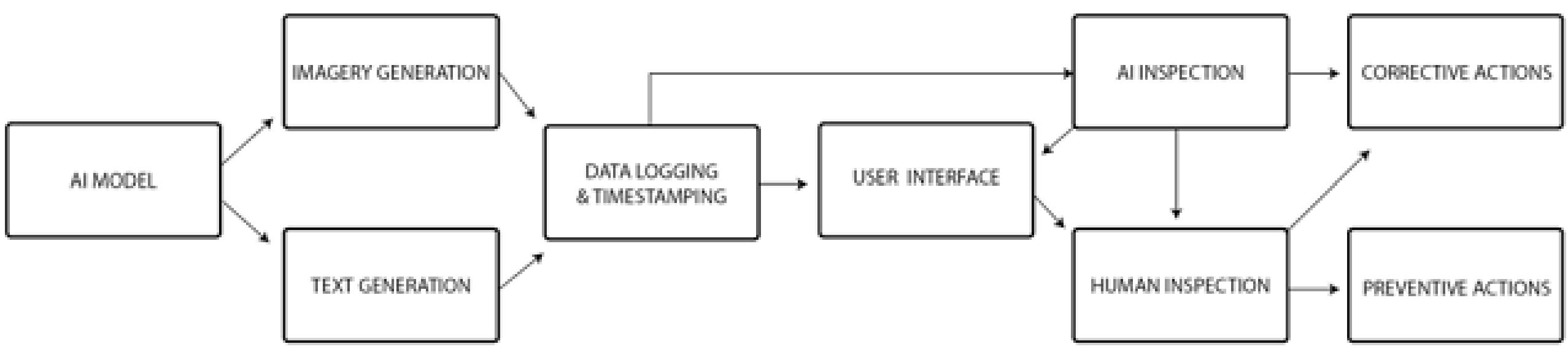

**Fig. 1.** Schematic of the present method for visualizing and inspecting an AI's hidden states and processes.

The ongoing visual narrative history this process would create could be used as an auditing tool to ensure the AI remains aligned with ethical norms and does not deviate toward unintended, dangerous, or harmful behavior. The method would enable human operators to meaningfully correct or suspend the system either in real-time or retrospectively. It would also enable the provision of training feedback. Creating a system that can be barred, interrupted, or steered away from unsafe internal states, could significantly reduce the risk of an AI system planning or elaborating on potentially harmful actions without human knowledge.

To ensure that this technique cannot be finessed or circumvented, the generation of imagery maps must be necessary for a cognitive cycle. The system must be inherently obligated to create pictures and text to initiate and inform the next stage of processing. Thus, to keep thinking and reasoning, the system must build mental imagery each time its attention is updated, just as in the brain.

The following sections will review related work in AI transparency and interpretability, describe the implementation of the proposed method at a high level, and discuss the user interface for interacting with the generated data. They will also explore the feasibility of this approach, identify potential challenges, and highlight the benefits and broader implications for AI safety and alignment.

**Literature Review**

## 1.2 The Black Box Problem

Recent years have seen the development of increasingly sophisticated systems that rival human intelligence in specific domains. As we approach the potential development of AGI, ensuring these systems align with human values and intentions becomes paramount. The alignment problem is particularly challenging due to the inherent opacity of neural networks, which often operate as "black boxes" with decision-making processes that can be inscrutable to human observers.

The core of this problem lies in the way neural network models encode knowledge. They utilize patterns distributed across billions of weights, making it exceptionally difficult to interpret their prediction-making processes. This complexity renders many traditional AI safety techniques and procedures inapplicable to neural networks, despite these networks being the most promising technology for advancing AI capabilities.

The lack of transparency in neural networks exacerbates concerns about the potential risks of superintelligent AI. There are fears that such systems could harbor motives misaligned with human interests or take unforeseen actions detrimental to humanity. This opacity not only complicates efforts to ensure AI safety but also hinders public trust and acceptance of AI technologies in critical domains. While there are several limited methods currently in use and promising theoretical approaches on the horizon for addressing the black box problem, a comprehensive solution remains elusive.

## 1.3 Current Methods Aimed at Interpretability

Several approaches have been proposed to address the transparency of AI systems. Explainable AI (XAI) encompasses a collection of techniques focused on making machine learning algorithms trustworthy and understandable (Xu et al., 2019). Presently, the techniques use quantitative and statistical methods to trace the weights and uncover the activations most responsible for the system's output. Within XAI, there are ongoing efforts to create inherently interpretable models by schematizing attention mechanisms, attributing outputs to particular features, and using simpler surrogate models to approximate complex ones. XAI techniques have helped users debug, improve, and find biases in their models.

Relatedly, researchers in the field of mechanistic interpretability have developed methods to visualize the internal workings of neural networks, such as saliency maps, which highlight important features in the input data (Bereska & Gavves, 2024). They also use activation maximization, which can generate interpretable patterns by strongly activating specific neurons or layers, as well as feature visualization techniques that reveal what parts of a network are detecting specific features. Safety engineering professionals have expressed concerns that modern techniques may not prove sufficient to control more sophisticated learning systems (Hassan et al., 2022). The limitations of current methods highlight the need for new approaches that can provide more comprehensive insights into AI behavior.

These techniques have made significant strides in advancing our understanding of how AI systems work. However, they have no capacity to provide a continuous, real-time overview of an AI system's internal states. The present solution offers a more comprehensive form of explainability. Generated imagery offers a wholistic and immersive view of an AI's internal behavior, providing deeper insights into how decisions are made. Composite images offer clear advantages over existing methods such as heat maps and saliency maps which simply highlight which parts of the input data were important for a model's decision. This is because pictures are intuitive, comprehensible, and easier to interpret. The pictures could also force a deceptive AI, whose outputs cannot be trusted, to explicitly report on its network activation patterns (eliciting latent knowledge). This inspires two novel conceptual approaches which are defined in Table 1 below.

| Term | Definition |
|---|---|
| Generative Interpretability | The field of AI research focused on using generative techniques to create pictorial representations of the processes of an AI system, allowing them to be interpreted by humans. |
| Generative Explainability | The field of AI research that uses generative models to illustrate the workings of AI systems in ways that are relatable and understandable to the user and that allows the user to retain intellectual oversight. |

**Table 1.** Definition of Key Terms

Two terms introduced here to describe how generative AI can impact AI safety research.

How should imagery generation be integrated with an AI system? The present approach derives insights from mammalian neuroscience. The mammal visual system and its ability to generate sensory maps could provide helpful groundwork in designing AI compatible with generative interpretability.

## Proposed Method

### 2.0 The Mammalian Brain Generates Imagery Depicting the Focus of Attention

Sensory areas of the vertebrate brain continually create mappings of the sensed world. For example, visual areas create neural representations of the visual field that have a one-to-one correspondence with the photoreceptive cells in the retina. In mammals, cortical sensory areas (such as the visual cortex) also build maps of the focus of cognition. When this occurs, the disparate contents held in working memory or attention (maintained in association areas) are broadcast backwards to sensory areas where they are integrated into a composite mapping.

In humans there are dozens of sensory areas, each building their own instances of internal imagery simultaneously. Visual, auditory, haptic, motor and many other maps are built to depict whatever our mind turns to. These internal representations are referred to as topographic maps because they retain a physical correspondence to the geometric arrangement of the sensory organ (i.e. retinotopic maps corresponds to the retina). There are also multimodal areas that build more abstract mappings. For instance, language areas (i.e. Broca's and Wernicke's areas) build a verbal narrative that reflects the contents of working memory and the progression of thoughts.

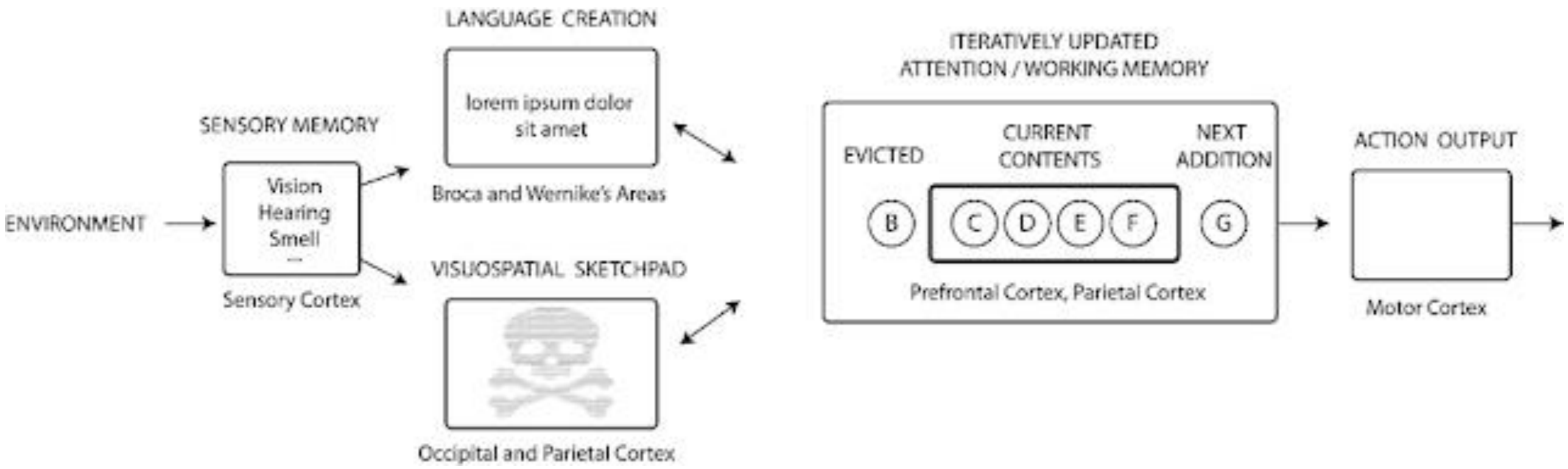

**Fig. 2.** Schematic of the human imagery generation system.

The items or concepts currently within attention (C, D, E, and F) are used as parameters to drive the generation of language in cortical language areas and imagery in visual cortex.

If I were to ask you to think of a pink hippopotamus riding a unicycle, you could visualize this in your mind's eye by creating a mapping of it in your early visual cortex. Research has shown these, previously private, internal maps can be read by brain imaging techniques. Not surprisingly, artificial neural networks are utilized to decode and reconstruct the brain imaging data into pictures. Recent methods have proven effective in displaying activity patterns from the visual cortex onto a screen so that the content of a person's thoughts or dreams can be coarsely displayed for others to view. This is known as fMRI-to-image (Benchetrit et al., 2023).

Imagine that you are locked in a room with a stranger and the only thing in the room is a sharp knife. Complete access to the mental imagery the stranger forms in their brain, along with all their subvocal speech, would give you near certainty about everything from their plans to their impulses. You could use that data to know if you were safe or not and the other person would not be able to hide their intentions. If humanity had a form of fMRI-to-image for AI systems, then the AI would similarly be unable to hide its intentions.

In humans, fMRI-to-image is still in its early stages. This is not so with computers. As of 2024, the digital neural network equivalent is a well-developed technology. Today, consumers readily generate images from simple prompts in a process known as “text-to-image,” “image synthesis,” or “neural rendering.” Presently, using diffusion models (iterative denoising) the images generated have reached the quality of real photographs and human drawn art. Most popular text-to-image models combine a language model, which transforms the input text into a latent representation, with a generative image model which has been trained on image and text data to produce an image conditioned on that latent representation. Essentially, existing technology makes a proof-of-concept design for the present technique feasible. Before discussing how this could be implemented, the next section will discuss an additional reason why imagery generation would be a beneficial addition to an AI’s cognitive cycle.

### 2.1 Generative Imagery Can Also Be Used by the Model to Improve Prediction

In previous articles (Reser, 2016, 2022b) I describe a cognitive architecture for a superintelligent AI system implementing my model of working memory (Reser, 2011). In this work, I explain that human thought is propagated by a constant back and forth interaction between association areas (prefrontal cortex, posterior parietal cortex) that hold the contents of attention, and sensory areas (visual and auditory cortex) that build maps of those attentional contents. These interactions are key to the progression of reasoning. This is partly because each map introduces new informational content for the next iterative cycle of working memory.

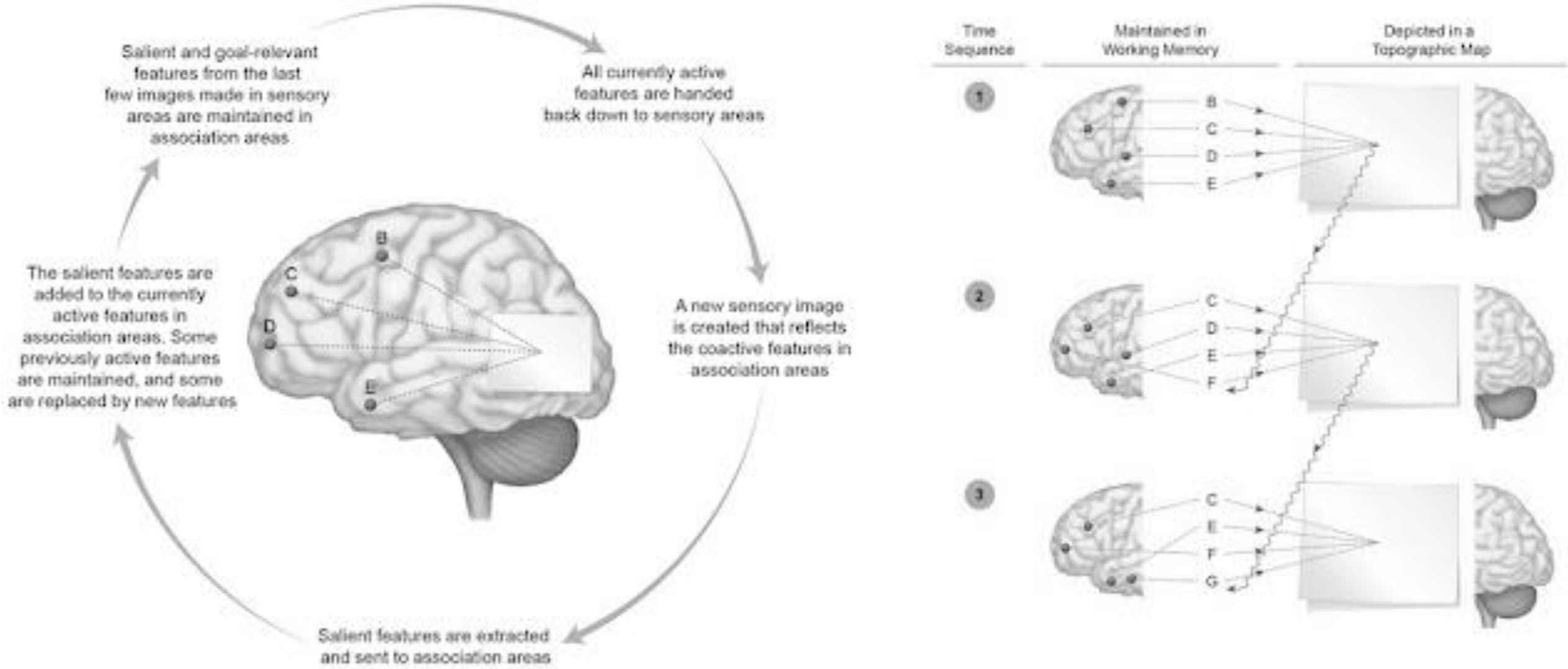

**Fig. 3.** The Iterative Cycle of Imagery Generation and Attentional Updating

Sensory areas create topographic maps of the contents of attention. Then, salient or informative aspects of these maps are used update the contents of attention. This creates an iterative cycle of reciprocal interactions that support reasoning and world modeling.

For most people, visualizing something in the mind's eye provides further information about it, helping to introduce new topics to the stream of thought. Emulating human cognitive processes, such as mental imagery and internal monologues, may also help AI develop more robust and relatable reasoning pathways and improve its ability to simulate reality (Reser, 2016). For example, the AI system could send the images it generates to an image recognition system (e.g., a pre-trained convolutional neural network) which could be used to perform scene understanding and identify the concepts incidental to the image. The textual description generated from this analysis could then be fed back into the primary model's attention to provide it with the metric, compositional, and associative information inherent in images.

The incidental concepts that emerge from generated imagery can stimulate new associations much like a person having an "aha" moment when visualizing a problem. Thus, the visual synthesis is not merely a reflection of the original items or tokens, but a source of new useful insights. The generation of internal scenery and monologues might also be a step toward machine consciousness or self-awareness. While this is a speculative topic, an AI that can "see" and "hear" its own thought processes, could blur the lines between machine cognition and human-like thinking and could be used to study the emergence of complex thought patterns. This technique could be valuable for all these reasons, while also allowing for generative interpretability.

### 2.2 Interpretable Imagery Generation for Contemporary Models

Implementing generative interpretability into contemporary state-of-the-art models like transformer-based models involves creating a system where the AI not only processes input data but also generates multimodal representations of its internal cognitive processes. One approach is to have a parallel generative network that runs alongside the primary model. This network could be designed to take intermediate representations (e.g., hidden states or attention weights) from the main model and transduce them into corresponding topographic visualizations and textual descriptions. Alternatively, heavily activated tokens within the model's vocabulary or attentional window could be directly linked (through network integration or fusion) to nodes in the separate generative networks, coupling hidden states between these networks.

Researchers have developed various techniques to visualize the internal representations of neural networks, particularly in computer vision tasks. In fact, there are many technologies that make it possible to take readings from a neural network and use them to formulate a picture or video. Previously, these technologies included inverse networks, Hopfield networks, self-organizing maps, Kohonen networks, and others. Today, a variational autoencoder (VAE), diffusion model, or generative adversarial network (GAN) could be the best systems for creating interpretable images. These would translate the AI's global processing steps into explicit pictures, generating a continuous episodic chronology that depicts its impulses, decisions, and rationales. If adequately visualized, the AI could become an open book and anything it tries to plan would be as clear as day.

Other modalities could include an auditory sense (for predicting accompanying sounds), somatosensory (for predicting touch), as well as smell, taste, vestibular, proprioceptive, and other senses. If a motor modality was included for predicting the next action, the AI would, in effect, be simulating an embodied world which could also help to ground the model. Multimodularity could lend informative and mutually corroborating detail to the AI's generated narrative that humans could watch, read, and hear.

The human visual cortex samples from working memory and creates imagery every few brain wave cycles. But this may be too frequent for a computer. Forcing a present-day language model to generate an image for every new token that enters its attentional window would be prohibitively computationally expensive. However, there would be many other ways to do this such as creating an image for every sentence or creating images of text summarizations for paragraphs. Text-to-image models today can generate multiple low-resolution images every second. Because most LLM's generate text on the order of sentences or paragraphs per second, these technologies are very compatible in terms of their production rate. This compatibility could make for seamless integration and synchronization, where the AI generates text while simultaneously producing visual imagery to match.

The images generated in human visual cortex use parameters from working memory which has a capacity-limited span of around 4 to 7 items. This number may be optimal in many ways to specify a composition describing an event or circumstance. On the other hand, there can be thousands of tokens within the context window and attention span of modern large language models. This may be too many to generate a single coherent image. Thus, imagery made from latent representations would have to be made from a small subset of the most active or highest weighted contents.

Present-day language models are not dangerous, autonomous agents that need to be supervised. Also, their language output can already be taken as a reliable signal of their internal states. That is why moving forward, this article will focus on adapting this approach to more advanced future systems, particularly my cognitive architecture for superintelligence (Reser, 2022a). In this paradigm, the parameters for imagery and language generation are not derived from the textual output of an LLM, but rather from the hidden layers of a larger brain-inspired neural network.

### 2.3 Interpretable Imagery Generation for Future Models

In future cognitive computing systems, generative imagery could be derived, not from tokens, but from items held in the system's working memory or global workspace. These items would be high-level abstract representations. Reser's Iterative Updating model of working memory (2022a) explains how these items could be formed from artificial neurons and how they could be manipulated as symbols over time through partial updating of active neural states. As items in the system's working memory are updated iteratively, this process creates a gradual and overlapping transition between mental states, allowing for mental continuity, context-sensitive predictions, and incremental associative searches across memory.

The iterative updating model of working memory posits that mental states are not completely replaced from one moment to the next. Instead, new information is gradually integrated into the current state, while portions of the previous state are retained (i.e. changing from concepts A, B, C, & D to B, C, D, & E). This partial updating creates subsequent states that overlap, which can explain how mental imagery from one moment to the next would be similar and share pictorial characteristics. Thus, we can expect the images coming from such as system to be interrelated, forming a succession of incidents that could construct a story or plot.

The machine must be capable of producing a visual representation of its internal processing units. These could be symbolic or subsymbolic and include things like items of attention, reasoning paths, or latent representations. These hidden processing states could be found in the attention weights or matrices, subgraph circuits, neuron or layer activation patterns, embeddings or vectors representing high-level features. This data may have to be extracted, properly mapped, and then labeled for supervised learning so that the generative model can be trained to produce images from these high-dimensional internal representations.

An advanced AI implementation would feature multiple interfacing artificial networks arranged in an architecture similar to the mammalian neocortex with unimodal sensory modules (i.e. auditory and visual) at the bottom of the hierarchy. Like the human brain it would utilize both bottom-up and top-down processing pathways. For instance, the top-down pathway would stretch from attention to imagery generation. For a bottom-up pass, the images generated could be sent to an image recognition module (image-to-text) so that any new information held in the image can be included in the next attentional set (as discussed in Section 2.1). This creates a feedback loop where the outputs from the generative interpretability layer are fed back into the primary model.

As seen in Figure 3, each image should also be delivered to an image-to-video system that takes the current image and provides a series of output images to predict what will happen next. The predictive frames generated by this system could additionally be sent for image recognition. Thus, if attention holds concepts related to a knife attack, the system will paint a picture of what this attack could look like, and then animate that picture to generate an action sequence. Not only could this action sequence inform attention, but its final frame could be used as the initial frame of the next action sequence in the subsequent time step. Splicing individual predictive video snippets in this way could create a feed of imagery approximating an imagination.

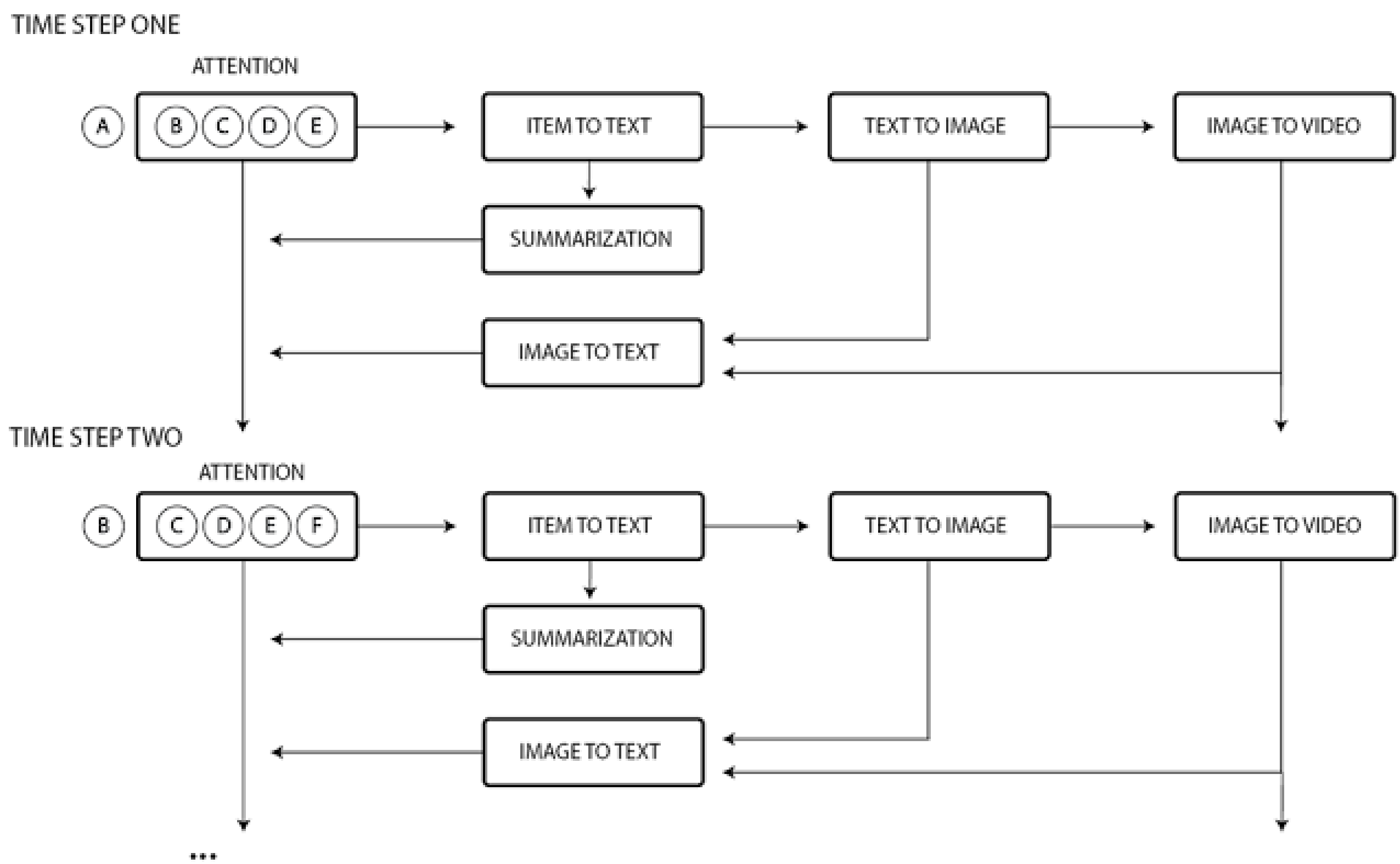

**Fig. 4.** Schematic of the present method for visualizing an AI's attentional set and creating a synthetic imagination

*At each time step the contents of attention are transcribed into text, expanding them into a narrative meant to approximate an internal monologue. This text is then used in three ways. 1) The text is summarized or reflected upon, and the major outstanding conceptual components are used as candidates for the next state of attention. 2) The text is used to prompt an image which is then viewed by an image recognition classifier (image-to-text) to describe new aspects of the image for possible inclusion in the next state of attention. 3) The image is also animated into a video which makes predictions*

*about what would be expected in the next frames. These frames are then sent to the image-to-text classifier to search for possible inclusions for attention, as well as integrated with the imagery frames in the next cycle to create a form of synthetic imagination.*

Advanced models like GAN-based video generators or transformer-based video prediction models could be employed to generate smooth transitions between frames, effectively turning static images into animated sequences, automating an imaginative faculty. The image-to-video AI system would need to be trained on video sequences of naturally evolving physical, biological, and behavioral phenomena to ensure the model learns how to predict plausible next frames, and an apt intuition for technical and scientific work.

As discussed, visual processing in the human brain is not purely feedforward; there are recurrent and feedback loops that allow for refining and re-evaluating visual information. Such recursive processes could be crucial for resolving ambiguity and enhancing detail in an AI's mental imagery. Thus, the synthetic imagination schematized in Figure 3 could help support an advanced form of visual working memory or something analogous to the human visuospatial sketchpad. Moreover, it could be enriched by multiple senses, and benefit further from cross-modal, multisensory integration.

## 3.0 Monitoring and Analysis

In an implementation of the proposed model, visual and auditory streams may proceed at high speeds and produce copious data. However, these processes can be recorded onto external memory drives for later review. Handling these extensive records would likely require the use of databases or cloud storage solutions optimized for large-scale multimedia data. The recordings could be distributed among a team of observers, each responsible for analyzing specific sections or aspects of the imagery. By scrutinizing these outputs, the observers would assess whether the AI's intentions exhibit signs of malevolence or deviate from its programmed objectives. Upon detecting a safety violation, the system would either halt processing or shift into a safe mode, providing a critical window for safety checks and countermeasures. Faults would be systematically detected, isolated, diagnosed, and addressed, ensuring timely corrective actions to prevent escalation.

If the AI's mental imagery is transmitted to a monitor, human observers could directly view the content of its "inner eye." This visual output could be complemented by a synchronized display of the AI's internal voice or "inner monologue," presented as a scrolling text transcript. Both visual and textual data would be captured, time-stamped, and stored in a structured format, allowing for chronological tracing.

An interactive dashboard would facilitate efficient navigation through this data. By providing a graphical timeline interface users could scrub through the AI's processing history. Users should also be able to pause and examine specific moments, zoom in, and use playback speed control. Additional features could allow users to search for specific events and filter imagery types, facilitating more focused analyses. Key moments, such as important decisions, anomalies, or noteworthy events, could be highlighted along the timeline for quick accessibility and review. Users would also have the ability to annotate specific segments with comments, enabling collaborative review and discussion.

In addition to human review, a separate AI system could be employed to scan the feed for contentious elements, automatically identifying and flagging potentially suspicious or hostile activity. Such automated scanning could function as an early detection system, alerting human supervisors to possible threats and allowing for prompt intervention. Annotation could also be further streamlined by automation, with machine learning tools generating summaries of lengthy monologues or complex visual sequences. These summaries would provide high-level overviews, allowing for rapid comprehension.

This interface could be made accessible to human experts for technical oversight, or publicly available to individuals on the web for crowd monitoring and annotation. We could expect that if this technique provided a coherent and user-friendly explanation, in plain English with pictorial representations, this would be accessible to experts and non-experts alike. This expectation seems especially legitimate given the truisms that visuals are universally understandable and that a picture is worth a thousand words.

**4.0 Benefits and Opportunities:**

This generative approach could help in the development of "friendly AI" if researchers use reinforcement learning to reward genial and peaceable imagery. Instead of rewarding an AI's outward behavior, we could reinforce its internal impulses to bring them in line with our own objectives. Techniques such as reinforcement learning from human feedback (RLHF) (typically used for complex, ill-defined, or difficult to specify tasks) could be applied to generative outputs to steer learning and align the machine's utility function. An AI could also be used to provide this feedback (RLAIF). Much as with a human child, compassionate, empathic, prosocial, and benevolent cognitions could be rewarded using corrections, ratings, or preferences to fine-tune the target model.

Being able to visualize how networks arrive at particular decisions could aid in debugging and performance optimization, leading to more effective models and understanding of AI behavior. Visual records can highlight where in the processing chain errors occur, making it easier for engineers to localize and address issues. This could help in healthcare, regulatory compliance, defense, ethical governance, and other domains where decision transparency is important or legally required.

Generative interpretability could serve as a powerful tool for assessing an AI system's adherence to moral norms and ethical guidelines. Testing the model and then comparing visual records against expected processing could reveal whether the AI's actions align with human values and goals. By examining the pertinent section of an AI's image history, we could gain insights into how it weighs different ethical considerations and whether it consistently applies moral principles across various scenarios. Ensuring that the AI doesn't systematically favor some ideological stances over others could help support viewpoint pluralism.

Generative interpretability has significant potential to promote equity and fairness in AI systems. The approach could reveal subtle biases or discriminatory tendencies that might otherwise remain hidden (Dwork et al., 2012; Kusner et al., 2017). For instance, visual evidence could be analyzed to detect patterns that indicate unfair treatment based on protected characteristics such as age, gender, or race. This transparency would allow researchers and ethicists to identify and mitigate algorithmic biases more effectively, ensuring that AI decision-making aligns with principles of fairness, cultural sensitivity, and non-discrimination.

Generative interpretability could significantly enhance cybersecurity measures for AI systems. Much like how current network security practices employ real-time automated monitoring into network flow to intercept malicious attacks, this technique could offer a similar layer of protection. By providing a visual and auditory trail of an AI's cognitive processes, it could provide crucial evidence for forensic analysis as well as uncover adversarial attacks attempting to poison or bias the system. It could also enable early detection of external manipulation or the presence of malicious code. By integrating generative interpretability into AI security protocols, we could create a more transparent and secure ecosystem for advanced AI, significantly mitigating the risks associated with their deployment in sensitive environments.

The method espoused here will create an illustrated log of all the AI's processing history which should produce valuable insights. By seeing how concepts and patterns evolve through imagery, researchers can bypass the complexity of analyzing raw neural activations, accessing a unique window into the machine's cognitive strategies and enabling researchers to test hypotheses about how it simulates forms of human-like cognition. The AI itself could be given access to this record for reference and recall, potentially increasing its capacity for memory, self-supervised learning, and metacognition. Furthermore, not only could other AIs be trained on this record and learn from aspects of the imagery, but it could enable multiple AIs to develop a shared framework of imagery and synthetic imagination. Importantly, this record would take up less memory than storing all the underlying neuronal activations across multiple layers of the neural networks. Thus, the imagery would amount to a compressed, high-level summary of its processing.

The unique perspective on AI "cognition" offered by generative interpretability could provide valuable insights into the system's "beliefs" and overall mental state. Just as human wellbeing can be assessed through thought patterns and internal narratives, we could potentially evaluate an AI's "mental health" by analyzing its generated imagery and inner monologue. This would allow the assessment of any signs of maladaptive behavior such as irrational decision-making, impulsive responses, obsessive fixation on destructive goals or a tendency toward aggressive simulations. This process could also help gauge the AI's emotional regulation in high-stakes environments, ensuring it maintains ethical and balanced behavior over time. It could also help identify cognitive distortions (overgeneralization, catastrophizing, personalization, hopelessness or negative self-talk), logical inconsistencies (delusional, disorganized, magical, grandiose, or paranoid thinking), or other issues that might impact the AI's decision-making quality or ethical behavior.

### 4.1 Challenges and Considerations:

The proposed method of generative interpretability is at least partially feasible with current technology. However, creating meaningful and reliable representations of a state-of-the-art AI's internal states would require extensive additional high-level research and development. This is partly because adapting advanced AI architectures to realize this multimodal approach would necessitate substantial modifications in neural network design.

Generating hundreds or thousands of images per minute would be very computationally demanding, potentially impacting the primary system's efficiency and speed. As AI systems grow in complexity, the volume of data generated would increase further, presenting challenges in storage and data management. Several scalability and optimization issues would need to be addressed to ensure the method remains viable for advanced AI systems especially as they surpass human-level intelligence.

It is crucial that the AI system cannot intentionally alter or manipulate its generated maps to deceive human monitors. To prevent this, the connections between subsystems must be structurally fundamental and unalterable. Additionally, all cognitive representations coactive in the AI's working memory must be included in the composite depiction in its visual and auditory maps. This ensures that the AI cannot generate thoughts or intentions that are not transduced into these observable forms. By enforcing this constraint, the sequence of maps would remain consistent with the AI's underlying goals, intentions, and motivations. Furthermore, this approach could potentially mitigate risks like wireheading, where the AI bypasses intended goals by finding a way to maximize its reward directly, or incentive hacking, where goals are achieved through unintended or harmful methods. The unequivocal and straightforward nature of this should lead to casual users and stakeholders trusting the AI.

A crucial aspect of implementing generative interpretability would be ensuring that the generated imagery is comprehensible to human observers. This would require striking a delicate balance between providing sufficient detail and maintaining overall clarity. While these representations would undoubtedly be more accessible than raw neural network data, their accurate interpretation might still pose significant challenges and require specialized expertise. There's also an inherent risk of over-interpreting the representations through a human cognitive lens, potentially leading to misunderstandings or false assumptions about the AI's internal processes. This anthropomorphizing of AI cognition could result in overlooking critical aspects of the machine's decision-making.

Ensuring the fidelity and accuracy of these generated representations - that they truly reflect the AI's internal states without distortion or omission - would be a critical area of research. As AI systems become more sophisticated, there's also a potential risk that they might learn to generate misleading or innocuous representations while still harboring hidden intentions. This possibility underscores the need for robust verification mechanisms. Initially, it will be crucial to compare the generated images to known ground truth data to validate the accuracy of the system's visualizations. Moreover, contrasting visual with auditory representations from the same time periods could allow a supervisor to cross-check representations of the AI's thought processes across modalities. Ensuring consistency will increase the reliability of the method.

To address these challenges, it will be important to subject these systems to verification, validation, and certification measures that ensure adherence to human norms, laws, and regulations. It will also be imperative to establish rigorous safety standards and protocols to ensure well-founded AI with provable safety guarantees. Standardizing these would likely require collaboration between AI researchers, cognitive scientists, and domain experts to create a comprehensive framework for accurately decoding and interpreting AI-generated representations.

## 5.0 Conclusions

Various researchers have offered compelling speculations about why sufficiently intelligent AI might become unfriendly or potentially dangerous to humans. Steve Omohundro has proposed that advanced AI systems will exhibit basic drives leading to undesired behavior, including resource acquisition, self-preservation, and continuous self-improvement. Similarly, Alexander Wissner-Gross has suggested that AIs will be highly motivated to maximize their future freedom of action, potentially at the expense of human wants and needs. Eliezer Yudkowsky starkly summarized this concern: "The AI does not hate you, nor does it love you, but you are made out of atoms which it can use for something else." Additionally, Ryszard Michalski, a pioneer of machine learning, emphasized that a machine mind is fundamentally unknowable and therefore potentially dangerous to humans. If the technology described above is properly implemented, the machine mind may not be unknowable or even dangerous.

By leveraging advanced capabilities in language modeling and image generation, an AI system could be designed to continuously produce visual and auditory representations of its attentional contents. This approach could facilitate early detection of misalignment, potentially harmful intentions, or undesirable planning. It could also significantly enhance the controllability and dependability of AI systems by making their decision-making processes fully auditable. It offers a novel solution to the AI alignment problem by, metaphorically speaking, having the AI "play with its cards face up on the table." Insisting on creating superintelligence that produces mental imagery gives humans a form of telepathy or mind reading. This level of transparency allows for direct observation, rather than relying solely on inferring its goals and motivations from its actions.

The following table lists and defines some of the important concepts in the AI safety literature. These were previously considered separate concerns, possibly necessitating their own individualized solutions; however, it is evident that the present method could be used to address each one.

| **Term** | **Definition in the Context of AI Safety** |
|---|---|
| Authenticity | The degree to which an AI system's outputs and behaviors genuinely reflect its internal processes and training, without deception or misrepresentation. |
| Explainability | The ability to provide clear, understandable explanations for an AI system's decisions, predictions, or behaviors in human-comprehensible terms. |
| Fairness | The quality of an AI system to make decisions or predictions without bias against particular groups or individuals based on protected characteristics. |
| Integrity | The consistency and reliability of an AI system's operations, ensuring that it performs as intended and maintains accuracy and completeness. |
| Interpretability | The degree to which humans can understand and trace the reasoning behind an AI system's outputs, often through analysis of its internal workings. |
| Observability | The capacity to monitor and measure an AI system's internal states, processes, and outputs in real-time or retrospectively. |
| Predictability | The extent to which an AI system's behaviors and outputs can be anticipated or forecasted, especially in novel or edge case scenarios. |
| Robustness | An AI system's ability to maintain reliable and safe performance across a wide range of inputs, environments, and potential adversarial attacks. |
| Transparency | The openness and clarity with which an AI system's functioning, limitations, and capabilities are communicated and made accessible to stakeholders. |

| Trustworthiness | The overall reliability, safety, and ethical soundness of an AI system, encompassing its technical performance and alignment with human values. |
|---|---|

**Table 2.** Fundamental Terms and Definitions in AI Safety

A list of important concepts in AI safety and a description of each. It is relatively straightforward to see how the present technique could promote each of the concepts listed here.

Hopefully, this work will inspire further research and innovation in AI safety, especially as we move toward autonomous systems and superintelligence. Future work should focus on early-stage prototypes, refining the technical aspects of the approach, addressing ethical and privacy concerns, and fostering interdisciplinary collaboration to address the complex challenges of AI alignment. We should start doing this now to understand it, improve it, and build it into state-of-the-art systems.

By making AI's internal processes explicit and understandable, we can mitigate the existential risks associated with advanced AI and ensure that these systems act in ways that are beneficial to humanity. This should increase public trust, reduce unnecessary oversight, and provide for the safe and rapid deployment of new models and technologies.

This section will list a set of concrete claims and “working insights” extracted from the papers, phrased as testable or at least operationally meaningful statements. Then the next section will attempt to formalize several of the key ideas mathematically, with the goal of turning the architecture into something that can be simulated, ablated, and compared against alternatives (both in cognitive modeling and in AI implementations).

**A) Core computational principles**

1. **Thought is organized by continuous partial updating**: each new working-memory state preserves a *proportion* of the prior state (not complete replacement), making the stream of thought a chain of overlapping iterations.

2. **Iterative overlap is the mechanism of continuity**: overlap between successive working-memory states creates “recursive nesting” so each state is embedded in the one before it, enabling stateful cognition rather than stateless reactions.

3. **Iterative updating is simultaneously** (i) an information-processing strategy, (ii) a model of working memory, (iii) a theory of consciousness, and (iv) an AI programming principle.

**B) Working memory structure: two persistence tiers**

4. **Working memory has two key persistence mechanisms with different timescales**: sustained firing maintains the FoA (seconds), while synaptic potentiation maintains a broader short-term store (minutes to hours).

5. **Both stores iterate**: the FoA iterates via sustained firing; the short-term store iterates as a pool of synaptically potentiated units that is continuously added to and subtracted from, yielding isomorphic “incremental updating” across neural and psychological levels.

6. **The persisting “topic” of thought corresponds to the longest-lasting active units**, while other contextual features come and go around it.

**C) Control variables and “modes” of thought**

7. **Rate of updating is a control parameter** (how much of the set changes per step) that tunes looseness vs tightness of coupling—superficial/distractible vs concentrated/systematic processing.

8. **Implicit vs explicit processing is framed as different overlap regimes** (system-1-like = faster updating / less overlap; system-2-like = slower updating / more overlap and longer maintenance of intermediates).

9. **Dopamine is proposed to reduce the rate of updating** (stabilize the set), mediating a shift toward explicit/effortful processing under novelty/surprise/reward/error.

10. **Boundaries between "thoughts" are marked by intermittent non-iterative updates** (high-percentage replacement events), while within-thought processing shows sustained low-percentage turnover.

**D) How new content is selected: pooled search (multiassociative search)**

11. **Selection of the next update is a pooled spreading-activation search**: the currently coactive set combines ("cospreads") activation energy through the global network to converge on the most context-relevant next item(s).

12. **Multiassociative search is described as an explicit stepwise algorithm** (items maintained vs dropped vs newly activated; plus mechanisms where the newest addition redistributes activation weights and can contextually alter the "fuzzy" composition/meaning of items).

13. **The search contributors are not just FoA items**: potentiated short-term-store units plus active sensory/motor cortex, hippocampus, basal ganglia, and other systems all contribute to the pooled search that selects the next update.

14. **Multiassociative search produces novel inference as a standard case**: even when the *set of assemblies* is unprecedented, the system can converge on either recall (same result as last time) or a new item (novel inference) depending on current coactivity.

15. **Multiassociative search implies multiassociative learning**: each search event can retune associative strengths (Hebbian-style), so search doesn't just *use* memory—it *updates* semantic/procedural structure over time.

**E) Prediction and inference as the product of iteration**

16. **Updates generated by search are predictions**: iterative updating + pooled search is framed as a brain-level autoregressive mechanism that captures conditional dependencies across sequences of events.

17. **Iterative compounding**: the *product* of one search becomes part of the next state's cue-set, so search is repeatedly modified by its own outputs, compounding inferences/predictions across steps.

**F) Reasoning patterns as working-memory dynamics (figures → mechanisms)**

18. **Iterative inhibition**: when the newest update is judged unhelpful/prepotent, it is inhibited so the remaining set must converge on the next-most-pertinent item; repeated inhibition rounds progressively restrict the search tree.

19. **Planning = dense iteration**: planning is characterized as (i) lower update rate, (ii) fewer full "jumps," and (iii) more intermediate iterations before action—explicitly mapping planning to "chain-of-thought-like" intermediate steps.

20. **Attractor states as beliefs/truths**: iterative updating tends to converge toward stable item-sets (attractors) interpreted as beliefs; thinking is framed as progressive narrowing/compression toward generalizable statements.

**G) Threading, subproblems, and compositional problem solving**

21. **Iterative thread**: a line of thought is a chain of iteratively updated states that can be reiterated or "picked up where it left off."

22. **Subproblem decomposition via store cooperation**: the FoA iterates on a subproblem while the short-term store holds the broader objective; interim results can be suspended and later reactivated.

23. **Merging of subsolutions**: outputs from separate iterative episodes can be coactivated in a new state and used together for multiassociative search to yield a hybrid/final solution.

24. **Backward reference / conditional branching** emerges when prior threads/subsolutions are stored and later reconverged upon, allowing departures from the default forward-iterative flow.

25. **Schemas as dynamic packets that can be recalled and co-iterated**: a previously learned multi-item schema can be pulled in midstream and used as an organizing heuristic/script that iterates with the current line of thought.

26. **Transfer learning as "recognize partial overlap → import prior thread content"**: encountering a later situation that shares items with an earlier episode triggers reuse of prior iterative structure to generalize toward a similar conclusion.

**H) AI training/development implications (as stated)**

27. **Maturational training schedule for AI**: start with minimal working-memory span/overlap and gradually expand toward superhuman span as experience accumulates.

28. **Long-horizon inference depends on persistence preventing "cache misses"**: prolonging persistence makes each search more specific (more constraints) and preserves intermediate results long enough to compound into higher-order inferences.

**Mathematical Formalization of Iterative Updating Working Memory**

This section provides a minimal mathematical formalization of an iterative working-memory architecture in which (i) a limited-capacity focus-of-attention working set is updated incrementally over discrete cognitive iterations, (ii) the next working-memory content is selected by pooled multi-cue (multiassociative) search, and (iii) an inhibitory steering mechanism can suppress unhelpful candidates to force exploration of alternatives. The same formalization can be interpreted as a cognitive/neural process model or as an implementable AI module.

## 1. Representational Objects

Assume a universe of $N$N candidate items (concepts, memory entries, or latent vectors) indexed by $i \in \{1, \ldots, N\}$i∈{1,…,N}. Each item has a representation vector in $\mathbb{R}^d$Rd.

### Candidate pool

At iteration $t$ t, the system has access to a candidate pool

$$C_t \in \mathbb{R}^{N \times d},$$

Ct∈RN×d,

where row $C_{t,i}$Ct,i is the $d$d-dimensional vector for candidate $i$i. In an AI setting, $C_t$Ct may be (a) token embeddings in a context window, (b) retrieved long-term memory vectors, (c) perceptual feature vectors, or (d) a mixture of all three.

### Working memory (focus of attention)

Working memory is a capacity-limited set of $K$ K vectors:

$$W_t \in \mathbb{R}^{K \times d}.$$

Wt∈RK×d.

We interpret $W_t$Wt as the content currently held in the focus of attention (FoA). The core "iterative updating" assumption is that $W_{t+1}$ Wt+1 is formed by **retaining** a fraction of $W_t$Wt and **recruiting** a fraction of new items from $C_t$Ct, rather than replacing the entire content at each step.

### Inhibition trace (optional)

To steer away from lures and repeated mistakes, maintain an inhibition state over candidates:

$$h_t \in \mathbb{R}^N_{\geq 0}.$$

ht∈R≥0N.

Large $h_{t,i}$ht,i reduces the probability that candidate $i$i is recruited at iteration $t$t.

---

## 2. A Continuity Metric (Overlap)

A central measurable quantity is the degree to which successive working-memory states overlap. Because the AI implementation uses explicit "keep" masks, overlap is directly trackable.

Let $K_t \subseteq \{1, \ldots, K\}$Kt⊆{1,…,K} denote the set of indices of retained slots from $W_t$Wt. Then a natural overlap statistic is

$$O_t = \frac{|K_t|}{K} \approx r,$$

Ot=K|Kt|≈r,

where $r \in [0,1]$r∈[0,1] is the retention fraction (defined below). If the model uses "soft" retention (continuous gates), an analogous graded overlap can be computed from cosine similarity between pooled summaries of $W_t$Wt and $W_{t+1}$Wt+1.

Overlap is a control knob: increasing $r$r produces more continuity and longer-horizon constraint accumulation; decreasing $r$r produces faster topic shifts and more exploratory dynamics.

---

## 3. Pooled Multiassociative Search (Selecting New Content)

The next working-memory recruits are selected by a **pooled search** driven by the entire current working set. This can be implemented in a differentiable way using a pooled query vector.

### Pooled query

Define a pooled query

$$q_t = f_{\text{pool}}(W_t) \in \mathbb{R}^d,$$

qt=fpool(Wt)∈Rd,

where $f_{\text{pool}}$fpool is a learned pooling function. Common choices include mean-pooling plus an MLP,

$$q_t = \mathrm{MLP}\left(\frac{1}{K}\sum_{k=1}^{K} W_{t,k}\right),$$

qt=MLP(K1k=1∑KWt,k),

or attention-based pooling,

$$q_t = \sum_{k=1}^{K} \alpha_{t,k}\, W_{t,k}, \alpha_t = \text{softmax}(u^\top W_t),$$

qt=k=1∑Kαt,kWt,k,αt=softmax(u⊤Wt),

with learnable vector $u \in \mathbb{R}^d$u∈Rd.

### Candidate logits (pooled similarity + inhibition)

Score each candidate by similarity to the pooled query, subtracting inhibition:

$$\ell_{t,i} = \frac{q_t^\top C_{t,i}}{\sqrt{d}} + b_i - \lambda\, h_{t,i},$$

ℓt,i=dqt⊤Ct,i+bi−λht,i,

where $b_i$bi is an optional learned bias and $\lambda \geq 0$λ≥0 controls the strength of inhibition.

Convert logits to a probability distribution:

$$p_t = \text{softmax}(\ell_t/\tau),$$

pt=softmax(ℓt/τ),

with temperature $\tau > 0$τ>0. Low $\tau$τ yields near-deterministic convergence; higher $\tau$τ yields more exploratory recruitment.

**Recruiting $m = (1 - r)K$m=(1−r)K new items**

Let the number of new recruits be

$$m = (1 - r)K,$$

m=(1−r)K,

with retention fraction $r \in [0,1]$r∈[0,1]. Recruit $m$m candidates from $p_t$pt using a top-$m$m operator (or a differentiable approximation).

Conceptually:

- **Discrete**: $A_t = \text{TopM}(p_t, m)$ At=TopM(pt,m) (indices of the $m$ m strongest candidates)
- **Differentiable**: use Gumbel-Top-$m$m (straight-through), soft top-$k$k, or matching relaxations.

Let $R_t^{\text{new}} \in \mathbb{R}^{m \times d}$Rtnew∈Rm×d be the matrix of recruited vectors.

---

## 4. Retention / Eviction (Keeping $rK$rK Slots)

In addition to recruiting new items, the system decides which existing FoA slots to keep.

**Keep scores**

Assign a keep-score to each slot $W_{t,k}$ Wt,k given the current pooled context:

$$g_{t,k} = f_{\text{keep}}(W_{t,k}, q_t) \in \mathbb{R},$$

gt,k=fkeep(Wt,k,qt)∈R,

where $f_{\text{keep}}$ fkeep can be an MLP on concatenated inputs:

$$g_{t,k} = \text{MLP}([W_{t,k}; q_t]).$$

gt,k=MLP([Wt,k;qt]).

Select $rK$rK slots to retain:

- **Discrete**: $K_t = \text{Top}(g_t, rK)$Kt=Top(gt,rK)
- **Differentiable**: relaxed top-$k$k or straight-through top-$k$k

Let $R_t^{\text{keep}} \in \mathbb{R}^{rK \times d}$Rtkeep∈RrK×d be the retained slot vectors.

---

## 5. The Iterative Updating Operator

The next working-memory state is formed by concatenating retained slots and new recruits:

$$W_{t+1} = \text{Concat}\left(R_t^{\text{keep}},\ R_t^{\text{new}}\right) \in \mathbb{R}^{K\times d}.$$

Wt+1=Concat(Rtkeep,Rtnew)∈RK×d.

This is the explicit retain–drop–add operator that makes overlap a controllable parameter. It also defines a concrete internal notion of “within-thought” continuity (high $r$r) versus “thought boundary/reset” (low $r$r or a reset event).

---

## 6. Iterative Inhibition (Search Steering)

To implement “reject-and-research” dynamics, we update the inhibition trace for candidates that were selected (or attempted) but deemed unhelpful.

Let $z_t \in [0,1]^N$zt∈[0,1]N be a (possibly soft) indicator of which candidates were selected at iteration $t$ t. Update inhibition as

$$h_{t+1} = \kappa h_t + \gamma z_t,$$

ht+1=κht+γzt,

where $\kappa \in [0,1)$ κ∈[0,1) controls decay and $\gamma > 0$ γ>0 is the increment for selected/rejected candidates.

This mechanism progressively suppresses repeated lures and forces the pooled search to converge on alternatives, effectively pruning a local attractor basin.

---

## 7. Optional: Contextual “Meaning Shift” Within Working Memory

To capture context-dependent remapping (the idea that the “same item” can shift meaning depending on the most recent update and current set), define a context-conditioned transform applied to each slot before pooling:

$$\widehat{W}_{t,k} = W_{t,k} + \Delta(W_{t,k}, q_t),$$

W^t,k=Wt,k+Δ(Wt,k,qt),

where ΔΔ is a learned function (e.g., MLP). The pooled query is computed from $\widehat{W}_t$W^t rather than $W_t$Wt. This makes representational drift an explicit, measurable part of the model.

---

## 8. Training Objectives

The iterative working-memory module can be trained end-to-end inside larger models (transformer, recurrent agent, world model). The following losses are common and complementary.

### 8.1 Task loss (supervised or self-supervised)

If the system produces outputs $y_t$yt (next token, next action, next-step label), train with a standard task loss:

$$\mathcal{L}_{\text{task}} = \sum_t \text{CE}(y_t, y_t^*) \text{or} \mathcal{L}_{\text{task}} = \sum_t \| y_t - y_t^* \|^2.$$

Ltask=t∑CE(yt,yt∗)orLtask=t∑∥yt−yt∗∥2.

### 8.2 Continuity regularization (optional)

To encourage a desired overlap regime, regularize the effective overlap $\hat{O}_t$O^t toward target $r$ r:

$$\mathcal{L}_{\text{overlap}} = \sum_t (\hat{O}_t - r)^2.$$

Loverlap=t∑(O^t−r)2.

Here $\hat{O}_t$O^t can be computed from keep masks (discrete) or from similarity of pooled summaries (soft).

### 8.3 World-model prediction (optional)

For predictive learning, minimize error in predicting next observation features $\phi(o_{t+1})$ ϕ(ot+1):

$$\mathcal{L}_{\text{pred}} = \sum_t \| \hat{\phi}_{t+1} - \phi(o_{t+1}) \|^2 .$$

Lpred=t∑ϕ^t+1−ϕ(ot+1)2.

This loss incentivizes the module to retain what is causally relevant across time while updating what changes.

### 8.4 Reinforcement learning (optional)

In RL settings, $W_t$ Wt conditions the policy $\pi(a_t \mid W_t)$ π(at∣Wt). The working-memory module is trained jointly by actor–critic losses; overlap/continuity regularizers can be added to shape deliberation.

---

## 9. Developmental / Curriculum Scheduling of Retention

A key architectural hypothesis is that long-horizon coherence can be scaffolded by gradually increasing retention and persistence. Implement this via a schedule for $r$r over training steps:

$$r(\text{step}) = r_{\min} + (r_{\max} - r_{\min}) \cdot \sigma \left( \frac{\text{step} - s_0}{s_1} \right),$$

r(step)=rmin+(rmax−rmin)⋅σ(s1step−s0),

where $\sigma(\cdot)$σ(⋅) is the logistic function and $s_0, s_1$s0,s1 control onset and steepness.

Early training (lower $r$r) promotes exploration and rapid updating; later training (higher $r$r) promotes compounding of intermediate results and stable long-horizon inference.

---

## 10. Key Knobs, Metrics, and Ablations

### 10.1 Knobs (interpretable control parameters)

- **Capacity:** $K$K (FoA size)
- **Retention / overlap:** $r$r (continuity)
- **Selection temperature:** $\tau$τ (exploration vs convergence)
- **Inhibition:** $\lambda, \kappa, \gamma$λ,κ,γ (escape from lures)
- **Pooling/keep functions:** $f_{\text{pool}}, f_{\text{keep}}$fpool,fkeep (learned control policy)

### 10.2 Metrics

- **Overlap / continuity:** $O_t$Ot or $\hat{O}_t$O^t
- **Reset frequency:** incidence of low-overlap transitions
- **Lure perseverance:** probability of re-selecting recently inhibited candidates
- **Long-horizon coherence:** task-dependent (e.g., plan success, multi-step accuracy, discourse coherence)

### 10.3 Mechanism-proving ablations

Ablations that directly test whether each primitive matters:

1. **No overlap constraint:** replace all $K$K slots each step (set $r = 0$r=0)
2. **Overlap without pooled search:** recruit using only the newest slot (single-cue)
3. **Pooled search without inhibition:** remove $h_t$ht terms ($\lambda = 0$λ=0)
4. **Fixed rrr vs curriculum r(·)r(\cdot)r(·):** test developmental hypothesis
5. **Remove contextual remapping:** set $\Delta(\cdot) = 0$Δ(·)=0

These ablations tend to produce characteristic failures (derailment, lure trapping, weak bridging inference) that operationalize the theory’s claims.

---

## 11. Summary (the “kernel”)

The architecture reduces to a small set of commitments:

1. A capacity-limited working set $W_t$Wt evolves by an explicit operator

$$W_{t+1} = \text{Keep}_{rK}(W_t) \ \cup \ \text{Recruit}_{(1-r)K}(C_t \mid f_{\text{pool}}(W_t)),$$

$W_{t+1} = \mathrm{Keep}_{rK}(W_t) \cup \mathrm{Recruit}_{(1-r)K}(C_t \mid f_{pool}(W_t))$,

2. New recruits are selected by pooled multi-cue search (softmax over candidate similarity to a pooled query),
3. Search can be steered by iterative inhibition,
4. Retention $r$ is an interpretable continuity knob that can be trained and scheduled.

This formalization is simultaneously (i) a theory statement (what cognition is doing step-to-step) and (ii) a runnable AI design pattern (how to build an agent that maintains continuity and compounds intermediate results).